\documentclass[preprint,5p,twocolumn,authoryear]{elsarticle}
\usepackage{amssymb}
\usepackage{color}
\usepackage[]{overpic}
\usepackage{wasysym}
\usepackage{multirow}
\usepackage{lineno}
\modulolinenumbers[5]
\usepackage[T1]{fontenc}		% gives the eth for Steinn's name, broken for these bib styles
\usepackage{url}

\def\lsim{\mathrel{\rlap{\lower 3pt \hbox{$\sim$}} \raise 2.0pt \hbox{$<$}}}
\def\gsim{\mathrel{\rlap{\lower 3pt \hbox{$\sim$}} \raise 2.0pt \hbox{$>$}}}
\def\chandra{{\it Chandra\/}}
\def\xmm{{XMM-{\it Newton\/}}}
\def\swift{{\it Swift\/}}
\def\nustar{{\it NuSTAR\/}}
\def\athena{{\it Athena\/}}
\def\erosita{{\it eROSITA\/}}
\def\lynx{{\it Lynx\/}}
\def\axis{{\it AXIS\/}}
\def\spica{{\it SPICA\/}}
\def\gaia{{\it Gaia\/}}
\def\arcsec{\hbox{$^{\prime\prime}$}}
 
\newcommand{\lum}{\rm erg~s$^{-1}$}
\def\gtrsim{\mathrel{\hbox{\rlap{\hbox{\lower4pt\hbox{$\sim$}}}\hbox{$>$}}}}
\newcommand{\ltsima}{$\; \buildrel < \over \sim \;$}
\newcommand{\simlt}{\lower.5ex\hbox{\ltsima}}
\newcommand{\gtsima}{$\; \buildrel > \over \sim \;$}
\newcommand{\simgt}{\lower.5ex\hbox{\gtsima}}

\def\cm2{{cm$^{-2}$}}
\def\nh{{N$_{\rm H}$}}

\def\micron{\hbox{$\mu$m}}

\newcommand{\msun}{{\rm M}_\odot}% Proceedings of the SPIE

\journal{New Astronomy Reviews}

\begin{document}

\begin{frontmatter}

%with \documentclass[preprint,12pt,authoryear]{elsarticle}
%\title{The Quest for Multiple Supermassive Black Holes:\\ A Multi-Messenger View}
%with \documentclass[final,5p,twocolumn,authoryear]{elsarticle}
%\title{The Quest for Multiple Supermassive Black Holes: A Multi-Messenger View}
\title{The Quest for Dual and Binary Supermassive Black Holes: A Multi-Messenger View}

%\author[Address1,Address2]{Author1}
%\author[Address2,Address3]{Author2}
%\address[Address1]{Address1}
%\address[Address2]{Address2}
%\address[Address3]{Address3}

\author[IAPS]{Alessandra De Rosa}
\author[Ubologna,INAF_OAS]{Cristian Vignali}
\author[Gtech]{Tamara Bogdanovi\'c}
\author[Uzurich]{Pedro R. Capelo}
\author[Caltech]{Maria Charisi}
\author[Ubicocca,INFN-Mi]{Massimo Dotti}
\author[MPIA]{Bernd Husemann}
\author[Udurham,Ufirenze,OAA]{Elisabeta Lusso}
\author[Uzurich]{Lucio Mayer}
\author[JIVE]{Zsolt Paragi}
\author[Umichigan,Unashville]{Jessie Runnoe}
\author[Ubicocca,Ubirmingham]{Alberto Sesana}
\author[Umunchen]{Lisa Steinborn}
\author[Uromatre]{Stefano Bianchi}
\author[Ubicocca]{Monica Colpi}
\author[IAP]{Luciano Del Valle}
\author[Konkoly_obs]{S\'andor Frey}
\author[Konkoly_obs,MTA-ELTE,Lor\'and U]{Krisztina \'E. Gab\'anyi}
\author[CSIC]{Margherita Giustini}
\author[ESA-ESTEC]{Matteo Guainazzi}
\author[Ucolumbia]{Zoltan Haiman}
\author[Ubochum]{Noelia Herrera Ruiz}     
\author[ESO-Chile,Barcelona]{Rub\'en Herrero-Illana}
\author[ICREA]{Kazushi Iwasawa}
\author[NAOC]{S. Komossa}
\author[SRON,Uradbound]{Davide Lena}
\author[ESA-ESAC]{Nora Loiseau}
\author[IAA-CSIC]{Miguel Perez-Torres}
\author[OAR]{Enrico Piconcelli}
\author[IAP]{Marta Volonteri}

\address[IAPS]{INAF/IAPS - Istituto di Astrofisica e Planetologia Spaziali, Via del Fosso del Cavaliere I-00133, Roma, Italy}
\address[Ubologna]{Dipartimento di Fisica e Astronomia, Alma Mater Studiorum, Universit\`a degli Studi di Bologna, Via Gobetti 93/2, 40129 Bologna, Italy}
\address[INAF_OAS]{INAF -- Osservatorio di Astrofisica e Scienza dello Spazio di Bologna, Via Gobetti 93/3, I-40129 Bologna, Italy}
\address[Gtech]{Center for Relativistic Astrophysics, School of Physics, Georgia Institute of Technology, 837 State Street, Atlanta, GA 30332-0430}
\address[Uzurich]{Center for Theoretical Astrophysics and Cosmology, Institute for Computational Science, University of Zurich, Winterthurerstrasse 190, CH-8057 Z$\ddot{u}$rich, Switzerland}
\address[Caltech]{TAPIR, California Institute of Technology, 1200 E. California Blvd., Pasadena, CA 91125, USA}
\address[Ubicocca]{Universit\`a degli Studi di Milano-Bicocca, Piazza della Scienza 3, 20126
Milano, Italy}
\address[INFN-Mi]{INFN, Sezione di Milano-Bicocca, Piazza della Scienza 3, 20126 Milano,
Italy}
\address[MPIA]{Max-Planck-Institut f\"{u}r Astronomie, K\"{o}nigstuhl 17, D-69117 Heidelberg, Germany}
\address[Udurham]{Centre for Extragalactic Astronomy, Durham University, South Road, Durham, DH1 3LE, UK}
\address[Ufirenze]{Dipartimento di Fisica e Astronomia, Universit\'a degli Studi di Firenze, Via. G. Sansone 1, I-50019 Sesto Fiorentino (FI), Italy}
\address[OAA]{INAF - Osservatorio Astrofisico di Arcetri, Largo E. Fermi 5, 50125 Firenze, Italy}
\address[JIVE]{Joint Institute for VLBI ERIC, Oude Hoogeveensedijk 4, 7991 PD, Dwingeloo, The Netherlands}
\address[Umichigan]{Department of Astronomy, University of Michigan, Ann Arbor, MI 48109, USA}
\address[Unashville]{Department of Physics and Astronomy, Vanderbilt University, Nashville, TN 37235, USA}
\address[Ubirmingham]{School of Physics and Astronomy and Institute of Gravitational Wave Astronomy, University of Birmingham, Edgbaston, Birmingham B15 2TT, United Kingdom}
\address[Umunchen]{Universit\"ats-Sternwarte M\"unchen, Scheinerstr.1, D-81679 M\"unchen, Germany}
\address[Uromatre]{Dipartimento di Matematica e Fisica, Universit\`a degli Studi Roma Tre, via della Vasca Navale 84, 00146 Roma, Italy}
\address[IAP]{Sorbonne Universites et CNRS, UMR 7095, Institut d'Astrophysique de Paris, 98 bis bd Arago, F-75014 Paris, France)}
\address[Konkoly_obs]{Konkoly Observatory, Research Centre for Astronomy and Earth Sciences, Konkoly Thege M. \'ut 15-17, H-1121 Budapest, Hungary}
\address[MTA-ELTE]{MTA-ELTE Extragalactic Astrophysics Research Group, ELTE TTK P\'azm\'any P\'eter s\'et\'any 1/A, H-1117 Budapest, Hungary}
\address[Lor\'and U]{Department of Astronomy, E\"otv\"os Lor\'and University, P\'azm\'any P\'eter s\'et\'any 1/A, H-1117 Budapest, Hungary}
\address[CSIC]{Centro de Astrobiolog\'ia (CSIC-INTA), Departamento de Astrof\'isica; Camino Bajo del Castillo s/n, Villanueva de la Ca\~nada, E-28692 Madrid, Spain}
\address[ESA-ESTEC]{ESA - European Space Agency. ESTEC, Keplerlaan 1 2201AZ Noordwijk, The Netherlands}
\address[Ucolumbia]{Department of Astronomy, Columbia University, New York, NY 10027, USA}
\address[Ubochum]{Astronomisches Institut, Ruhr-Universit\"at Bochum, Universit\"atsstrasse 150, 44801 Bochum, Germany}
\address[ESO-Chile]{European Southern Observatory, Alonso de C\'ordova 3107, Vitacura, Casilla 19001, Santiago de Chile, Chile}
\address[Barcelona]{Institute of Space Sciences (ICE, CSIC), Campus UAB, Carrer de Magrans, E-08193 Barcelona, Spain}
\address[ICREA]{ICREA \& Institut de Ci\`{e}ncies del Cosmos (ICCUB), Universitat de Barcelona (IEEC-UB), Barcelona 08028, Spain}
\address[NAOC]{National Astronomical Observatories,
Chinese Academy of Sciences, Beijing 100012, China}
\address[SRON]{SRON, Netherlands Institute for Space Research, Sorbonnelaan 2, 3584~CA, Utrecht, The Netherlands}
\address[Uradbound]{Department of Astrophysics/IMAPP, Radboud University,
P.O.~Box 9010, 6500 GL, Nijmegen, The Netherlands}
\address[ESA-ESAC]{XMM-Newton SOC, ESAC/ESA, E-28692 Villanueva de la Ca\~nada, Madrid, Spain}
\address[IAA-CSIC]{Instituto de Astrof\'isica de Andaluc\'ia (IAA-CSIC), E-18008, Granada, Spain}
\address[OAR]{INAF - Osservatorio Astronomico di Roma, via Frascati 33, 00040 Monte Porzio Catone (Roma), Italy}

\begin{abstract}
The quest for binary and dual supermassive black holes (SMBHs) at the dawn of the multi-messenger era is compelling.
Detecting dual active galactic nuclei (AGN) -- active SMBHs at projected separations larger than several parsecs -- and binary AGN -- probing the scale where SMBHs are bound in a Keplerian binary -- is an observational challenge. The study of AGN pairs (either dual or binary) also represents an overarching theoretical problem in cosmology and astrophysics. The AGN triggering calls for detailed knowledge of the hydrodynamical conditions of gas in the imminent surroundings of the SMBHs and, at the same time, their duality calls for detailed knowledge on how galaxies assemble through major and minor mergers and grow fed by matter along the filaments of the cosmic web. 
This review describes the techniques used across the electromagnetic  spectrum to detect dual and binary AGN candidates and proposes new avenues for their search. The current observational status is compared with the state-of-the-art numerical simulations and models for formation of dual and binary AGN. 
Binary SMBHs are among the loudest sources of gravitational waves (GWs) in the Universe. The search for a background of GWs at nHz frequencies from inspiralling SMBHs at low redshifts, and the direct detection of signals from their coalescence by the Laser Interferometer Space Antenna in the next decade, make this a theme of major interest for multi-messenger astrophysics. 
This review discusses the future facilities and observational strategies that are likely to significantly advance this fascinating field.
\end{abstract}


\end{frontmatter}
\onecolumn
\tableofcontents{}
\twocolumn

\part*{Introduction}

Supermassive black holes (SMBHs) with mass of $\sim$10$^6$--10$^9$ M$_\odot$ are ubiquitous in ellipticals, in the bulges of disk galaxies and in at least a fraction of dwarf galaxies. 
The tight correlation between the black hole mass and the bulge stellar velocity dispersion \citep{ferrarese&merritt00,gebhardtetal2000,Kormendy13} suggests that SMBHs are likely to affect the evolution of the host galaxy over cosmological time-scales.
In contemporary astrophysics, a large variety of physical phenomena concur to establish the close connection between the formation and evolution of galaxies and of their central SMBHs \citep{silk&rees98,DiMatteo_et_al_2005}.
Galaxy mergers may be a way through which SMBHs form by direct collapse of gas at the center of protogalaxies \citep{Begelman2006,Mayer2010,Mayer19}. Furthermore, there is growing evidence that major mergers trigger the most luminous AGN  (e.g.,  \citealt{treisteretal12,fan_etal2016,goulding+2018}), although this result is still debated. Numerical simulations show that galaxy collisions are conducive to episodes of major gas inflows that feed the central SMBH, thus powering accretion and nuclear activity (e.g.,  \citealt{DiMatteo_et_al_2005}). Models of structure formation can reproduce the observed large-scale properties of quasars (e.g., their environment and clustering clustering) if their bright and short-lived active phases are most likely  triggered by mergers (see \citealt{KauffmanHaehnelt2000,2012NewAR..56...93A} and references therein). However, not all nuclear activity is triggered by galaxy collisions as SMBH growth can occur also through secular processes \citep{2018MNRAS.476.2801M,Ricarte2019} with mergers triggering an initial rapid growth phase \citep{2018MNRAS.481.3118M}.  In the last decade,  quasar pairs at sub-Mpc (projected) separations have raised interest  as these systems could possibly trace regions of systematic large-scale overdensities of galaxies (e.g.,  \citealt{hennawi2006,hennawi2010,2014MNRAS.444.1835S,2017MNRAS.468...77E,2018MNRAS.474.4925S,lusso2018}). 

The connection between AGN triggering and galaxy mergers is not clear yet.  
To discuss this connection, in this paper we define {\it dual} AGN those interacting galaxy systems containing two active nuclei powered by accretion onto SMBHs that are nested inside their host  but that are not mutually gravitationally bound. If only one SMBH in active, we refer to this system as {\it offset} AGN. Similarly, {\it binary} AGN at sub-pc separations are defined as active SMBHs which are gravitationally bound, forming a Keplerian binary.\footnote {Hereafter, the separations between observed AGN and galaxy pairs are meant as projected if not stated otherwise.} The (projected) separation ranges between $\sim 1$~pc and $\sim 100$~kpc in dual AGN, while it lies in the pc--sub-pc range in binary systems (pc-scale binaries could fall into either category depending on the parameters of the system). 

A number of studies in different wavebands show evidence for a higher fraction of dual AGN in galaxies with a close companion, suggesting that galaxy interactions play a role in the AGN triggering process \citep[e.g.][]{Ellison2011,Koss_et_al_2012, Silverman_et_al_2011,satyapaletal14,kocevskietal15,kossetal2018}. 
However, there are investigations revealing no enhanced  AGN activity in mergers compared to 
a matched control sample of inactive galaxies \citep[e.g.][]{Cisternas:2011,Mechtley:2016}. These conflicting results may be a consequence of different sample selection criteria (such as merger stage of interacting galaxies and/or AGN luminosity) and observational biases  (e.g., nuclear obscuration and AGN variability). 
In light of these uncertainties, the detection and characterisation of dual and binary SMBHs is fundamental if we want to understand the formation and accretion history of SMBHs across cosmic ages.

A further issue which makes the study of dual and binary AGN one of the forefront topics in modern astrophysics is that these systems are the natural precursors of coalescing binary SMBHs, which are strong emitters of low-frequency gravitational waves (GWs).  Detecting the GW signal from the inspiral, merger and ringdown of binary black holes with LISA, the ESA's Laser Interferometer Space Antenna \citep{LISA17}, will let us unveil the rich population of SMBHs of $\sim$10$^{4-7}~\msun$ forming in the collision of galaxy halos, out to redshifts as large as $z\sim 20$ \citep{Colpi2020}. Note that LISA will detect the ``light'' SMBHs at the low-mass end of their mass distribution. Thus, LISA will unveil the origin of the first quasars, carrying exquisite measurements of the black hole masses and spins, and providing the first census of this yet unexplored population of sources. Soon, Pulsar Timing Array (PTA) experiments at nHz frequencies will detect the cosmic GW background radiation from inspiraling SMBH binaries (SMBHBs) of $\sim$10$^9$~$\msun$ at $z\sim 1$ \citep{Kelley2020}.
Discovering binary massive black holes over such a wide range of masses will shed light on the deep link between SMBHs and galaxies, their growth, evolution and assembly.

This review, far from being exhaustive on the topic of multiple AGN systems, surveys  the main results that emerged during a workshop organized at the Lorentz Center in Leiden ``The Quest for Multiple Supermassive Black Holes: A Multi-Messenger View'' (\texttt{https://www.lorentzcenter.nl}). The Workshop was dedicated to multi-messenger studies of multiple SMBHs, with emphasis on the best strategies for their detections with current and upcoming observatories, and 
to the most updated and comprehensive numerical simulations and theories on the pairing and merging of SMBHs.
As such, this review may guide the community to answer to the most compelling questions posed during the meeting: \\
1) Can state-of-the art simulations of interacting galaxies provide key indicators that will let us discover dual, obscured AGN systems?\\
2) How can we unveil the missing population of 10--100~pc dual black holes?\\ 
3) Which observational strategies should we envisage in order to demonstrate the existence of the binary SMBHs at separations close to and below a milli-parsec?\\
4) Is there a population of transient AGN that may be candidate counterparts of  inspiralling or/and merging SMBHs, and how can this population be unambiguously identified?

The review is organized in two main bodies describing the systems depending on their separation: Sect.~\ref{sec:kpc_scale} is focused on AGN and multiplets with kpc-to-pc scale separation, while Sect.~\ref{sec:BH_binaries} on gravitationally bound SMBHs with sub-pc separations.
Within each section we describe the datasets already available, the observational results and the predictions resulting from models and state-of-the art numerical simulations. We then discuss the most effective ways to define, observe, analyse and interpret the wealth of data collected in different wavebands.

Finally, in Sect.~\ref{sec:Future_perspective}, we present first the techniques and strategy developments in this field, describing the future electromagnetic observing facilities and numerical set-up. 
Then, we explore the low-frequency GW Universe, the new window which will provide a complementary view of SMBHs by detecting the GW signal from those black holes coalescing in binaries. Measuring the BH masses and spins across cosmic ages will let us uncover their nature, growth and yet unknown origin.

For the calculations presented in this review, a concordance cosmology with $H_0 =70$~km~s$^{-1}$~Mpc$^{-1}$, $\Omega_{\rm M}=0.3$ and $\Omega_{\Lambda}=0.7$ \citep{Wright06} has been adopted, and magnitudes are in the AB system; when 
literature results are reported, we refer to the original papers for the 
correspondingly adopted cosmology. 

\part{AGN pairs and multiplets with kpc-to-pc scale spatial separations}
\label{sec:kpc_scale}

The triggering of AGN activity in galaxy mergers has been extensively studied both from an observational and a theoretical point of view. The varied observational and numerical techniques adopted thus far are highly complementary, since different bands and/or numerical setups and recipes have inherently distinct limitations.

Different methods have been proposed to identify good candidates of dual and binary AGN. Most of the AGN candidates and pairs with pc-to-kpc separations have been identified either through extensive optical, radio, mid-infrared and hard X-ray surveys, and through pointed observations mainly in the high-energy domain; all these technique will be discussed in the following sections. Nevertheless, dual AGN systems are rare, and most of their detection has been serendipitous.

A challenge in this kind of studies is the need for a statistically significant sample of dual and multiple AGN covering a wide dynamical range in spatial separations, from pc- to kpc-scale  separation. While a number of AGN pair candidates and merging galaxies have been discovered over the past several years, only a handful of these have eventually been confirmed, usually through intense and observationally costly multiband follow-ups.
An example is the increasing availability of SDSS optical spectral data where dual AGN are identified through doubled-peaked O[III] line, but only about 2 per cent of these candidates are finally confirmed with multi-wavelength follow-up. 
This is due to the fact that this signature (i.e., the presence of a doubled-peaked profile) is not unique, indicating other possible effects originated nearby a single AGN (e.g., matter outflows). 

Similar arguments can be applied to observations at other frequencies: if not sensitive enough, it might be challenging to distinguish AGN from star formation processes in radio-emitting regions using only radio data; besides, only about 10 per cent of AGN are radio emitting. Furthermore, the detection of X-ray emission associated with a source is not {\em per se} an indication of the presence of an active nucleus, since also star-forming galaxies and weakly accreting black holes (e.g., LINERs, low-ionization nuclear emission-line regions, and AGN with inefficient disks) can produce some level of X-ray emission. 
Finally, mid-IR all-sky surveys can detect a number of obscured systems and then complement the optical search of AGN. 
Overall, the best observing strategy requires the search for good candidates to be then confirmed through appropriate follow-up programs, and each wavelength is pursuing its own quest in this direction, contributing to the final detection/confirmation.
\\
On the numerical side, there are two main avenues of research: cosmological and isolated simulations of mergers. Galaxy mergers and systems of multiple AGN can be produced through cosmological simulations.  This research provides important counterpart to observations, allowing the measure of the dual AGN fraction with respect to single AGN.
A variety of hydrodynamic cosmological simulations exists; however, only a few are able to produce AGN pairs with separations down to kpc scales, since combining large volumes with relatively high resolution is computationally very expensive. 
On the other hand, idealized merger simulations are able to resolve sub-kpc scales, that is fundamental to follow the dynamics of SMBHs. Nevertheless, simulations of isolated galaxy mergers cannot provide any prediction on the fraction of AGN pairs out of the total number of AGN, although it is possible to compute an ``activity-dual time'' normalized to the time when at least one BH is active. \\
In this chapter, we review the current observational constraints on the existence of merger-triggered single and dual AGN (i.e. AGN pairs that are not gravitationally bound)  observed at different wavelengths (Sect.~\ref{ssec:BH_Pairs_Observations}). In each section we will describe the best strategy adopted in a given band to select good candidates and follow-up observations to confirm or disprove them. We will also show that a multi-wavelength approach is the only viable way to improve the detection rate of dual AGN systems. 
We then summarize our most up-to-date theoretical understanding of this process (Sect.~\ref{ssec:BH_Pairs_theory}) describing mergers in larger but coarser cosmological simulations and mergers between idealised, isolated galaxies from higher-resolution simulations. We also review the physical processes that may produce SMBH pairs stalling at 1--100~pc spatial separations, possibly detectable as tight dual AGN.
The predictions from both these sets of simulations, such as the dual AGN fraction,  nuclear environment (e.g., gas reservoir) and system properties (e.g., BH mass ratio, separation) will then be compared with the available and planned observations.

\section{Observations of dual AGN}
\label{ssec:BH_Pairs_Observations}

The detection of spatially-resolved dual AGN at projected separation of a few tens of kpcs, in optical and mid-infrared surveys, mostly from the {\it Sloan Digital Sky Survey} (SDSS, \citealt{YorkAA00,AbazajianAMC09}), the {\it Baryonic Oscillation Spectroscopic Survey} (BOSS, \citealt{2012ApJS..203...21A}), and the {\it Wide-field Infrared Survey Explorer} (WISE, \citealt{Wright2010}), reinforces the idea that gas-rich mergers may trigger the active nuclear phase in both galaxies (e.g.,  \citealt{2008ApJ...678..635M,2009ApJ...693.1554F,satyapaletal14,satyapaletal17,Weston2017}). 
Bright AGN pairs with separations less than a few hundreds of kpc observed at similar redshifts are ideal probes of both the small-scale ($\simlt$100~kpc) structure of the intergalactic medium (e.g., \citealt{2017Sci...356..418R}) and the large-scale (order of Mpc) rich environment where mergers are more likely to happen (e.g.,  \citealt{Djorgovski2007,2011ApJ...736L...7L,2013MNRAS.431.1019F,deane2014,Hennawi2015,lusso2018}).  

One method of detecting/observing dual AGN consists of time-expensive follow-up observations of previously identified candidates, selected through wide-field surveys or large catalogues.
Another method relies on using the data from survey fields (e.g., SDSS in optical, GOALS in mid-IR, COSMOS and Chandra Deep Field-North and South in X-rays) to select dual AGN candidates and define also their properties. 
These two approaches are highly complementary, as discussed in the following. 
We present in this section the photometric and spectroscopic surveys and integral-field spectroscopy outcomes and their success in identifying the closest candidates for dual and multiple AGN. These candidates should then be confirmed by pointed follow-up observations.
We also show the investigations carried out through optical, X-ray, mid-infrared, and radio techniques used to detect and characterize spatially resolved and unresolved dual AGN systems with kpc- to pc-scale separation.

\subsection{Optical surveys of dual AGN at sub-kpc/kpc separation: the search for candidates}
\label{sssec:optical_kpc}

In the optical, the photometric and spectroscopic SDSS and BOSS surveys are vast databases that have been explored to search for pair candidates (e.g., \citealt{hennawi2006, hennawi2010, 2008ApJ...678..635M, Shen2010,2014MNRAS.444.1835S,2018MNRAS.474.4925S,lusso2018,LenaPE18}). 
Yet, one of the main limitations of such surveys in the search for pair candidates is mostly due to the \textit{fiber collision limit}: fibers cannot be placed closer than 55$^{\prime\prime}$ for SDSS and 62$^{\prime\prime}$ for BOSS. For a target located at $z = 0.008 - 0.7$ (or a luminosity distance D$_{L} \approx 35 - 4300$~Mpc), the SDSS fiber encompasses between 0.5 and 21~kpc (given the fiber size of 3$^{\prime\prime}$ and a 2$^{\prime\prime}$ in  SDSS and BOSS, respectively). 
For the galaxies which do not fall within a single fiber, this limit implies a minimum physical distance between pairs of about 500~kpc at $z \simeq 2$. One way to overcome the fiber collision limit is through overlapping plates (e.g.,  \citealt{2003AJ....125.2276B}). However, only $\sim30$  per cent and $\sim$ 40 per cent of the sky observed by SDSS and BOSS, respectively, is covered by such overlap.  

Another possibility to select suitable close pair candidates, also at higher redshifts, is to mine the photometric quasar catalogues such as the XDQSO (\citealt{2011ApJ...729..141B,2012ApJ...749...41B,2015MNRAS.452.3124D}), which can then be followed up spectroscopically (e.g., see \citealt{hennawi2006,2008ApJ...678..635M,hennawi2010,2017MNRAS.468...77E} for details).
To build the XDQSO catalogue, \citet{2011ApJ...729..141B} segregate quasars from stars selected from the photometric SDSS/BOSS catalogues of $10^6$ sources with $r<22$ using machine learning algorithms. 
This work uncovered $\sim$300 quasar pairs with projected separations $R_\perp < 1$\,Mpc. Only 60 of them have similar redshifts (with redshift differences around 5,000--10,000 km/s at $z\sim2$). 
The probability to find a quasar pair against pairs of stars is (QSO pairs)/(star pairs)\,$\sim10^{-4}$, with a success rate of pair confirmation of $\sim$25 per cent. We refer to \citet{findlay2018} for the most updated release of pair candidates at large separation with spectroscopic follow-up  and for further statistical details on the chance probability of finding a quasar pair over the single quasar population (see also, e.g., \citealt{2012MNRAS.424.1363K,Eftekharzadeh2019}).
One should note, however, that these rare quasar pairs at kpc-scale projected separation are biased towards being very blue (i.e., SDSS selected). Thus these surveys  miss interacting, much closer dust-obscured nuclei hosted in dual AGN.

In the following subsections, we will discuss spectrophotometric and spatial techniques to identify dual AGN candidates. These observational methods are not mutually exclusive but can be ideally used in combination to improve the selection of dual AGN candidates and possibly their further confirmation.

\subsubsection{Optical spectroscopy of unresolved dual AGN candidates: double-peaked narrow-line emitters}
\label{sssect:obs_kpc_opt}

Once two AGN become closer than the spatial resolution of the imaging data, it is nearly impossible to identify them as independent sources. Depending on the redshift of the source, for ground-based observatories this already happens at kpc scales. Spectroscopy can overcome this limitation by detecting several kinematic components in velocity space through emission lines produced by AGN ionization, e.g., so-called double-peaked emitters. This technique is based on the assumption that  each AGN in a pair carries its own Narrow Line Region (NLR) -- the clouds at a scale of $\sim$1~kpc from the AGN core, which trace the systemic velocity of the AGN as they move in their common gravitational potential. The identification of double-peaked emitters is only limited by the spectral resolution of the spectrograph with respect to the width and flux ratio of the emission lines.
The same technique has been applied to the BLR as observational evidence of sub-pc SMBHBs systems; this is extensively discussed in Sect.~\ref{sssec:subpcspec2}.

SDSS has provided the largest database of extra-galactic spectra and has been commonly used for the identification of double-peaked narrow-line emitters. In particular, the [OIII] $\lambda\lambda4960,5007$ emission lines have been investigated for signatures of double peaks, since these are expected to be the brightest rest-frame optical emission lines in case of AGN photoionization \citep{rosario2010}. Other high-ionization lines, such as [NeV] $\lambda 3426$ or [NeIII] $\lambda3869$, can be be used in a similar way in the higher-redshift sources. High-ionization double-peaked emission lines have been identified serendipitously in individual sources \citep{Gerke:2007, Xue:2009,Barrows:2012,Benitez:2013} or through systematic studies of large spectroscopic databases such as SDSS \citep{Wang:2009,Smith:2010,Liu:2010,Pilyugin:2012, Ge:2012, Barrows:2013,Lyu:2016}, LAMOST \citep{Shi:2014} or DEEP2 and AGES \citep{comerford2013}. Those systematic studies revealed that roughly one per cent of the AGN population exhibit double peaks in the forbidden high-ionization lines. However, not all double-peaked emitters are really dual AGN, due to the complex internal kinematics of the NLR in luminous AGN. AGN-driven outflows (e.g., \citealt{2015ARA&A..53..115K}), compact rotating gas disks (e.g., \citealt{2015AJ....149...92V}) or illumination of interacting companion galaxies (e.g.,  \citealt{Xue:2009, 2016ApJ...818...64S}) may lead to similar signatures in the line profiles without having a dual AGN origin.\footnote {A further potential challenge in ground-based spectroscopic searches is the seeing, which may project the emission-lines of a single AGN onto its companion galaxy, thus mimicking an active pair. } Furthermore, single-peaked AGN in SDSS may turn out to be intrinsically double-peaked AGN when higher spectral resolution observations are employed \citep[e.g.][]{Woo:2014}. Hence, neither the parent sample of double-peaked narrow-line emitters based on SDSS is complete nor does their number count directly relate to the dual AGN fractions, without further observations.

\begin{figure*}[t]
\center $
\begin{array}{c}
\includegraphics[trim=0cm 10.2cm 0cm 0cm, clip=true, scale=0.8]{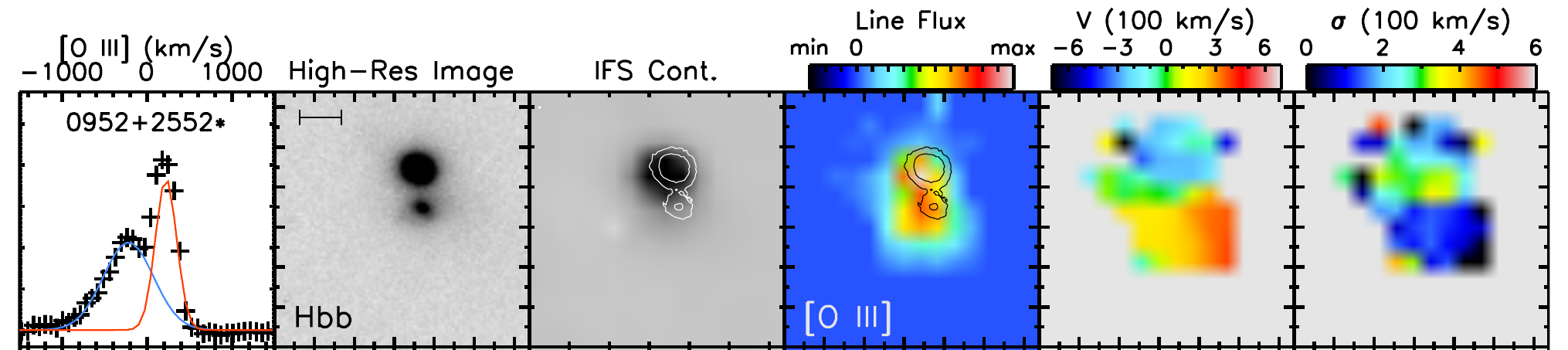}\\
\includegraphics[trim=0cm 3.23cm 0cm 23.65cm, clip=true, scale=0.965]{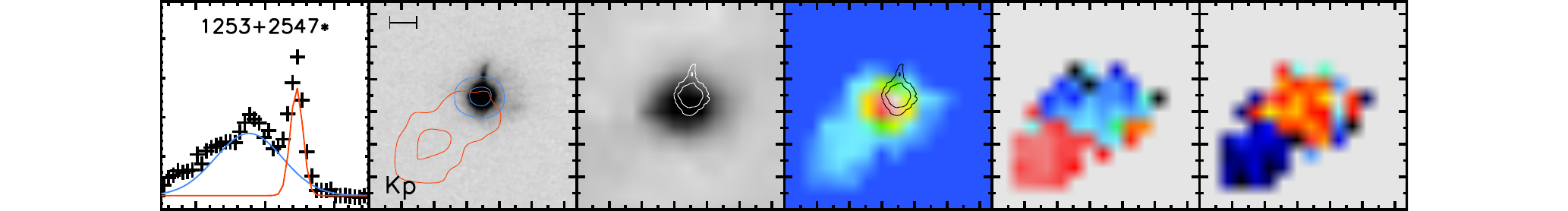}\\
\end{array}$
\caption{Top: candidate dual AGN where the double-peaked emission lines are due to the relative orbital motion of two galaxies. Bottom: single AGN with extended narrow-line region producing a double-peaked emission-line profile. From left to right: SDSS [OIII]$\lambda$5007 emission line profile with the two-component fit overplotted, high-resolution broad-band image, emission-line--free continuum from the IFS data, emission-line intensity map, velocity map, and velocity dispersion map. SDSS object designations are labeled in the first column. The filter for the broad-band images are labeled, and the scale bar indicates a transverse separation of 5~kpc. Blue and red contours on the high-resolution image of the single AGN indicate the kinematically-resolved blue-shifted and red-shifted components. Contours in the third and fourth column are from the broad-band images in the second column, which have been spatially aligned with the datacubes. The measured emission line is labeled in the fourth column. N is up and E is to the left; major tickmarks are spaced in $1^{\prime\prime}$. Adapted from figure 1 and figure 4 of \cite{FuYM12}. 
}
\label{fig: ifs_fu}
\end{figure*}

\subsubsection{From candidates to detection: follow-up studies for the closest, sub-kpc pairs, and the power of integral-field spectroscopy}

As specified above, a wide range of phenomena might concur in producing the shape of the continuum and the profile of the spectral lines. The underlying stellar population is distributed in a combination of morpho-kinematical features such as bulges, disks, bars, and tidal streams; neutral and ionised gas experiences a combination of virial (e.g., rotation in the plane of the galaxy) and non-virial kinematics (i.e. inflows and outflows powered by AGN and/or star formation activity).
As a result, it is clear that the detection of double-peaked [OIII] emission lines remains, in the great majority of cases, the starting point for a deeper investigation. 

Probing the activity level of galactic nuclei is nontrivial, especially in the case of obscured AGN pairs, and the task requires the examination of multi-wavelength evidence, often from both the spectroscopic and imaging side \citep[e.g.][]{Komossa2003,MazzarellaIV12,LiuCS13,gabanyi2016,LenaPE18}.

Within the optical-to-IR realm, a valuable tool in the investigation of galaxies is integral-field spectroscopy (IFS), a technique providing, simultaneously, imaging and spatially resolved spectroscopy (e.g., \citealt{Lena15}). With a single observation one can obtain a two-dimensional view of the gaseous and stellar kinematics, of the emission-line flux distributions, and of the emission line ratios, a diagnostic for the dominant ionisation mechanism (AGN, star formation, shocks) at play at different locations within the galaxy \citep{Baldwin81}.
Indeed, IFS has been used by a number of authors to follow-up dual AGN candidates \citep[e.g.][]{McGurkMR11,FuYM12, McGurk_2015}, to gain further insights on confirmed dual AGN \citep[e.g.][]{KosecBS17}, and to produce serendipitous discoveries \citep[e.g.][]{Husemann2018}. Thanks to the recent availability of IFS surveys, such as SDSS MaNGA \citep{Bundy2015}, IFS data have also been used as a starting point for the identification of new dual candidates \citep{EllisonSM17}. They represent a particularly promising avenue for discovery and characterization of closer, sub-kpc separation AGN pairs.

The work presented by \cite{FuYM12} is illustrative: the authors performed high-resolution imaging plus optical and NIR IFS for a sample of approximately 100 double-peaked AGN selected from SDSS-DR7 \citep{AbazajianAMC09}. Imaging was performed to achieve a spatial resolution of approximately 0.1$^{\prime\prime}$; towards this goal they used archival Hubble Space Telescope images, H-band and K$^{\prime}$-band images obtained with OSIRIS \citep{LarkinBK06} and NIRC2 on the Keck telescope (see also \citealt{rosario2011,fu2011,McGurkMR11,McGurk_2015,liux2018} for similar experiments). NIR IFS data assisted by adaptive optics were obtained with the OSIRIS integral field unit \citep{LarkinBK06,WizinowichCJ06}; for the majority of the targets, seeing-limited IFS in the optical and NIR was obtained with the 2.2m Hawaii Telescope instrument SNIFS \citep{AlderingAA02,LantzAA04} at a spatial resolution of approximately 1.2$^{\prime\prime}$. The combination of high-resolution imaging and spatially-resolved spectroscopy allowed the authors to distinguish between candidates where the double-peaked emission-lines arise from the relative orbital motion of two galaxies (possibly both active), and single AGN mimicking the spectral signatures of potential dual AGN, for example because of emission from an extended narrow-emission line regions (both examples are shown in Fig.~\ref{fig: ifs_fu}). They showed that, for 98 per cent of their targets, the double-peaked emission lines could be explained as the result of gas kinematics within a single AGN, with only 2 per cent of the double-peaked profiles being produced by the relative velocity of merging systems. Still, a conclusive result on the nature of the four dual AGN candidates selected with this investigation could not be achieved, as the scenario of a single AGN illuminating gas in two merging galaxies remained a viable option. As it is often the case, high-spatial resolution X-ray or radio observations were invoked to dissipate the residual ambiguities (see Sect.~\ref{sssec:xray} and future perspectives in Sect.~\ref{sssec:future_xray}).

To summarise, IFS is a powerful technique which allows a clear view of the kinematics and ionisation mechanisms at play within a galaxy. It brings with it the strengths inherent to the optical and NIR wavelength range (that is medium/high spatial and spectral resolution); however, it also bears its limitations (mostly the effects of dust obscuration, and the fact that not every AGN displays optical signatures). Moreover, the drastic reduction in the number of candidates that this technique allows to achieve comes at the cost of using a wealth of data which often are not publicly available and that can be obtained only through the access to highly competitive facilities.

\subsection{X-ray observations of dual AGN}
\label{sssec:xray}

\begin{figure}[t]
\center 
\includegraphics[
% trim=0cm 5.2cm 0cm 0cm, clip=true, 
scale=0.35]{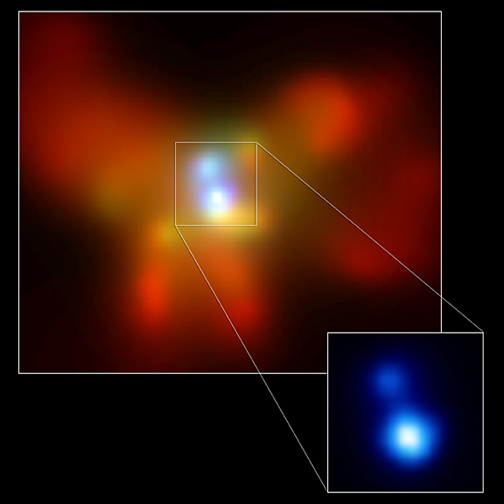}\\
\caption{\chandra\ image of NGC~6240 and its pair of accreting SMBHs. In this colour-coded image, red represents soft X-rays, and blue hard X-rays. [Image credit: NASA/CXC; \cite{Komossa2003}].
}
\label{fig: n6240_komossa}
\end{figure}

The high penetrative power of hard X-rays provides a unique and often ultimate tool in the hunt for multiple active nuclei in a galaxy, being less affected by contamination from stellar processes and absorption, though still limited by spatial resolution.
Indeed, it was X-ray imaging spectroscopy which led to the  identification of the first pair of accreting SMBHs in the galaxy NGC\,6240 (see the next section). 
A large (above $\sim10^{42}$~\lum) point-like luminosity is more likely ascribed to an AGN that to a starburst and/or emission from low-mass X-ray binaries (see e.g.,  \citealt{fragosetal2013,lehmeretal2016}).
 
However, often both a nuclear and an extended component can contribute to the X-ray spectrum of a dual system, requiring a high spatial resolution and a low background to properly disentangle the contribution of a weak AGN emission from a diffuse component. 
In the end, X-ray  observations represent an efficient way to detect accretion-powered processes in low-to-moderate obscured sources. Although heavy obscuration prevents us from providing a complete census of the AGN population, deep exposures, coupled with hard X-ray coverage, may mitigate this issue. 

%There are two possible ways to discover and/or characterize dual AGN in the X-ray band, provided that the X-ray data have the required spatial resolution and sensitivity to achieve this goal. In this regard,
The best angular resolution currently available in the X-ray band is that of \chandra, of the order of one arcsec for on-axis observations. This translates into $\approx$20~pc as the minimal angular scale probed at $z=0.001$, which becomes $\approx$1.8~kpc at $z=0.1$ and $\approx$8~kpc at $z=1$. From these numbers it appears clear that at present X-rays can reasonably probe only dual AGN at kpc-scale separations, and this will hold true until the launch of a mission with a \chandra-like resolution but with the support of a much larger effective area (e.g., \axis, a NASA probe-class mission, and  \lynx, a NASA large mission, currently under study, see Sect.~\ref{sssec:future_xray}). 
Needless to say that many of the \chandra\ observations suffer from low photon-counting statistics, enabling the detection/characterization of sources only in nearby galaxies. Despite this, \chandra\ provided X-ray images of remarkable detail and quality for the first time, thanks to the very low intrinsic background and sharp Point Spread Function (PSF). Limited photon statistics for more distant sources can be overcome by increasing the exposure time. Alternatively, \xmm\ can overcome the issue of photon statistics due to its much higher effective area, but its on-axis PSF FWHM ($\approx$6\arcsec, i.e. about six times larger than \chandra) limits the study of dual AGN to even larger-separation systems than \chandra. 

Besides using X-ray imaging spectroscopy for the discovery of AGN pairs, one method of detecting/observing dual AGN in X-rays consists of follow-up observations of previously identified multiple AGN systems (e.g., pre-selected in optical band, see previous Sect.~\ref{sssec:optical_kpc}). This technique is presented in Sect.~\ref{ssubsec:pointed X-rays}. Another method relies on using survey fields (e.g., COSMOS, the \chandra\ Deep Field North and South), taking advantage of the plethora of multi-wavelength data typically available in such fields, needed to select dual AGN candidates and define their properties; this technique is further discussed in Sect.~\ref{ssubsec:surveys X-rays} and Sect.~\ref{cgoals}. The advantage of the first method is clearly the possibility of targeting the system with a ``proper'' instrumental set-up, chosen to maximize the possibility of detecting/separating the AGN on the one side and defining their properties on the other side. If the search for dual AGN is pursued using the second method - where this investigation is essentially one of the multiple science goals that an X-ray survey can fulfill - the entire field of view can be used to reveal dual AGN, although the decrease in sensitivity and the broadening of the PSF at off-axis positions can limit their effectiveness. The latter approach can benefit from the typically already available spectroscopy in X-ray survey fields, which is of the order of 54--65 per cent in e.g. COSMOS-Legacy and CDF-S \citep{Marchesi2016, Luo2017}. Candidate dual AGN can be selected using photometric redshifts in all the cases where multi-band accurate photometry, coupled with proper galaxy (or galaxy$+$AGN) templates, is available. This technique, although simple, can be used at zeroth order also to identify large-scale structures to which dual AGN may belong. 
Clearly, spectroscopic follow-up observations are needed to finally confirm that the two AGN have comparable redshifts, hence being a ``certified'' pair.

\subsubsection{Discovery and follow-up observations of dual AGN and candidates} 
\label{ssubsec:pointed X-rays}

\begin{figure*}[t]
\centering
\includegraphics[scale=0.5,angle=0]{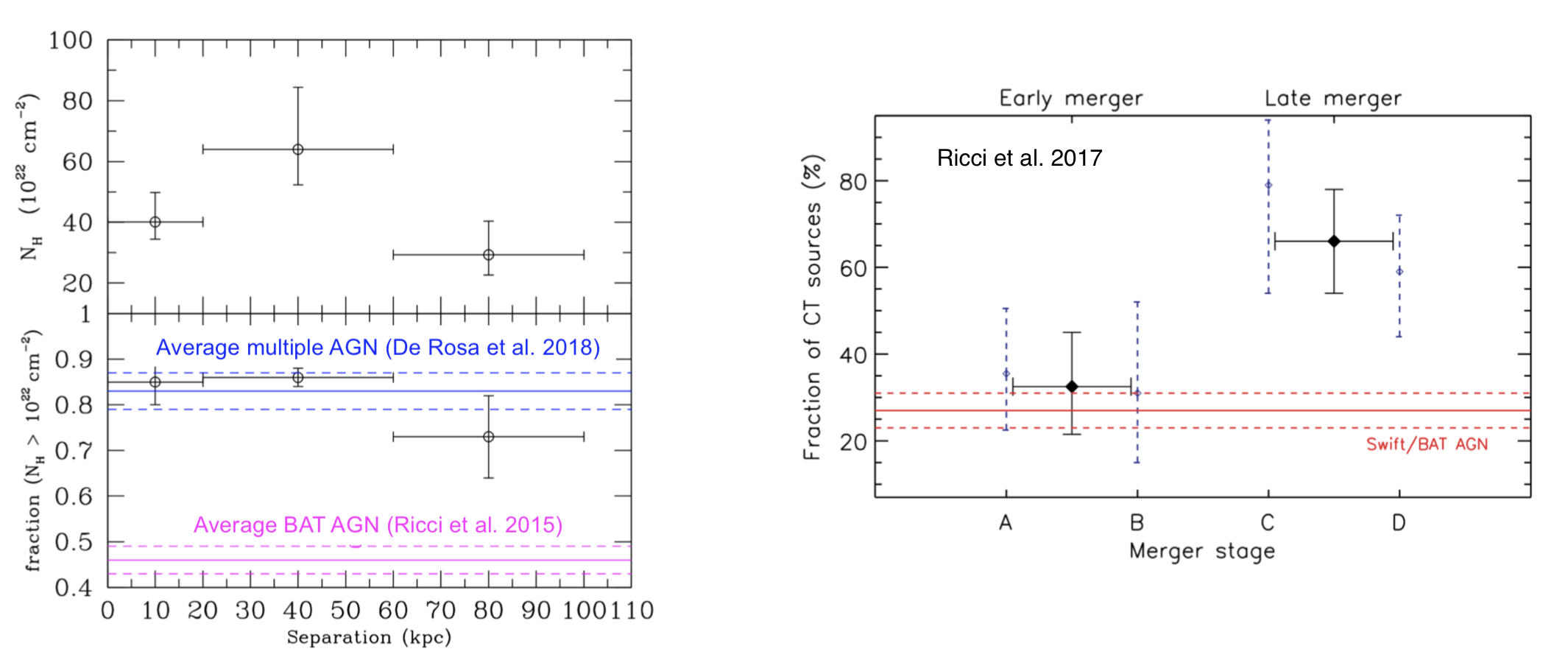}
\caption{{\it Left panel}: Absorption properties of a samples of dual systems observed in X-rays but selected in different ways  (optical, IR, X-rays) presented in \cite{derosaetal2018}. 
The upper panel shows the values of the absorption column density averaged in 3 ranges of projected galaxy separations: 0--20, 20--60 and 60--100~kpc, while the lower panel reports the fraction of AGN in dual/multiple systems with \nh\ above 10$^{22}$ \cm2 as a function of the projected separation.  Blue lines represent the average value as obtained for the large sample investigated in \cite{derosaetal2018} (61 sources) when 90 per cent errors on \nh\ are taken into account. Magenta lines represent the \swift/BAT average values and 1$\sigma$ error as reported in \citet{riccietal15} in the same bin of 2--10~keV luminosity. {\it Right panel:} Fraction of Compton-thick AGN versus merger stage as investigated in \cite{riccietal17}. These authors analysed a sample of infrared-selected local luminous and ultraluminous infrared galaxies in different merger stages and followed them up in the X-ray band with \nustar\ (up to $\sim$40~keV). The empty blue diamonds represent the values for the four merger stages separately, while the filled black diamonds are the values for early and late mergers.  The red continuous line represents the intrinsic fraction of Compton-thick AGN measured by Swift/BAT \citep{riccietal15} with associated 1$\sigma$ uncertainty.
}
\label{fig:Xray_abs}
\end{figure*}

The first spatially resolved SMBH pair was identified in X-rays, and was based on a dedicated \chandra\ observation which targeted the nearby ultraluminous IR galaxy (ULIRG) NGC~6240. \chandra\ imaging spectroscopy has revealed that both galaxy cores emit luminous point-like X-rays (Fig.~\ref{fig: n6240_komossa}), and show similar X-ray spectra which are flat, heavily absorbed, both exhibiting a strong neutral iron line \citep{Komossa2003, Nardini2017}. These are the tell-tale signs of heavily obscured AGN. %
The nuclei are at $\sim$ 1~kpc separation and are likely both Compton thick (i.e., with absorption column density along the line of sight \nh \gtsima 10$^{24}$ \cm2).
Recently, \cite{Kollatschny19}, using VLT/MUSE data at a resolution of 75 milli-arcsec, have reported the discovery of a third nucleus in this galaxy: the Southern component appears to host two distinct nuclei, separated by 198~pc only. The lack of a radio counterpart for the newly detected nucleus suggests that it could be not active; furthermore, no $^{12}$CO(2-1) emission at its position is detected by ALMA \citep{Treister2020}, in stark contrast with the two previously known nuclei.

In X-ray targeted observations, new dual AGN  have been discovered serendipitously.
The clearest cases are those for which both double nuclei show emission up to highest energies, above 10~keV: Mrk~739, a kpc-scale separation (3.4~kpc) at high ($\approx0.7$) Eddington ratio \citep{Koss2011}, selected in the hard-X ray domain by \swift/BAT and followed-up with the highest spatial resolution  available (\chandra); the ultra-luminous IR galaxy Mrk~463  (3.8~kpc separation; see \citealt{Bianchi2008}), for which \nustar\ broad-band analysis has recently confirmed the presence of Compton-thin obscuration in both nuclei \citep{yamada_etal2018}; Mrk~273, in the C-GOALS IR-selected sample, at 1~kpc separation: also in this case both nuclei are heavily obscured, one likely in the Compton-thick regime according to broad-band \chandra\ and \nustar\ observations \citep{Iwasawa2018}.

In the search and detection of dual AGN at kpc-scale separation with \xmm\ we can mention IRAS~20210+1121, hosting a pair of Type~2 AGN at 11~kpc separation \citep{Piconcelli2010}, the early-stage merging system ESO~509-IG066 (d=10.5~kpc, \citealt{Guainazzi2005}) whose double nuclei are both mildly obscured and with a possible variable absorption column density \citep{KosecBS17}, and Arp~299, where the presence and full characterization of two AGN in the companion galaxies IC~694 and NGC~3690 (separation of 4.6~kpc) was possible thanks to the combination of \chandra\ and \xmm\ data \citep{Ballo2004}. 
As said, at ~pc scale separation, X-ray observations have the limitation due to the PSF. The presence of two active, heavily obscured nuclei in the spiral galaxy NGC~3393 \citep[at $\sim$150~pc separation;][]{Fabbiano2011}, has been questioned by \cite{Koss2015} using deeper \chandra\ data combined with adaptive optics near-IR data and radio interferometric observations, resulting in a totally different interpretation (i.e., a strongly obscured AGN with a two-sided jet). This result shows that even X-ray observations alone are sometimes not conclusive (see, e.g., the recent results by \citealt{hou2019} on an [OIII]-selected AGN candidate pair sample, but see also \citealt{Comerford2015}, where the invaluable strength of X-rays to pinpoint and characterize the AGN emission in [OIII]-selected dual AGN is fully exploited) or can be misinterpreted if multi-wavelength data of comparable/higher resolution and quality are not available. 
We also note that sophisticated analysis techniques may sometimes help in identifying dual AGN \citep{foord19} at relatively close separations. 

One peculiarity of many of the dual systems mentioned so far is that some members of these systems show no (or very weak) explicit AGN evidence in their optical/near-infrared spectra. A rather common property of these AGN pairs is that they are heavily dust-enshrouded, with strong indications of being also heavily obscured in the X-rays. 
Investigating a sample of infrared-bright AGN (the GOALS sample, see more details about sample properties below), \cite{riccietal17} found the  fraction of Compton-thick AGN hosted in late-merger galaxies is higher than in local hard X-ray selected AGN (65$^{+12}_{-13}$ $vs$ 27 $\pm$ 4 per cent, see right panel in Fig.~\ref{fig:Xray_abs}). 
More generally, all AGN of their sample in late-merger galaxies are characterized to have N$_{\rm H} \geq$ 10$^{23}$~\cm2. Further evidence for high absorption, with column densities above 10$^{23}$~\cm2, in an IR (WISE)-selected sample of interacting/merging galaxies has been presented by \cite{pfeifle19} (see Sect. \ref{sssec:mid_IR_AGN_pairs}).
 
Similar results have been recently found by \cite{derosaetal2018},  where four Seyfert-Seyfert AGN systems at low redshift ($z<0.08$),  selected from SDSS, have been followed-up by \xmm. They find that the X-ray emission of three out of the eight AGN is consistent with being absorbed by Compton-thick material, and further three AGN are in the Compton-thin regime.
\cite{derosaetal2018} also compared the absorption properties in their dual AGN with those in larger samples observed in X-ray but selected in different ways (optical, IR and hard X-rays).
They found that the fraction of obscured (N$_{\rm H} \geq$ 10$^{22}$~\cm2) AGN is in the range of 64--90 per cent up to large-scale separations ($\sim$ 100~kpc, see left panel in Fig.~\ref{fig:Xray_abs}), i.e. higher than in isolated AGN observed by BAT \citep[43--49 per cent;][]{riccietal15}. 

This behaviour is in agreement with prediction of numerical simulations of isolated black holes (e.g.,  \citealt{Capelo_et_al_2017,Blecha_et_al_2017}; see discussion in Sect.~\ref{sssec:BH_Pairs_Theory_Isolated_Mergers}). In particular, numerical simulations of AGN in galaxy mergers show an increase of \nh\ as the distance decreases. This effect is due to merger dynamics; the median value of \nh\ was about 3$\times$10$^{23}$~\cm2, in good agreement with the value found in the large collection of dual systems presented by \cite{derosaetal2018}.
However, these numerical simulations only probe absorption on relatively large scales (above $\sim$ 50~pc, then absorption from the torus on typical pc-scale is not considered), therefore the \nh\ value should be considered as a lower limit.

It is, however, worth mentioning also the recent \chandra\ observations of eight  radio-selected dual AGN in the Stripe~82 field ($\sim4-9$~kpc separations,  \citealt{gross2019}) and seven optically selected dual AGN targeted in X-rays as part of a complete optical sample of dual systems with small transverse separations (15--27~kpc, \citealt{Green2011}; see also \citealt{Green2010}).
Of these quasars, only one object is obscured in X-rays; this result is probably due to the original selection from the SDSS, being biased towards systems with low intrinsic extinction. 
Furthermore, recent studies of X-ray properties of an SDSS-selected sample of galaxy pairs, where only one member of the pair exhibits spectroscopic signatures of AGN activity, indicate a non-negligible fraction of X-ray unobscured AGN (Guainazzi et al., in prep.). 
Intriguingly, galaxy pairs hosting  only one active member show a similar absorption fraction with respect to isolated AGN, i.e. much lower with respect to AGN in dual systems (as shown in Fig.~\ref{fig:Xray_abs}, left panel).

\subsubsection{Searching for dual AGN in X-ray survey fields} 
%\subsubsection*{Searching for dual AGN systems in X-ray survey fields} \
\label{ssubsec:surveys X-rays}
Regarding surveys,  \cite{Silverman_et_al_2011} selected kinematic pairs (galaxies) in the zCOSMOS field (i.e., through the optical spectroscopy carried out in COSMOS), and the X-ray coverage was used to probe the fraction of such galaxies showing nuclear activity, likely triggered by close galaxy encounters. This fraction seems to increase at separations below 75~kpc (projected distance) and line-of-sight velocity offsets less than 500~km~s$^{-1}$. This result is not dissimilar to what has been found using SDSS data by other groups (\citealt[and references therein]{satyapaletal14,Ellison2011})

It is worth mentioning the case of the X-ray COSMOS source CID-42 at z=0.36 that was found to host two  compact nuclei at close separation ($\approx$0.5 arcsec, corresponding to about 2.5~kpc; see \citealt{Civano2010}). Optical spectroscopy indicates the presence of a large $\approx$1200~km~s$^{-1}$ velocity offset between the narrow and broad components of the H$\beta$ emission line (see also \citealt{Comerford2009}). The X-ray spectra show an absorbing feature (with an inverted P-Cygni profile) at the energy corresponding to iron line, variable in energy and intensity on year-timescale, indicative of a high-velocity inflow (v/c$\approx$0.02-0.14;  \citealt{Civano2010, Lanzuisi2013}). 
Two scenarios may explain these results: either a GW recoiling black hole (e.g., \citealt{Peres1962,Campanelli2007,Blecha2011}) or a dual AGN system, where one of the two is recoiling because of slingshot (i.e., a 3-body BH scattering) effect. No firm conclusion can be drawn even after follow-up observations, although high-resolution X-ray data \citep{Civano2012} have shown that most of the X-ray emission is coming from only one nucleus (the South-Eastern one), the other being most likely a star-forming region, thus potentially favouring the first hypothesis. 
JVLA 3-GHz data have found that the emission is associated with the  nucleus emitting in X-rays; the radio spectrum, built up using also broad-band literature data, is suggestive of a Type~1, radio-quiet flat-spectrum nucleus \citep{Novak2015}. 
Overall, the current observational picture in the radio band is still consistent with the recoiling black hole hypothesis but cannot rule out the presence of an obscured and radio-quiet SMBH in the North-Western component. 

The study of the triggering mechanism in dual AGN systems requires complete samples, unbiased towards obscuration as much as possible (as discussed in the previous subsection). In this regard, the fraction of dual AGN has been estimated by \cite{Koss_et_al_2012} in a sample of local ($z<0.05$), moderate-luminosity AGN selected from the  all-sky  hard-X ray \swift/BAT survey. At scales $<$100~kpc, the dual AGN frequency is 10 per cent, where small-separation ($<30$~kpc) systems appear to be more common among the most luminous AGN, and 50 per cent of the AGN with a very close ($<$15~kpc) companion are dual. The fact that the X-ray luminosity of both AGN increases strongly with decreasing galaxy separation further suggests that merging (as shown by signs of disruption in the host galaxies by \citealt{Koss_et_al_2010}, and of galaxy interactions from morphological studies, \citealt{Cotini2013}) is probably the key in powering both active nuclei. In this regard, X-rays provide better insight into the nuclear activity, whereas  extinction and dilution by star formation strongly limit the search for accretion-related activity in the optical band. 

It worth mentioning that in the study of a large sample of galaxy pairs selected in the mid-IR, the fraction of AGN, relative to a mass-, redshift- and environment-matched control sample, increases with decreasing projected separation \citep[see discussion in Sect.~\ref{sssec:mid_IR_AGN_pairs}]{Ellison2011,satyapaletal14}.
This evidence further suggests the role of (close) galaxy encounters in driving gas towards the inner regions of AGN, thus efficiently fueling both AGN. As an example, see the case of the dual AGN system Mrk~739 as detected with \chandra\ \citep{Koss2011}.

Recently, \cite{kossetal2018} have preformed follow-up high-resolution infrared observations of hard-X-ray-selected black holes from \swift/BAT. Their study shows that obscured luminous SMBHs (with L$_{\rm bol}$ above 2$\times$10$^{44}$ \lum) show a significant excess of late-stage nuclear mergers (17.6 per cent) compared to a sample of inactive galaxies (1.1 per cent). The link between mergers and obscuration was also found in the COSMOS field by \cite{Lanzuisi2018} using \chandra\ and HST data; in particular, they found that the fraction of Compton-thick AGN in mergers/interacting systems increases with luminosity and redshift.

\subsubsection{C-GOALS: X-ray follow-up of IR-bright galaxies}
\label{cgoals}

\citet{iwasawa11} presented the X-ray properties of the GOALS sample \citep{armus09}, based on the X-ray \chandra\ data on the high infrared-luminosity complete sample (C-GOALS). Whilst the whole GOALS sample consists of $\sim 200$ far-IR selected nearby LIRGs, they focused on a complete sample of 44 systems with log $\log(L_{\rm ir}/L_{\odot})> 11.73$ in erg~s$^{-1}$.

The $L_{\rm ir}$ of the 44 systems range from 11.73 to 12.57 with a median $\log(L_{\rm ir}/L_{\odot}) = 12.0$. Given the selection method, they were not {\it a priori} known to be galaxy mergers, but the frequency of galaxy mergers turned out to be high. The number of systems in which clearly separated galaxy nuclei or galaxies are seen is 26. However, the unresolved galaxies all show tidal features hinting that they also underwent a recent galaxy merger. All these LIRGs are dusty objects and sometimes near-IR imaging was needed to reveal close pairs (with the nuclear separations smaller than 1~kpc).

The projected nuclear separation spans from 0 to 65~kpc \citep{2013ApJ...768..102K}, with a median value of 1.6~kpc. When limited to the 26 resolved systems, the median separation is 6.7~kpc. Out of the 44 systems, 43 X-ray detections were recorded. The AGN fraction is 37 per cent when only the X-ray diagnostics were applied (X-ray colour and Fe K$\alpha$ emission line), but this value increases to 48 per cent when supplemented by the mid-IR line diagnostics (e.g., [Ne {\sc v}] $\lambda$14.3$\mu m$, \citealt{petricetal11}). This AGN fraction is higher than that found in the lower-IR luminosity GOALS sample \citep{torres-alba_etal2018}. Among the detected AGN, 9 are Compton-thick AGN. The AGN frequencies in three categories of nuclear separation are found as follows: 1) unresolved and $<1$~kpc: 2/4 and 11/18; 2) 1--10~kpc: 7/15; and 3) $>10$~kpc: 2/7. As already discussed above, there might be a possible tendency of increasing AGN occurrence at smaller separations in the widened GOALS sample \citep{riccietal17}. Double AGN within a single galaxy 
are rare: in this category we have only NGC~6240 \citep{Komossa2003} and Mrk 273 \citep{Iwasawa2018}, both with sub-kpc nuclear separations. The X-ray luminosity of AGN (as observed) spans from $\log L_{2-10 \rm keV}=40.8$ to 43.1 in units of erg/s, with a median value of 41.65, which are relatively low for these IR-luminous systems. Absorption correction ranges from a factor of $\sim 2$ for a moderately absorbed source to $\sim 30$ for a Compton-thick AGN, and the average absorption correction is about 1 dex. When further applying the standard bolometric correction (10--20, e.g., \citealt{marconietal04,2012MNRAS.425..623L}), the AGN bolometric luminosity would be estimated to be of the order of 10$^{44}$ erg/s, i.e. less than 10 per cent of the total bolometric luminosity of ULIRGs.

\subsubsection{High-z X-ray observations of multiple AGN systems}
The situation at $z>1$ has been poorly investigated thus far, since detecting distant, possibly obscured AGN is challenging for \chandra\ observations of moderate depth. In this context, we may mention the triple quasar system at z=2.05 LBQS~1429-0053 (aka QQ1429-008; \citealt{Djorgovski2007}), for which a $\approx21$~ks \chandra\ pointing allowed the clear detection of the two brightest members of the system.
Although complete samples of dual quasars at high redshift are not easy to construct (see \citealt{hennawi2010, findlay2018}) and their complete follow-up 
in X-rays would probably require very large projects (of the order of several-Ms exposures), the few X-ray observations thus far available have allowed to check whether and how quasars in the golden quasar epoch are influenced by the surrounding environment (i.e., the presence of companions). In this regard, recently \cite{Vignali2018} have selected two systems of dual quasars at z=3.0--3.3 drawn from the \cite{hennawi2010} sample, hence from the SDSS-DR6, at separations of 
6--8\arcsec\ (43--65~kpc), providing a first-order source characterization for both quasars of each pair thanks to two \chandra\ 65~ks observations. All of these quasars have intrinsically high rest-frame 2--10~keV luminosities ($2\times10^{44}-5\times10^{45}$~erg~s$^{-1}$), with signs of obscuration in those quasars showing indications for broad absorption lines (BAL) in their optical spectra (hence being classified either as mini-BAL or BALQSOs). In particular, one of these four quasars has an order of magnitude higher optical luminosity than the median value of SDSS quasars at z=3.3; the possibility that this is due by chance (and, thus, not because of the presence of a companion) has been estimated to be of the order of 3 per cent. 
In the same work, \cite{Vignali2018} found X-ray emission from both quasar members of a pair at z=5.0 (projected separation of 21\arcsec, corresponding to 136~kpc), for which \xmm\ archival data ($\approx$~80~ks) allowed the authors to detect the two components separately, finding unobscured X-ray emission, consistent with their optical classification. To our knowledge, this represents the highest-distance quasar pair ever detected in X-rays. 

Still focusing on the quest of multiple AGN systems at high redshift, an astonishing and very recent field of research is given by the discovery of enormous ($\gtrsim200$~kpc) Ly$\alpha$ nebulae (ELANe) around $\simlt$10 per cent of radio-quiet quasars (e.g., \citealt{Hennawi2015},  following \citealt{Hennawi2013}; see also \citealt{Cantalupo2014, Cai2017}). Up to now, all of them show the presence of multiple companion AGN. Additional recent detections of ELANe at high redshift \citep{Fumagalli2017, Arrigoni-Battaia2018, Arrigoni-Battaia2019} indicate that these structures are almost ubiquitous found around luminous high-redshift quasars (mostly thanks to VLT/MUSE, Keck/KCWI, and the ongoing statistical surveys targeting quasars; this will be extensively discussed in Sect.~\ref{sssec:future_opticalspec}). 
For comparison, we note that other less extended and lower-luminosity nebulae \citep{Borisova2016} and Ly$\alpha$ blobs (LABs) reported in literature do not contain multiple AGN. Among ELANe, the Jackpot Nebula at $z=2$ was recently observed with \chandra\ (140~ks pointing; Vignali et al., in prep.). This nebula comprises an extended ($\approx$310~kpc) Ly$\alpha$ emission region, four AGN in the inner $\approx$200~kpc plus three additional AGN on larger scales, identified by \chandra, likely at similar redshift, a factor $\approx$20 overdensity of Ly$\alpha$ emitters, and several submm galaxies probably at the same redshift. The \chandra\ detection of extended emission at the 5.4$\sigma$ level in the inner region further supports the interpretation of the field as hosting a protocluster with a very high incidence of AGN (Vignali et al., in prep.). 
ELANe could help identify high-density regions at high-redshifts, such as prototclusters, where enhanced multiple AGN activity can be favored by halo assembly bias or by physical interactions in a dense galaxy environment.

\subsection{AGN dual detection in near- and mid-infrared}
\label{sssec:mid_IR_AGN_pairs}

It has been known since the publication of the IRAS Catalog of Bright Galaxies that most LIRGs and ULIRGs are  interacting/mergers. 
Later on it was proven that the fraction of LIRGs/ULIRGs hosting an AGN is higher than for non-IR bright galaxies; in fact, at the highest luminosity ($L_{\rm ir}$ > $10^{12}$ L$_\odot$), nearly all objects appear in advanced mergers powered by both AGN and circumnuclear starbursts, which are fueled by an enormous concentration of molecular gas that has been funneled into the merger nucleus \citep{Sanders1988,1996ARA&A..34..749S}. 
\cite{mateos17} found that the majority of luminous, rapidly accreting SMBHs at $z\leq$1 reside in highly obscured nuclear environments; most of these sources are so deeply enshrouded that they have so far escaped detection in $<$10~keV X-ray wide-area surveys. 

WISE \citep{Wright2010} all-sky mid-IR data proved to be extraordinarily fruitful in the characterization of mid-infrared properties of AGN, showing the power of mid-IR colours as AGN diagnostic tools (e.g., \citealt{Jarrett_et_al_2011, stern2012, mateos2012, Rovilos2014, satyapal2018}). %
The hot dust surrounding AGN produces a strong mid-infrared continuum, whose emission can be easily disentangled from that related to stellar processes via SED-fitting techniques in both obscured and unobscured AGN. 
Through a large mid-infrared study of AGN in mergers and galaxy pairs selected by matching SDSS and WISE data, \citet{satyapaletal14} found that there is an enhanced fraction of IR-selected AGN with respect to optically selected AGN in advanced mergers. They also found that the fraction of obscured AGN increases  with the merger stage, with the most obscured AGN becoming more prevalent in the most advanced mergers, where also star formation rates are the highest \citep{Ellison2015}.
There are additional observational works presenting the close link between mid-IR selection of AGN and mergers \citep{Donley2007,Urrutia2008,Veilleux2009,Assef2013,Glikman2015,fan_etal2016,goulding+2018}, highlighting the fact that a large number of objects are missed by optical classification because of heavy extinction and calling for a synergic approach in AGN selection involving different wavelengths. 

In a similar context, \cite{Weston2017} investigated the connection between merging/interacting galaxies and dusty obscured AGN using SDSS and WISE data at $z\simlt0.08$.
The analysis showed a clear link between the presence of major mergers/interactions and red [3.4] -- [4.6] colour, at least at high ($>10^{10}\ \msun$) stellar masses; non-merging galaxies in a similar mass range are less likely to have such red mid-IR colour. Thanks to  optical spectroscopy diagnostics (e.g., BPT diagram, \citealt{Baldwin81}), they found that one-quarter of Seyferts show redder [3.4] -- [4.6] colours than $\sim$99 per cent of non-Seyferts. From their sample, AGN are five times more likely to be obscured when hosted by a merging galaxy.  

\begin{figure}[t]
\centering
\includegraphics[scale=0.3]{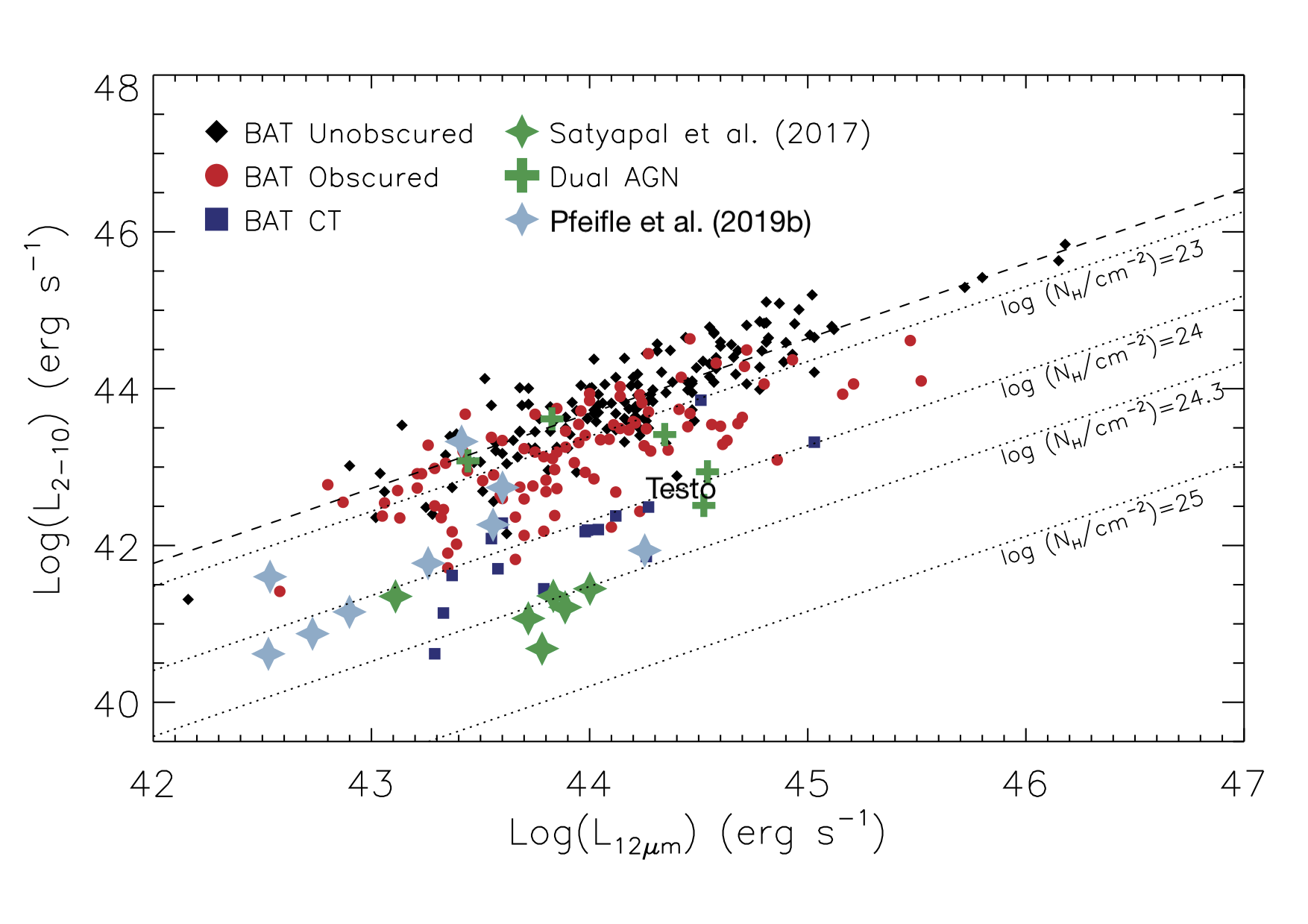}
\caption{2--10~keV observed luminosity $vs$ mid-IR (12~\micron) luminosity for the mergers included in the program described in Sect.~\ref{sssec:mid_IR_AGN_pairs} (i.e., X-ray follow-ups of WISE-selected mergers). The plot also includes the sources already confirmed as dual AGN in the literature. The samples of hard X-ray selected AGN with the \swift/BAT survey are also shown; for these AGN, a direct measurement of the absorbing column density N$_{\rm H}$ is possible \citep{riccietal15,riccietal17}. The dual AGN candidates derived from the WISE selection are the most heavily absorbed. Dotted lines represent the effect of different absorption column densities on the unabsorbed best-fit linear relation from the \swift/BAT sample (dashed line). Adapted from \cite{pfeifle19}.}
\label{fig:midIR}
\end{figure}

Findings in line with these were recently reported by \cite{pfeifle19} and \cite{satyapaletal17} for samples of mid-IR selected AGN with signatures of ongoing interactions. 
In particular, these authors analysed \chandra\ and \xmm\ and near-IR (Large Binocular telescope, LBT) data of a sample of advanced mergers with nuclear separations below 10~kpc, which were pre-selected through mid-IR colours using WISE. 
The combined X-ray, near-infrared, and mid-infrared data were able to discover the presence of  dual/multiple AGN systems. 
\citet{satyapaletal17} detected the presence of the AGN in a least four out of the six mergers, with four systems possibly hosting dual AGN. \citet{pfeifle19} detected at least one nuclear X-ray source in all 15 mergers analysed in their work; eight of these systems exhibit two sources suggestive of dual (or triple) AGN. The high spatial resolution X-ray, near-IR, and optical spectroscopic diagnostics allowed the discovery of the triplet system SDSS~J084905.51+111447.2 at z = 0.077  \citep{Pfeifle2019b}.
It is worth noting that none of these analysed systems show evidence for AGN in their optical spectra, probably due to the large extinction. Support to this interpretation comes from the measured large nuclear obscuration revealed in X-rays (N$_{\rm H}$ above $10^{24}$~cm$^{-2}$).
The absorbed 2--10~keV luminosity in all these systems is much lower than the mid-IR luminosity (see Fig.~\ref{fig:midIR}). 
These findings suggest that the pre-selection in the mid-IR provides an efficient way to detect dual AGN in late-mergers, although data at these wavelengths are limited in terms of angular resolution and may require information at other frequencies (e.g., in X-rays) to eventually provide conclusive results on the presence of nuclear activity. A significant step forward in terms of angular resolution, hence in the process of confirmation and characterization of close systems, will be provided by the mission James Webb Space Telescope, {\it JWST} (see Sect.~\ref{sssec:future_mir} for more details on future mid-IR observations). 
In this regard, \cite{Imanishi_Saito2014} used high spatial resolution K- and L$\sp{\prime}$- band observations to search for dual AGN in kpc-scale merging systems, previously identified in the optical as double-peaked; they found dual AGN in four systems among the 29 analyzed. This result can be interpreted assuming that the active phase is not simultaneous in the two merging galaxies.

\subsection{Dual AGN in the radio waveband}
\label{sssec:Dual_radio}

 One of the advantages of observations in radio bands is that the propagation of radio waves is not affected by dust obscuration. Therefore, radio surveys are an essential element of multi-wavelength studies. At centimetre wavelengths, the radio continuum emission from extragalactic sources is dominated by non-thermal (synchrotron) emission from either AGN or star-forming activity. In the former case, the radio emission originates from ultrarelativistic electrons spiralling around the magnetic field lines in the vicinity of SMBHs. In the case of ``normal'' galaxies without AGN, the origin of radio emission is partly by relativistic electrons accelerated by supernova remnants and travelling through the magnetic field of the galaxy, or free-free emission from HII regions \citep[for a comprehensive review of radio emission from galaxies, see][]{condon1992}.

Distinguishing between AGN and star formation related radio-emitting regions requires observations with sufficiently high angular resolution. While the radio emission from AGN is compact and localised to the innermost central part of the host galaxy, star-forming activity is distributed over much larger volumes. 

Connected-element radio interferometers, with maximum baselines of tens of km such as the Karl G. Jansky Very Large Array\footnote{\url{http://www.vla.nrao.edu/}} (VLA) or the Australia Telescope Compact Array\footnote{\url{https://www.narrabri.atnf.csiro.au/}} (ATCA) routinely provide angular resolutions of 1--10 arcsec at cm wavelengths. The electronic Multi-Element Remotely-Linked Interferometre Network\footnote{\url{http://www.e-merlin.ac.uk/}} (eMERLIN) in the UK, with maximum baselines of a few hundred km, provides angular resolutions as small as 40 and 150 milliarcsecond (mas) at 6 and 20 cm, respectively.

Such angular scales allow to detect and characterize not only the compact sources corresponding to the AGN themselves, but also extended emission in their surroundings, such as radio jets or jet-driven outflows. Furthermore, multi-frequency observations give useful spectral information to discern them: AGN typically show flat spectra, with a few exceptions \citep{1998PASP..110..493O}, while jet-driven structures present steeper profiles. This makes radio observations with arcsec resolution a powerful tool to confirm the nature of multiple AGN systems (e.g., \citealt{mullersanchez2015});  this is particularly important given that gas photoionization from one AGN can create a spurious signature of a second AGN in optical data, as described in Sect.~\ref{sssec:optical_kpc}.
The highest angular resolution in radio astronomy is offered by very long baseline interferometry (VLBI). 

By this technique, the radio telescopes do not need to be physically connected, so the distance between them can reach thousands of km. The mas angular resolution achievable by VLBI at cm wavelengths makes it possible to identify very compact radio sources and indicates that the brightness temperature of the detected source exceeds $\sim$10$^6$\,K. Compact continuum radio sources with such high brightness temperatures are often associated with AGN, being the brightness temperature of star-forming galaxies $\lesssim\,10^{5}$\,K for frequencies above $\sim 1$\,GHz \citep{condon1992}. Star-forming activity, gamma-ray bursts, radio supernovae or their complexes can occasionally reach this limit. However, transient events are rather rare at radio wavelengths -- only $\sim$50 supernovae have been detected at radio frequencies after $\sim$30 years of observations \citep{lien2011} -- and the luminosity of star formation quickly drops below detection limits when the redshift of the host galaxy is larger than $\approx$0.1. Therefore, based on luminosity and brightness temperature arguments, if a radio source is detected with VLBI in an extragalactic object at redshift $z\approx0.1$ or beyond, it is certain that the emission is AGN-related \citep{middelberg2013} -- with the single exception of hyperluminous IR galaxies at high redshift, some of which have exceptionally high star-formation rates and can be partly powered by hyperluminous nuclear starbursts. For these reasons, VLBI is a powerful tool to separate AGN from star-forming galaxies above relatively low redshifts ($z\gtrsim\,0.1$). In other words, while connected-element interferometers are sensitive to both starburst-related and AGN radio emission spatially blended in the same source,
VLBI with its longer baselines acts as an efficient spatial filter for very compact non-thermal AGN radio emission at $z\gtrsim\,0.1$. It is therefore a unique observing tool to confirm the existence of jetted (radio) AGN.{\footnote{In some cases, however, multiple knots in a radio jet can mimic two AGN.}} 

In the context of dual AGN, the typical mas-scale angular resolution achievable with VLBI networks at cm wavelengths allows us to directly resolve pairs with projected linear separation as small as about a pc in the local Universe and $\sim 10$\,pc at any redshift. The technique is thus ideally suited for observing kpc-scale dual AGN \citep[e.g.][]{2018RaSc...53.1211A}. However, a severe limitation of its applicability is that only a minority of AGN, less than $10$ per cent of the population, are strong (i.e. at least mJy-level) radio emitters \citep[e.g.][]{ivezic2002}. Therefore, based on VLBI observations alone, it is only possible to prove unambiguously the AGN nature of a candidate dual source if both companions are radio AGN. However, it must be also remarked that some of the AGN classified as ``radio-quiet'' do show compact radio emission that can be detected with sensitive VLBI observations \citep[e.g.][]{herreraruiz2016}. 

\begin{figure}[t]
\center 
\includegraphics[
scale=0.45]{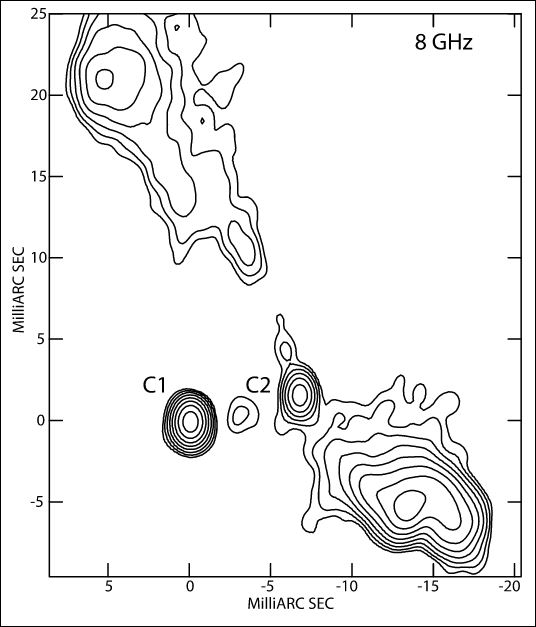}\\
\caption{ Radio VLBA image contours of the system 0402+379 at 8 GHz. Components C1 and C2 correspond to the two radio nuclei at projected separation of 7.3\,pc.
From \citealt{rodriguez2006}.}
\label{fig:radio0402}
\end{figure}

Because of their narrow (typically arcsec-scale) fields of view, the use of VLBI networks as ``blind'' survey instruments is also limited, unless wide-field observing and multi-phase-centre correlation techniques are employed \citep[e.g.][]{chi2013,herreraruiz2017}. A more typical application of VLBI is to study individual objects or small samples of dual AGN candidates that are selected based on independent indications of duality. It comes as no surprise that kpc-scale dual AGN cases confirmed with VLBI observations are still rare. 
A systematic study of the most luminous radio AGN in the multi-frequency VLBI archive \citep{burkespolaor2011} has produced a single case of a double flat-spectrum core, the nearby radio galaxy 0402+379 ($z=0.055$), with two nuclei separated by just 7.3\,pc (see Fig.~\ref{fig:radio0402}, \citealt{rodriguez2006}).

Based on the analysis of VLBA (the network of ten radio antennae located across the United States) data spanning 12~yr, recently \cite{bansal2017} claimed the detection of relative motion of the companion AGN. If this is due to orbital motion, and assuming a circular orbit, the authors could derive an orbital period of about $3 \times 10^{4}$~yr. While the VLBI technique is capable to detect positional changes of compact components on sub-mas scales, over the currently available time baselines of several decades, only a minor part of an assumed orbit can be sampled in a binary with pc-scale separation. Consequently, the derivation of the orbital parameters is challenging. 

There are practical difficulties in deciding which sources can be regarded as reliably ``confirmed'' dual AGN. Ideally, at least two different methods, based on observing at multiple wavebands, would be desirable. In this regard, radio observations also have their role (although with the already mentioned disadvantage that only a small fraction of AGN are expected to be strong radio emitters). Care must also be taken when conducting only low-resolution or single-frequency measurements, since the observed properties could be reconciled with several different explanations. 

For example, based on VLA observations, the quasar J1023+3243 ($z=0.127$), showing double-peaked narrow optical emission lines, was found to be a compelling case for a dual AGN through radio observations \citep{mullersanchez2015}. Conversely,  higher-resolution VLBI observations, filtering out extended non-AGN emission, failed to reveal even a single compact radio AGN in J1023+3243 \citep{gabanyi2016}. This is consistent with the fact that arcsec-resolution radio imaging alone is not decisive in confirming AGN duality, as described above. 

There are also cases when the interpretation of VLBI observations is not straightforward. The quasar J1425+3231 ($z=0.478$) was found as a kpc-separation dual AGN candidate based on double-peaked narrow [OIII] emission lines in its optical spectrum \citep{peng2011}. Follow-up VLBI observations of this radio source revealed two compact components separated by 2.6\,kpc \citep{frey2012}, one of them with a flat and another with a steep radio spectrum. While both components must be AGN-related synchrotron sources, they are not necessarily associated with AGN cores. Indeed, sensitive intermediate-resolution ($\sim 50-200$\,mas) e-MERLIN radio images later made it clear that the steep-spectrum feature is not an AGN core but instead a hot spot in a kpc-scale lobe. 
Therefore, this quasar is apparently not a dual AGN system, and the double-peaked optical emission lines are likely related to a jet-induced symmetric outflow from the vicinity of a single central SMBH \citep{gabanyi2017}.    

An example of the debated interpretation of VLBI results is the case of J1502+1115 ($z=0.390$). Multi-frequency VLA observations by \citet{fu2011} showed that two radio components could be detected at the core positions separated by $7.4$\,kpc previously revealed by optical and near-infared measurements. \citet{deane2014} reported the surprising result that, based upon their high-resolution VLBI observations, one of the two radio AGN contains two flat-spectrum components separated by just 0.14\,kpc and interpreted this as J1502+1115 being a triple radio AGN system. The picture with a close-pair binary was also supported by the helical pattern of the large-scale radio jet. The scenario was, however, challenged by \citet{wrobel2014} who argued that the closely-separated radio components are instead double hot spots energized by a single SMBH in a compact symmetric object. Sensitive high-resolution multi-frequency VLBI observations would be needed to give a final answer to this debate.

Probably the most spectacular example of a confirmed dual radio bright AGN is 3C\,75 ($z=0.023$). The double nuclei of the source were first mapped in the radio by \citet{owen1985} using the VLA. Later, observations conducted at various wavebands from X-rays to radio revealed that the source is indeed a pair of radio-emitting AGN separated by $\sim 8$\,kpc \citep[][and references therein]{hudson2006}. The radio interferometric images of the jets of the two AGN provide a unique opportunity to study their interaction with each other, and with the intracluster medium. 

\section{Theory on dual AGN systems: cosmological and isolated simulations of mergers}
\label{ssec:BH_Pairs_theory}

In this section, we extensively discuss cosmological (Sect.~\ref{sssec:BH_Pairs_Theory_Cosmological_Simulations}) and idealized (Sect.~\ref{sssec:BH_Pairs_Theory_Isolated_Mergers}) merger simulations on the formation of  dual AGN. We also describe  physical processes occurring in circumnuclear
disks (CNDs; Sect.~\ref{sssec:stallingBH_2}) surrounding BH pairs and discuss their pairing at ~sub-kpc-scale separations. Whereas large-scale cosmological simulations (e.g., modelling of gas inflows from filaments on evolving galactic halos) enable direct comparisons with surveys, idealized merger simulations, which typically reach a much higher numerical resolution, allow to follow in greater detail the dynamics and growth of the SMBHs, and to study systematically the evolution of their parameters (e.g., mass, spin, and merger times).

\subsection{Cosmological Simulations}
\label{sssec:BH_Pairs_Theory_Cosmological_Simulations}

In the last few years, several cosmological hydrodynamic simulations have been carried on with the aim at producing  large samples of simulated galaxies (e.g., Magneticum: \citealt{Hirschmann_et_al_2014}; Horizon-AGN: \citealt{2014MNRAS.444.1453D}; Illustris: \citealt{Vogelsberger_et_al_2014}; EAGLE: \citealt{Schaye_et_al_2015}; MassiveBlack-II: \citealt{Khandai_et_al_2015}; \textsc{BlueTides}: \citealt{Feng_et_al_2016}; MUFASA: \citealt{Dave_et_al_2016}; Romulus: \citealt{Tremmel_et_al_2017}; IllustrisTNG: \citealt{Springel_et_al_2018}), including more and more detailed physical processes introduced at sub-grid level, e.g. star formation \citep[e.g.][]{Springel_Hernquist_2002, Schaye_DallaVecchia_2008}, stellar evolution, supernova feedback, chemical enrichment \citep[e.g.][]{Tornatore_et_al_2003, Tornatore_et_al_2007, Pillepich_et_al_2018}, magnetic fields \citep[e.g.][]{Pakmor_et_al_2011, Pakmor_Springel_2013}, growth of SMBHs and their associated AGN feedback \citep[e.g.][]{Springel_et_al_2005, Fabjan_et_al_2010, 2012MNRAS.420.2662D, Rosas_Guevara_et_al_2015, Steinborn_et_al_2015, Weinberger_et_al_2017}.
The overall properties of the simulated galaxy samples, i.e. the stellar mass function, the star formation main sequence, the mass-size relation, and the mass-metallicity relation, do in general agree surprisingly well with observations, given the fact that they are produced self-consistently \citep[e.g.][]{Hirschmann_et_al_2014, Steinborn_et_al_2015, Remus_et_al_2017, Dolag_et_al_2017, Weinberger_et_al_2018}.

Simulations including SMBHs do also reproduce the observed BH mass function as well as observed scaling relations between BHs and their host galaxies relatively well, including the relation between the BH mass $M_\bullet$ and the stellar mass $M_*$ or the $M_{\bullet}$-$\sigma$ relation \citep[e.g.][]{Sijacki_et_al_2015, Hirschmann_et_al_2014, Steinborn_et_al_2015,Volonteri_et_al_2016, Weinberger_et_al_2018}.
In these simulations, the BH growth is commonly modelled using the Bondi formalism to compute the accretion rate, limited to a maximum value of $x \cdot \dot{M}_{\mathrm{Edd}}$, where $x$ is customarily set to $x=1$ \citep{Springel_et_al_2005}:
\begin{equation}
\dot{M}_\mathrm{\bullet} = \mathrm{max}\left(\frac{4 \pi \alpha G^2 M_\bullet^2 \langle \rho_{\infty} \rangle}{(\langle v \rangle ^2 +
  \langle c_{\mathrm{s}} \rangle ^2)^{3/2}}, x \cdot \dot{M}_{\mathrm{Edd}}\right),
\label{accretion_rate}
\end{equation}
where $\langle \rho_{\infty} \rangle$, $\langle c_{\rm s} \rangle$, and $\langle v \rangle$ are the mean density, the mean sound speed, and the mean velocity of the accreted gas relative to that of the BH, respectively. The empirical boost factor $\alpha$ has been introduced by \citet{Springel_et_al_2005} to compensate for the fact that the resolution is too low to properly apply the Bondi model. This parameter was originally set equal to $\alpha=100$, but it has been modified later-on such to account for the dependence  on the gas pressure \citep{Vogelsberger_et_al_2013}, on its angular momentum content \citep{Rosas_Guevara_et_al_2015,Tremmel_et_al_2017}, or on the gas temperature \citep{Steinborn_et_al_2015}.

However, despite the variety of existing hydrodynamic cosmological simulations, only a few of the currently used BH models are capable of producing AGN pairs with separations down to kpc scales within a cosmological simulation.
One reason is that very large volumes are required at relatively high resolutions, making such simulations computationally very expensive. Another reason is that most of the currently used BH models are not capable of tracing BHs down to the resolution limit during a galaxy merger.
Specifically, in most cosmological simulations BHs are pinned to the gravitational potential minimum \citep{Blecha_et_al_2016, Springel_et_al_2005} to avoid that the BHs become artificially dislocated from the galaxy centre.
Consequently, during a galaxy merger, the two central BHs merge instantaneously as soon as the two merging galaxies are identified as only one galaxy by the halo finder.
Since this occurs relatively early during the merger, it is not possible to directly trace the dynamics and growth of two BHs during the merger.
Making predictions about the number density of SMBH pairs using such simulations is only possible indirectly, for example using models like in \citet{2017MNRAS.464.3131K} and in \citet{Salcido_et_al_2016}, making predictions about GWs from SMBH mergers adding delay times in post-processing.

However, some authors developed methods to avoid that the BHs have to be pinned to the potential minimum.
In the cosmological zoom-in simulations from \citet{Bellovary_et_al_2010}, for example, dark matter (DM) particle masses are set similar to the gas particle masses in the high-resolution region.
\citet{Okamoto_et_al_2008} artificially drag the BHs along the local density gradients. \citet{2012MNRAS.420.2662D} include dynamical friction from unresolved gas, while 
\citet{Hirschmann_et_al_2014}, \citet{Tremmel_et_al_2015}, and \citet{2019MNRAS.486..101P} use a similar approach by accounting for a dynamical friction force, from stars and DM, caused by small- and large-scale perturbations of the surrounding particles (see, e.g., \citealt{Tremmel_et_al_2015} and references therein for a detailed description).
Due to these perturbations BHs are, equivalently to stars, decelerated in the direction of their motion \citep{Chandrasekhar_1943}:
\begin{equation}
\Delta \vec{v} = \delta \vec{v} (\Delta t) - \eta \vec{v} \Delta t.
\end{equation}
Following \citet{Chandrasekhar_1943}, and assuming that $M_{\bullet}$ is much larger than the masses of the surrounding particles, $m_i$, the dynamical friction coefficient $\eta$ is given by
\begin{equation}
\eta = 4 \pi m_i M_{\bullet} G^2 \frac{\vec{v_{\bullet}}}{v_{\bullet}^3} \mathrm{ln}\left( \frac{D_0 <v^2>}{G M_{\bullet}} \right) \int_0^{v_{\bullet}} N(v_i) \mathrm{d} v_i,
\end{equation}

\noindent where $D_0$ is the average distance between the BH and the surrounding particles,
$\vec{v}_{\bullet}$ is the velocity of the BH, $\vec{v}_{i}$ are the velocities of the neighboring particles, $<v^2>$ is the mean square of these velocities, and $N(v_i)$ is the according velocity distribution.
Note that only particles with $v<v_{\bullet}$ are considered in this formula.
Such a friction term, however, is only required when the resolution is too low to model dynamical friction self-consistently within the simulation \citep{2018MNRAS.478..995B}.
Alternatively, in the EAGLE simulations,  BHs are positioned in correspondence to the potential minimum only below a mass threshold $M_{\bullet}<100 m_{\mathrm{gas}}$, where $m_{\mathrm{gas}}$ is the initial mass of gas particles \citep{Rosas_Guevara_et_al_2019}.
BH mergers are generally allowed when the separation and the relative velocity between the two BHs decrease below certain thresholds.

\subsubsection{Fraction of AGN pairs}
%\subsubsection*{Fraction of AGN pairs}
Due to the large size of the simulations required to produce AGN pairs, there is either lack of available snapshots or poor spatial resolution, or too small simulated volumes to produce AGN pairs within the same time step. \citet{Steinborn_et_al_2016} use a simulation with a very large volume of (128~Mpc/h)$^3$ and a spatial resolution of roughly 2~kpc. However, the simulation, which is taken from the Magneticum Pathfinder simulation set, ran only down to $z=2$.
The Horizon-AGN simulation \citep{Volonteri_et_al_2016} has a smaller simulation volume of (100~Mpc/h)$^3$ and a spatial resolution of 1~kpc, but it ran down to $z=0$.
The Romulus simulation \citep{Tremmel_et_al_2017,Tremmel_et_al_2018} also ran down to $z=0$. Its spatial resolution is about four times higher than in the Magneticum simulation and twice that in the Horizon-AGN simulation. However, the simulation volume is relatively small with (25~Mpc/h)$^3$.
To compensate for the small volume, \citet{Tremmel_et_al_2017} do not use single snapshots, but all galaxy mergers across cosmic time.

In order to compare different simulations, we need to adopt the same selection criteria, in particular the same spatial separations between two AGN (typically $<10$~kpc) and the same galaxy and AGN luminosity threshold. Often, AGN are simply selected through their bolometric luminosity, and a typical threshold is $L_\mathrm{bol} > 10^{43}$~\lum. When comparing simulations with observations, one has to be aware of the fact that such a selection does not include any obscuration effects. In addition, simulations make different assumptions on the modelling of the galaxies and their SMBHs, leading to very different predictions. State-of-the-art simulations of galaxy mergers are still in their infancy in tracking the dynamics and growth of SMBHs realistically, and it is still challenging to reproduce the observed properties of AGN and SMBHs as a function of redshift. 

\begin{table}
\centering
\begin{tabular} {l|llll}
{} & \multicolumn{2}{l}{\multirow{2}{*}{$M_* > 10^{10} M_\odot$}} & \multicolumn{2}{l}{\multirow{2}{*}{$M_* > 10^{11} M_\odot$}} \\
{} & \multicolumn{2}{l}{\multirow{2}{*}{}} & \multicolumn{2}{l}{\multirow{2}{*}{}} \\
{} & $z=0$ & $z=2$ & $z=0$ & $z=2$ \\
\hline
{} & {} & {} & {} & {} \\
{Horizon-AGN} & {0.1} & {2 } & {0.9 } & {11 } \\
%{Magneticum} & {-} & {0.013} & {-} & {1.2} \\ %for <10ckpc
{Magneticum} & {-} & {0.19} & {-} & {1.7} \\ %for <30pkpc
\end{tabular}
\caption{Fraction of dual AGN (per cent) with respect to the total number of galaxies above the given mass threshold and at the given redshift, for the Magneticum simulation and the Horizon-AGN simulation.
        }
\label{cosm_sim_comparison}
\end{table}

To demonstrate this, Table~\ref{cosm_sim_comparison} shows a comparison of the dual AGN fractions in the Magneticum simulation run by \citet{Steinborn_et_al_2016} and the Horizon-AGN simulation by \citet{Volonteri_et_al_2016}.
At $z=2$, the Horizon-AGN simulation \citep{Volonteri_et_al_2016} predicts a dual AGN fraction of 2 per cent (11 per cent) with respect to all galaxies above $M_* > 10^{10} M_\odot$ ($M_* > 10^{11} M_\odot$). 
 
The Magneticum simulation run predicts much lower dual AGN fractions: 0.19 per cent and 1.7 per cent for galaxies above $M_* > 10^{10} M_\odot$ and $M_* > 10^{11} M_\odot$, respectively. There might be several reasons for these discrepancies, most of them being related to selection criteria, as specified above, for dual AGN (such as the threshold for BH mass, galaxy mass, AGN luminosity, and Eddington ratio). A minor effect is the slightly different definition of dual AGN. For the Magneticum simulation result we include only AGN pairs with proper separations $< 30$~kpc, whereas \citet{Volonteri_et_al_2016} include also AGN pairs with larger separations. However, only a very small fraction of their dual AGN have separations above 30~kpc, making this point negligible. 
More importantly, Horizon-AGN selects dual AGN using a threshold for galaxy mass, whereas Magneticum uses a threshold on BH mass (equal for both AGN, $M_\mathrm{\bullet} > 10^{7} M_\odot$); assuming the same criteria on BH mass in Horizon-AGN, the dual AGN fraction is closer to the one reported in Magneticum (0.6 per cent at $z=2$).
Also, imposing the same BH mass threshold as in the Magneticum simulation will select systems with mass ratio close to one, that will increase the paucity  of the selected dual AGN systems.
The number density of AGN in general is much larger in the Horizon-AGN simulation than in the Magneticum simulation. Specifically, the faint luminosity end of the luminosity function from the Horizon-AGN simulation is over-predicted, compared to observations from \citet{Hopkins_et_al_2007}. In contrast, the Magneticum simulation under-predicts the number of luminous AGN, but agrees with \citet{Hopkins_et_al_2007} at the low-luminosity end.
Since faint AGN are the majority of all AGN, this leads to a much larger number density of AGN and, thus, also of dual AGN.
One reason for these differences might be the differences in the AGN feedback model.
In particular, the model from \citet{Steinborn_et_al_2015}, which is used in the Magneticum simulation run, suppresses BH growth much earlier than in the feedback model used in the Horizon-AGN simulation.
Particularly at $z=2$, this leads to a larger fraction of quiescent galaxies with respect to star-forming galaxies.
Assuming that AGN activity and star formation activity are driven by the same mechanisms, this implies a generally smaller fraction of AGN.
However, since observations also predict very different AGN luminosity functions \citep[e.g.][]{Lacy_et_al_2015}, it is still unclear which simulations give better predictions.
Thus, any predictions of dual AGN fractions with respect to the overall galaxy sample, from simulations as well as from observations, have to be taken with great caution.

\begin{figure}
\centering
\includegraphics[trim = 10mm 10mm 10mm 10mm,width=\columnwidth,angle=0]{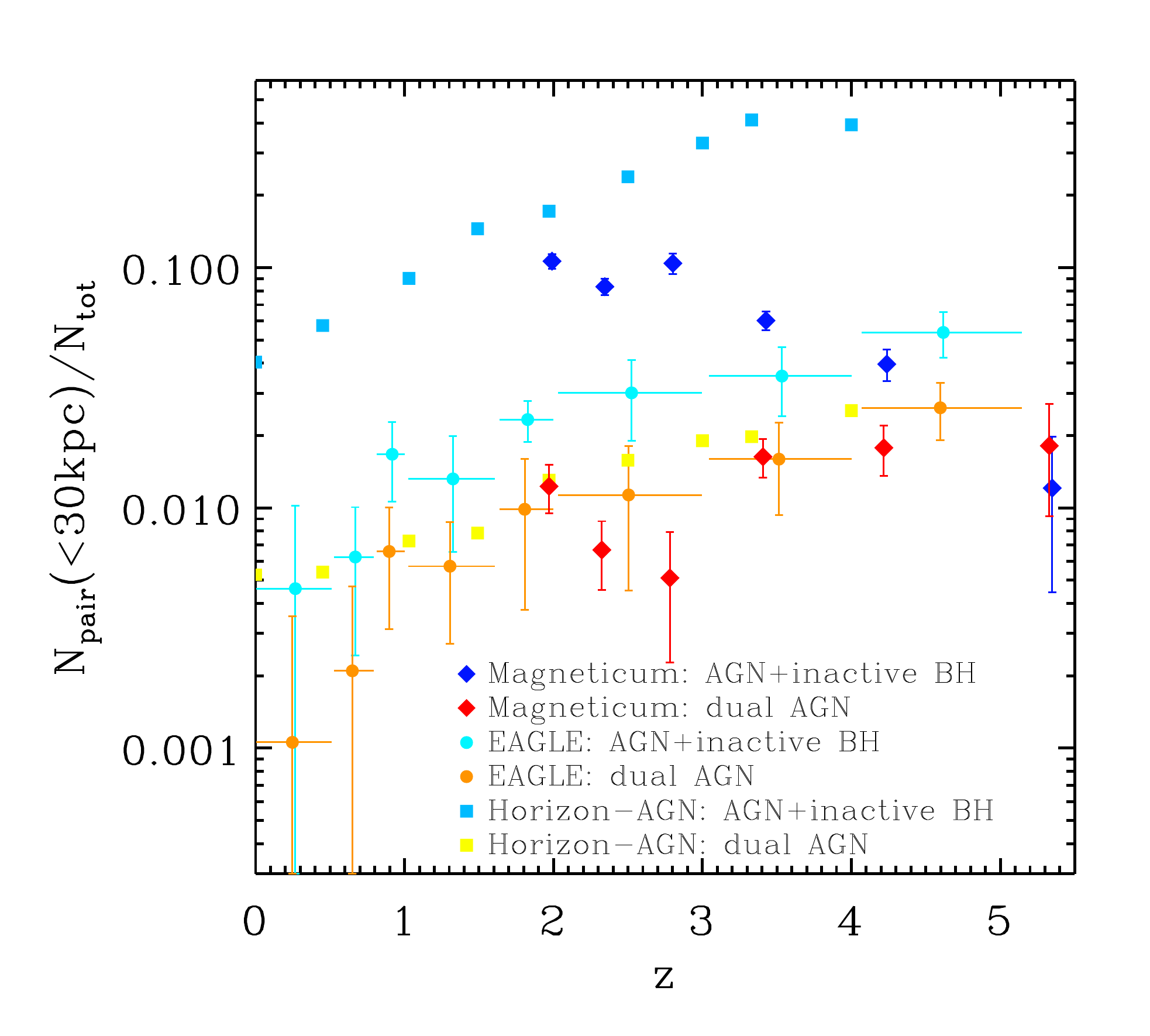}
\caption[]{Fraction of dual AGN (red, orange, and yellow datapoints) and BH pairs with only one AGN (dark blue, cyan, and light blue datapoints) with respect to the total number of AGN, for Magneticum (red and dark blue diamonds), EAGLE (orange and cyan circles, taken from \citealt{Rosas_Guevara_et_al_2019}), and Horizon-AGN (yellow and light blue squares). For all simulations we include only AGN pairs with $L_\mathrm{bol} > 10^{43}$~\lum and a proper separation $<30$~kpc, while for Magneticum and Horizon-AGN we also apply a BH mass cut of $M_\mathrm{\bullet}>10^7 M_\odot$ for one BH.}
\label{fig:frac_cosmsims}
\end{figure}

Next, we compare the redshift evolution of the dual AGN fraction and the fraction of BH pairs with only one active BH of the Magneticum simulation run presented in \citet{Steinborn_et_al_2016} with Horizon-AGN and with the largest EAGLE simulation run (Fig.~\ref{fig:frac_cosmsims}). 
The datapoints from \citet[][EAGLE]{Rosas_Guevara_et_al_2019} are shown as cyan (one AGN) and orange (dual AGN) filled circles and
the Magneticum simulation results are represented by dark blue (one AGN) and red (dual AGN) diamonds. The datapoints for Horizon-AGN are shown in light blue (one AGN) and yellow (dual AGN). To be able to compare the two simulations, we again adopt the same selection criterion for the Magneticum simulation like in \citet{Rosas_Guevara_et_al_2019},  which is a proper separation of $<30$~kpc. For Magneticum and Horizon-AGN, we apply the same BH mass cut of $M_\mathrm{\bullet}>10^7 M_\odot$. While the dual AGN fractions predicted by the simulations agree very well with each other, the fraction of BH pairs with one AGN is generally increasing in the EAGLE and Horizon-AGN simulations, whereas in the Magneticum simulation it is decreasing, at least above $z \sim 3$. For $z \lesssim 3.5$ the Magneticum simulation has a significantly larger fraction of such pairs than the EAGLE simulation, but similar to Horizon-AGN.

We stress here the importance of the exact definitions of both numerator and denominator in defining these fractions and their redshift evolution. As noted above, imposing a galaxy mass cut is not the same as imposing a BH mass cut, since the choice, when coupled with a luminosity threshold, impacts the Eddington ratio distribution of the selected population. For instance, in the case of Horizon-AGN at $z=2$, the dual AGN fraction is 0.8 per cent taking only a $L_\mathrm{bol} > 10^{43}$~\lum threshold and it is basically constant with redshift; the fraction becomes 2.4 per cent if applying also a galaxy mass cut of $M_* > 10^{10} M_\odot$ and it increases strongly from low to high redshift. Applying the luminosity cut and a BH mass cut of $M_\mathrm{\bullet}>10^7 M_\odot$ to both BHs in the pair the fraction at $z=2$ is 0.6 per cent and it decreases somewhat at $z>2$, finally when applying the same BH mass cut to one AGN only the fraction at $z=2$ is 1.3 per cent and it increases weakly from $z=0$ to $z=4$ (see fig. \ref{fig:frac_cosmsims}).

Furthermore, when comparing with observations, it is imperative to adopt the same approach. For instance, applying the same luminosity threshold to the sample, as done in these examples, differs from selecting all AGN above a luminosity threshold and then looking at which of these have a companion that can be somewhat fainter. While it is understandable that the choice of definition is sometimes forced by the parent sample properties, one needs to check carefully whether the same definitions are used, before performing a comparison.

However, predictions have to be taken with caution. For instance, to avoid spurious scattering, the Magneticum simulation uses two masses, a ``real'' mass and a ``dynamical mass''. The ``real'' mass is used to calculate gas accretion and feedback, whereas the ``dynamical mass'' is used for gravity as long as the BH mass is not large enough to avoid wandering caused by numerical noise (see \citealt{Hirschmann_et_al_2014} and \citealt{Steinborn_et_al_2015} for details). In that way they have a relatively smooth transition between the galaxy without and, then, with a BH hole, which would otherwise cause too much AGN feedback right after seeding the BH. However, since the ``real'' mass is different from the ``dynamical mass'', which determines the motion of gas near the BH, for these BHs it is uncertain whether they would also be active if their mass had been larger and exerted a stronger feedback (see \citealt{Steinborn_et_al_2016}).
Higher redshifts, where many BHs are seeded, should be affected more than lower redshifts. This could explain the difference between EAGLE and Magneticum in Fig.~\ref{fig:frac_cosmsims}.

In contrast to \citet{Volonteri_et_al_2016}, \citet{Steinborn_et_al_2016}, and \citet{Rosas_Guevara_et_al_2019}, \citet{Tremmel_et_al_2018} concentrate on haloes with Milky Way-like masses at $z=0$. They find that multiple SMBHs within 10~kpc from the galaxy centre are very common in such low-mass haloes: on average, they host $5.1 \pm 3.1$
SMBHs, independent of the merger history and morphology. This implies that SMBH pairs and multiplets might be common \citep[for early predictions giving similar numbers for Milky Way-size galaxies,  see][]{2005MNRAS.358..913V,Bellovary_et_al_2010}. However, AGN activity and, in particular, dual AGN activity is very rare in these haloes since off-center BHs tend to avoid the gaseous disk.
One explanation for the large amount of BH pairs and multiplets \citep{Tremmel_et_al_2018} is that SMBHs often spend a long time getting to the centre \citep{Tremmel_et_al_2017}.
The results from \citet{Tremmel_et_al_2018} are very different from those of \citet{Steinborn_et_al_2016}, who find only 35 BH pairs in total.
One reason might be the different redshifts, which would imply that the number of offset SMBHs must increase strongly between $z=2$ and $z=0$, probably due to mergers.
Furthermore, \citet{Tremmel_et_al_2018} and \citet{Steinborn_et_al_2016} use different mass ranges due to the different resolutions and simulation volumes.
More likely, however, the discrepancies originate from the different resolutions, seeding criteria and the slightly different dynamical descriptions of SMBHs.
This, as widely discussed here, demonstrates again that current cosmological simulations have to be taken with caution when making predictions for the amount of SMBH pairs and thus, for the fraction of dual AGN.

\subsubsection{Conditions for the formation of dual AGN}
Although cosmological simulations are not able to make reliable predictions for the fraction of AGN pairs, they are very useful for making predictions about general trends regarding dual AGN, for example in comparison either to inactive SMBH pairs or to SMBH pairs with only one active BH.
In particular, \citet{Steinborn_et_al_2016} and \citet{Volonteri_et_al_2016} agree in the following points, both being in agreement with simulations of isolated mergers (Sect.~\ref{sssec:BH_Pairs_Theory_Isolated_Mergers}):
\begin{itemize}
\item The probability for dual AGN activity increases with decreasing distance between the BHs.
\item The BH mass ratio of dual AGN is close to unity.
\end{itemize}

Cosmological simulations do not only produce AGN pairs self-consistently in a cosmological context, but they also contain more information than observations. Most importantly, the time evolution can give insights into the formation and trigger mechanisms of dual AGN.
In particular, \citet{Steinborn_et_al_2016} conclude:
\begin{itemize}
\item Dual AGN do, on average, require a larger gas reservoir than single AGN. They are not only driven by internal gas of the host galaxy, but also by gas which has recently been accreted by the galaxy, for example from cold gas filaments or clumpy accretion.
\item Only when both SMBHs have similar BH masses, they are both able to be active. 
As soon as one BH is significantly more massive than its counterpart, its gravitational potential accumulates and heats the gas even in the vicinity of the smaller BH, suppressing its accretion which could otherwise be accreted by the less massive BH. 

\end{itemize}

Both these conclusions agree with those obtained from idealized simulation (see Sect.~\ref{sssec:BH_Pairs_Theory_Isolated_Mergers} on BH activity) and with observing data that reveal a more obscured environment in dual AGN with respect to isolated systems (see Sect.~\ref{sssec:xray}). 

Although the specific dynamical description of BHs, as well as the underlying AGN models used in cosmological simulations have to be refined, current state-of-the-art cosmological simulations already have important implications for the interpretation of observations as well as of simulations of isolated mergers.
In particular, dual AGN activity is not simply a tracer of galaxy mergers, since many inactive BH pairs and multiplets with kpc-scale separations exist, which are not necessarily linked to a recent merger \citep{Tremmel_et_al_2018}.
Furthermore, the results from \citet{Steinborn_et_al_2016} suggest that having a closer look onto the detailed morphology as well as onto the environment of galaxies which host dual AGN might give further insights into their driving mechanisms. Since the accretion of gas onto the galaxy seems to play a major role \citep{Steinborn_et_al_2016}, it will be particularly interesting to have a look at how the gas is accreted by the galaxy and how it is driven towards the centre.

\subsection{Isolated Mergers}

\label{sssec:BH_Pairs_Theory_Isolated_Mergers}

\begin{figure*}[!t]
\centering
\vspace{3pt}
\begin{overpic}[width=0.67\columnwidth,angle=0]{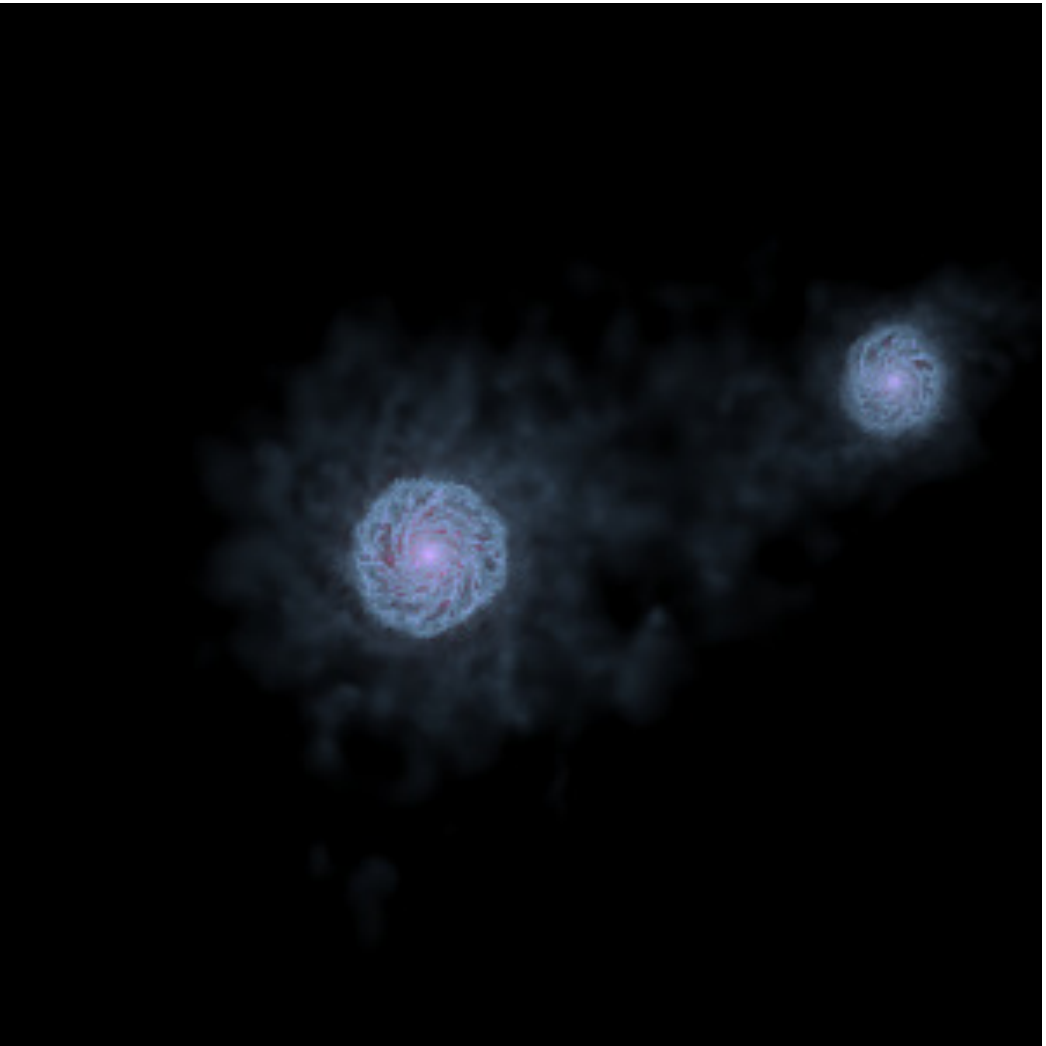}
\put (3,90) {\textcolor{white}{1}}
\end{overpic}
\begin{overpic}[width=0.67\columnwidth,angle=0]{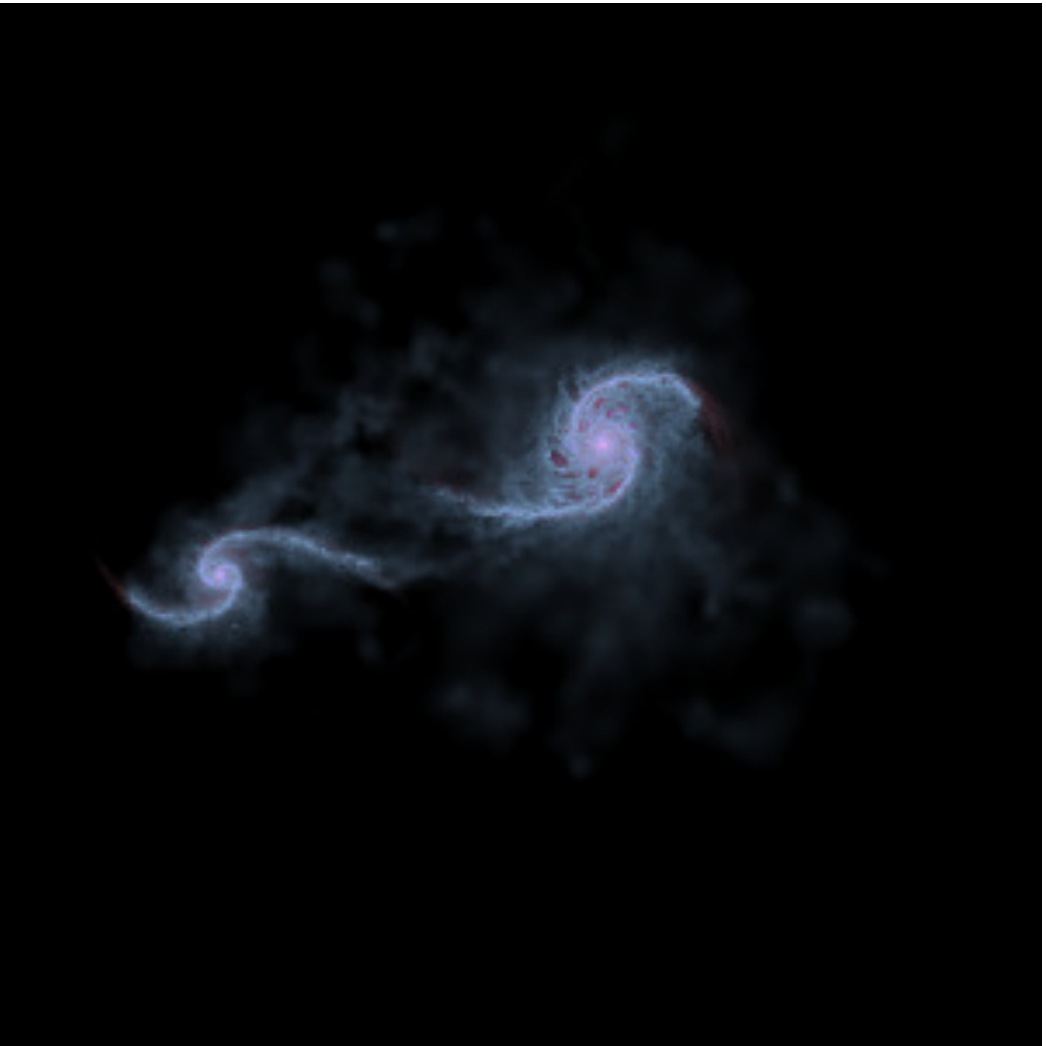}
\put (3,90) {\textcolor{white}{2}}
\end{overpic}
\begin{overpic}[width=0.67\columnwidth,angle=0]{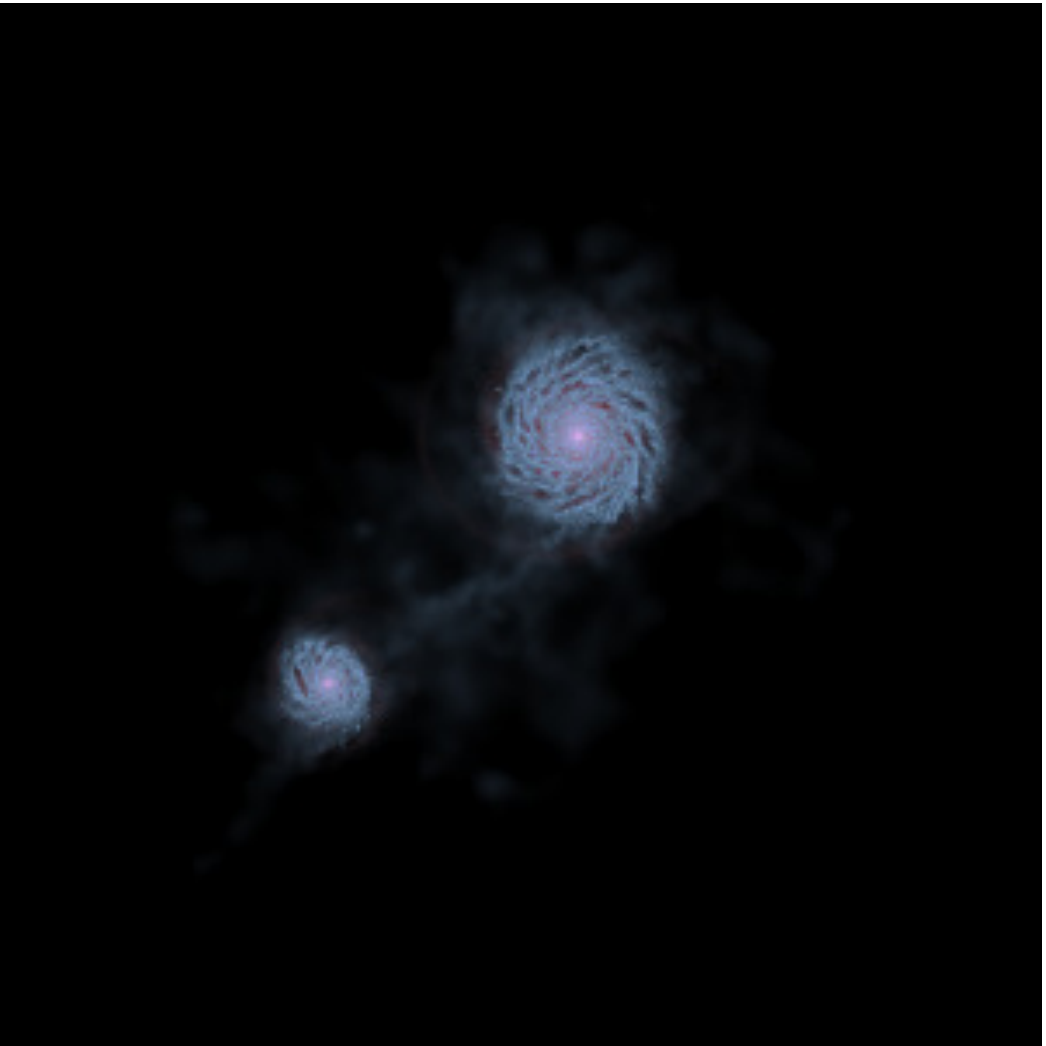}
\put (3,90) {\textcolor{white}{3}}
\end{overpic}
\vskip 0.5mm
\hspace{1.3mm}\begin{overpic}[width=0.67\columnwidth,angle=0]{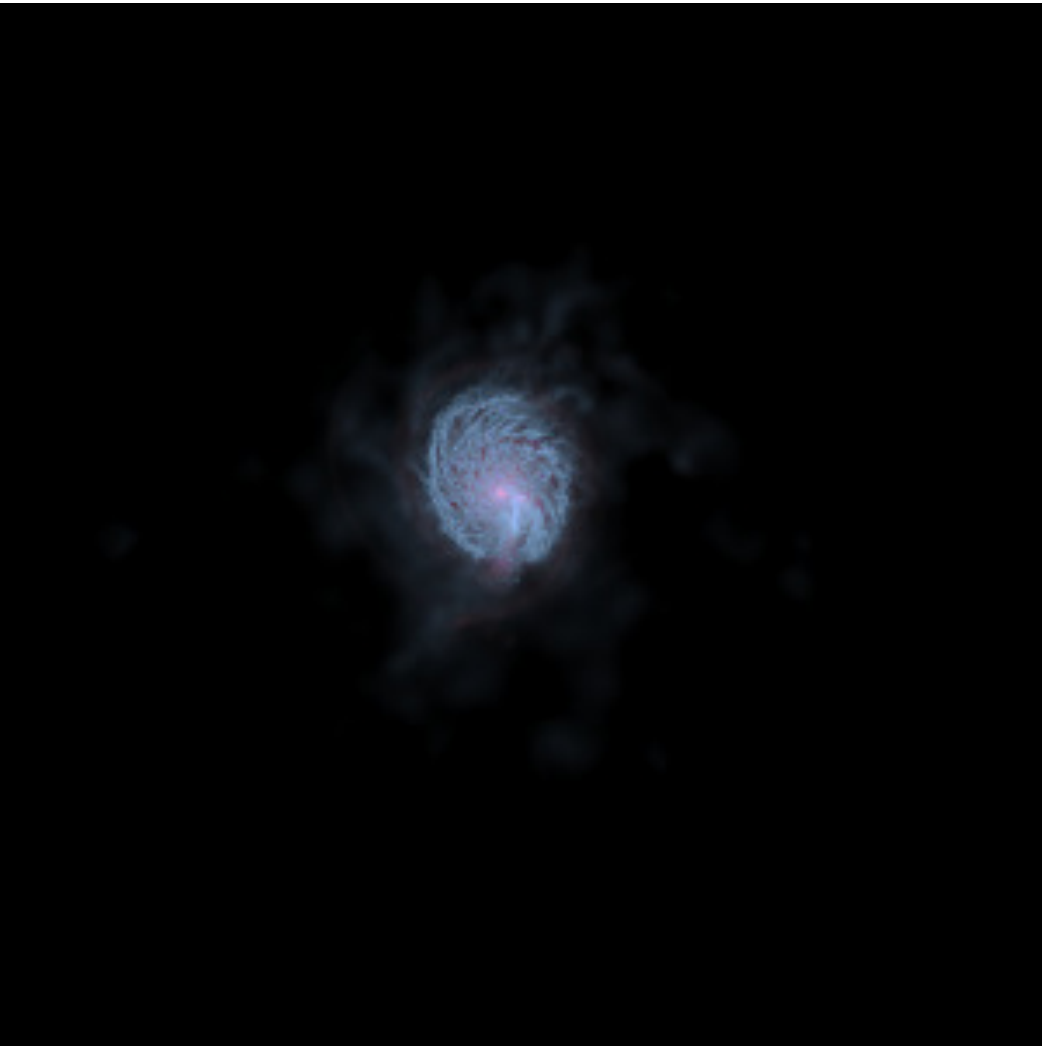}
\put (3,90) {\textcolor{white}{4}}
\end{overpic}
\begin{overpic}[width=0.67\columnwidth,angle=0]{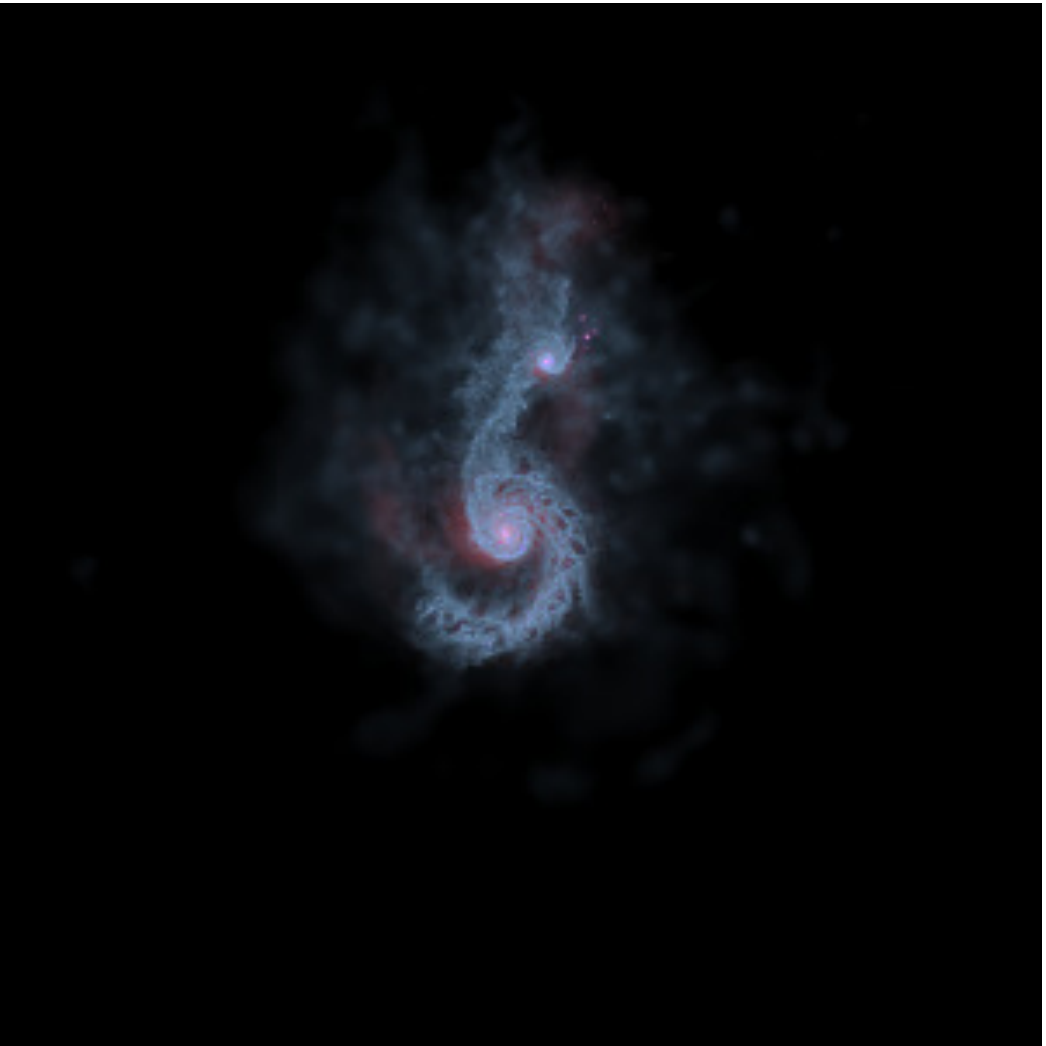}
\put (3,90) {\textcolor{white}{5}}
\end{overpic}
\begin{overpic}[width=0.67\columnwidth,angle=0]{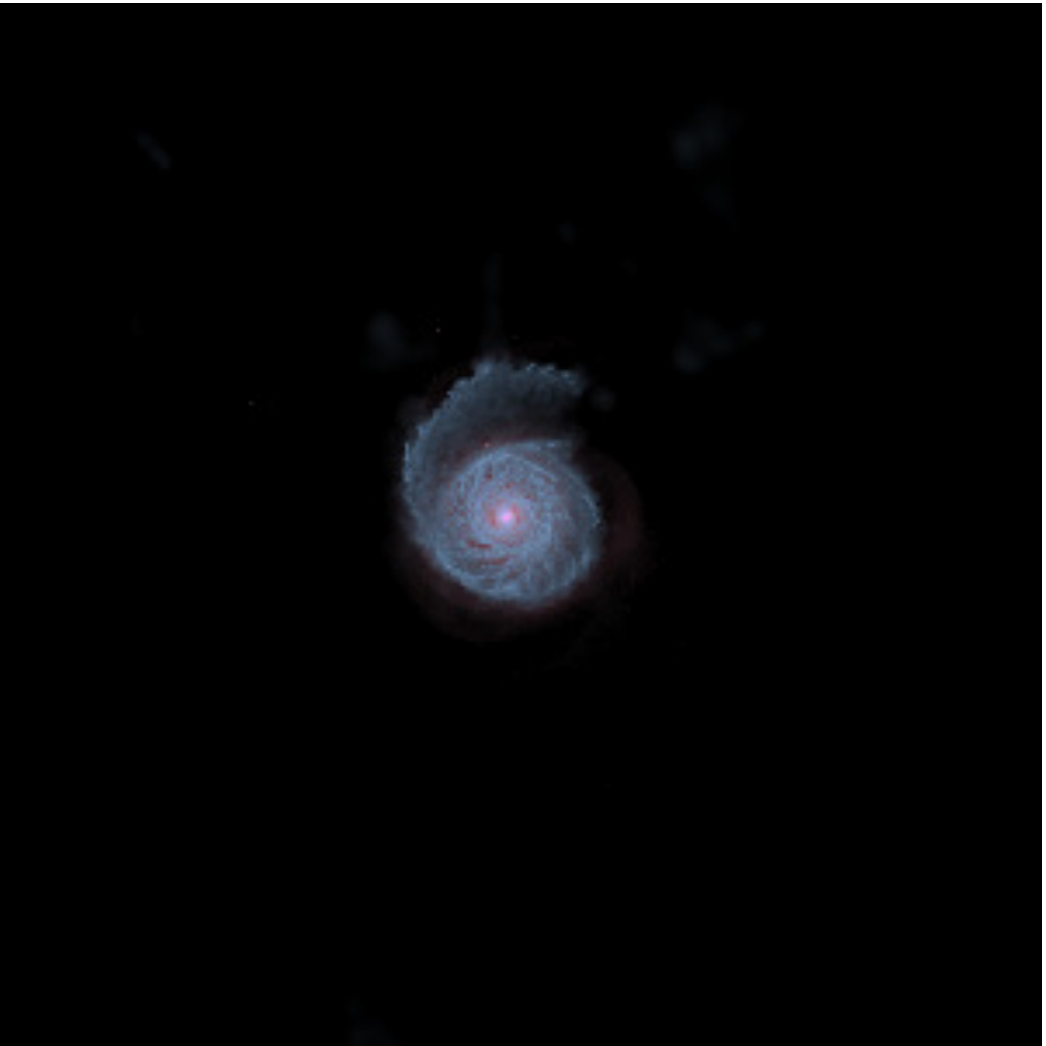}
\put (3,90) {\textcolor{white}{6}}
\end{overpic}
\vspace{0pt}
\caption[Density snapshots]{Stellar (red) and gas (blue) density snapshots (viewed face-on) at representative times of a 1:4 coplanar, prograde--prograde merger (first described in \citealt{Capelo_et_al_2015}; Run~07): (1) 0.196, (2) 0.391, (3) 0.880, (4) 0.973 (second pericentric passage), (5) 1.055, and (6) 1.561~Gyr. The primary (secondary) galaxy starts the parabolic orbit on the left (right) of the first snapshot, moving right (left) wards. In order to make the gas more visible, gas density was overemphasized with respect to stellar density. Each image's size is $70 \times 70$~kpc. Adapted from \citet{Capelo_et_al_2015}.}
\label{fig:1to4_density_snapshots}
\end{figure*}

\begin{figure*}[!t]
\centering
\vspace{5.0pt}
\includegraphics[width=1.02\columnwidth,angle=0]{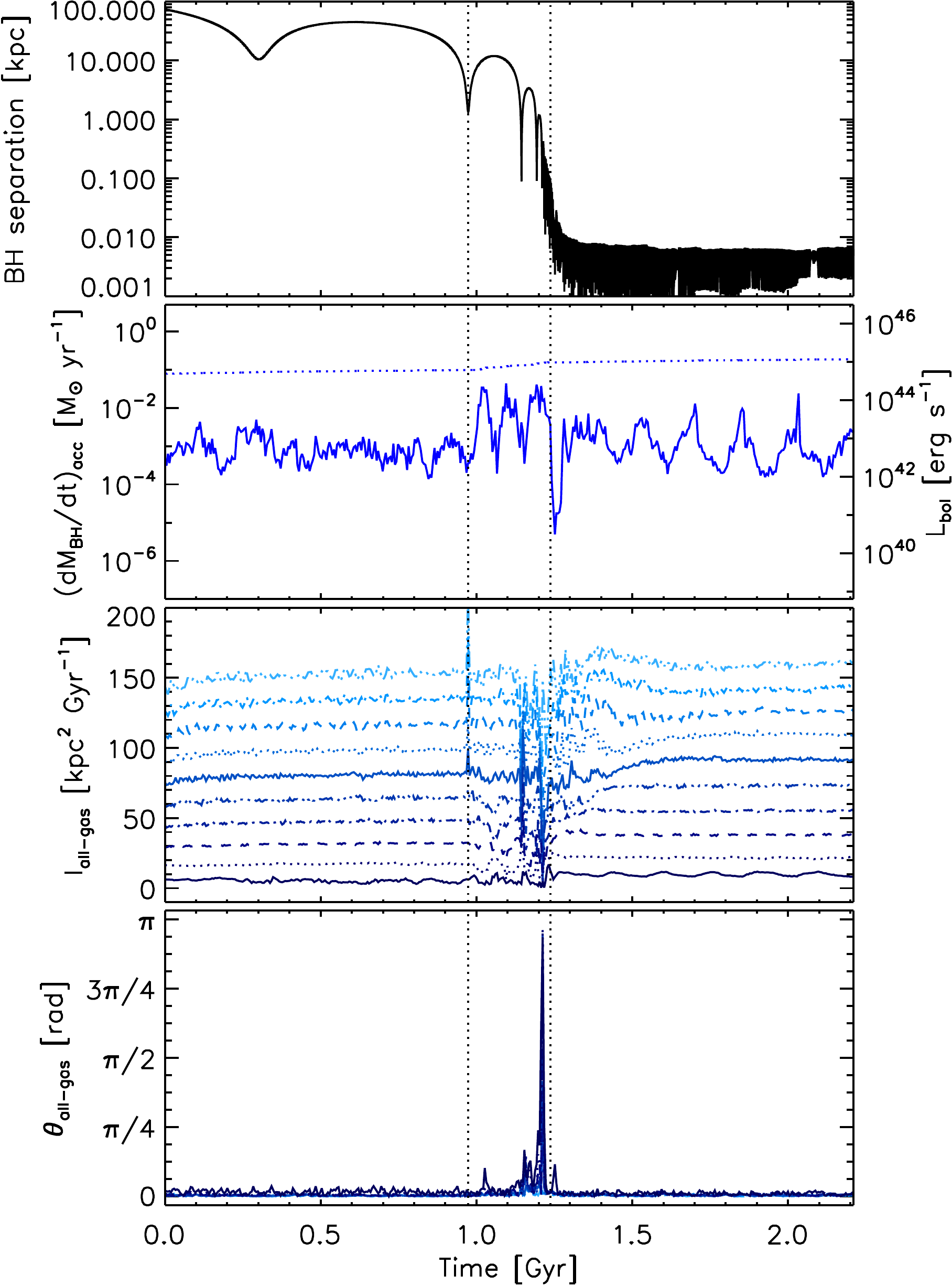}
\includegraphics[width=1.02\columnwidth,angle=0]{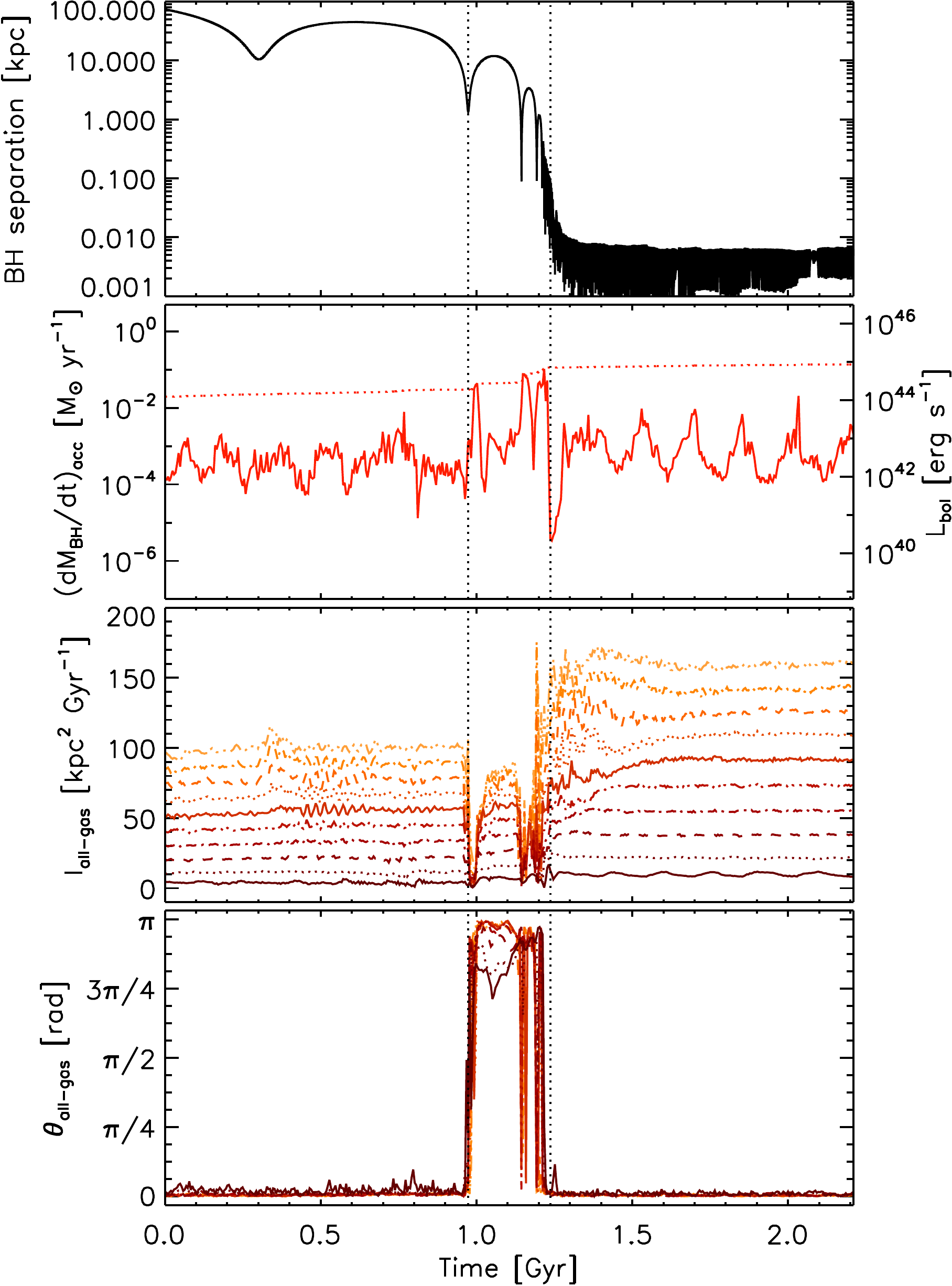}
\vspace{-0.0pt}
\caption[]{Temporal evolution of several quantities of the primary (left-hand panels) and secondary (right-hand panels) galaxy of a 1:4 coplanar, prograde--prograde merger (see also Fig.~\ref{fig:1to4_density_snapshots}). In all panels, the vertical, dotted, black lines show the separation between the stochastic, the merger, and the remnant stage. First panel: separation between the two BHs. Second panel: BH accretion rate (solid line) and BH Eddington accretion rate (dotted line). Third panel: magnitude of the gas specific angular momentum $l$ in 10 concentric shells of 100-pc thickness around the local centre of mass near the BH, in the inner kpc; $l$ grows monotonically with radius at the beginning and at the end of the run. Fourth panel: same as the third panel, but for the polar angle of the specific angular momentum vector. Adapted from \citet{Capelo_et_al_2015}.
}
\label{fig:1to4_temporal_evolution}
\end{figure*}

\begin{figure*}[!t]
\centering
\vspace{5.0pt}
\begin{overpic}[width=0.50\columnwidth,angle=0]{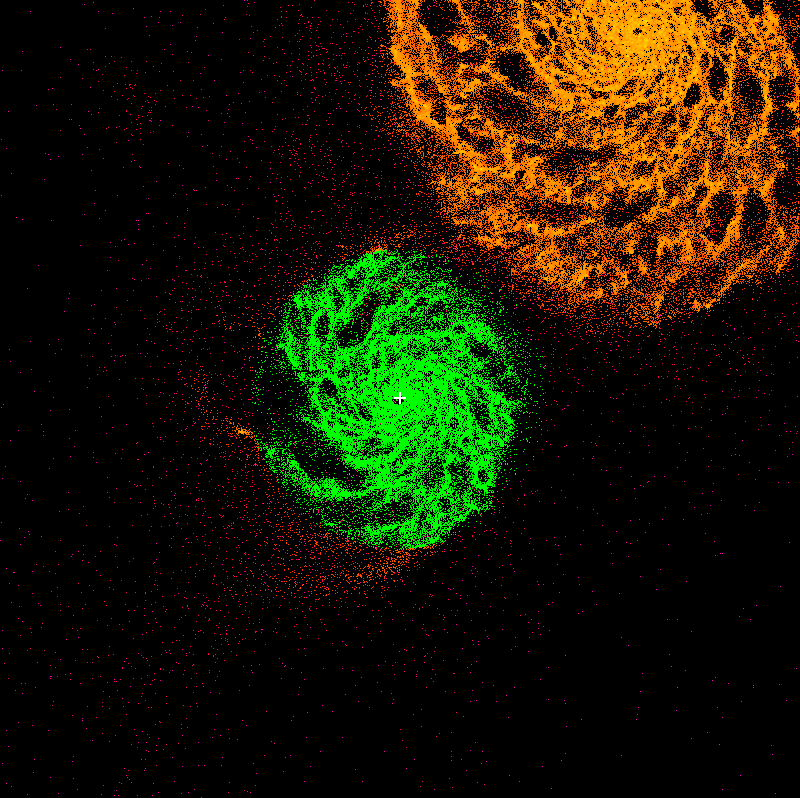}
\end{overpic}
\begin{overpic}[width=0.50\columnwidth,angle=0]{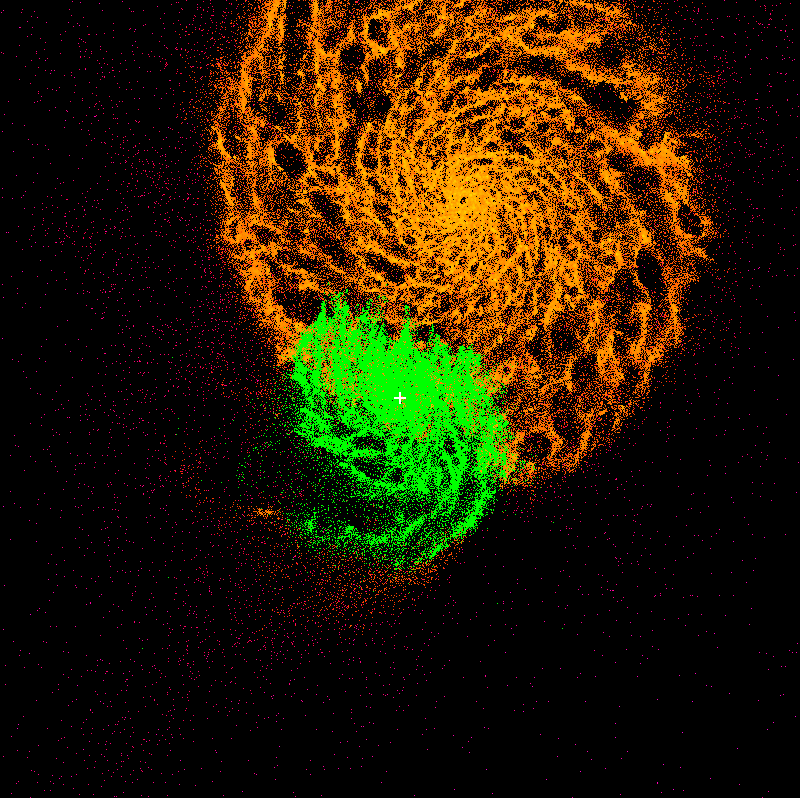}
\end{overpic}
\begin{overpic}[width=0.50\columnwidth,angle=0]{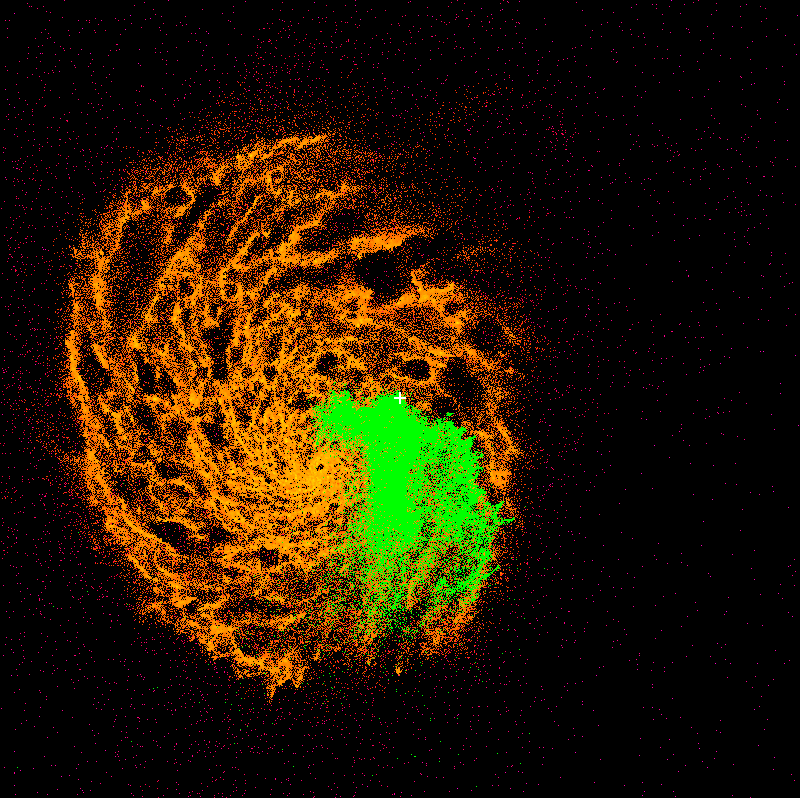}
\end{overpic}
\begin{overpic}[width=0.50\columnwidth,angle=0]{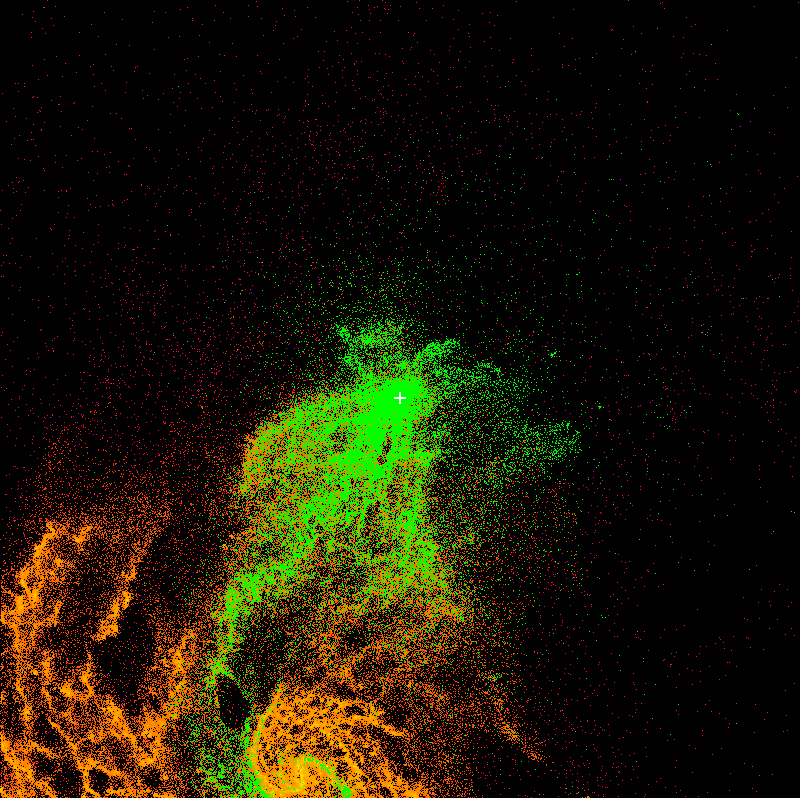}
\end{overpic}
\vskip 0.5mm
\begin{overpic}[width=0.50\columnwidth,angle=0]{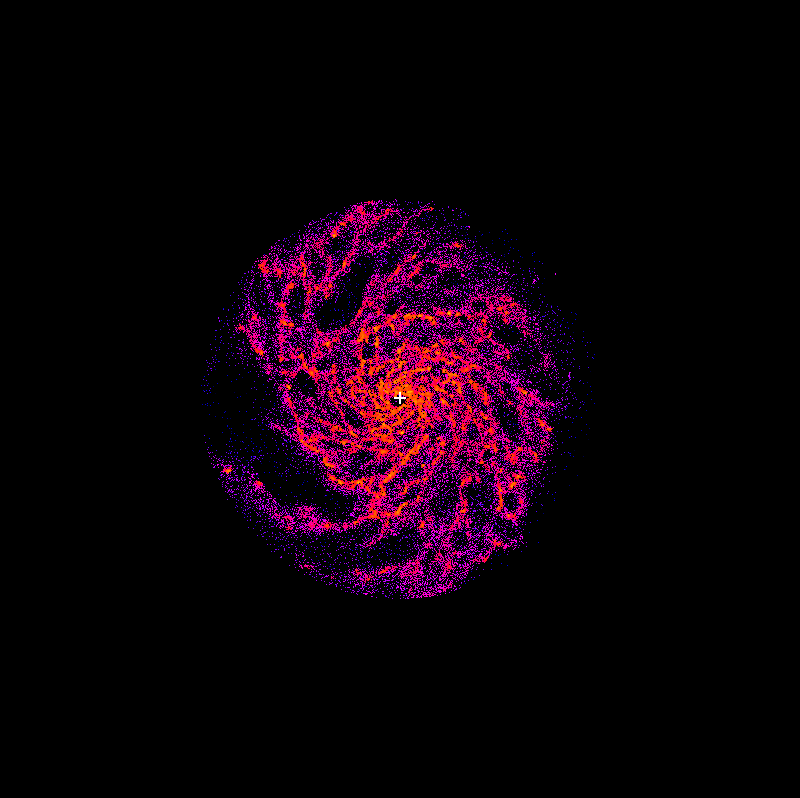}
\end{overpic}
\begin{overpic}[width=0.50\columnwidth,angle=0]{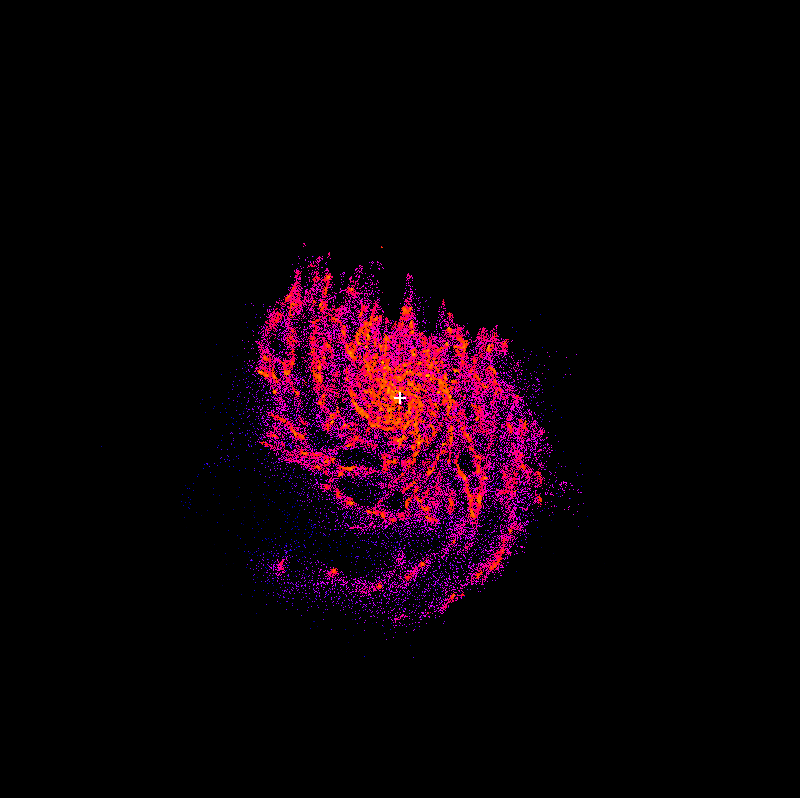}
\end{overpic}
\begin{overpic}[width=0.50\columnwidth,angle=0]{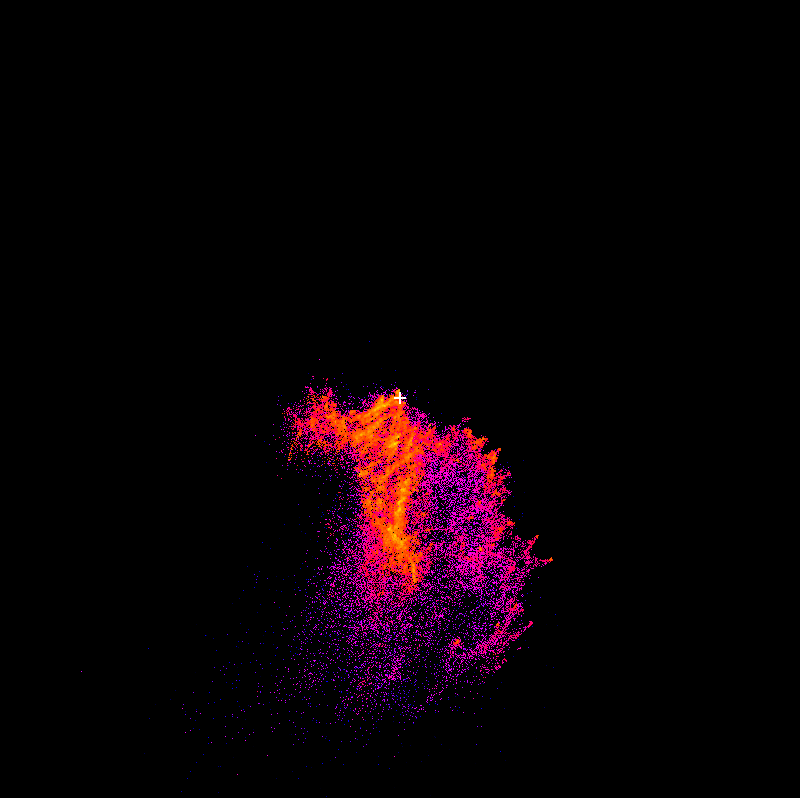}
\end{overpic}
\begin{overpic}[width=0.50\columnwidth,angle=0]{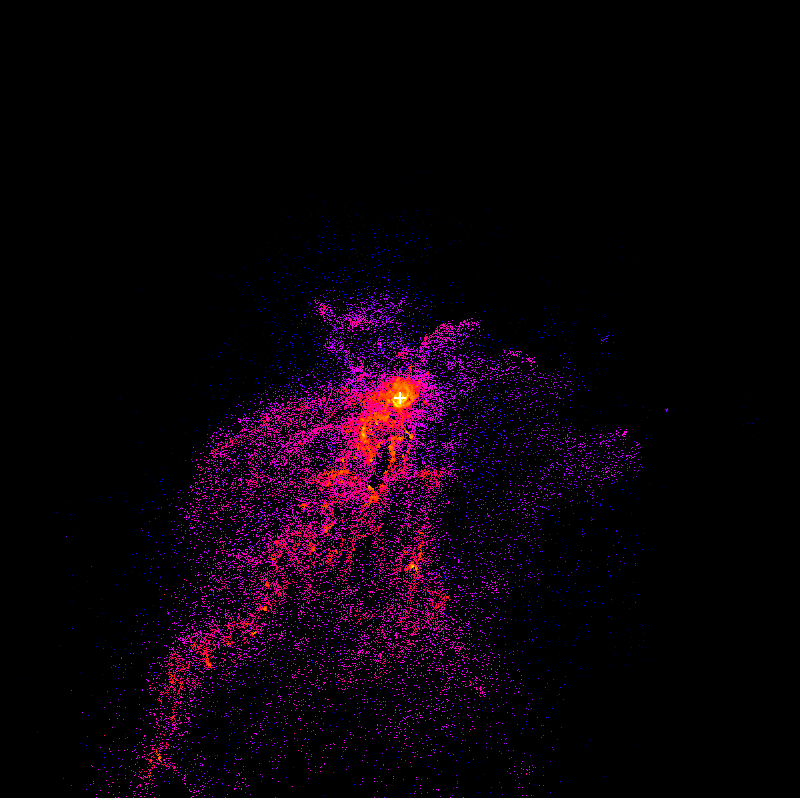}
\end{overpic}
\vskip 0.5mm
\begin{overpic}[width=0.50\columnwidth,angle=0]{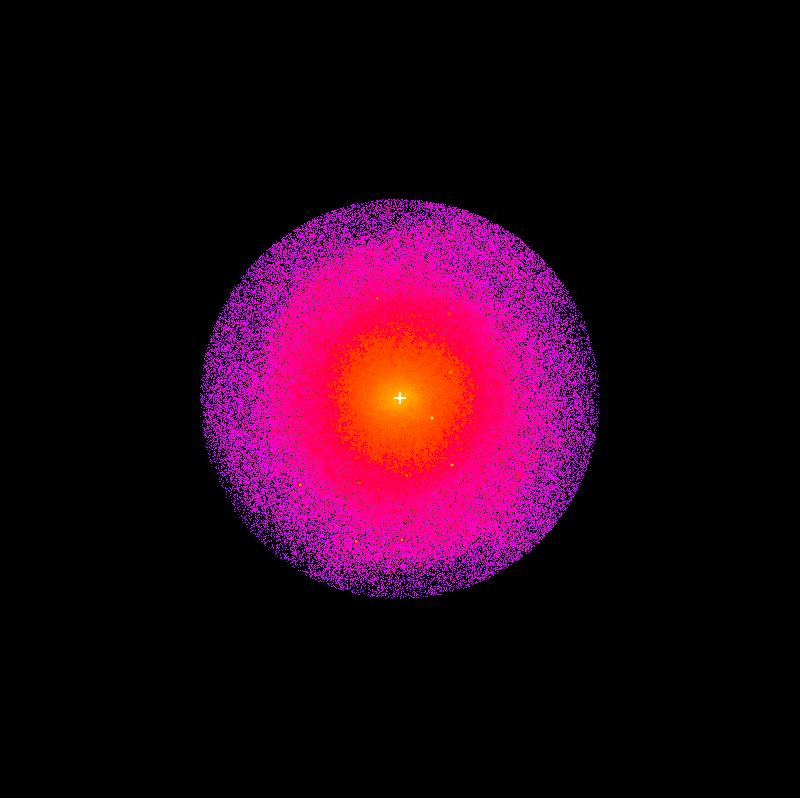}
\end{overpic}
\begin{overpic}[width=0.50\columnwidth,angle=0]{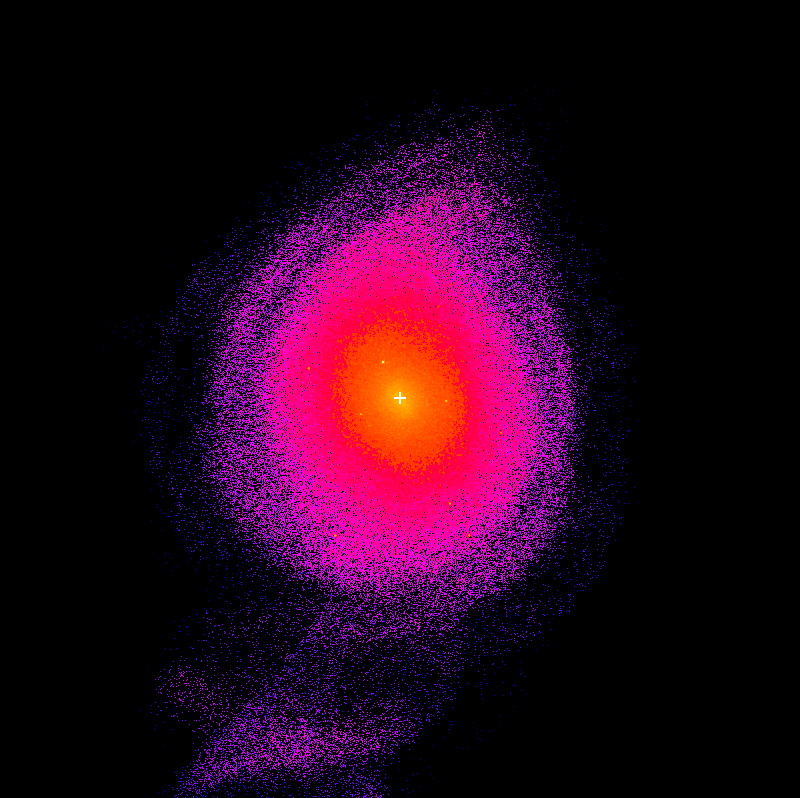}
\end{overpic}
\begin{overpic}[width=0.50\columnwidth,angle=0]{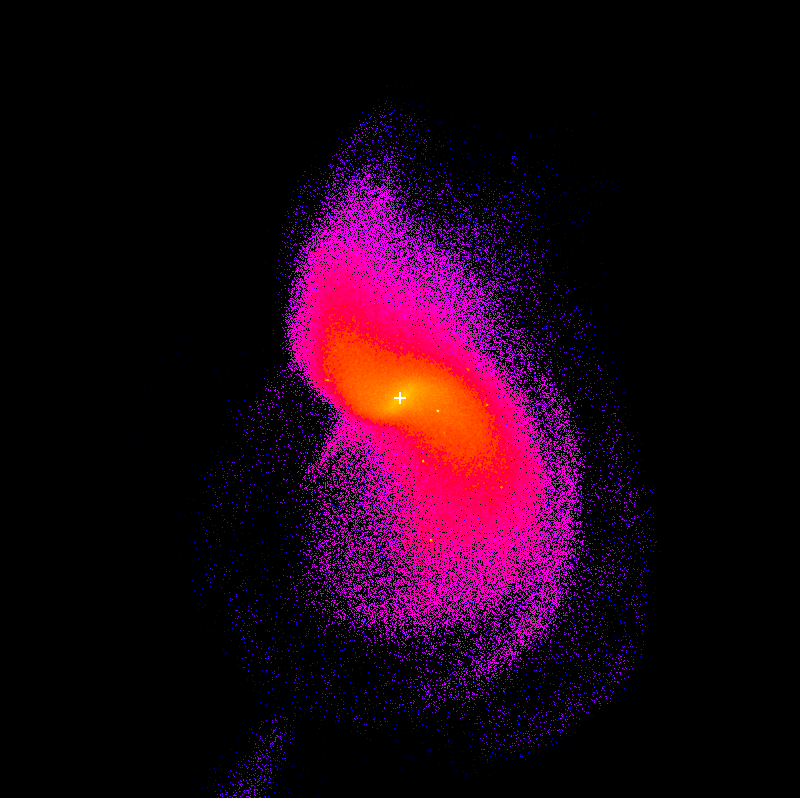}
\end{overpic}
\begin{overpic}[width=0.50\columnwidth,angle=0]{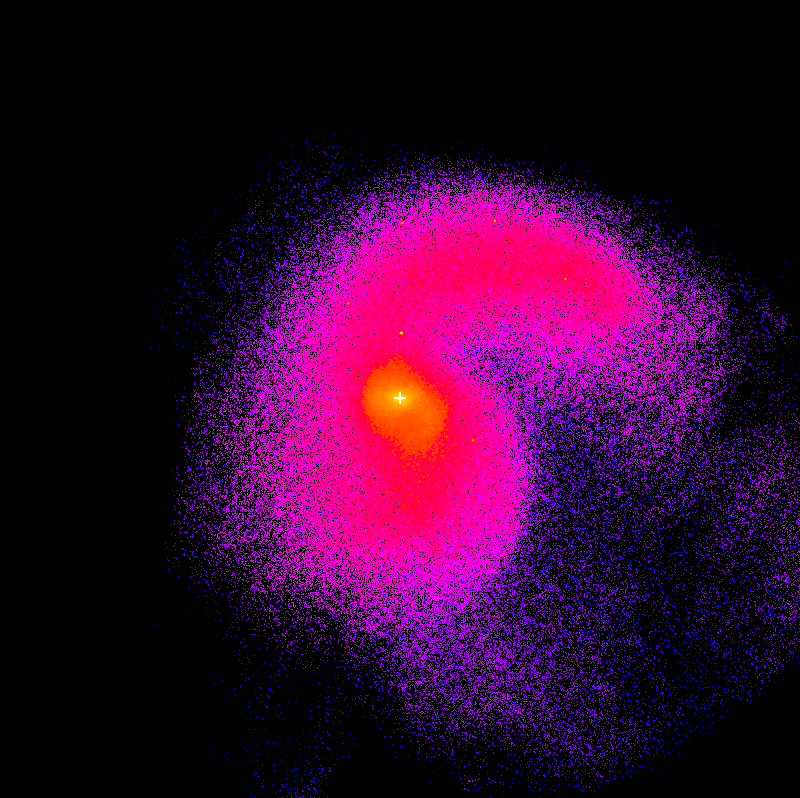}
\end{overpic}
\vskip 0.8mm
\hspace{1.0mm}\begin{overpic}[height=0.6cm,width=2.01\columnwidth,angle=0]{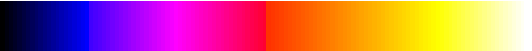}
\put (1,1) {\textcolor{white}{$10^1$}}
\put (23.5,1) {$10^3$}
\put (49,1) {$10^5$}
\put (74.5,1) {$10^7$}
\put (96.5,1) {$10^9$}
\end{overpic}
\vspace{0pt}
\caption[Density snapshots -- Shock]{Detailed evolution of a 1:4 coplanar, prograde--prograde merger (see also Fig.~\ref{fig:1to4_density_snapshots} and Fig.~\ref{fig:1to4_temporal_evolution}) at times close to the second pericentric passage (from left to right: 0.948, 0.963, 0.978, and 1.002~Gyr). Top row: positions of the gas particles of both galaxies, with the gas originally within 3~kpc from the secondary galaxy's centre marked in green. Middle row: density map of the gas originally within 3~kpc from the centre of the secondary. Bottom row: same as the middle row, but for the stars. The colour bar shows the (logarithmic) density scale in units of $2.2 \times 10^5$~M$_{\odot}$~kpc$^{-3}$ for the middle and bottom rows. Adapted from \citet{Capelo_Dotti_2017}.}
\label{fig:1to4_density_snapshots_shock}
\end{figure*}

The dynamics of isolated galaxy mergers have been presented in a plethora of studies, from the early works of, e.g.,  \citet{Holmberg_1941} and \citet{Toomre_Toomre_1972}, to the extensive research in the 1990's \citep[e.g.][]{Barnes_Hernquist_1991,Barnes_1992,Barnes_Hernquist_1996} and 2000's \citep[e.g.][]{Cox_et_al_2008}. 
Only in the past two decades did merger simulations start to include the dynamics and/or growth of massive BHs \citep[e.g.][]{DiMatteo_et_al_2005,Hopkins_et_al_2006,Younger_et_al_2008,Johansson_et_al_2009,Hopkins_Quataert_2010,Callegari_et_al_2009,Callegari_et_al_2011,VanWassenhove_et_al_2012,VanWassenhove_et_al_2014,Hayward_et_al_2014,Colpi2014,Capelo_et_al_2015,Volonteri_et_al_2015a,Volonteri_et_al_2015b,Gabor_et_al_2016,Capelo_et_al_2017,Pfister_et_al_2017,Khan_et_al_2018}. In this subsection, we limit our discussion to the few studies that focused on the theoretical predictions of observational signatures of single or double AGN activity.

\subsubsection{Link between merger dynamics and AGN fuelling}

As an illustrative case, we discuss the dynamical evolution of one of the mergers presented in \citet{Capelo_et_al_2015}, in which the initial mass ratio between the two galaxies as well as between the two BHs is 1:4, the merger is coplanar (i.e. both disk planes lie on the merger orbital plane), and the internal angular momenta of the two disks are both aligned with the merger angular momentum (1:4 coplanar, prograde--prograde merger; Run~07 in the original paper). The simulation starts when the two galaxies are on a parabolic orbit and separated by $\sim$70~kpc, corresponding to the sum of the two galaxy virial radii. Six snapshots highlighting the main phases of the merger are shown in Fig.~\ref{fig:1to4_density_snapshots}: the first panel shows a very early stage, before the first encounter, whereas in the second panel the tidal perturbations onto the two galaxies are clearly visible after the first pericentre. The fourth panel shows the extremely close second pericentre, when the two galaxies actually pass through each other, drastically changing their own morphologies, with the by far larger effect being observable on the gas component of the secondary galaxy (as shown at the second apocentre in the fifth panel). The sixth panel shows the galactic remnant $\sim$1.5~Gyr after the beginning of the run, when the galaxy merger has terminated.

The simulation \citep[as well as the other runs in][]{Capelo_et_al_2015} has a high mass and spatial resolution ($3.3 \times 10^3$ M$_\odot$ and 10~pc for stellar particles, $4.6 \times 10^3$ M$_{\odot}$ and 20~pc for gas) and has been designed to save a full output of the run with an extremely high cadence, allowing for a detailed analysis of the impact of the merger on the stellar and gaseous components of the two systems. Fig.~\ref{fig:1to4_temporal_evolution} presents a direct comparison between the BH separation (upper panels), the evolution of the BH mass, and the dynamical properties of the gas in the primary (left-hand panels) and secondary (right-hand panels) host galaxies. The accretion rates onto both the primary and secondary BH (second row of panels) evolve from an initial stage of relatively low values with small fluctuations \citep[{\it stochastic phase}, as dubbed in][]{Capelo_et_al_2015} to a second  bursting phase ({\it merger phase}), starting immediately after the huge perturbation occurring at the second pericentre (see Fig.~\ref{fig:1to4_density_snapshots}) and lasting until the merger completion. The accretion burst is correlated with strong variations in the gas angular momentum (see the third row of panels in Fig.~\ref{fig:1to4_temporal_evolution}). This is especially true for the secondary galaxy, where a clear angular momentum loss is observable along with a 90-degree re-orientation of the gas angular momentum. The end of the merger phase shows a second orbital flip in the secondary, after which the run enters the {\it remnant phase}, in which the separation between the two BHs is poorly resolved and the two objects accrete from the same gas reservoir.

The main physical process responsible for the violent evolution of the gas angular momentum and, as a consequence, of the burst of accretion 
onto the secondary BH is ram-pressure, as discussed in \citeauthor{Capelo_Dotti_2017} (\citeyear{Capelo_Dotti_2017}; see also \citealt{Barnes_2002} and \citealt{Blumenthal_Barnes_2018} for similar larger-scale results) and shown in Fig.~\ref{fig:1to4_density_snapshots_shock}. The four columns refer to four different times centred around the beginning of the merger phase, from immediately before to immediately after the second pericentre. It can be noted that the distribution of stars belonging to the secondary galaxy is tidally perturbed, being elongated in the radial direction toward the primary galaxy, and develops clear tidal tails after the interaction (bottom row). The evolution of the gas component is, however, very different, as shown in the top and middle rows. A strong hydrodynamical interaction is observable as soon as the two gas components interact, creating a clear shock front in the secondary gas and braking it (second column). Such deceleration of the gas (with respect to the stellar component of its host) is even more visible at the pericentre, and results in the gas being dynamically decoupled from the stars. A fraction of the gas unbinds from the secondary galaxy and gets accreted by the primary galaxy (upper-rightmost panel), but the majority of the gas remains bound to the secondary, with a significantly smaller angular momentum, forming a dense nuclear disk that is counter-rotating with respect to the stars.

\begin{figure}[!t]
\centering
\vspace{4.0pt}
\includegraphics[width=0.98\columnwidth,angle=0]{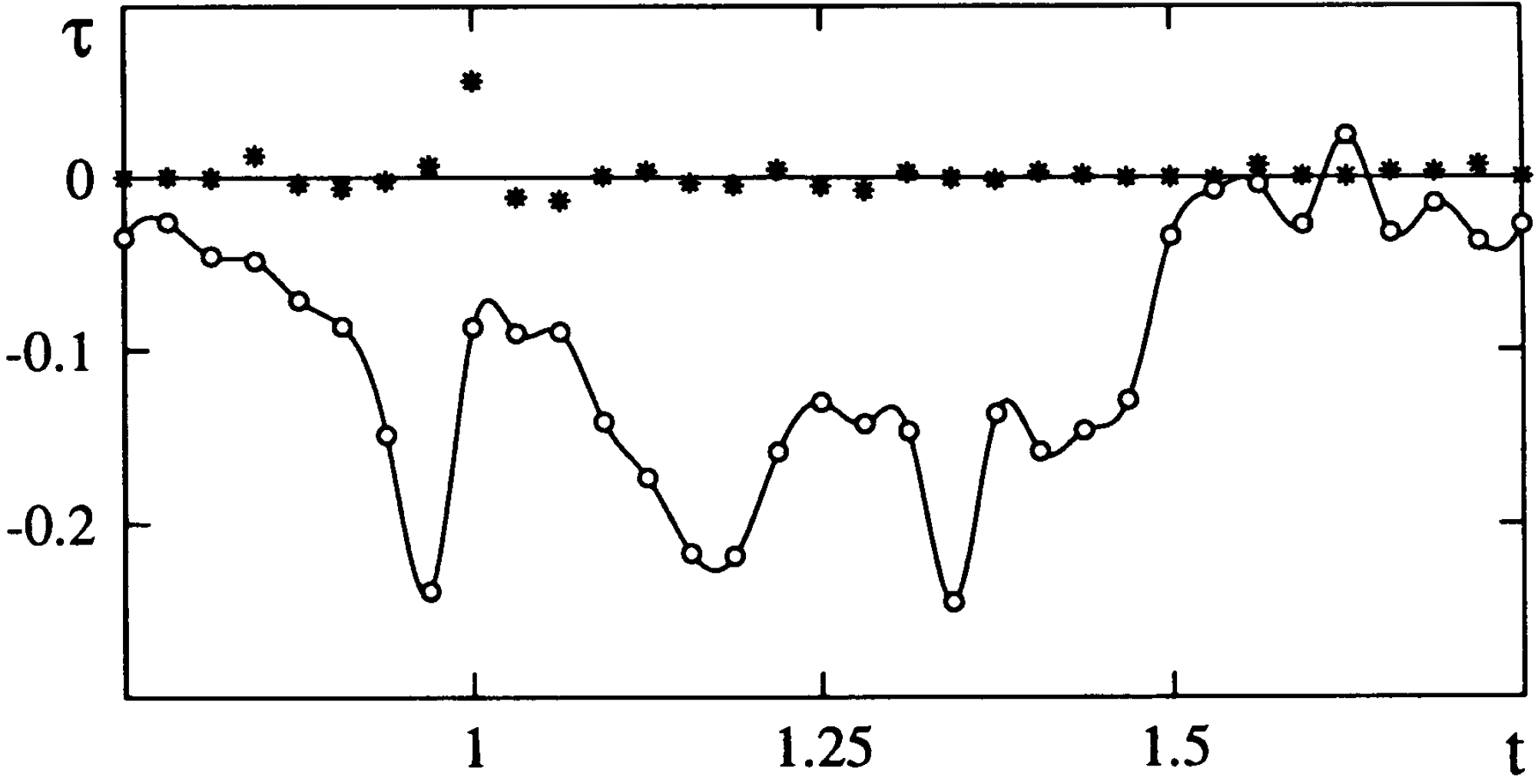}
\vspace{-0.0pt}
\caption[]{Specific torques acting on the gas at the centre of the face-on disk in a gas-rich equal-mass merger between two disk galaxies, as a function of time (with the units of length, mass, and time being roughly 40~kpc, $2.2 \times 10^{11}$~M$_\odot$, and 250~Myr, respectively). In this relatively low-resolution simulation, the gravitational torques exerted on the gas by the rest of the system (open circles) are much stronger than those due to hydrodynamic forces (star symbols), as opposed to the high-resolution case described in Figs~\ref{fig:1to4_density_snapshots}--\ref{fig:1to4_density_snapshots_shock}. From  \citet{Barnes_Hernquist_1991}.}
\label{fig:Barnes_Hernquist_1991_Fig2}
\end{figure}

As discussed in \citet{Capelo_Dotti_2017}, such strong perturbation depends on the dynamical properties of the encounter, and a complete flip of the gas orbital plane is not expected to be ubiquitous. In addition to this process, the above-mentioned tidal torques can lead to deviations from axisymmetry in the stellar components, further supporting the gas inflow \citep[e.g.][]{Hernquist_1989,Barnes_Hernquist_1991,Barnes_Hernquist_1996,Mihos_Hernquist_1996}. 
These early works, in fact, showed that the role of hydrodynamical torques is negligible (see, e.g.,  Fig.~\ref{fig:Barnes_Hernquist_1991_Fig2}), possibly due to a combination of lower resolution and the less accurate treatment of hydrodynamics in earlier versions of the smoothed particle hydrodynamic (SPH) methods. The different relative importance of ram pressure with respect to tidal torques in driving nuclear activity will depend on the different galaxy mass ratios, impact parameters, etc., and a broad parametric study aiming at  quantifying the relative impact of the two processes has not been presented to date.

\subsubsection{Dual AGN properties and occurrence}

\begin{figure*}[!t]
\centering
\vspace{-6.0pt}
\includegraphics[width=0.99\columnwidth,angle=0]{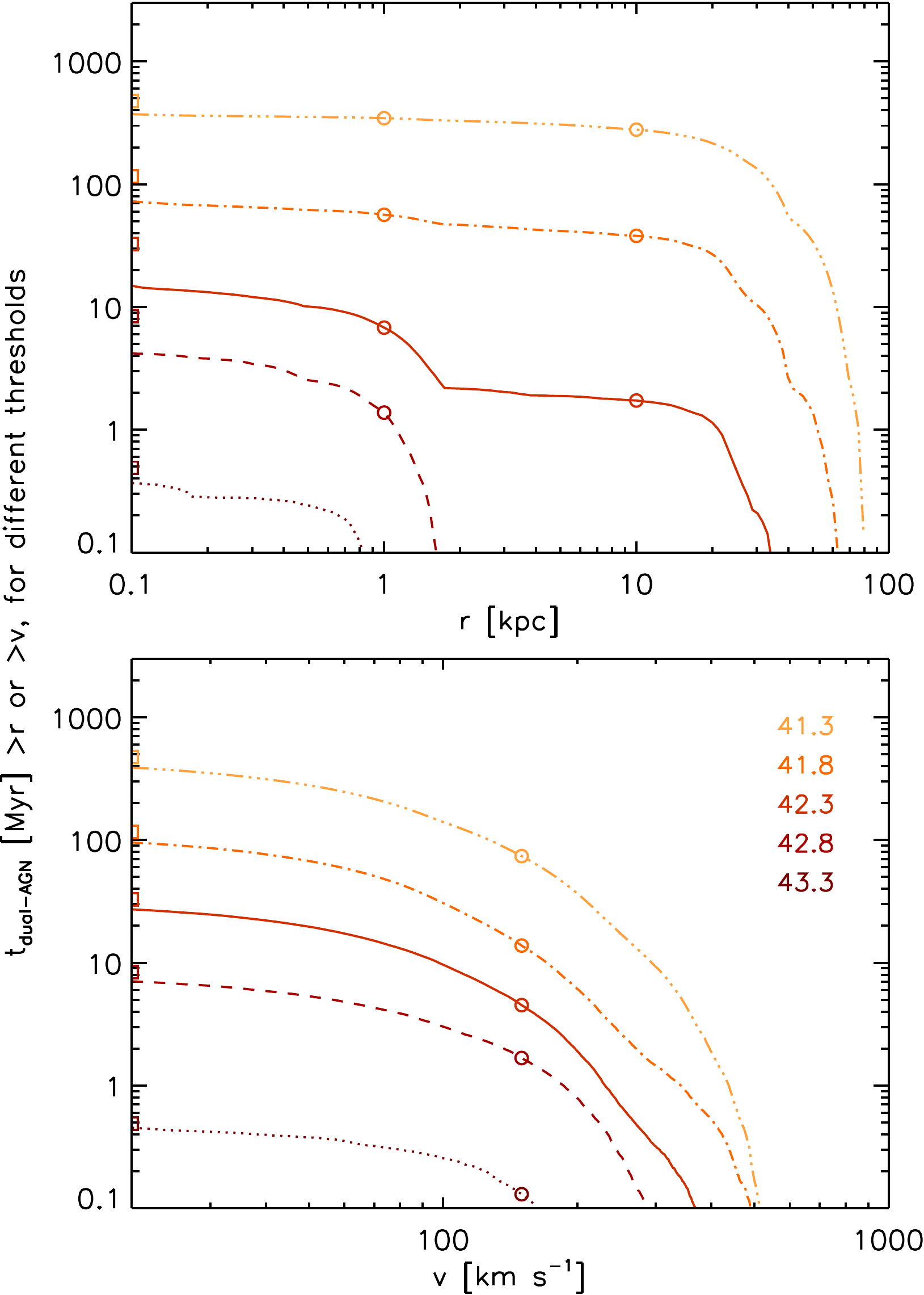}
\includegraphics[width=1.05\columnwidth,angle=0]{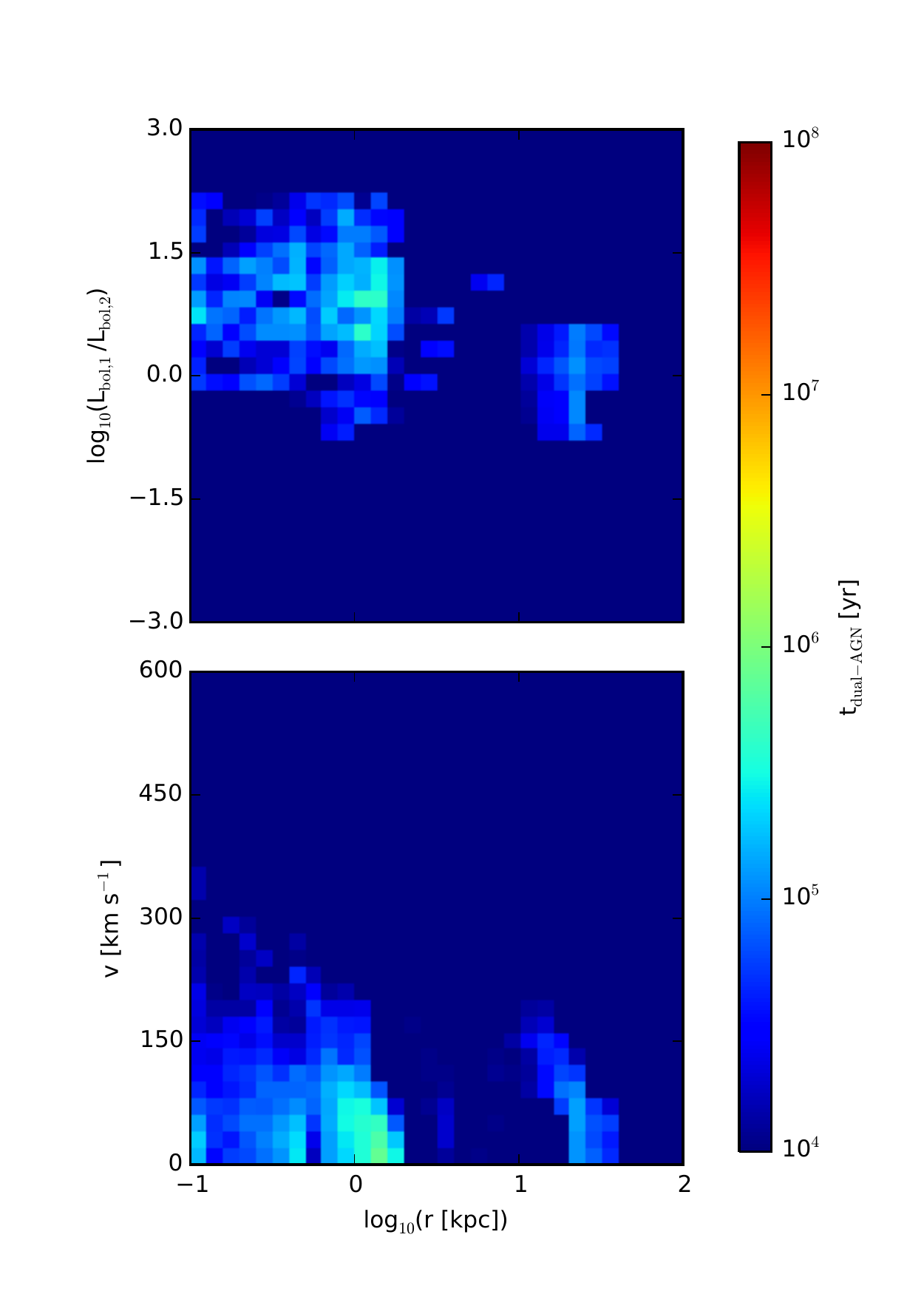}
\vspace{-0.0pt}
\caption[]{Dual-activity time for a 1:2 coplanar, prograde--prograde merger (first described in \citealt{Capelo_et_al_2015}; Run~02). Left-hand panels: time above a given projected separation $r$ (top panel) and velocity difference $v$ (bottom panel) between the two BHs, for several 2--10~keV luminosity thresholds [$10^{41.3}$ (dash-triple-dot), $10^{41.8}$ (dash-dot), $10^{42.3}$ (solid), $10^{42.8}$ (dash), and $10^{43.3}$ (dot) erg~s$^{-1}$]. The squares indicate the time regardless of any $r$ or $v$ filter (i.e. $r = v = 0$), whereas the circles indicate $r = 1$ and 10~kpc, and 150~km~s$^{-1}$. Right-hand panels: time as a function of projected separation and velocity difference (bottom panel), and bolometric-luminosity ratio between the two BHs (top panel), for the luminosity threshold $L_{2-10\, {\rm keV}} = 10^{42.3}$~erg~s$^{-1}$. Each side is divided in 30 bins ($\Delta \log_{10}(r\, {\rm [kpc]}) = 0.1$, $\Delta v = 20$~km~s$^{-1}$, and $\Delta \log_{10} (L_{\rm bol,1}/L_{\rm bol,2}) = 0.2$). Adapted from \citet{Capelo_et_al_2017}.
}
\label{fig:1to2_dual_activity}
\end{figure*}

\begin{figure}[!t]
\centering
\vspace{4.0pt}
\includegraphics[width=0.98\columnwidth,angle=0]{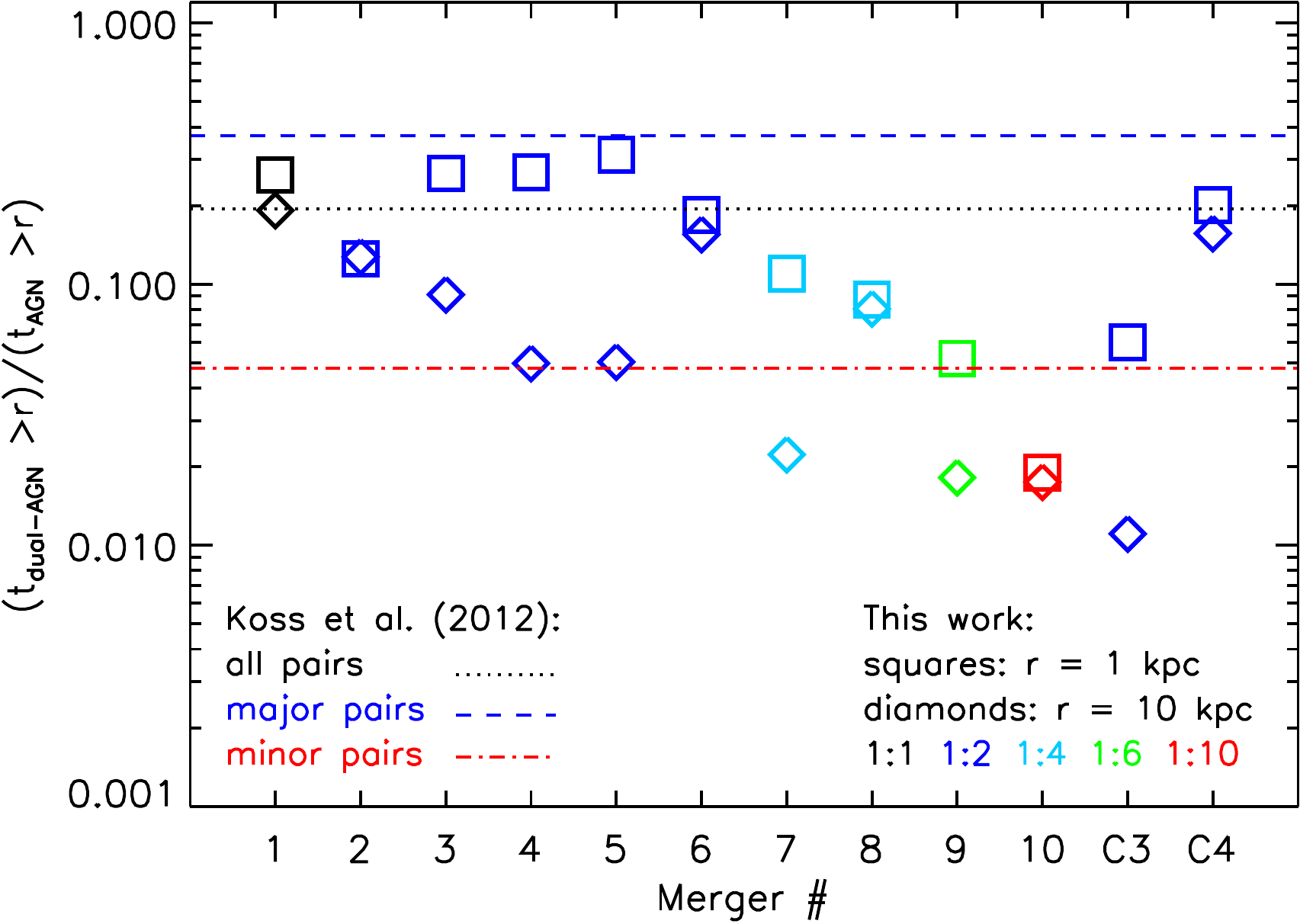}
\vspace{-0.0pt}
\caption[]{Dual-activity time above a given projected separation (1~kpc -- squares -- and 10~kpc -- diamonds) divided by the activity time above the same threshold, assuming a bolometric luminosity threshold of $10^{43}$~erg~s$^{-1}$, for all the mergers in the suite of \citet{Capelo_et_al_2017}. The merger number in the $x$-axis is the same as in Tables~1--2 of \citet{Capelo_et_al_2017}. The colours refer to the initial mass ratio: 1:1 -- black, 1:2 -- blue, 1:4 -- cyan, 1:6 -- green, and 1:10 -- red. The horizontal lines show the observed fraction of dual AGN systems out of interacting systems (in which at least one system has an AGN) by \citet{Koss_et_al_2012}, for their full sample of \swift-BAT AGN pairs (black, dot), their major pairs (blue, dash), and their minor pairs (red, dash-dot) with projected separation 1--100~kpc. Major mergers with 0.1 per cent BH feedback efficiency (squares of Runs~01--06, and C4) should be compared to the blue, dashed line, whereas minor mergers (squares of Runs~07--10) should be compared to the red, dash-dotted line. Adapted from \citet{Capelo_et_al_2017}.
}
\label{fig:dual_activity_koss}
\end{figure}

Since merger-driven tidal torques and ram-pressure shocks efficiently funnel gas into the central regions of the merging galaxies, where the BHs lie, we expect some level of dual-AGN activity associated with galaxy mergers. This, however, is difficult to quantify, given the extreme complexity of such systems, wherein one needs to take into account many physical processes (e.g., DM, stellar, and gas dynamics; star formation and stellar feedback; radiative cooling; BH accretion and feedback; etc.). High-resolution numerical simulations are therefore needed to study this process, in order to understand how, when, and why and to predict when two BHs may accrete at the same time. In this respect, suites of high-resolution isolated merger simulations are particularly useful, as one can appreciate the importance of any given BH, merger, or galactic property by varying one parameter at the time.

Different numerical investigations have been performed to test the effects of galaxy morphology \citep[disk versus elliptical; e.g.][]{VanWassenhove_et_al_2012}, bulge-to-disk stellar mass ratio \citep{Blecha_et_al_2013, Blecha_et_al_2017}, orbital configuration, BH mass, and BH feedback efficiency \citep{Capelo_et_al_2017}, and primary-to-secondary-galaxy and gas-to-stellar mass ratios (all of the above references). They all agree on dual AGN activity being more efficiently promoted during the last stages of galaxy mergers, when the two BHs are separated by less than 1--10~kpc. This last stage is, however, shorter than the initial galaxy pairing phase, when the average BH luminosities are lower and most often not simultaneous. To compute the chances of observing an AGN pair, one has to consider all of the above-mentioned effects. Moreover, these chances strongly depend on the observational window that is used to search for AGN pairs, as detailed in the following.

In order to have a more meaningful comparison with observations, \citet{Capelo_et_al_2017} used the results of their set of simulations to predict hard X-ray luminosities. This observational band is way less affected than the optical one by obscuration \citep[e.g.][]{Koss_et_al_2010}. Indeed, \citet{Capelo_et_al_2017} checked for the effect of obscuration on resolvable scales ($>$100~pc, i.e. neglecting any effects from nuclear tori), and found it to be negligible for high-redshift ($z = 3$) galaxies and moderate for local galaxies, producing a change in the dual-activity time-scales by a factor of $\sim$2. They also computed projected (rather than 3D) BH separations and relative velocities, to make the comparison with observations more direct.\footnote{The usage of projected quantities is crucial, as dual-AGN time-scales can differ by factors of up to $\sim$4 at a given separation/relative velocity.}

The link between simulated merger history and dual-AGN activity is shown in Fig.~\ref{fig:1to2_dual_activity} for one major merger of the suite of \citet{Capelo_et_al_2017} -- the 1:2 coplanar, prograde--prograde merger with 30 per cent disk gas fraction and 0.1 per cent BH feedback efficiency (Run~02 in the original paper). In the left-hand panels, we show the time the two BHs spend above a given projected separation (top panel) and velocity difference (bottom panel), assuming several 2--10~keV luminosity thresholds, from $2 \times 10^{41}$ to $2 \times 10^{43}$~erg~s$^{-1}$ (from top to bottom). In the low-luminosity cases, the BHs are both active for a significant fraction of the encounter and, as a consequence, the curves simply follow the orbital history of the BHs. In the high-luminosity cases, dual activity occurs only during the merger stage, at low (projected) separations. The solid line corresponds to the case when $L_{\rm 2-10\, keV} = 10^{42.3}$~erg~s$^{-1}$, which is typically used as the threshold to define an AGN \citep[][]{Silverman_et_al_2011}. This luminosity case is further exemplified in the right-hand panels of Fig.~\ref{fig:1to2_dual_activity}, where the dual-AGN time is shown as a function of luminosity ratio, projected separation, and projected velocity difference, and in which the higher-density time-clouds are related to the apocentric passages, where the residence time is longer.

We stress that the quoted  relative velocities in \citet{Capelo_et_al_2017} should be considered as a broad estimate more than an actual prediction of the velocity shift measurable through optical spectroscopy, as these were obtained by projecting the 3D relative velocity between the two BHs along the line of sight. Such assumption does not take into account the size and dynamics of the NLRs. These effects were considered by \citet{Blecha_et_al_2013}  through a post-processing analysis of a merger simulation suite, who found that only a minority of double-peaked NLR were directly linked to the relative motion of massive BHs.  

Simulations of isolated galaxy mergers (see also recent studies by \citealt{Solanes_et_al_2019} and \citealt{Yang_et_al_2019}) cannot provide any prediction on the fraction of AGN pairs out of the total number of AGN (for that, large-scale cosmological simulations are needed; see Sect.~\ref{sssec:BH_Pairs_Theory_Cosmological_Simulations}). However, it is possible to compute the dual-activity time, normalised by the activity time (defined as the time when at least one BH is active), and compare this quantity to observed fractions of dual AGN systems out of interacting systems in which at least one system is known to host an AGN. This is shown in Fig.~\ref{fig:dual_activity_koss}, where such normalised dual-activity time for all the mergers in the suite of \citet{Capelo_et_al_2017} is compared to data from the all-sky \swift-BAT survey \citep[][]{Koss_et_al_2012}. Remarkably, both major and minor simulated mergers are consistent (to within a factor of $\sim$2 in time) with the observations.

Moreover, it is also possible to compute the dual-activity time, divided by the total merger time, and compare it to cosmological simulations which provide the fraction of dual AGN out of the total number of BH pairs. In this case, when comparing, for example, major-merger dual-activity times of isolated mergers \citep[][]{Capelo_et_al_2017} to the cosmological run by \citeauthor{Steinborn_et_al_2016} (\citeyear{Steinborn_et_al_2016}; see Sect.~\ref{sssec:BH_Pairs_Theory_Cosmological_Simulations}), they are within a factor of $\sim$2 from each other.

Although less direct than the X-ray based selections, infrared studies  of dual AGN activity allow to build significantly larger samples, as current X-ray surveys are mostly comparatively shallower. However, mid-IR bands can be contaminated by star formation, often enhanced in mergers, and particular care has to be taken in characterizing the IR (dual-)AGN selection criteria \citep{2016ApJ...832..119H}.  

\citet{Blecha_et_al_2017} performed eight hydrodynamic simulations of isolated mergers with resolution and gas fraction comparable to the ones discussed above, and a mass ratio of 1:2 (as the case shown in Fig.~\ref{fig:1to2_dual_activity}), except for two runs exploring the extreme cases 1:1 and 1:5, varying the bulge-to-total stellar mass ratio in the range 0--0.2 (slightly lower with respect to the 0.22 mass ratio in the two runs mentioned above). They coupled the galaxy evolution to dust radiative transfer in post processing, to quantify the mid-IR observability of (single and) dual AGN. The post-processing analysis misses the contribution of structures on unresolved scales \citep[similarly to the case in ][]{Capelo_et_al_2017}, but we stress that the input SED for the AGN they used includes a mid-IR contribution by a sub-resolution dusty torus.

\begin{figure}[!t]
\centering
\vspace{-10pt}
\begin{overpic}[width=1.04\columnwidth,angle=0,trim={0 15.0cm 0 0},clip]{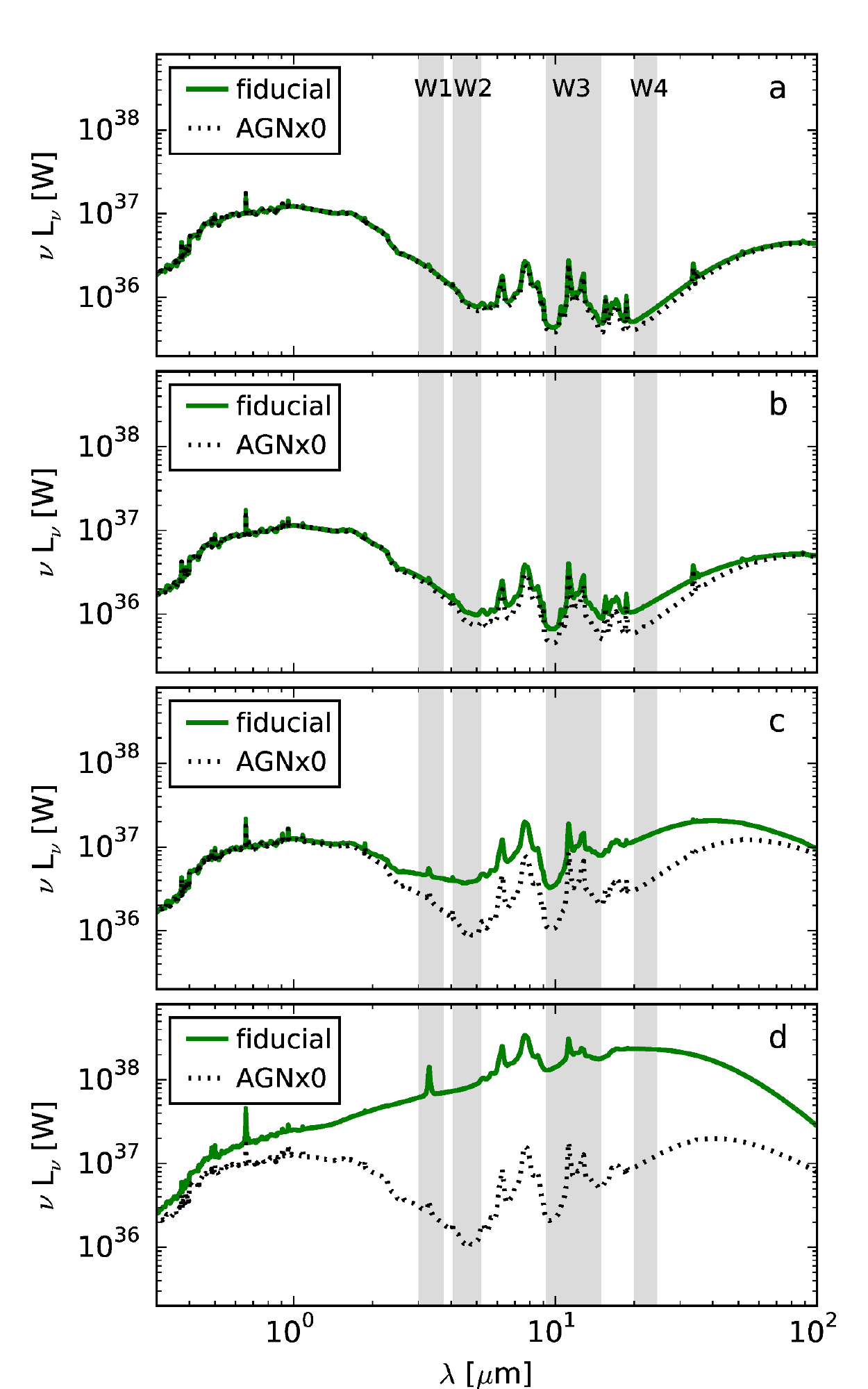}
\put (88,30.49) {\textcolor{white}{\LARGE{$\CIRCLE$}}}
\put (87,30.49) {\textcolor{white}{\LARGE{$\CIRCLE$}}}
\end{overpic}
\begin{overpic}[width=1.04\columnwidth,angle=0,trim={0 0 0 14.6cm},clip]{figures_sect2/fig2new.pdf}
\put (88,39.41) {\textcolor{white}{\LARGE{$\CIRCLE$}}}
\put (87,39.41) {\textcolor{white}{\LARGE{$\CIRCLE$}}}
\end{overpic}
\vspace{0pt}
\caption[]{Optical to IR SED of a 1:2 gas-rich merger in \citet{Blecha_et_al_2017}, the first time at which the separation between the two BHs drops below 10~kpc (upper panel), and immediately after the final pairing of the two BHs (lower panel). The solid line refers to the fiducial analysis, whereas the dotted line shows the SED obtained neglecting the input from any AGN activity. The shaded areas indicate the four WISE bands. Adapted from \citet{Blecha_et_al_2017}.}
\label{Blecha_SED}
\end{figure}

Figure~\ref{Blecha_SED} shows the optical to IR SEDs of a major (mass ratio 1:2), gas-rich ($M_{\rm gas}/M_{\star, \rm disc}$ = 0.3, for both galaxies) merger involving two massive bulge-less galaxies \citep[Run~A1A0 in][]{Blecha_et_al_2017}, for the cases in which the contribution from the AGN SED was included (solid line) or not (dotted line). The emission in the upper panel (corresponding to the first time at which the separation between the two BHs drops below 10~kpc) is clearly dominated by star formation, and the AGN contribution is negligible. A clear increase in the IR emission due to the AGN contribution and the reddening of the mid-IR colours are evident immediately after the final pairing of the BHs (lower panel). We stress that our discussion applies to low-redshift ($z<0.5$) mergers, where the optical/near-IR peak of star formation has a limited impact on the W1 and W2 WISE bands (see Sect.~\ref{sssec:mid_IR_AGN_pairs} and \citealt{Blecha_et_al_2017} for a discussion about the limitations of the IR selection at higher redshifts).

\begin{figure}[!t]
\centering
\vspace{0.0pt}
\includegraphics[width=1.03\columnwidth,angle=0]{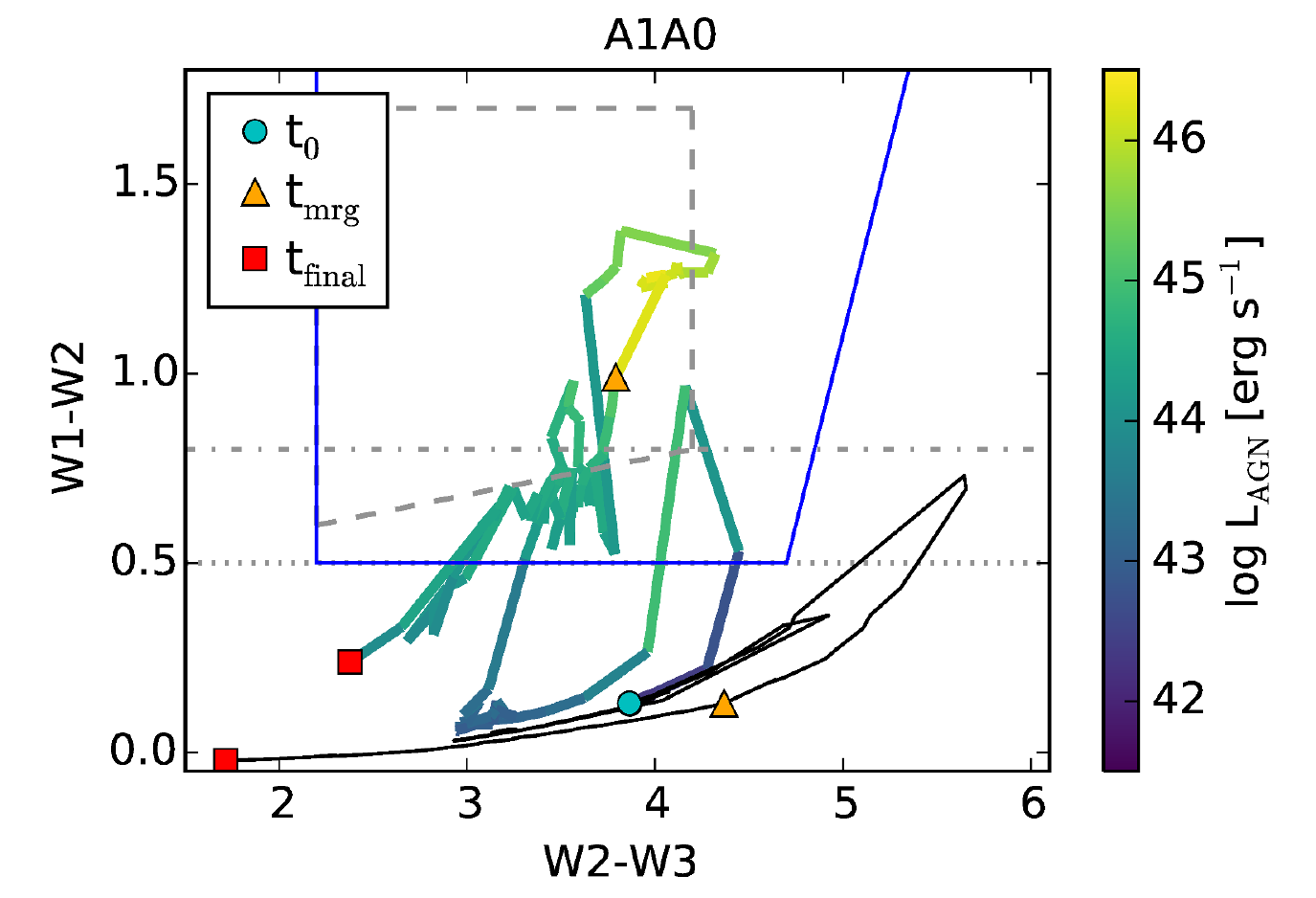}
\vspace{-0.0pt}
\caption[]{Time evolution of the same merger shown in Fig.~\ref{Blecha_SED} in the WISE colour-colour space. The thick coloured line refers to the fiducial analysis, whereas the thinner black line refers to the analysis not including any AGN component. The circles, triangles, and squares pinpoint the beginning of the run, the time of BH merger, and the end of the run, respectively. The horizontal lines refer to the commonly used thresholds for AGN selection (W1-W2~$>0.5$ and W1-W2~$>0.8$). The gray dashed line refers to the two-colour selection criterion proposed in \citet{Jarrett_et_al_2011}, whereas the blue solid line indicates the selection criterion proposed by \citet{Blecha_et_al_2017}. Adapted from \citet{Blecha_et_al_2017}.
}
\label{Blecha_colour_evolution}
\end{figure}

Blecha and collaborators used their results to provide an a-priori estimate of the completeness and purity of AGN samples selected with different IR colour criteria. As shown in Fig.~\ref{Blecha_colour_evolution} for the run discussed above, a single-colour criterion either misses a significant fraction of AGN (when the W1-W2 threshold is too high), or it is contaminated by star formation (for lower W1-W2 thresholds). A two-colour selection criterion such as the one proposed by \citet{Jarrett_et_al_2011} performs better, but does not maximise completeness and purity at the same time. The results of the different runs have then been used to engineer a new simulation-informed criterion, shown with the blue solid line in Fig.~\ref{Blecha_colour_evolution}.  

The analysis performed by \citet{Blecha_et_al_2017} suggests that mid-IR selection can be very effective at identifying AGN hosts in mergers (see also Sect.~\ref{sssec:mid_IR_AGN_pairs}). Using the same simulations, they estimated the expected fraction of dual AGN in each merger stage, based on the fraction of time at each separation when both BHs are simultaneously active, and found that up to $\sim$80 per cent of the mergers with a projected separation $<$3~kpc could host a pair of accreting BHs, making the mid-IR-based samples of dual-AGN candidates the ideal ballpark for X-ray follow-ups and the perfect targets for future sub-kpc imaging and spectroscopy with the {\it JWST} (see Sect.~\ref{sssec:future_mir} for more details on future mid-IR observations).

In conclusion, we are still far from a detailed picture of the connection between mergers and (dual) AGN activity. AGN obscuration and variability on scales shorter and smaller than the typical temporal and spatial resolution of idealised merger simulations cannot be taken into account. This, together with the fact that dust obscuration from the torus in the vicinity of the BHs is also not currently resolved, likely leads to an over-estimate of the X-ray dual-AGN time, possibly resulting in an uncertain one-to-one association between mid-IR-selected double AGN and detected X-ray pairs. Moreover, these results may be sensitive to which numerical method is employed \citep[e.g.][]{Hayward_et_al_2014,Gabor_et_al_2016} and to the specific implementation of BH dynamics \citep[e.g.][]{Lupi_et_al_2015a,Lupi_et_al_2015b,Biernacki_et_al_2017}, accretion \citep[e.g.][]{Debuhr_et_al_2010,Hopkins_Quataert_2011}, and feedback \citep[e.g.][]{Newton_Kay_2013}. Finer resolution and improved subgrid recipes will be helpful to gain a better understanding of these complex systems.

\subsubsection{Stalling BH pairs as tight dual AGN}\label{ssec:stallingBH}

The above mentioned studies \citep[e.g.][]{Capelo_et_al_2017,Blecha_et_al_2017} have performed a detailed analysis of only one particular type of  merger (with gas rich, though non-clumpy, massive disk galaxies). Even  though a systematic study expanding upon this region of parameters has  not been done, it is interesting to note that the dynamics and possible  detection of BHs can vary significantly if one studies different types  of galaxies, e.g., if there is little (or much more) gas, or if the structure  of the galaxy is significantly different.

In particular, some processes can produce BH pair stalling at sub-kpc separations, and such stalling BH pairs could be detectable as {\it tight dual AGN}. Such processes are more numerous in gas-rich environments \citep[e.g.][]{Fiacconi_et_al_2013,Tamburello_et_al_2017a, SouzaLima_et_al_2017}, contrary to previous expectations that gas-rich media would be more conducive to fast decay and binary formation \citep[e.g.][]{Mayer:2013:MBHBGasRev}. Additionally, low-mass/dwarf galaxies offer a special environment where slow orbit decaying or even stalling can occur, albeit for reasons different from those playing a role in normal galaxies. Below we list some of the regimes that are relevant. In each of these regimes, a quantity of interest, which has clear observational implications, is the {\it characteristic  residency time-scale} in the stalling phase, namely the typical time-scale the two BHs will spend at a separation of 10--100~pc. In all these regimes, a {\it stalled BH pair} would arise when the final separation of the two BHs is too large for a bound binary to form for a wide range of BH masses. Note that a long characteristic residency time, for example significantly longer than the orbital time, implies a higher probability to observe a dual AGN source, if both BHs are accreting, and/or an offset AGN, if only one is accreting.

\bigskip

{\it (a) Inefficient SMBH decay in minor mergers}. If the secondary galaxy is tidally disrupted far from the centre of the primary, it can deliver a stalling BH at a few hundred pc separations that will hardly decay efficiently afterwards \citep[e.g.][]{Callegari_et_al_2009, Capelo_et_al_2015}. The observability of dual AGN from tens of pc to $>$1~kpc separations in major as well as minor mergers (down to 1:10 mass ratios) has been studied thoroughly in \citet{Capelo_et_al_2017}. The general trend, at all separations, was found to be that the dual activity time is much shorter in 1:10 mergers than in nearly equal mass mergers, by more than an order of magnitude (see Fig.~\ref{fig:dual_activity_koss} and figures~6--7 of the original paper). We note that, for separations above 1~kpc, the results of \citet{Capelo_et_al_2017} are in agreement with the observational findings of \citet{Koss_et_al_2012}, that the fraction of dual AGN increases with decreasing separation and increasing mass ratio. Still, even at the smallest separations considered (20--40~pc), not yet probed in observational samples, the dual activity time of 1:10 mergers is always the smallest. The fraction of dual AGN in the latter minor mergers is thus predicted to be down to a few per cent relative to a single AGN sample. Whether this is true also for even smaller separations and/or in the other regimes discussed here is yet to be explored. When interpreting the mapping between galaxy mass ratio in the merger and BH mass ratio, we recall that, despite low dual activity time, one BH can accrete proportionally more or less than the other, depending on the initial mass ratio, resulting in an increase (decrease) of the BH mass ratio in minor (major) mergers \citep[][]{Capelo_et_al_2015}.

\bigskip

{\it (b) Stochastic SMBH decay in clumpy galaxies}. If the BH pair is evolving in a clumpy galaxy such as the massive star-forming galaxies  detected at $z \sim 1$--3 \citep[][]{ForsterSchreiber_et_al_2011,Tacconi_et_al_2013,Wisnioski_et_al_2015}, scattering from giant clouds/stellar clusters can cause a random walk of the orbit of the BHs, leading to stalling. Here the stalling of SMBHs can happen at separations as large as a kpc, but typically of the order of a few hundred pc. \citet{Tamburello_et_al_2017a} considered the decay of two massive BHs that are already embedded in a gas-rich massive galactic disk subject to fragmentation into massive star-forming clumps. They considered also the effect of BH accretion and feedback, as well as star formation and feedback from supernova explosions. They explored a large suite of simulations, with galactic hosts having different masses and gas fractions, and choosing different eccentricities for the orbit of the secondary BH (the mass ratio of the two BHs was fixed to 5:1, which should be statistically representative, as the same mass ratio is most typical for halo/galaxy mergers). The main result is that the lighter BH, with mass in the range $10^7$ to $4 \times 10^7$~M$_{\odot}$, experiences repeated gravitational scattering by the most massive clumps as well as by strong spiral density waves. This often results in its ejection from the disk plane, slowing down its orbital decay by at least an order of magnitude, as dynamical friction is suppressed in the low-density envelope around the disk midplane. The suppression of orbital decay is exacerbated when AGN feedback is included, as both dynamical friction and disk torques are weaker even when the secondary BH is still in the disk plane, due to local heating of the gaseous background. In this case, since it sinks more slowly, the secondary can be ejected when it is further away from the primary. The secondary, following an ejection, can still wander at separations larger than 1~kpc for Gyr time-scales. Without considering eventual subsequent dynamical perturbations, at face value this would imply a complete abortion of the coalescence process. However, since \citet{Tamburello_et_al_2017a} used galaxy models that lack an extended bulge/spheroid component, dynamical friction was underestimated away from the galactic disk plane. By computing analytically the extra drag resulting from a \citet{Hernquist_1990} spheroid with realistic structural parameters, it was found that, if no more ejections occur, wandering secondaries should coalesce on time-scales of order 1~Gyr. This is still a very long time-scale, comparable to, if not longer than, the typical merging time-scales of galaxies at $z \sim 2$, opening the possibility that triplets and their mutual dynamical interactions \citep{Bonetti+2018} might be crucial to ascertain the final BH pair/binary state, and hence the predictions for dual AGN activity.

\bigskip

{\it (c) Inefficient decay/pairing of IMBHs/MBHs in merging dwarf galaxies}. The number of candidate central BHs in dwarf galaxies \citep[e.g.][]{Greene_Ho_2004,Greene_Ho_2007,Reines_et_al_2013,Moran_et_al_2014,Lemons_et_al_2015,Baldassare_et_al_2015} is increasing, owing to multi-wavelength surveys  such as the \chandra\ COSMOS Legacy Survey \citep[e.g.][]{Mezcua_et_al_2018}. These candidates are in the mass range $10^4$--$10^6$~M$_{\odot}$, hence bridging to the mass window that is often referred  to as intermediate-mass BHs (IMBHs). The interest towards these lightweight central BHs is motivated by the fact that, as pointed out in Sect.~\ref{sssec:LISA}, their mass is the ideal target mass for LISA \citep[][]{LISA17,gwverse_paper}.

Here the effect of the DM distribution is important, at odds with more massive galactic hosts, because dwarf galaxies are DM-dominated at nearly all radii. A novel numerical study finds that the pairing of IMBHs in merging dwarfs is affected significantly by the DM mass distribution inside the hosts, and can be rather inefficient \citep[][]{Tamfal_et_al_2018}. In the latter work, it appears that only in the ideal case in which the DM profile is a cuspy CDM profile \citep[e.g.][]{NFW96} does the BH decay continue unimpeded (until the resolution limit is reached; see \citealt{Biava_et_al_2019} for an analytical study of the later stages of the merger). Instead, when the DM density profile is flatter, either because of the nature of the DM particle [e.g., self-interacting DM (SIDM) or fuzzy DM as opposed to CDM], or because baryonic-feedback effects alter the mass distribution of both baryons and dark matter, dynamical friction becomes very inefficient already at $\sim$100~pc separation and results in stalling of the BH pairing process before a bound binary can form. Stalling at such separation is found to persist for several Gyr, although we caution that the stalling and inspiral time-scales could be affected by the presence of gas, which was not included in the collisionless simulations by \citet{Tamfal_et_al_2018}. Albeit with lower resolution, a statistical study of SMBH dynamics in galaxies with SIDM halos, assuming a large self-interaction cross-section that produces kpc-scale constant density cores, versus galaxies in CDM halos, has shown a similar trend of BH pairing suppression \citep[][]{DiCintio_et_al_2017}. Namely, most SMBHs end up wandering at few hundred pc to few kpc separations in low-mass galaxies. The dynamics of a sinking perturber, such as a BH binary, in a shallow/cored mass distribution is expected to end up with stalling for two reasons: a lower density, and a non-Maxwellian velocity distribution with many dark matter particles on N-horn periodic orbits \citep[][]{Cole_et_al_2012,Petts_et_al_2015,Petts_et_al_2016}, the latter being orbits such that lumps move faster than the BHs, accelerating them rather than dragging them. A similar stalling result was shown also in the cosmological zoom-in simulations presented in \citet{Bellovary_et_al_2018}, in which about half of the massive BHs in dwarf galaxies are found wandering within a few kpc of the centre of the galaxy. 

\subsection{From kpc- to pc-scale separations: SMBH pairs in circumnuclear disks}
\label{sssec:stallingBH_2}

As already discussed, there are very few observations of SMBH pair candidates on 1--100~pc scale, e.g NGC~3393 (see Sect.~\ref{ssubsec:pointed X-rays};  \citealt{{Fabbiano2011}}) or 0402+379 (see Sect.~\ref{sssec:Dual_radio} and Fig.~\ref{fig:radio0402}). In Sect.~\ref{ssec:stallingBH}, we have shown that, while state-of-the-art galaxy merger simulations do provide a wealth of information on the dynamics of BH pairs down to 10--100~pc separations, their resolution is not sufficient to
capture the complex multi-phase structure of the interstellar medium (ISM) on nuclear scales, and its effect on the evolution of SMBH pairs. In this sub-section, we review physical processes that arise on these, intermediate spatial scales and discuss how they affect the rate of gravitational pairing of SMBHs, that occurs at separations of $\sim$1~pc.

The multi-phase ISM on these scales consists of cold star-forming clouds and clumps embedded in a warm-hot
diffuse medium, spanning decades in density as well as in temperature (see, e.g., \citealt{downes_solomon1998}).  The hot medium is
likely the result of powerful stellar and supernova feedback, as many of these gas-rich nuclei host starbursts or have features typical of post-starburst environments. In particular, dense multi-phase CNDs of gas and stars, with sizes of a few hundred to a few tens of pc, dominate the mass distribution of resolved galactic nuclei at low and intermediate redshift 
in merger remnants as well as in the central regions of Seyfert galaxies \citep[][]{Medling_et_al_2014,Izumi_et_al_2016}, which are structurally typical late-type spiral galaxies. A common feature of these CNDs in observations is that, while they show clear evidence of rotation in their gas and stellar kinematics,
the velocity dispersion of both gas and stars is almost comparable to rotation, akin to the high-redshift clumpy galaxies discussed in the previous section [paragraph
(b)]. The dense clumpy phase comprises a large fraction of the mass in CNDs, while the diffuse phase dominates in terms of volume, providing
significant pressure support (Downes \& Solomon 1998). The picture that emerges is that of a thick turbulent rotationally supported CND, very
different from the larger scale thin, kinematically cold disk that dominates the mass distribution of late-type galaxies at kpc scales.
Since clouds and clumps of atomic and molecular gas have masses of $\sim$ 10$^3$--10$^6$~M$_\odot$ and sizes of 1--10~pc, an appropriate numerical hydrodynamical model 
requires a mass resolution much higher than that in galaxy mergers,  and a spatial resolution
of order pc or better. For this reason, small-scale simulations that model specifically BH pairs embedded in CNDs about a hundred pc in size have been carried out by
a number of authors.

The clumpiness of the ISM, due to the presence of cold clouds, is an important factor for the late stage of BH pair decay, at 10~pc separations and below, near the phase in which a bound binary forms. \citet{Fiacconi_et_al_2013} performed the first study of BH pairs in clumpy CNDs, documenting the dramatic effect of gravitational scattering by clumps and spiral density waves that we already illustrated for the case of clumpy high-redshift galaxies in the previous section. They considered BHs right in the LISA window ($10^5$--$10^7$~M$_{\odot}$). They also found that in some cases the decay can be accelerated due to the capture of a massive gas cloud (of mass $>$10$^6$~M$_{\odot}$) by the secondary. They proposed the general notion that {\it orbital decay is a stochastic process in a clumpy CND}, with orbital decay time-scales down to $\sim$1~pc separations ranging from a few Myr to as much as 100~Myr. The shortest time-scales agree with previous results for smooth CNDs \citep[see][]{escala05,Dotti_et_al_2006,dotti07,Mayer:2013:MBHBGasRev}. Statistically, though, a larger fraction of the simulations resulted in long time-scales, in line with the same findings of \citet{Tamburello_et_al_2017a}, especially for BHs on eccentric orbits.

Early CND simulations were quite idealized, missing feedback mechanisms which should play a major role in CNDs, such as to generate the warm-hot diffuse phase, as these are often hosting a starburst or at least elevated star formation activity \citep[][]{Medling_et_al_2015}. \citet{delValle_et_al_2015} and \citet{SouzaLima_et_al_2017} recently carried out similar studies but included a much richer inventory of physical processes. \citet{delValle_et_al_2015} included star formation and a weak form of supernova feedback, whereas \citet{SouzaLima_et_al_2017} employed a popular effective sub-grid model of supernova feedback, the blast-wave feedback \citep[][]{Stinson_et_al_2006}, which has been shown to produce galaxies with realistic structural properties in galaxy formation simulations \citep[][]{Guedes_et_al_2011,Munshi_et_al_2013,Tollet_et_al_2016,Sokolowska_et_al_2017}. \citet{SouzaLima_et_al_2017} also added BH accretion and AGN feedback, using the same method implemented in \citet{Tamburello_et_al_2017a}, confirming the stochastic orbital-decay picture and the prevalence of suppressed orbital decay with time-scales in the range 50--100~Myr, whereas \citet{delValle_et_al_2015} found the suppression of binary formation to be a weaker effect in their simulations. Differences in the implementation of feedback, and the fact that the former consider eccentric orbits while the latter explore mostly circular orbits, might be at the origin of the discrepant conclusions. Long binary-formation time-scales were  also found in a complementary multi-scale study modelling all stages of the BH pair evolution in a realistic major galaxy merger \citep[][]{Roskar_et_al_2015}. The latter work also addressed the formation of the CND after the merger, but was restricted to evolve only one initial condition for the galaxy merger, due to the high computational burden introduced. 

Finally, \citet{SouzaLima_et_al_2017} uncovered another process that tends to suppress the BH decay and binary formation, this time caused by AGN feedback; this is the {\it wake evacuation effect}, by which the secondary BH carves a hole in the outer, lower-density region of the CND, as it launches a hot pressurized bubble resulting from AGN feedback. The result is that dynamical friction and disk torques are suppressed as the CND--BH dynamical coupling is temporarily stifled by the presence of a large cavity. Hence, the secondary BH decays more slowly even before a strong interaction with a clump or spiral wave occurs, often resulting in an ejection when it is still far from the center. As a result, in the runs with both a clumpy ISM and AGN feedback, the outcome is an even longer delay, of order of 1~Gyr (although we caution that different implementations of feedback could produce different results and should be investigated). The latter, however, should be regarded as an upper limit in the absence of an extended massive stellar spheroid (for the same argument, see \citealt{Tamburello_et_al_2017a} in the case of high-redshift galactic disks). When dynamical friction by an extended spheroid (with realistic parameters) is added, \citet{SouzaLima_et_al_2017} obtained a decay time-scale (to fraction of pc separation, namely close to the resolution limit) of order a few $10^8$~yr. This is still almost two orders of magnitude longer than the decay time-scale in a smooth CND with no feedback processes included \citep[see, e.g.][]{Mayer:2013:MBHBGasRev}. These effects are only some of those that are possible in a complex multi-phase CND. For example, recently \citet{Park_Bogdanovic_2017} have shown that radiation pressure can also produce an extra drag pulling opposite to the dynamical friction wake, which again slows down the orbital decay.

\part{Gravitationally bound SMBH binaries with sub-parsec separations}
\label{sec:BH_binaries}

As discussed in the previous section, the multi-wavelength searches for SMBH systems with kpc separations, corresponding to early stages of galactic mergers, have so far successfully identified a number of multiple, dual and offset AGN. 
Gravitationally bound SMBHs (SMBH Binary, SMBHBs) are representative of the later stages of galactic mergers in which the two SMBHs (with separation ranging from pc down to sub-pc scales) forms a Keplerian binary. 

As we describe in the remainder of this section, the key characteristic of gravitationally bound SMBHBs is that they are observationally elusive. At the time of this writing a few hundred SMBHB {\it candidates} have been described in the literature but their nature as true binaries is inconclusive and remains to be tested through continued multi-wavelength monitoring. For this reason, most of what is known about the formation, rate of evolution, physical properties, and expected observational signatures associated with gravitationally bound binaries hinges on theoretical arguments. In the remainder of this section, we follow this line of argument and first present expectations based on theoretical models and simulations of SMBHBs and then describe observational evidence for SMBHB candidates obtained with different observational techniques.

\section{Theoretical background}
\label{ssec:subpctheory}

A gravitationally bound binary forms at the point when the amount of gas and stars enclosed within its orbit becomes smaller than the total mass of the two black holes.  For a wide range of host properties and SMBH masses this happens at orbital separations $\lesssim 10$~pc \citep{mayer07,dotti07,khan12}. \citet{begelman1980} first described the subsequent series of physical processes that could remove energy and angular momentum from an SMBHB and allow it to spiral in.  In this picture, after the SMBHB hardens, scattering events with stars in the nucleus drive angular momentum loss and slowly bring the SMBHB to separations of $\sim$1~pc.  At separations of order of  10$^{-2}$ or 10$^{-3}$~pc, GW emission becomes efficient and the binary evolves to coalescence in $\lesssim$~a few$\times 10^8$~yrs. The exact rate of binary orbital evolution depends on the nature of gravitational interactions that it experiences and is still an area of active research. In what follows, we discuss the most important mechanisms that lead to SMBHB orbital evolution.

\subsection{Interactions with stars and other SMBHs}
It was not immediately clear whether SMBHB could transit between stellar scattering events and efficient emission of GWs within a Hubble Time. Three-body stellar scattering requires a supply of stars within a certain region of phase space (the so-called loss-cone), which may be depleted before the binary enters the GW-dominated regime. This possible slow-down in the orbital evolution of the parsec-scale SMBHBs, caused by inefficient interactions with stars \citep{mm01}, is often referred to as {\it the last parsec problem}. If the binary evolution is not efficient in this stage, a significant fraction of SMBHBs in the universe should reside at orbital separations of $\sim 1$~pc and would not yet be strong GW emitters. However, the loss-cone may be replenished \citep{yu05}, e.g., in galaxies with non-spherical potentials. Several  theoretical studies report that the binary evolution due to stellar scatterings continues unhindered to much smaller scales, when the stellar distribution possesses anisotropies  \citep{yu02,merritt04,berczik06, preto11, khan11,khan12a,khan13,vasiliev14,holley-bockelmann15,Vasiliev2015,Mirza2017}.  

In fact, mechanisms beyond three-body interactions with stars may also play a critical role. For instance, if stellar scattering were ineffective, and the binaries stalled, then a third (or fourth, etc.) merger with another SMBH, as a result of subsequent galaxy mergers, eventually tends to promote merger for a fraction of such stalled binaries~\citep{2003ApJ...582..559V,Hoffman2007,Bonetti+2018,Ryu+2018}. 
More specifically, a third SMBH ``intruder'' may end forming a hierarchical triplet,  i.e. a triple system where the hierarchy of orbital separation defines an inner and an outer binary, the latter consisting of the intruder and the centre of mass of the former. These systems may undergo Kozai-Lidov oscillations if the intruder is on a highly inclined orbit with respect to the inner binary \citep{Blaes2002}.  These oscillations increase the eccentricity of the inner binary accelerating its orbital decay driven by GW emission, and eventually driving it to coalescence. By performing rather extensive analyses, including the SMBH merger sequence and the dynamical evolution of triplets in a spherical galaxy potential, \cite{Ryu+2018} and \cite{Bonetti+2018}  find that  a fraction of at least 20-30 percent  of those SMBHBs that would otherwise stall are led to coalesce within a characteristic  timescale of 300 Myrs.

\subsection{Interactions with gas} 
The existence of gas in the central regions of the post-merger galaxy has also been proposed as a solution to the final parsec problem \citep{escala05}. The gas can not only catalyze the binary evolution, but also provide bright electromagnetic (EM) counterparts, significantly enhancing the chances of detecting the binary from its EM spectrum \citep{2002ApJ...567L...9A}. Thus, SMBHBs in predominantly gaseous environments are particularly interesting for this review and, in particular, for the binary signatures discussed below. The interaction of a binary with a circumbinary disk has been the topic of a number of theoretical studies \citep[see e.g.][for recent investigations]{an05, macfadyen08, bogdanovic08,  cuadra09, haiman09, hayasaki09,bogdanovic11, roedig12,shi12,noble12,kocsis12b, kocsis12a,dorazio13,farris14,rafikov16,Bowen18,Bowen19}. 
The phenomenology has been found to depend on the black hole mass ratio $q\equiv M_2/M_1$ and can be broadly divided in three mass ratio regimes: I) $q\ll 1$, II) $10^{-4} \lsim q \lsim 0.05$, and III) $0.1\lsim q\leq 1$.

\begin{itemize}
\item In the limit of small mass ratio ($q\ll 1$), significant work has been done to understand the linear interactions between the binary and the disk. This so-called Type I planetary migration is facilitated by linear spiral density waves launched at resonant disk locations
\citep[e.g.][]{GT80}. The rate and even the direction (inward or
outward) of Type I migration is sensitive to thermodynamics
\citep[e.g.][]{PM2006}.  

\item  The second regime ($10^{-4} \lsim q \lsim 0.05$) has
also been widely explored. The secondary opens a narrow annular gap in
the disk, resulting in so-called Type II migration
\citep[e.g.][]{Ward1997}.  The interaction is non-linear, generally causing inward migration on a time-scale comparable to the
viscous time-scale of the disk\footnote{The time scale on which the angular momentum is transported outward through the disk.} near the binary \citep[e.g.][]{LinPapa1986}. 
Deviations from this time-scale, however, can be significant, especially when the mass of the smaller BH exceeds that of the nearby disk, so that the disk cannot absorb the secondary's angular momentum, causing the migration to slow down~\citep[e.g.][]{SyerClarke1995,Ivanov99}.  Additionally, contrary to the simplistic view in which the secondary divides the disk into an inner and outer region and becomes, effectively, a particle accreting in lock-step with the background gas~\citep[e.g.][]{ArmitageReview}, the gas has been shown to be able to flow across the gap on horsehoe orbits \citep[e.g.][]{Duffell+2014,DK2017}.

\item The range of $0.1\lsim q\leq 1$ is numerically much more challenging because both black holes as well as the origin need to be on the coordinate grid, and the gas flows become much more violently disturbed by the binary. In this range, however, relatively little analogous work has been done regarding the rate and the direction of migration. There are several simulations in the literature of $q\sim 1$ binaries, including the compact GRMHD regime near merger. For example, \cite{Bowen18,Bowen19} simulated the full 3D GR magnetized mini-disks coupled to circumbinary accretion in a SMBHB approaching merger, and characterized the gas dynamics in this regime. However, the impact of gas on the binary orbit is not evaluated (and is likely insignificant at this late GW-driven stage). 
\cite{macfadyen08}, \cite{dorazio13} and \cite{Miranda+2017} used 2D grid-based hydrodynamical simulations to study SMBHBs embedded in thin $\alpha$ disks \citep{SS1973}.
However, these simulations excluded the innermost region surrounding the binary, by imposing an inner
boundary condition, neglecting potentially important gas dynamics and torque contributions from inside the excised region. \cite{farris14}, on the other hand, included the innermost region in their simulated domain, but their study did not focus on the binary-disk interaction, and did not present measurements of the gas torques from that region. Studies by \cite{Cuadra+2009} and \cite{roedig12} followed the interaction between SMBHBs with $q=1/3$ and a self-gravitating circumbinary disk, using 3D SPH simulation, and generally found gas disk torques to be strong enough to drive a sub-pc
binary into the GW-driven regime in a Hubble time (except perhaps the
most massive systems). 

 However, recent work \citep{Tang+2017,Munoz+2019,Moody+2019} has called this conclusion
into question, because in the quasi-steady state and the much higher
spatial resolutions reached in these works, the torques appear
dominated by gas very close to the individual BHs, and cause outward, rather than inward-migration (see discussion presented below in this section \textit{Open Questions related to theory of SMBHBs with subparsec separations.}
\end{itemize}

\begin{figure*}[t]
\centering{
\includegraphics[trim=130 120 120 130, clip, scale=0.37,angle=90]{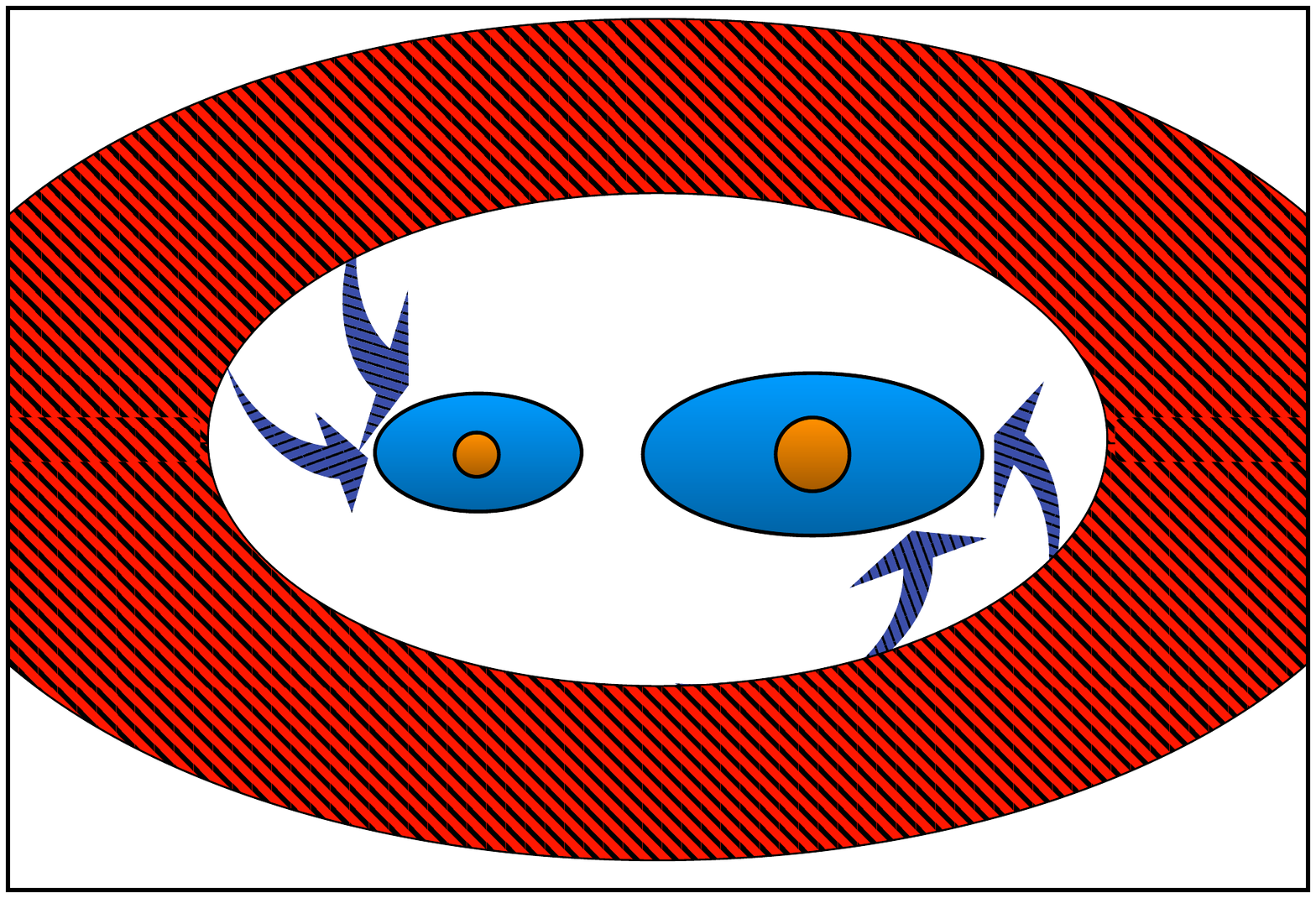} % b l t r
\includegraphics[trim=260 355 260 350, clip, scale=1.65,angle=0]{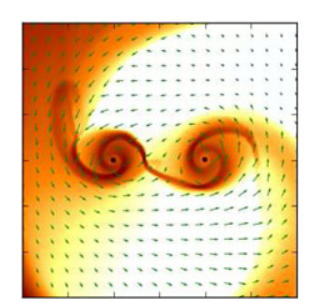} %  l b r t
}
\caption{Illustration of the geometry of circumbinary region after the binary has created a low density region in the disk. {\it Left:} Simulations indicate that SMBHBs in this evolutionary phase can accrete by capturing gas from the inner rim of the circumbinary disk and can maintain mini-disks bound to individual holes. Adapted from \citet{bogdanovic15}. {\it Right:} Snapshot of gas surface density from a 2D hydrodynamic simulation of a circumbinary disk around an SMBHB with mass ratio 1:4. Orbital motion is in the counterclockwise direction. Superposed green arrows trace the fluid velocity. Adapted from \citet{farris14}. }
\label{fig_cmbd}      
\end{figure*}

The $0.1\lsim q\leq 1$ regime is of particular interest to SMBHBs, as such systems may be naturally favored by the binary formation process (see Sect.~\ref{ssec:stallingBH}) and could result in two AGN of comparable luminosity and the loudest GW sources.
Fig.~\ref{fig_cmbd} illustrates the geometry of the circumbinary region for SMBHBs in this mass ratio range.  Hydrodynamic simulations of prograde binaries (rotating in the same sense as the circumbinary disk) indicate that in unequal mass binaries, the binary expels the gas from the central region, creating a low-density cavity. Additionally, accretion occurs preferentially onto the smaller of the two SMBHs, which orbits closer to the inner edge of the circumbinary disk \citep{artymowicz1994, gunther02, hayasaki07, Roedig11, farris14}. The proximity to the circumbinary disk and a smaller velocity relative to it (when compared to the primary SMBH) allow the secondary to capture more gas and result in a higher accretion rate. This phenomenon of ``accretion inversion'' is of practical interest because it implies that the AGN powered by the lower mass SMBH may appear more luminous than the primary counterpart. 

As the binary orbit decays, the inner rim of the circumbinary disk follows it inward until the timescale for orbital decay by gravitational radiation becomes shorter than the viscous timescale of the disk \citep{an05}. At that point, the rapid loss of orbital energy and angular momentum through gravitational radiation can cause the binary to detach from the circumbinary disk and to accelerate towards coalescence; this phase is known as ``decoupling''.
The evolution differs for retrograde binaries, which due to the cancellation or orbital angular momentum do not open a low density cavity in the disk, and for the same reason tend to shrink faster \citep{roedig_sesana2014}.

\subsection{Open questions related to theory of SMBHBs with sub-parsec separations} Thermodynamic properties of the circumbinary disk (which define the viscous time-scale) are also expected to play an important role in the binary evolution. These are uncertain, since they are still prohibitively computationally expensive to model from first principles and are not constrained by observations. More specifically, the thermodynamics of the disk is determined by the binary dynamics as well as the presence of magnetic field and radiative heating/cooling of the gas. While the role of magnetic fields in circumbinary disks has been explored in some simulations \citep{giaco12,shi12,noble12,farris12,gold14,Kelly2017}, a fully consistent calculation of radiative processes is still beyond computational reach. 
 
On top of this, significant uncertainties related to the SMBHB orbital evolution timescales still remain. In particular, a recent study which resolved the minidisks and gas flows around the individual BHs at much higher resolution~\citep{Tang+2017} found that the torques are even stronger and are dominated by the gas very close to the BHs 
(near or even inside their tidal truncation radii, $R_H\approx 0.3q^{1/3}a$, where $q$ is the mass ratio and $a$ is the semi-major axis of the binary; gas can formally be bound to individual BHs inside this radius).

The net torque arises from a small
asymmetry in the shape of the gas distribution near the edges of the
minidisks ahead and behind the BHs.  \cite{Tang+2017} found that 
for a disk density corresponding to an accretion rate normalized to
0.3 times the Eddington value, the binary inspiral time is $\approx
3\times10^6$ years, independent of BH mass and binary separation.
However, given the dependence of this result on the small asymmetry in
the gas distribution near the BHs, one has to keep in mind that 3D
effects, radiation pressure, winds, etc. are likely to modify this
timescale significantly.  
Likewise, apart from additional physical effects, the subtlety of the torques also make them susceptible to numerical issues.  For example, \cite{Tang+2017} found that the numerical sink speed affects the torques, and the torques turn positive (causing outspiral) for rapid sinks. \citet{Munoz+2019} found a positive torque with a similar setup, resulting in widening, rather than shrinking of the SMBHB orbit.  Recent work by \citet{Moody+2019} has also found positive torques, even with an apparently slower sink prescription.

The next generation of simulations involving circumbinary disks must be improved, also adding the effect of radiation and feedback and increasing the number of orbits considered for the evolution of the system. In addition, a more realistic cosmological setting may provide more physically motivated conditions for these (so far local) simulations and help to put them in a broader context of a galaxy merger (see Sect.~\ref{sssec:BH_Pairs_Theory_Cosmological_Simulations}).  Similarly, the simulations of binary-stellar interactions should be also improved,  expanding to realistic timescales and galactic environments. 

Observations of the orbital properties of SMBHBs are key to constraining and understanding binary evolution. This is because the frequency of binaries as a function of their orbital separation is directly related to the rate at which binaries evolve towards coalescence. Theoretical models predict that the exchange of angular momentum with the ambient medium is likely to result in SMBHB orbits with eccentricities $\gtrsim 0.1$, with the exact value depending on whether gravitationally bound SMBHs evolve in mostly stellar or gas-rich environments \citep{Roedig11,sesana11,holley-bockelmann15}. Known semi-major axes and eccentricity distributions would therefore provide a direct test for a large body of theoretical models.  

The key characteristic of gravitationally bound SMBHBs is that they are observationally elusive and expected to be intrinsically rare. While the frequency of binaries is uncertain and dependent on their unknown rate of evolution on small scales (see  Sect.~\ref{sssec:BH_Pairs_Theory_Cosmological_Simulations}), theorists estimate that a fraction $<10^{-3}$ of AGN at redshift $z<0.7$ may host detectable SMBHBs \citep{Volonteri09}. Similar fractions have been found with different approaches \citep{2012MNRAS.420..860S,2019MNRAS.485.1579K,krolik2019} This result has two important implications: (a) any observational search for SMBHBs must involve a large sample of AGN, making the archival data from large surveys of AGN an attractive starting point and (b) the observational technique used in the search must be able to distinguish signatures of binaries from those of AGN powered by single SMBHs. 

\section{Observational evidence for SMBH binaries with sub-parsec separations.}
\label{ssec:subpbobs}

Observational techniques used to search for SMBHBs systems have so far largely relied on direct imaging, photometry, and spectroscopic measurements \citep[see][for a review]{bogdanovic15, komossa_zensus2016}. They have recently been complemented by observations with PTAs. We summarize the outcomes of these different observational approaches in this section and note that in all cases SMBHBs have been challenging to identify because of their small angular separation on the sky (spatially unresolved in most cases), as well as the uncertainties related to the uniqueness of their observational signatures.

\subsection{Radio imaging searches for SMBHBs with the VLBI}
\label{ssubsec:subpcradio}

As described in Sect.~\ref{sssec:Dual_radio}, radio-emitting binaries can be imaged to redshift $z\sim0.1$ on pc scale with VLBI, and 10-pc scale separations are resolvable practically at any redshift.
We have already mentioned the well-known example of an SMBHB system with pc-separation, discovered and studied with VLBI in the nearby radio galaxy 0402+379 (see Fig.~\ref{fig:radio0402}, \citealt{rodriguez2006}). 
Recently, \cite{kharb2017} reported on a sub-pc separation candidate binary system in the Seyfert galaxy NGC\,7674 ($z=0.0289$). Since the assumed two cores were only detected at a single frequency, the radio spectral indices that would provide the strongest proof of their optically thick nature, could not be derived.
 
It is worth mentioning again that low-resolution data should be taken with care.
For example, SDSS\,J113126.08$-$020459.2 and SDSS\,J110851.04$+$065901.4, two double-peaked [O III] emitting AGN (see Sect.~\ref{sssec:optical_kpc}) identified as candidate kpc-scale pairs AGN using optical and near-IR  observations \citep{Liu:2010,2010ApJ...715L..30L}, were observed with the European VLBI Network. Those observations yielded a detection of only one AGN in the source SDSS\,J113126.08$-$020459.2, and none in the case of SDSS\,J110851.04$+$065901.4 \citep{2016A&A...588A.102B}. 
Another interesting example is J1536+0441 which was originally proposed as a tight ($\sim 0.1$\,pc separation) binary SMBH by interpreting its 
unusual optical emission line systems \citep{boroson2009}. Soon after, radio observations with the VLA instead found two components separated by 5.1\,kpc \citep{wrobel2009} whose compact AGN nature was proven by high-resolution VLBI imaging \citep{bondi2010}.

\citet{burkespolaor2011} systematically analysed archival multi-frequency VLBI data available for the most luminous radio AGN known -- more than 3000 objects -- and found only a single case with double flat-spectrum cores. This previously known object, the nearby radio galaxy 0402+379 ($z=0.055$), has two nuclei separated by just 7.3\,pc (see Fig.~\ref{fig:radio0402}, \citealt{rodriguez2006}).

The superior angular resolution is not the only benefit radio interferometric observations offer for studying binary AGN. The presence of a binary companion can be imprinted on the the symmetric pair of relativistic jets emanating from the vicinity of one of the SMBHs. 
In this model, the SMBH that is producing the jet has periodically changing velocity (due to the orbital motion in the binary) which may lead to an observable modulation on an otherwise straight jet \citep[e.g.][]{gower1982,kaastra1992,roos1993,hardee1994}. 
Such periodic perturbations may result in rotationally symmetric helical S-shaped radio structures \citep[e.g.][]{begelman1980,roos1988,lobanov2005,kun2014,deane2014}. 

The presence of precessing jets does however does not necessarily indicate modulation caused by binary SMBHs. For example, precessing jets can also form due to tilted accretion disks in single AGN \citep{liska2018}. Internal plasma instabilities in the jet, or the interaction of a jet with the ambient medium could also produce ``wiggling'' jets. Therefore semi-periodic patterns in radio jets alone cannot be considered a definitive observational evidence of SMBHBs. However, in cases where multiple indications are available, the precessing jet model could constrain the SMBH masses, separation and orbital period. Jet studies are often invoked for tight binary candidates that are radio sources but not directly resolvable with interferometers, to seek indirect supportive evidence for binarity \citep{kun2015,mohan2016,mooley2018}. These targets are selected e.g., on the basis of their periodic optical variability; this will be discussed in detail in Sect.~\ref{sssec:subpc_photometry}.

From SMBHB candidates separated well below the resolution limits of radio interferometers, an extensively studied example is the blazar OJ~287. Its quasi-periodic optical light curve shows double peaks about every 12~yr \citep{sillanpaa1988}; see also the recent review by \cite{dey19}. These variations have been explained in the context of a number of different binary SMBH scenarios. The best-explored of these requires an SMBHB on an eccentric orbit with a semi-major axis of about 0.05~pc \citep{valtonen2007} crossing the accretion disk of the primary \citep{valtonen2008} twice during each orbit. 

However, connecting the radio morphology to the detected periodicities in the optical light curve is not straightforward. The complex VLBI jet of OJ~287 has been described with a helical model \citep{valtonen2013} in a way that is consistent with the binary orbital motion. Most recently, \cite{Britzen_et_al_2018} presented an extensive study of the jet emission of OJ~287 at cm wavelengths, combining VLBI and single-dish radio data, and found that precession of an accretion disk around a single SMBH can explain the observed periodic behaviour of the jet morphology and radio flux density changes (but not the optical long-term lightcurve and polarimetry;
\citealt{dey19}). 
A binary is still likely needed in that model for explaining the disk precession itself.
 
In addition, \cite{Agudo_et_al_2012}, who studied the jet at higher resolution at mm wavelengths, detected erratic, wobbling movement of the jet. It was explained by variable accretion leading to fluctuating plasma injection to the jet.
The most recent modelling of OJ~287 in the context of the SMBHB model of Valtonen and collaborators is based on 4.5 order post-Newtonian dynamics. It requires a binary of mass ratio $q\simeq 0.01$, a massive primary of 
$1.8 \times 10^{10} M_{\odot}$, and an eccentric orbit of the secondary with $\epsilon = 0.7$ (review by \citealt{dey19}). The orbit is subject to General Relativity (GR) precession of the pericenter of $\Delta \phi = 39$ deg/orbit. Independent supporting evidence for a large mass of the primary SMBH comes from optical imaging and scaling arguments \citep{wright1998,valtonen2012}.

Another remarkable example is the BL Lac object PG~1553+113 at z=0.5, extensively monitored in radio, which shows evidence of a possible quasi-periodic trend in gamma-ray emission (at energies above 100 GeV), with main peaks occurring over a period of 2.18 years (observer frame, corresponding to 1.5 year in the source frame) and strengthened by correlated oscillations observed in radio and optical fluxes \citep{ackermannetal15,tavanietal18}. The periodicity has been interpreted as an SMBHBs system with a total mass of $\sim10^8\,\msun$ and a milliparsec separation. Jet nutation from the misalignment of the rotation SMBH spins or  Magnetohydrodynamic (MHD)-kinetic tearing instabilities in the jet of the more massive BH due to the stress of the smaller BH at the periastron could be the source of the periodicity (see \citealt[and references therein]{tavanietal18,cavaliereatal17}).

On larger scales, far away from the central engine, the jets/lobes may provide indications of past merging events. X-shaped (or winged) radio sources are radio galaxies where, besides the usual pair of lobes, a second pair of low-surface brightness radio-emitting wings can be detected \citep{Leahy_Parma_1992}. \cite{Merritt_Ekers_2002} proposed that a rapid change in the direction of the BH spin (spin-flip) in a merger event can lead to the formation of these structures. \cite{Roberts_Saripalli_2015} used the existing data on X-shaped radio galaxies to estimate the GW background. They found that most sources in the sample are likely not associated with spin axis flips but the X-shaped structures are rather due to backflows or axis drifts. The number of genuine X-shaped radio sources is at most $\sim$20 per cent \citep{2015ApJS..220....7R} of the initial sample of candidates selected by \cite{2007AJ....133.2097C}, and only a fraction of these could be caused by merger-induced spin flips 
(see also \citealt{Liu+2003} for theoretical considerations).  
 
The inferred rate for major galaxy mergers results in SMBH coalescence is much lower in comparison, and less than 0.13 per Gyr per radio galaxy host. This suggests that most of X-shaped radio sources cannot be produced as a consequence of SMBHB mergers. Recently, \cite{Saripalli_Roberts_2018} presented a detailed morphological study of more than 80 X-shaped radio galaxies. The data suggest that phenomena related to the central SMBH spin axis cause most of the morphologies seen. While binary black holes are a strong contender in causing axis changes, other (disk-related) explanations are not excluded.

\subsection{Spectroscopic searches for SMBHBs}

\smallskip
\subsubsection{Broad emission-line velocity shifts as signatures of sub-pc SMBHBs} 
\label{sssec:subpcspec1}

Another approach to searching for SMBHBs is to make analogy to spectroscopic binary stars \citep{Komberg1968,begelman1980} and look for the radial velocity signature of orbital motion in the emission lines of AGN and quasars \citep{Gaskell1983}.  This scenario corresponds to a physical picture where one or both of the SMBHs is active \citep[e.g.,][as described in Sect.~\ref{ssec:subpctheory}]{Hayasaki2007,Cuadra+2009}, but their broad-line regions (BLRs) are distinct and suffer at most mild truncation due to their membership in the binary (however, see the next paragraph for a detailed discussion of the effect of BLR truncation). 

This can occur at a limited range in SMBHB separation.  At close separations comparable to or smaller than the BLR size, the BLRs become substantially truncated \citep[][]{Roedig+2014,Runnoe2015} and eventually merge \citep{krolik2019}.  At wide separations, the broad-line velocity offset or acceleration due to orbital motion will be imperceptible \citep[e.g.,][and see discussion below]{Eracleous2012,Shen2010, Popovic2012}.  
 
Thus, the assumption is that the BLR is not dissimilar to that of a normal AGN: it is a flattened distribution of clouds originating in the outer parts of the accretion disk, at $\sim10^3\,r_g$ ($r_g = GM/c^2$ and $M$ is the SMBH mass), gravitationally bound to its SMBH with a predominantly Keplerian velocity field \cite[e.g.,][]{Wills1986,Koratkar1991,Peterson2000,Denney2010,Grier2013}. A disk-like emission-line profiles are seen in AGN at lower luminosity, while in Seyfert~1 galaxies an additional narrower component (FWHM $\sim 1000-2000\,{\rm km\,s^{-1}}$) is observed, probably due to a non disk BLR gas, usually moving at lower velocities \citep{Storchi-Bergmann2017}.  The hypothesis is then that bulk orbital motion of the SMBH and BLR is observable as a time-dependent velocity shift in the broad lines of some AGN relative to the narrow emission lines that are emitted from larger size scales in the host galaxy.   

There are limits on the type and properties of SMBHBs likely to be uncovered by spectroscopic searches. \citet{Pflueger2018} studied this question using an analytic model to determine the likelihood for detection of SMBHBs by ongoing spectroscopic surveys. The model combines the parameterized rate of orbital evolution of SMBHBs in circumbinary disks with the selection effects of spectroscopic surveys and returns a multivariate likelihood for SMBHB detection. Based on this model, they find that spectroscopic searches with yearly cadences of observations are in principle sensitive to binaries with orbital separations less than a ${\rm few} \times 10^4\,r_g$ (where $r_g$ is now defined in terms of the binary mass), and for every one SMBHB in this range, there should be over 200 more gravitationally bound systems with similar properties, at larger separations. Furthermore, if spectra of all SMBHBs in this separation range exhibit the AGN-like emission lines utilized by spectroscopic searches, the projection factors imply five undetected binaries for each observed $10^8\,M_\odot$ SMBHB with mass ratio 0.3 and orbital separation $10^4\,r_g$ (and more if some fraction of SMBHBs is inactive). Assuming a binary mass of $10^7-10^8$~M$_{\odot}$, these orbital separations translate to $\sim0.01-0.1$~pc and orbital periods of order a few tens of years to few centuries. Notably, observational monitoring campaigns cannot hope to observe many cycles of radial velocity curves from such systems, so the signature of a binary will rather be a monotonic increase or decrease in the observed velocity of an emission line.  

\begin{figure*}[!t]
\centering
\vspace{0.0pt}
\includegraphics[width=0.97\columnwidth,angle=0]{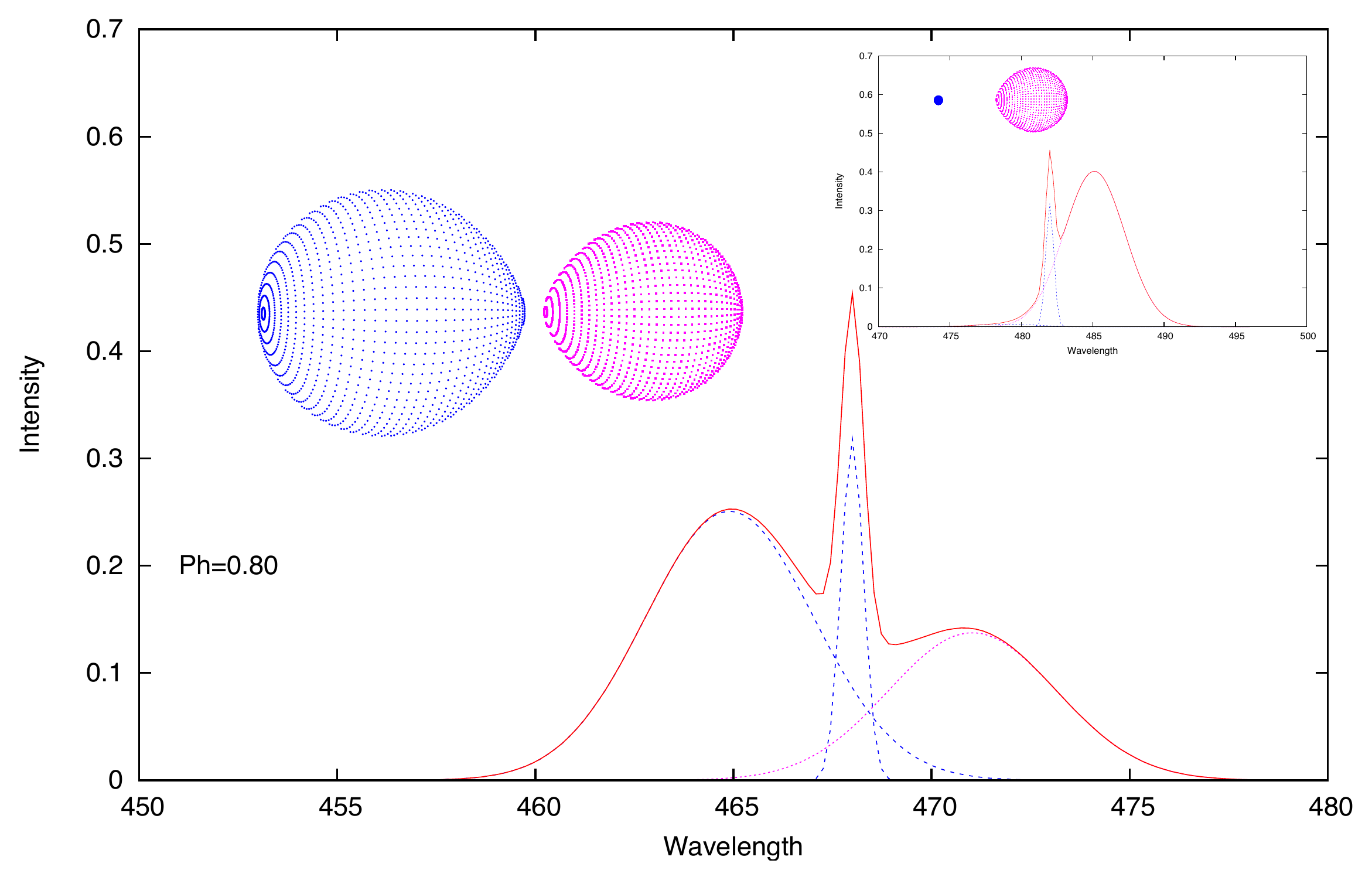}
\hskip2cm
\includegraphics[width=0.8\columnwidth,angle=0]{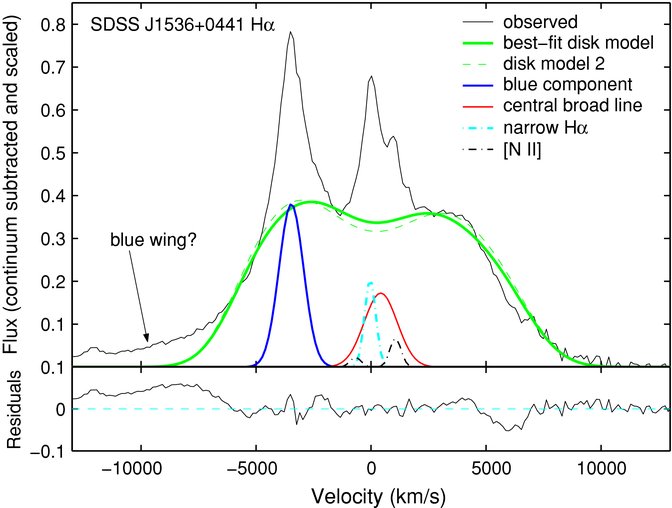}
%\textcolor{blue} 
\caption[]{Left panel: H$\beta$ emission-line profile emitted from the SMBHB system when both black holes have the BLR, while in the inset the line profile when only one component has the BLR is shown. Adapted from \cite{Popovic2012}. 
Right panel: H$\alpha$ emission line (once the continuum spectrum is removed; solid black line) fitted by the disk model (solid green line). There are additionally two relatively broad peaks, reproduced using two Gaussian profiles (blue and red line), while the cyan dash-dotted line is the Gaussian fit to the narrow H$\alpha$ component. The black dash-dotted lines are the Gaussian fits to [NII]$ \lambda$6548, 6583 lines, with fixed FWHM given by fitting [OIII] $\lambda$4959, 5007. 
The dashed green line shows the disk model line with outer radius 8000 r$_g$, and the same inner radius and inclination as the best-fit solid green line. The black solid line in the lower panel shows the residuals. From \cite{Tang2009}.}
\label{fig:dpeprofiles}
\end{figure*}

\begin{figure}[!ht]
\centering
\vspace{0.0pt}
\includegraphics[width=0.9\columnwidth,angle=0]{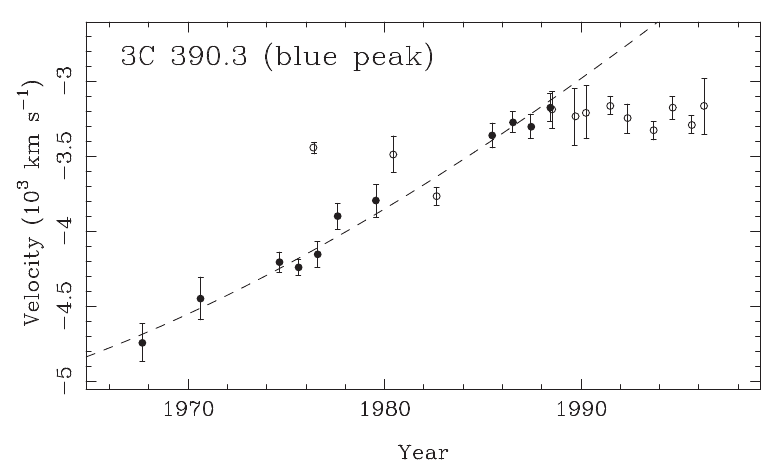}
\vspace{-0.0pt}
\caption[]{The radial velocity curve for one shoulder of the broad H$\alpha$ emission line in the double-peaked emitter 3C~390.3.  This family of AGN was considered as the observational consequence of an SMBHB embedded in a gaseous disk with both black holes active and having distinct BLRs.  The dashed line shows the expected behavior of the velocity offset of the peak in this scenario based on the data prior to 1988. As described in Sect.~\ref{sssec:subpcspec1}, this is one among several lines of reasoning used to disfavor the SMBHB hypothesis for double-peaked emitters. Filled and open points differentiate between error bars calculated using error in the mean and rms dispersion, respectively. Adapted from \citet{Eracleous1997}.}
\label{fig:dpevar}
\end{figure}

{\it Velocity shifts in double-peaked emitters.} In the physical model of an SMBHB 
embedded in a gaseous disk, either one or both SMBHs can be active. Quasars with 
double-peaked broad emission lines \citep{Eracleous1994,Eracleous2009} have been  
hypothesized as the observational consequence of the latter scenario 
\citep{Gaskell1983,Gaskell1984,Gaskell1988,Popovic2012}; see Fig.~\ref{fig:dpeprofiles}. 
This hypothesis has been tested and is no longer favored based on a number of theoretical and observational results.  First, the velocity separated peaks observed in the broad lines of these systems  are not considered to be the signature of an SMBHB \citep{Chen1989}.  From Kepler's laws, the orbital velocity should not exceed the dispersion of the gas around a single BH \citep[e.g.,][]{Eracleous1997}, so velocity splitting of the line profile is seen in only a very small fraction of orbital configurations \citep{Shen2010}.  Second, reverberation mapping campaigns for individual double-peaked emitters \citep[e.g., 3C~390.3,][]{Dietrich1998,OBrien1998} show that both shoulders of the broad line respond to changes in the photoionizing continuum at the same time.  Such a straightforward response is not expected for a system with two central sources and two BLRs.  Finally, the long-term radial velocity curves of the broad lines are inconsistent with the SMBHB hypothesis (e.g., see Fig.~\ref{fig:dpevar} taken from \citealt{Eracleous1997} and see also \citealt{Gezari2007,Lewis2010,JLiu2016}).

Although double-peaked broad lines are now thought to originate in the outer regions of the BLR disk \citep[e.g.,][]{Eracleous1994,Storchi-Bergmann2017}, the body of work testing the SMBHB hypothesis for this class of objects raises important points relevant for ongoing searches.  First, quasar variability can mimic the signal of an SMBHB on many timescales \citep{Eracleous1997,Halpern2000}. Thus, additional work in the area of time-domain spectroscopy of AGN is needed to identify unique signatures of orbital motion in an AGN. Second, in the interest of testing the nature of binary candidates, the exercise of ruling out a class of objects may be possible sooner than confirming a binary.  As an example, \citet{Eracleous1997} and, more recently, \citet{JLiu2016} have demonstrated for many double-peaked emitters that unphysically large black hole masses are required to explain the radial velocity curves in the context of the SMBHB hypothesis.
Finally, as already pointed out for double-peaked NLR emission lines (see Sect.~\ref{sssect:obs_kpc_opt}), also in this case it should be noted that the gas may be characterized by complex kinematics (i.e., inflows and outflows powered by AGN and/or star formation activity) that could produce the observed double-peaked profiles without necessarily requiring the presence of a pair of nuclei.

\begin{figure*}[ht]
\centering
\vspace{0.0pt}
\includegraphics[width=0.7\columnwidth,height=6.5cm, angle=0]{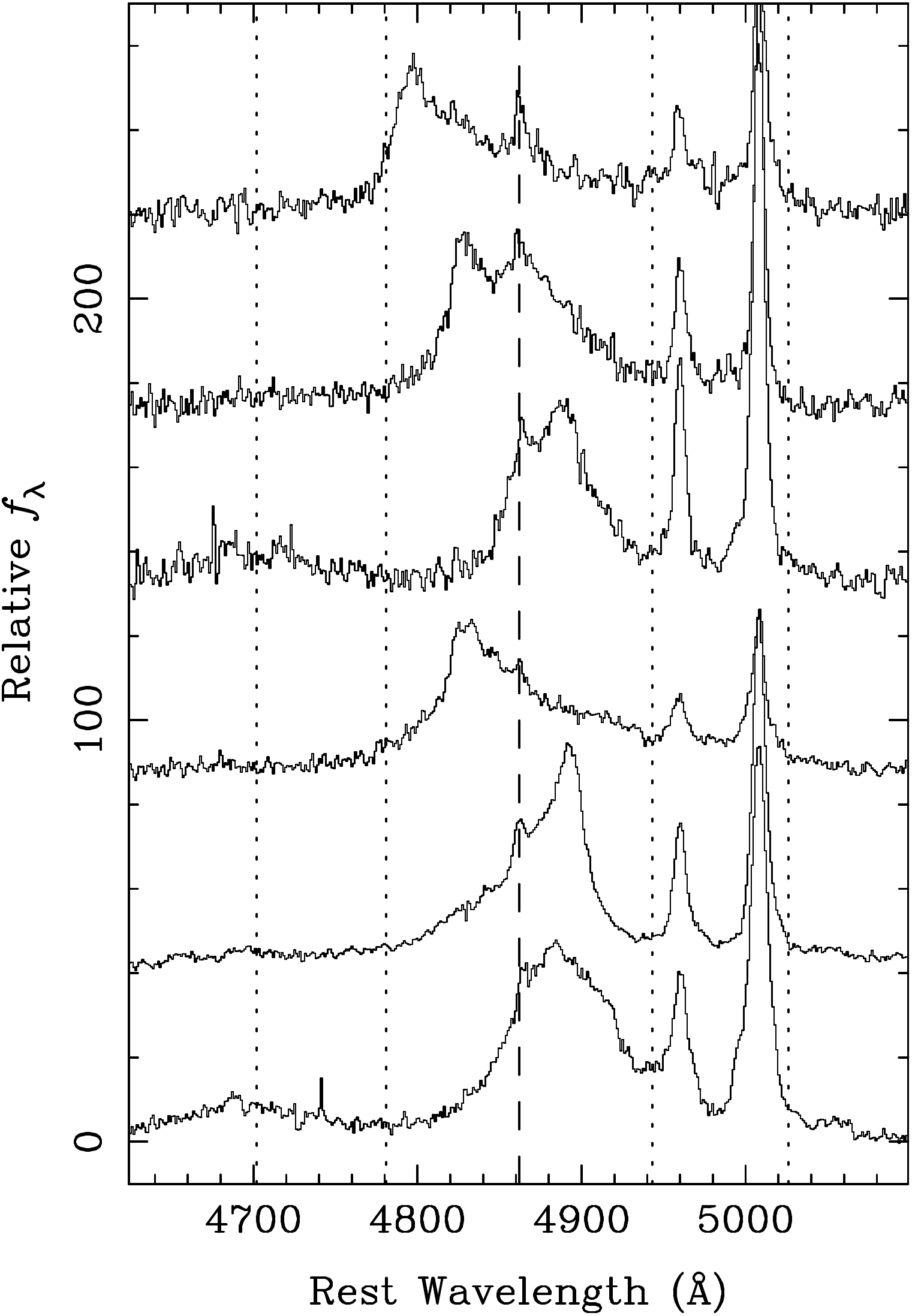}
\hspace{2cm}
\includegraphics[width=0.8\columnwidth,height=6.cm,angle=0]{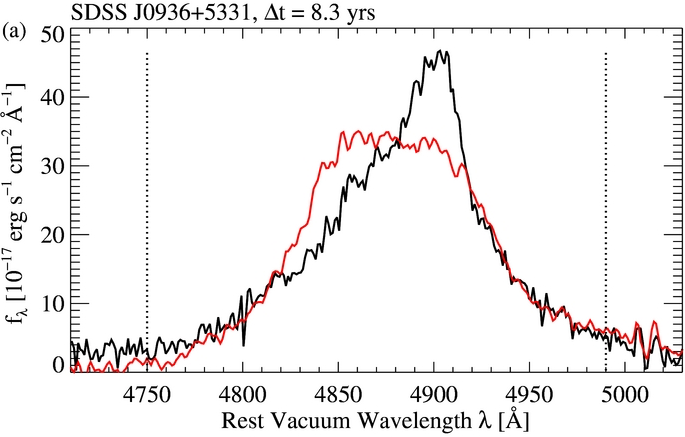}
%\vspace{-0.0pt}
%\textcolor{blue} {
\caption[]{SMBHB candidates identified through optical H$\beta$ offset. \textit{Left panel}:  these candidates were selected on the basis of the presence of broad H$\beta$ emission lines that are velocity shifted relative to the narrow  [O~\textsc{iii}]~$\lambda\lambda4959, 5007$ emission lines. The dashed line marks the rest wavelength of H$\beta$, and the dotted lines mark $\pm 5,000$ and 10,000~km~s$^{-1}$ relative to the dashed line.  Thus, the broad H$\beta$ peaks are offset by a few $\times 1000$~km~s$^{-1}$. This may be the observational consequence of a scenario where one of the BHs in a SMBHB is active and its orbital motion induces a periodic radial velocity shift in the broad H$\beta$ emission line relative to the narrow lines in the spectrum.  Such a system would likely have a sub-pc separation corresponding to a period of the order of decades to hundreds years. Adapted from \citet{Eracleous2012}.
\textit{Right Panel}: example of the cross-correlation analysis applied to measure the velocity shift of the broad H$\beta$ between two epochs. The broad H$\beta$ in the SDSS spectrum is shown in black, while the follow-up observation is in red. The spectral range of the cross-correlation analysis is marked by the dotted vertical lines. Negative values of the cross-correlation analysis mean that the emission line in the follow-up spectrum needs to be blue-shifted to match the SDSS spectrum. From \cite{liu14}.}
\label{fig:dvcan}
\end{figure*}

\begin{figure*}[t]
\centerline{
\includegraphics[scale=0.35]{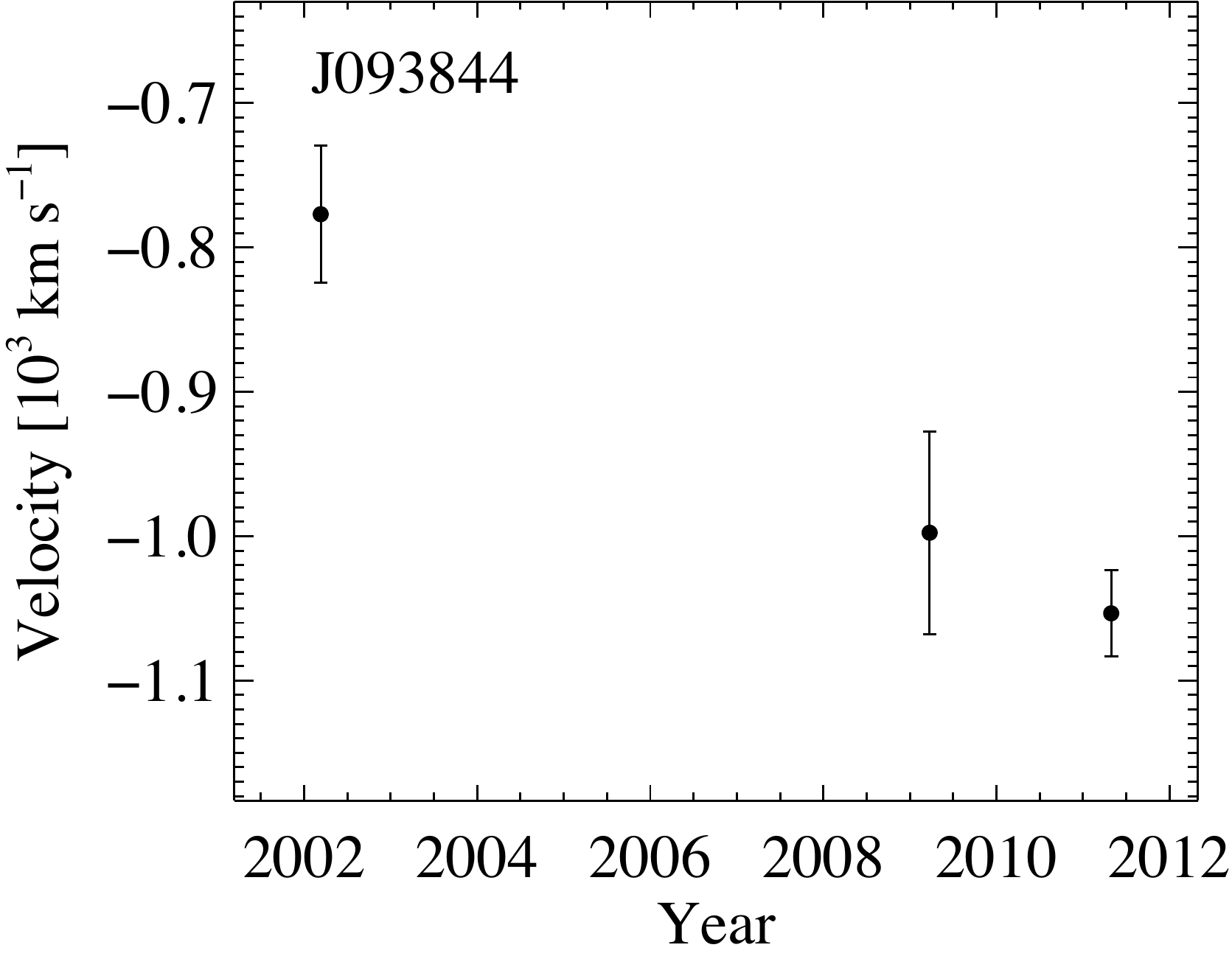}
\hspace{-0.6cm}
\includegraphics[scale=0.35]{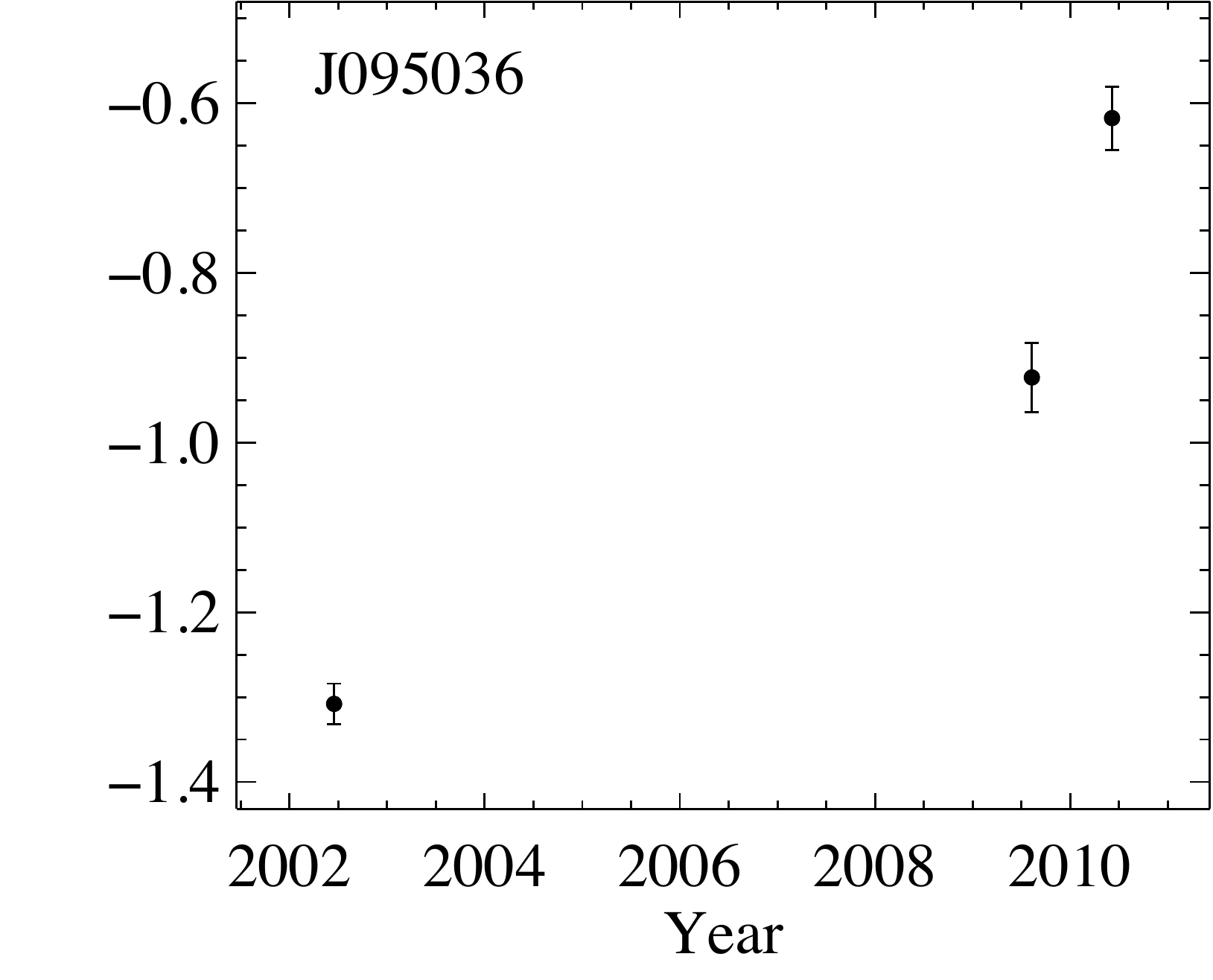}
\hspace{-0.6cm}
\includegraphics[scale=0.35]{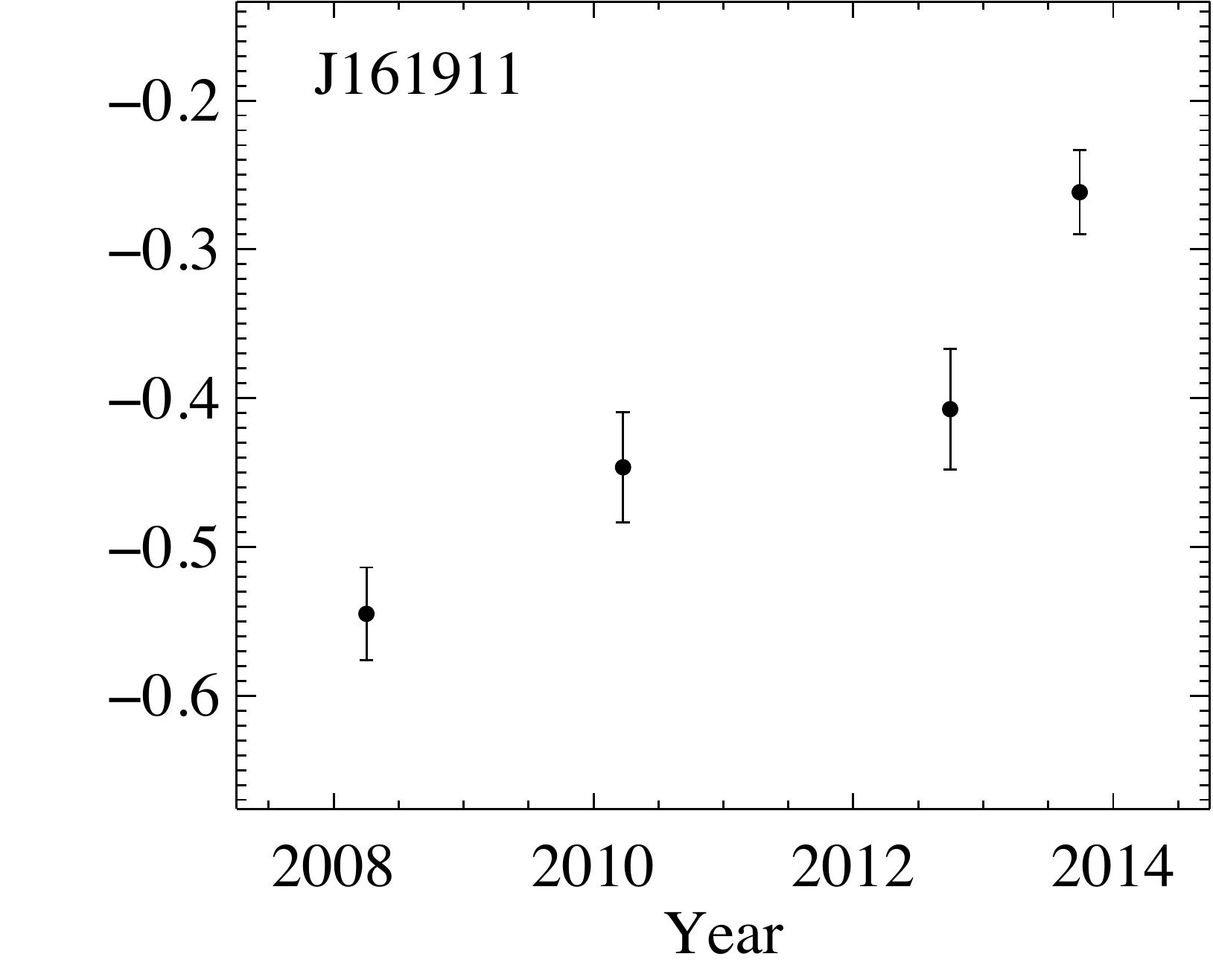}
}
\vskip -0.15truein
\caption{Radial velocity curves candidate SMBHBs at sub-parsec separation from \citet{runnoe17}.  Spectroscopic searches for SMBHBs are sensitive to systems with periods of the order of tens to hundreds of years. Thus, the expected signal of the binary is a monotonic increase or decrease of the radial velocity measurement over the duration of the monitoring campaign.  These are strong candidates in the sense that they display this signal and the SMBHB hypothesis cannot be ruled out based on current analyses.} 
\label{fig:bbhrv}
\end{figure*}

{\it Velocity shifts in single-peaked emitters.} The ``single-line spectroscopic binary'' case is still a viable avenue to search for SMBHBs among AGN with single-peaked broad emission lines. The first candidate SMBHBs were identified based on their velocity-offset broad emission line profiles in single-epoch spectra, starting with 3C~227 and 3C~668 \citep{Gaskell1983,Gaskell1984}.  Time-domain spectroscopic follow-up can then reveal the signal of orbital motion in the velocity shift of broad emission lines between two or more epochs of observation.  Recent investigations adopting this approach have monitored both AGN with velocity offset lines and also apparently normal AGN with broad lines located at their systemic velocity, corresponding to SMBHBs viewed in conjunction. 

There are of order $10^2$ SMBHB candidates identified via velocity offsets between broad H$\beta$ (see, e.g.,  Fig.~\ref{fig:dvcan}) and the host galaxy redshift that have been selected from SDSS quasar catalogs \citep{tsalmantza11,Eracleous2012,liu14}. Spectroscopic monitoring of these candidates has been used to measure or limit subsequent shifts of the broad lines \citep{decarli13} and place limits on the properties of the candidates \citep{liu14,Runnoe2015,Guo2019}.  \cite{runnoe17} presented radial velocity curves for the \citet{Eracleous2012} candidates probing timescales up to 12 years in the rest frame with 3--4 spectra per object.  They identified 3 candidates (SDSS J093844.45+005715.7, J095036.75+512838.1, J161911.24+501109.2) for sub-parsec separation SMBHBs which show the monotonic changes in their radial velocity curves (Fig.~\ref{fig:bbhrv}) that are expected for SMBHBs with periods of decades to hundreds of years.  They also adapt the methodology developed for the double peaked emitters of placing limits on the masses and separations of the hypothetical SMBHBs, but the limits they obtained were not yet restrictive.

Searches for SMBHBs among regular AGN (i.e. corresponding to SMBHBs seen in conjunction) have been conducted based on Mg~$\textsc{ii}$~$\lambda$2798 \citep{ju13,wang17} and H$\beta$ \citep{shen13}.  Using the distribution of observed accelerations measured for the 521 AGN in SDSS DR7 with multiple quality spectra and $z<0.9$, \cite{shen13} placed limits on the properties of SMBHBs among the quasar population.  If the accelerations are attributable entirely to SMBHBs, they infer that most of the quasars in their sample must be in binaries, the inactive black hole must usually be more massive, and that the separation is at most a few times the radius of the BLR \citep[which they estimate to be $0.01-0.1$~pc using the radius-luminosity relationship, e.g.,][]{bentz09}.  

Motivated by the described advances in observational searches for sub-parsec SMBHBs, \citet{nguyen16} have developed a semi-analytic model to describe spectral emission-line signatures of these systems. Based on this study, they find that the modeled profiles show distinct statistical properties as a function of the semimajor axis, mass ratio, eccentricity of the binary, and the degree of alignment of the triple disk system (including two mini-disks and a circumbinary disk). This suggests that the broad emission-line profiles from SMBHB systems can in principle be used to infer the distribution of these parameters and as such merit further investigation.  \citet{nguyen19,nguyen19b} also find that their modeled profile shapes are more compatible with the observed sample of SMBHB candidates \citep[drawn from the observations reported in][]{Eracleous2012, Runnoe2015, runnoe17} than with their control sample of regular AGN. Furthermore, they report that if the observed sample of SMBHBs is made up of genuine binaries, it must include compact systems (with orbital separations $\log(a/M) \approx 4.20\pm 0.42$) with comparable masses and misaligned mini-disks. If the considered SMBHB candidates are true binaries, this result would suggest that there is a physical process that allows initially unequal mass systems to evolve toward comparable mass ratios or point to some, yet unspecified, selection bias. Similarly, if upheld for confirmed SMBHBs, this finding would indicate the presence of a physical mechanism that maintains misalignment of the mini-disks (or causes them to be warped) down to sub-parsec binary separations.

However, if the H$\beta$ velocity shifts are the result of long-term quasar variability, then close, massive binaries are disfavored.  
Regular quasar spectral variability is the source of the biggest caveat to finding SMBHBs via this approach.  On short timescales of months, normal quasars can produce velocity shifts comparable to what is seen in SMBHB candidate samples \citep{barth15}. The long-term spectroscopic variability of regular quasars has not been well characterized, therefore this remains a critical uncertainty for finding SMBHBs with the radial velocity technique.  The baseline measurement of the spectroscopic variability for a control sample of regular AGN on long timescales is the promise of archival data and upcoming spectroscopic surveys (see Sect.~\ref{sssec:future_agn_variability}).  Because regular quasar variability can easily masquerade as the signal of orbital motion on timescales much shorter than the orbital period which is expected to be long, it is extremely important to conduct as many supporting tests as possible. 

\subsubsection{Peculiar broad emission-line ratios as signatures of sub-pc SMBHBs}
\label{sssec:subpcspec2} 

Close SMBHBs (with a semimajor axis $\lsim 0.1$~pc) have also been proposed to significantly alter the luminosity ratios between different pairs of broad emission lines (BELs produced in the BLR) by \cite{Montuori11}. They considered the scenario described in Sect.~\ref{ssec:subpctheory}, in which a binary is orbiting in the low-density gap region within a denser circumbinary disk. They further followed the common assumption that the secondary SMBH is more likely to be active, being closer to the circumbinary disk inner edge and having a lower relative velocity with respect to the outer gas  \citep{artymowicz1994, gunther02, hayasaki07, Roedig11, farris14}. As a first test, they assumed that the binary is wide enough to neglect any contribution of the outer circumbinary disk to the brightest BELs at optical-UV rest-frame wavelengths.

Within this simple framework, \cite{Montuori11} analyzed the tidal truncation that the primary exerts onto the secondary mini-disk, limiting the extent of the BLR. Whenever the gas responsible for the emission of a given Broad Emission Line (BEL) lies outside the secondary Roche lobe radius, it is ejected toward the circumbinary disk, decreasing the gas density in the vicinity of the active SMBH and therefore decreasing the BEL luminosity. This can result in an anomalous luminosity ratio between lines that are produced by gas with different spatial distributions, and is therefore expected to be more {\it effective} when comparing the observed fluxes of lines with significantly different ionization potentials. 

\cite{Montuori11} checked the effect of binaries of different masses, luminosities, and separations on two specific line ratios: $F_{\rm MgII}/F_{\rm H\beta}$ and $F_{\rm MgII}/F_{\rm CIV}$. They computed the BEL flux under the assumptions of the ``locally optimally emitting clouds'' model \citep[e.g.][]{Baldwin95}. Each cloud is then assumed to emit a line luminosity depending on the incoming AGN continuum flux and the local gas density, based on a grid of models computed using the photoionization code \textsc{cloudy} \citep{Ferland98}.\footnote{For a detailed description of the procedure, we point the reader to Sect.~3 of \cite{Montuori11}.}

The $F_{\rm M gII}/F_{\rm H\beta}$ ratio has the advantage of being measured in large optical surveys (such as SDSS, see Sect.~\ref{sssec:optical_kpc}) within the redshift range $0.4 \lsim z \lsim 0.8$, allowing for a cross search for shifted/asymmetric BELs and peculiar $F_{\rm MgII}/F_{\rm H\beta}$ values. On the other hand, the expected number of SMBHBs observable at low redshift is at most of the order of a few \citep{Volonteri09}. The analysis of the dependence of this ratio on the binary parameters demonstrated that peculiar values are expected only in an extremely limited range of the parameter space, due to the similar ionization potential of the  two elements, thus strongly limiting the utility of the $F_{\rm M gII}/F_{\rm H\beta}$ ratio.\footnote{A different and less constraining strategy in the redshift range $0.4 \lsim z \lsim 0.8$ is to search for Type I (i.e., unobscured) AGN having all the observable broad lines peculiarly faint and broad, as observed for example in the SMBHB candidate 4C+22.15, independently selected because of the shifted BELs \citep{Decarli10}.}

The $F_{\rm MgII}/F_{\rm CIV}$ line ratio can instead probe the AGN population observed in large optical surveys at $z \gsim 2$, allowing for the SMBHB search during the cosmic high noon, when most of the galaxies and SMBHs where at the peak of their growth. This second ratio is more affected by the tidal effect of the primary, since the two lines are preferentially emitted in spatially distinct regions. 

Indeed, the $F_{\rm MgII}/F_{\rm CIV}$ ratio was found to be reduced by up to an order of magnitude for close binaries, with respect to the $\sim$0.3--0.4 ratio expected for unperturbed BELs around isolated SMBHs. It must be noticed, however, that the $F_{\rm MgII}/F_{\rm CIV}$ ratio is not expected to keep decreasing indefinitely as the binary and the circumbinary disk shrink: at semimajor axes $< 0.01$~pc the inner edge of the circumbinary disk is close enough to the active secondary to emit efficiently the MgII BEL. This second effect has been described with a simple analytical model by \cite{Montuori12}. Since the structure of the circumbinary disk can in principle differ significantly from that of isolated BLRs, \cite{Montuori12} decided to take advantage of the numerical simulations of SMBHB--circumbinary disk systems presented in \cite{Sesana12}. The simulation results were post-processed with the code \textsc{cloudy} to constrain the relative contribution of the gas within the gap and in the outer disk. 

The results of the numerical analysis were in remarkable agreement with the simple analytical predictions: an order of magnitude reduction in $F_{\rm MgII}/F_{\rm CIV}$ is possible for semimajor axes $\sim(0.01-0.2)(f_{\rm Edd}/0.1)^{1/2}$~pc, for a mass of the secondary SMBH between $10^7$ and $10^9$~M$_{\odot}$, and a binary mass ratio $q = 0.3$.\footnote{$f_{\rm Edd}$ here corresponds to the bolometric luminosity of the accretion process normalized to the Eddington limit of the secondary.} These separations correspond to orbital periods in the range $\sim$(20-200)(f$_{\rm Edd}/0.1)^{3/4}$ yr, making the search for other SMBHB signatures such as BEL velocity variations and continuum variability studies possible but, due to the long monitoring required, extremely challenging.\footnote{All the above-mentioned estimates assume circular orbits. At fixed semimajor axis, an eccentric binary could have a different truncation radius, depending on the ability of the material within the secondary Roche lobe to readjust on the orbital period time-scale. We stress, however, that due to the existence of a limiting orbital eccentricity $e_{\rm crit} \approx 0.7$  for binaries co-rotating with the circumbinary disk \citep{Roedig11}, the upper limit of the semimajor axes quoted above increases by a factor 2--3 at most.}

\paragraph{Sub-parsec separations: optical-X-ray synergies}
\label{sssec:subpc_opt_xray}
A special case -- and a clear indication of the invaluable power of synergies at different wavelengths -- is that of MCG+11-11-032, a Seyfert~2 galaxy at z=0.036, originally selected from the SDSS spectrum showing a double-peaked emission-line profile in [OIII]. 
Its \swift/BAT 123-month light curve presents almost regular peaks and dips  every $\sim$25 months, while the \swift/XRT spectrum shows two narrow emission lines at rest-frame energies of $\sim$6.16~keV and $\sim$6.56~keV \citep{Severgnini2018}, interpreted as a double-peaked iron K$\alpha$ emission-line profile and possibly ascribed to a circumbinary accretion disk linked to two sub-pc scale SMBHs (see Sect.~\ref{ssec:subpctheory}). 
These results make the binary SMBH hypothesis highly plausible. 
Intriguingly, there is also a remarkable agreement between the putative SMBH pair orbital velocity derived from the \swift/BAT light curve and the velocity offset derived from the rest-frame energies of the two X-ray lines. Further support to this picture has recently come from a careful time-domain analysis of the \swift/BAT data (Serafinelli et al., in prep.). 
While firmly establishing with present and future facilities the presence of a sub-pc SMBH in this system will remain quite challenging, this kind of investigation reflects the capability of X-ray and optical data in unveiling SMBH pair candidates also in obscured sources and the potentialities of all-sky X-ray monitorings in this uncharted territory. 

\subsection{Photometric searches for sub-pc SMBHBs}
\label{sssec:subpc_photometry}

Another proposed method to search for compact active SMBHBs is to identify
quasars with photometric variability. One generic conclusion of the hydrodynamic simulations, discussed in Sect.~\ref{ssec:subpctheory}, is that the net accretion rate onto the SMBHs is significant and can produce quasar-like luminosities ~\citep{dorazio13,farris14,ShiKrolik2015,Ragusa+2016}. Additionally,  the mass accretion rate onto the SMBHs is modulated
periodically~\citep{Hayasaki2007,macfadyen08,roedig12,dorazio13,farris14,ShiKrolik2015}, which may be translated into periodic modulation of the brightness of the source.

While periodic modulations in the accretion rate appear inevitable, relativistic Doppler boost can also cause an SMBHB to exhibit periodic variability.  For example, a very unequal binary ($q\lsim 0.04$) may accrete steadily. However, if the binary is sufficiently compact, such that orbital velocities exceed $v/c\gsim$ few per cent, we expect periodic modulations from special relativity alone~\citep{Dorazio+2015}. More specifically, if the optical luminosity arises in gas bound to the moving BHs (e.g., in the mini-disks seen in hydrodynamical simulations), the binary appears blue-shifted (and brighter for typical values of optical spectral indices), when the more luminous SMBH (typically the less massive SMBH) is moving towards the observer, and vice versa. We note that this scenario is expected to produce smooth quasi-sinusoidal variability, whereas in the case of periodic accretion the periodicity may be more ``bursty''.

The advent of the modern time domain surveys, which systematically scan large areas of the sky, has allowed statistical searches for periodic variability in large samples of quasars. More specifically, \citet{Graham2015} analyzed 245,000 spectroscopically confirmed quasars from the Catalina Real-time Transient Survey (CRTS) and identified 111 candidates.
\citet{Charisi2016} analyzed a sample of 35,000 spectroscopically confirmed quasars from the Palomar Transient Factory (PTF) and identified 33 candidates with significant periodicity.

The two samples are somewhat complementary; the candidates from CRTS are relatively bright and have periods of $\sim 2-6.5$ yr, whereas the PTF candidates have fainter magnitudes (and higher redshifts) and periods of $\sim 150-800$ days, owing to the higher photometric precision and higher cadence of the survey. Accounting for selection effects in the CRTS sample (candidates were selected preferentially at the brightest end of the sample), the occurrence rate of SMBHB candidates would be similar in both samples, $\sim 1/1000$. Additionally, \citet{Charisi2016} examined the candidates as a population and found a preference for low-mass ratio binaries ($q=0.01$) in both samples (although this may be a selection effect, since both searches were sensitive to sinusoidal periodicity).

Several other individual candidates have recently appeared in the literature, either from the analysis of smaller samples or from serendipitous discoveries. For instance, \citet{Liu2015} reported one candidate with period of $\sim$ 550 days and a separation of 0.006~pc, from the PanSTARRS Medium Deep Survey. However, follow-up observations revealed that the periodicity of this candidate is not persistent \citep{Liu2016}. \citet{Zheng2016} identified an additional candidate in CRTS with two periodic components in the variability ($\sim 741$ days and $\sim 1500$ days) with the characteristic frequency ratio 1:2. This candidate was included in the sample of \citet{Graham2015}, but was not identified as periodic.
Additional candidates have emerged from the analysis of historical long-baseline light curves of bright AGN \citep{Li2016,Bon2016} and the analysis of blazar light curves from the Fermi gamma-ray telescope \citep{Sandrinelli_et_al_2016,Sandrinelli_et_al_2018}. Furthermore, \citet{Dorn_Wallenstein2017} claimed the detection of a periodic AGN in PTF, which was subsequently disputed by \citet{Barth2018}, who highlight the importance of interpreting correctly the null hypothesis simulation tests and and performing carefully the calculation of the false alarm probability.

The controversial nature of the above candidates illustrates that identifying quasars with periodic variability is challenging, mainly because the periods are relatively long compared to the available baselines and the underlying variability of quasars is stochastic. This was also clearly demonstrated in the case of PG~1302-102 -- the first (and brightest) candidate that emerged from the time domain surveys -- the statistical significance of which remains controversial.
\citet{Charisi2015} showed that the periodogram peak of PG~1302-102 is statistically significant at the 1 per cent level compared to a Damped Random Walk (DRW) model, but not significant compared with red noise variability.\footnote{This 1 per cent does not include the trial factors that are associated with large sample from which the source was chosen.} \citet{Dorazio+2015} showed that a sinusoidal model is preferred compared to pure DRW noise, whereas \citet{Vaughan2016}, with a similar analysis, reached the opposite conclusion. The main difference between the two is the choice of the DRW parameters. 
\citet{Liu2018} added more recent data from the ASAS-SN \citep{ASASSN_1,ASASSN_2} survey and suggested that the significance of the periodicity decreases, even though a sinusoidal model is still preferred compared to a DRW model.
	
\citet{Vaughan2016} also suggested that the incomplete knowledge of the underlying variability can lead to false detections and therefore the above samples are likely contaminated by false positives. This was also suggested by \citet{Sesana2018}, who calculated the GW background from the inferred population of binaries identified as periodic quasars. They found that, in order for the population to be consistent with the current upper limit from PTAs, the masses or the mass ratios of the binaries have to be unusually low. Therefore, as noted by \citet{Liu2016}, it is crucial to continue monitoring the candidates, in order to test the persistence of the periodicity and distinguish the genuine SMBHBs from the false detections.

An alternative approach is to search for independent lines of evidence for the binary nature of the candidates. These include: 1) \textit{Multiple periodic components in the optical variability with a characteristic frequency pattern}.
For $q\gsim 0.3$, several simulations have found that the inner cavity in the accretion disk becomes lopsided
\citep{macfadyen08,shi12,noble12,dorazio13,farris14}. In this case, the most
prominent period is 3-8 times longer than the orbital period. 
2) \textit{Multi-wavelength signatures of relativistic Doppler boost.} As mentioned above, the relativistic Doppler boost can explain the quasi-sinuosoidal variability in optical bands. Additionally, if the UV/X-ray luminosity also arises in gas bound to the BHs, their luminosity should vary in tandem with the optical, but with amplitudes which depend on the spectral curvature in the respective bands. Therefore, the relativistic Doppler boost scenario offers a robust prediction, which can be tested with multi-wavelength data \citep{Dorazio+2015,Charisi2018}. The possible application of this technique will be further discussed in Sect.~\ref{sssec:future_sims}. 3) \textit{Characteristic infrared echoes.} The optical/UV luminosity is produced in compact regions close to the central BH.  
Subsequently, it is reprocessed by the dusty torus, and it is re-emitted in IR. When the torus is illuminated anisotropically, e.g., if the source of the torus irradiation is the moving secondary BH in a binary, the IR light curve shows characteristic time lags and amplitudes that can help distinguishing binaries from a single central SMBH \citep{Dorazio2017,Jun2015}. 4) \textit{Periodic self-lensing flares.} If the binary is not too far from face-on (with inclination $\lsim 30$~deg), the accretion disk of one BH is lensed when it passes behind the other BH, which can produce bright X-ray/optical flares ~\citep{Haiman2017,DorazioDiStefano2018}.
{5) \textit{X-ray outburst from tidal disruption events.}
The tidal disruption of stars by SMBHs (tidal disruption events; TDEs) causes a characteristic X-ray outburst lightcurve, which declines as the stellar debris is accreted by the black hole (see \citealt{komossa_zensus2016} for a review). Lightcurves of TDEs which occur in SMBHB systems look characteristically different from TDEs of single black holes \citep{liu2009}. The lightcurves show dips and recoveries, as the second SMBH acts as a perturber and temporarily interrupts the accretion stream on the primary. Simulations by \citet{liu2014} have
shown, that the lightcurve of the TDE in the inactive galaxy SDSS\,J1201+3003 \citep{saxton2012} is consistent with a binary SMBH model with a primary mass of 10$^6 M_{\odot}$, a mass ratio $q=0.1$ and a semi-major axis of 0.6 milli-pc. This method is of particular interest, since TDE rates in galaxy major mergers are strongly enhanced by up to two orders of magnitude with respect to  single SMBH \citep{Lietal2017}, and it allows to identify SMBHBs in {\em non-active, quiescent} host galaxies; conversely, essentially all other SMBHB identification methods discussed in this review require that at least one, or both, SMBHs are active.}

\part{Future perspectives}
%\section{Future perspectives}
\label{sec:Future_perspective}

In this section we describe the observational strategies that will be available in the future to uncover dual and binary AGN, using ground- and space-based facilities operating in different energy bands.
As widely discussed in the previous sections, depending on the spatial separation, different techniques need to be applied in order to detect AGN dual systems, AGN in close binaries, and  SMBHs during the latest phases of the inspiral and  proper merger. We will explore several of these techniques, starting from systems at kpc separations down to $\mu$pc where the SMBHBs emit GWs. We will
describe recent advances in numerical simulations of circumbinary disks around SMBHBs and enter the realm of GW observations.

As discussed in Sect.~\ref{ssec:BH_Pairs_Observations}, so far, most of the dual- and multiple-AGN on kpc scales have been found serendipitously. In spite of the huge effort put into carefully constructing selected samples using a number of observing criteria, none of these appeared to be efficient in discovering  dual AGN so far. In the future our search strategy  will  shift toward reasonably well identified samples obtained by cross-matching various multi-waveband survey catalogs. Optical/NIR or MIR spectroscopy will be needed to infer the redshift of the host galaxies and confirm a physical association. In order to discover kpc-scale dual AGN, we describe below (without attempting to be comprehensive) planned surveys and future instruments dedicated to their search.  

To probe the 10--100~pc scales, as described in Sect.~\ref{ssec:stallingBH}, we need  radio interferometric observations with very high-resolution. Having contemporary AGN activity in the radio is much less likely than having 
dual AGN in the optical and X-rays, but at low accretion rates AGN become radio-loud, therefore sensitive radio instruments will be essential probes in this regime. 
In particular, two techniques can be mentioned: one relies on finding powerful transient radio sources -- signalling new AGN activity -- near an already known 
AGN. The other one consists of revealing dual activity comparing precision astrometry in the optical and the radio bands. 

The sub-pc regime presented in Sect.~\ref{ssec:subpbobs} will remain the most challenging, although AGN variability studies are quite promising. The breakthrough in this field will come with the advent of GW astronomy, that provides a direct way of detecting orbiting SMBHBs probing scales between $\sim 10\mu$pc and a few milli-pc, when the black holes are coalescing or in the verge of merging, years to centuries before the final plunge.

\section{Search for dual and multiple AGN in the era of surveys}
\label{ssec:future_surveys}
\subsection{Radio surveys in the cm waveband}
\label{sssec:radiofuture}

The Next Generation Very Large Array -- ngVLA and the Square Kilometre Array -- SKA (and its precursors: APERTIF, ASKAP, MeerKAT) are being (have been) designed to be able to carry out deep surveys that cover a large fraction of the sky. These interferometers will be complementary in various ways, and they will be synergistic with ESFRI-listed facilities (CTA, KM3Net and E-ELT) and other major observatories.

\begin{table*}[t]
\centering
\caption{Present and future high-resolution continuum radio surveys}\label{tabradiosrv}
\begin{tabular}{l|c|c|c|c|c} \hline \hline
Telescope & Survey & Resolution & Frequency  & Sensitivity & Sky      \\ 
% \hline
          &        & [arcsec]   & [GHz]      &rms [$\mu$Jy]& coverage \\ 
\hline
VLA	      & NVSS   & 45"        & 1.4        & 450         & $\delta>-40$~deg \\ % NVSS FoV: 4x4 deg.
VLA       & FIRST  & 5"         & 1.4        & 150         & $10^4$~deg$^2$ \\ % FIRST FoV: 46'.5 × 34'.5 images
VLA       & VLASS  & 2.5"       & 2--4       &  70         & $\delta>-40$~deg \\ % on-the-fly mosaicing
ASKAP     & EMU    & 10"        & 1.13--1.43 &  10         & $\delta<+30$~deg \\ % 36x1.2 deg FWHM beams: ~30 deg^2 FoV
MeerKAT   & MIGHTEE& 6"         & 0.9--1.67  &   1         & $20$~deg$^2$ \\
SKA1-MID* & SASS1  & 0.5--1"    & 0.95--1.76 &   4         & $\delta<+30$~deg\\ % visible 3 Pi
SKA2*     & SASS2  &$\sim 0.1"$ & 0.95--1.76 &   0.1       & $\delta<+30$~deg\\ % steradian
% 2.5 GHz band centred at ~10 GHz; Band 5 4.6--13.8 GHz
\hline \hline
\end{tabular}

*Indicative only. SKA surveys will have various tiers with different parameters.
\end{table*}
 
The VLA Sky Survey\footnote{\url{https://science.nrao.edu/science/surveys/vlass/}} 
is being carried out at 2--4~GHz, and therefore will have improved sensitivity and angular resolution compared to its predecessors at 1.4 GHz (NVSS and FIRST; see Table~\ref{tabradiosrv}). The large observing bandwidth will allow for spectral index measurements, making these data ideal to search for flat-spectrum dual-AGN cores for example. The VLASS potentials for AGN mergers with a separation less than 7~kpc have been described in a white paper by \citet{Burke-spolaor2014,Burke-Spolaor2018}. The ngVLA \citep{Murphy2018}
would greatly expand the instantaneous frequency range (1-116~GHz), the field of view, and the resolution of the array (with baselines at least 300~km). The ngVLA will be capable of probing active massive black holes and their feedback below $10^6~M_{\odot}$ \citep{Nyland2018}, entering a very interesting regime for dual-AGN activity studies.
 
ASKAP \citep{Johnston2007} and MeerKAT \citep{Norris2011} are SKA precursors employing different technologies, but both are very fast survey machines (Table~\ref{tabradiosrv}). They will survey the Southern sky down to $\sim \mu$Jy sensitivities, and produce catalogs for tens of millions of radio sources albeit with limited resolution.
 The mid-frequency telescope of the phase-I SKA (SKA1-MID) will have more power to distinguish between star formation and AGN activity (with baselines up to $\sim$150~km).
SKA1 Continuum Surveys \citep[SASS1;][and references therein]{Prandoni_Seymour2015} top priority science cases include the star formation history of the Universe, the role of black holes in galaxy evolution, gravitational lensing and more. These surveys will detect the bulk ($\sim$ 90 per cent) of the AGN population, the majority of which is missed in current radio surveys \citep{Smolcic2015}.

While SKA1-MID will have a resolution of $\sim 0.1-1$~arcsec, the core of the array can be coherently phased up to mimic a single radio telescope. By combining this very high sensitivity component with other radio telescopes around the world one may form a powerful very long baseline interferometry network, a concept known as SKA-VLBI \citep{Paragi2015}. This is particularly relevant for us, because very sensitive VLBI observations is the only way to directly address the (hitherto) missing population of dual-AGN with a separation of a few (tens of) parsec. SKA2 will have a sensitivity of at least an order of magnitude higher than SKA1, and its resolution will be about 20 times better, therefore it will be competitive in resolution with current VLBI arrays. The SASS2 1.4 GHz survey of the southern sky will detect some 3.5~billion radio sources at high angular resolution \citep{Norris2015}. 
These surveys will contain a huge number of dual-AGN candidates, that can be further down-selected based on radio spectral index, multi-band properties, and variability. This latter might reveal episodic accretion from an otherwise inactive pair to an already well established AGN, like the one recently discovered next to Cyg-A in the mid-IR \citep{Canalizo2003} and in the radio \citep{Perley2017}. 
A possible form of episodic accretion is represented by TDE, that are found to occur dominantly in galaxies with post-starburst star formation history and merger origin \citep{Zabludoff_et_al_1996,French2016,Pfister_et_al_2019}. The accretion rates are expected to be higher in black holes with lower masses, such as in the case of white-dwarf - intermediate-mass black hole encounters.
Radio (and in fact multi-band) observation of TDEs next to and already active nucleus will reveal exotic pairs of dual-AGN. Note that these IMBH+WD encounters may also lead to GW radiation that might be detectable by LISA in the Local Group \citep{Rosswog2009,Anninos2018}.

Finally, it is worth noting that VLBI offers down to $\sim 10~\mu$as level astrometric accuracy for bright radio AGN \citep{Fey2015}. This has been matched recently with the \gaia\ spacecraft in the optical, allowing for synergistic studies of AGN positions for a large overlapping sample in the radio and optical for the first time. Objects with significant optical--radio positional offsets may serve as natural candidates for dual AGN that warrant follow-up observations \citep{Orosz2013}. According to the most recent studies based on new \gaia\ data, the majority of significant (mas or sub-mas level) \gaia--VLBI offsets occur downstream or upstream of the AGN jet \citep{Kovalev2017,Plavin2019}. This suggests that in general strong optical jet emission is present at least on 20--50~pc scales from the central engine. In turn, optical--radio offsets with position angles significantly different from that of the jet may indicate either dual AGN containing a radio-loud and a radio-weak companion, or offset/recoiling AGN. Further \gaia\ data releases and improved VLBI astrometric solutions hold the potential for identifying more dual AGN candidates.

\subsection{X-ray surveys}
\label{sssec:future_xray}

In the next decades, new X-ray observatories will greatly enlarge the population of known AGN, in particular at high redshift. The key players are \erosita, \athena\ and, possibly, \axis\ and \lynx. 
Table~\ref{tab1xrs}
summarizes the main properties of their scientific payload relevant to the  observational study of dual AGN.

\begin{table*}
\centering
\caption{Performance of the X-ray payload discussed in Sect.~\ref{sssec:future_xray}. Legenda: HEW~=~High Energy Width, averaged over the FoV (field-of-view); sensitivity: in erg~s$^{-1}$~cm$^{-2}$ in the 0.5-2~keV energy range; the sky coverage is at the ``Sensitivity'' flux.}
\label{tab1xrs}
\begin{tabular}{l|c|c|c|c|c} \hline \hline
Payload & HEW & FoV & Energy range (keV) & Sensitivity & Sky coverage \\ \hline
{\it eROSITA}/SRG	& 28" & 1.03$^{\circ}$  dm. &	0.3--10 & $\approx$10$^{-14}$ &	All sky \\
WFI/{\it Athena} & 6" & 40'$\times$40' & 0.2-12 & $\approx$3$\times$10$^{-17}$ &	30~deg$^2$ \\
Lynx & 0.5" & 22'$\times$22' & 0.5-10 & $\approx$3$\times$10$^{-19}$ & 20~deg$^2$ \\
\hline \hline
\end{tabular}
\end{table*}

\begin{figure}[t]
    \centering
    \includegraphics[angle=90,width=0.47\textwidth]{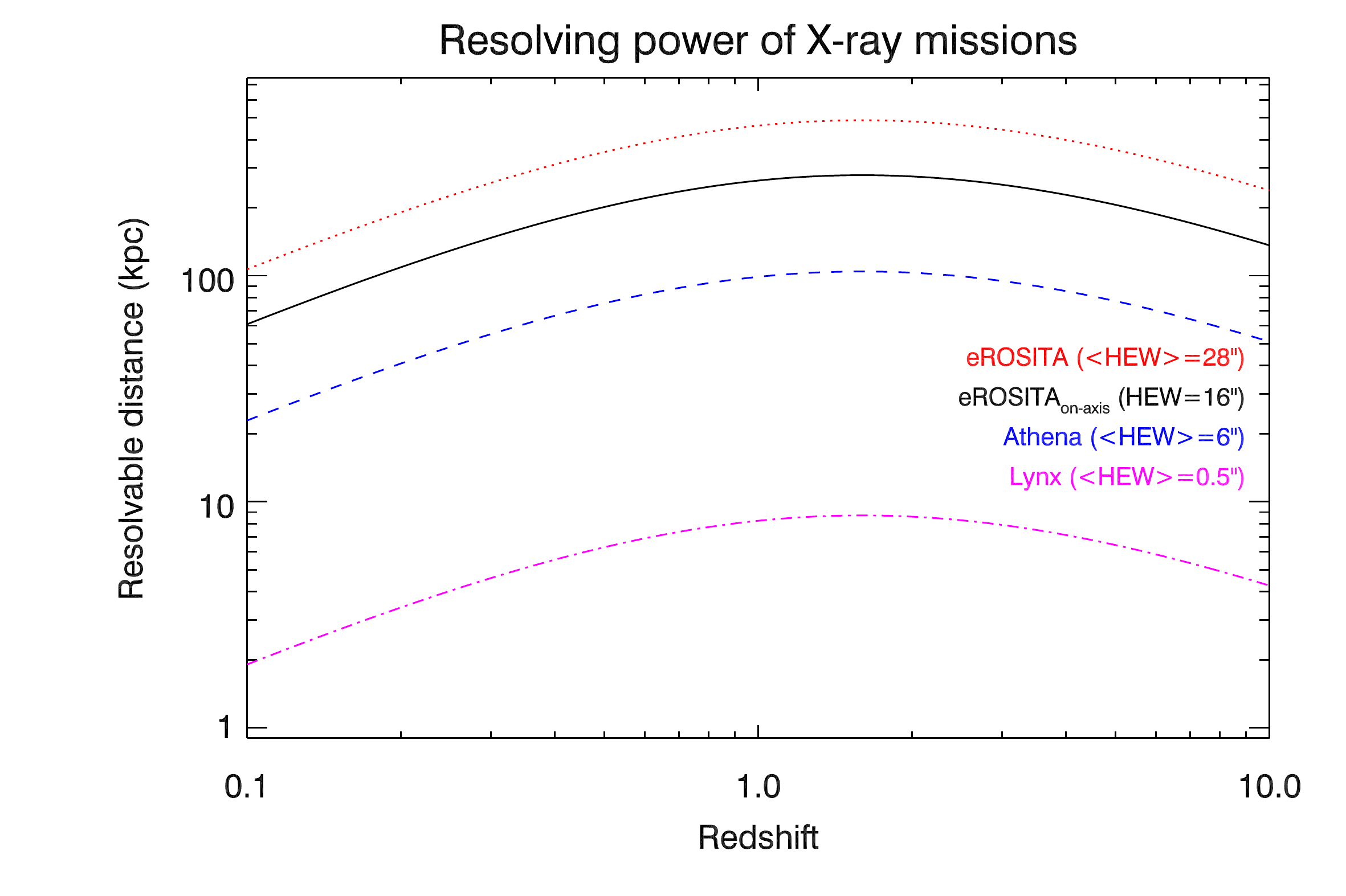}
    \caption{Resolvable distance (in kpc) versus redshift for the future X-ray facilities
    discussed in this paper. The quantity on the y-axis is twice 
    the angular resolution
    (in units of Half-Energy Width) following \cite{Ranalli2013}.
    From {\it top} to {\it bottom}: \erosita\ (average HEW), \erosita\ (on-axis), {\it Athena}, and {\it Lynx}.}
    \label{fig1xrs}
\end{figure}

\erosita\ \citep{merloni12_erosita} is the primary instrument on the Russian Spektrum-Roentgen-Gamma (SRG) mission. Successfully launched in July 2019, \erosita\ aims at performing a 4-year long survey of the X-ray sky. Building on the experience of the {\it ROSAT} All-Sky Survey \citep{Voges1999, Boller2016}, \erosita\ will be about 20 times more sensitive in the 0.5--2~keV energy band, while providing the first true imaging survey of the hard (2--10~keV) X-ray sky. Among the main scientific goals of \erosita, AGN studies feature prominently. \erosita\ aims at determining the accretion history of SMBHs by studying in unprecedented details the X-ray AGN luminosity function, in particular the still poorly understood luminosity-dependent fraction of obscured objects; studying the clustering properties of X-ray selected AGN at least up to $z \sim 2$; and identifying rare AGN sub-populations such as high-redshift, possibly highly obscured nuclei. This will be possible thanks to a large sample of about 3 million AGN to be detected during the nominal survey, of which several tens of thousands at redshift higher then 3, and a few thousands with bolometric luminosities larger than 10$^{46}$~erg~s$^{-1}$. The angular resolution will be comparable to that of \xmm, $\sim$16\arcsec\ Half-Energy Width (HEW) on-axis, and $\sim$28\arcsec\ averaged over the whole 1-degree diameter Field-of-View (FoV). This will limit the redshift range on which dual AGN with a separation $\le$100~kpc can be resolved to the local Universe ($z\le$0.3; see Fig.~\ref{fig1xrs}). The moderate spectral resolution ($\sim$130~eV at 6~keV, similar to present-day CCD-based instruments) will not allow efficient spectral separation of unresolved galaxy-AGN pairs. 
However, the \erosita\ observing cadence in the survey phase -- most of the sky will be revisited approximately every 6 months, with the poles being monitored with higher cadence -- will allow each location of the sky to get visited at least eight times in the four years of the mission, so detection of AGN via variability also in dual systems will be viable. Such a search will  however be possible only for relatively bright systems at the large separations allowed by \erosita\ spatial resolution, coupled with a proper spectroscopic identification of the variability-detected sources. 

\athena\ is a L-(Large) Class X-ray observatory in the Cosmic Vision Program of the European Space Agency \citep{Nandra2013}, due to launch in early 2030s. 
 \athena\ will combine a collecting area at 1~keV more than one order of magnitude larger than any existing or planned X-ray mission, and a 5\arcsec\ on-axis HEW mirror with a very gentle performance degradation over the field-of-view of the Wide Field Imager (WFI). This is a Silicon Active Pixel Sensor camera with a large FoV (40'$\times$40'), high-count rate capabilities, and CCD-like energy resolution (about 150~eV at 6~keV). This combination will allow a survey speed more than two orders of magnitude faster than \chandra\ and \xmm. Thanks to a multi-tier survey strategy covering almost 30~Ms during the 4-year nominal operational life, \athena\ will probe an AGN population more than two orders of magnitude fainter than the SDSS and Euclid QSOs. At the end of the nominal survey, {\it Athena} will have detected over 400,000 AGN, probing several thousand AGN at $z\ge$4, a few hundred at $z\ge$6, and several tens of Compton-thick AGN at the peak of the accreting black hole activity. Its  average angular resolution over the WFI FoV ($\sim$6\arcsec) will allow probing dual AGN down to a separation of a few tens of kpc (see Fig.~\ref{fig1xrs}). 
 A new observational window in X-rays will be opened by the other instrument onboard \athena, the X-ray Integral Field Units (X-IFU), thanks to its unprecedented spectroscopic capabilites (down to few eV). As outlined in \cite{McKernan_ford2015}, it will be possible, in nearby AGN, to search for the presence of binary systems (with $q\gtrsim0.01$ and moderate, $<0.1$, eccentricity) separated by several hundreds gravitational radii. 
 Radial velocity shifts will be imprinted on the broad iron K$\alpha$ line due to oscillations of a massive black hole around its barycenter, therefore multiple observations of the same object will eventually provide the imprints of the presence of a binary system close to merger. 
 Furthermore, if both AGN in a binary, obscured system are emitting fluorescence iron lines and if their difference in systemic velocity is larger than few hundreds km/s (as NGC~6240), it will be possible, for sufficiently long ($\sim$Ms) exposures, to spectrally resolve close systems through an analysis of the iron line energy peaks (Piconcelli et al., in prep.). 

On a longer time-scale, \lynx\footnote{{\tt https://wwwastro.msfc.nasa.gov/lynx/}} is a concept study for consideration by NASA in the context of the forthcoming 2020 Astrophysics Decadal Survey. The main technical advancement of this proposed mission is a densely-packed, thin grazing incidence mirror with an effective area of 2~m$^2$ at 1~keV, and sub-arc second angular resolution over a high-definition X-ray imager in the focal plane. \lynx\ is expected to push the quest for young SMBHs in the very early Universe by two orders of magnitude in intrinsic luminosity. Its exquisite angular resolution, well matching that of JWST and WFIRST, will allow to study binary AGN with a separation down to a few kpc over a very wide redshift range (Fig.~\ref{fig1xrs}). In particular, the large field-of-view of WFIRST Wide-Field Instrument, 0.8$\times$0.4~deg$^{2}$, coupled to its sensitivity, is ideal for surveys purposes, and will provide optical/near-IR counterparts to faint and/or obscured AGN pairs detected by \lynx. 

Along the pathway outlined by \lynx\ in terms of sub-arc second angular resolution and larger than \chandra\ effective area, we need to mention \axis, which is a probe-class NASA mission under study for the 2020 Decadal Survey \citep{mushotzky2018}.  
Although \axis\ is characterized by much lower sensitivity compared to \lynx\ (being, however, 10$\times$ better than \chandra\ over a 24$\times24$ arcmin$^{2}$ field-of-view), it will allow detection and characterization of kpc-scale AGN pairs up to high redshift, with a foreseen launch for 2028. 

Hard X-rays are able to trace accretion even in obscured AGN, with some limitations for the most heavily obscured systems. As such, they are an ideal tool in the hunt for multiple active nuclei in a galaxy, providing typically a higher contrast with respect to stellar-related emission processes than observations at other wavelengths. 
The capability of detecting nuclear, accretion-related emission in galaxies with limited (or negligible) indications of an active nucleus at other wavelengths is therefore one of the major advantages of X-ray observations, although the planned X-ray missions described above, \erosita\ and \athena, will have poor and moderate angular resolution, respectively, thus strongly limiting the possibilities of disclosing and properly characterizing close ($<$ kpc) dual systems if both members are active.  
 The high point-like luminosity, the non-thermal spectral shape of the continuum and the presence of a strong Fe line emission in the X-ray spectrum are clear signatures allowing to discriminate the presence of an AGN with respect to potential astrophysical ``contaminants'' such as hot shocked gas in region of intense nuclear star formation. However, the success of any strategies to identify AGN through X-ray surveys invariably hinges on a synergetic approach with facilities at other wavelengths. These synergies work bi-directionally, both trough X-ray follow-ups of candidates identified at other wavelengths, or by campaigns in optical/IR (OIR) to identify the AGN counterparts of X-ray sources. 

 An example of the former aspect is the potential synergy between SKA and \athena. The sky survey at 1.4~GHz of the first-phase SKA (SASS1 starting at early 2020's) will provide the sky map (3$\pi$ steradians) to an r.m.s. of about 4 $\mu$Jy/beam (1$\sigma$) with  a resolution of 1". SASS1 will therefore be a powerful detector of low-luminosity AGN ($\sim 10^{39}$ erg\,s$^{-1}$) even for the radio-quiet population\footnote{AGN with $\log(R_x) < -4.5$, where $R_x$ is radio loudness defined as the luminosity ratio between the luminosity densities at 5~GHz and in the 2--10~keV band; see \citealt{Terashima_Wilson2003}.}. As such, it will be invaluable resource for investigation of systems of dual and multiple AGN. At these faint flux densities, the main population consists of a mixture of both star-forming galaxies and radio-quiet AGN and different techniques are needed in order to distinguish among these two populations: radio morphology, spectral index, polarization, variability, radio-infrared ratio, optical and IR colours, SEDs, optical line ratios, X-ray power. 
 \athena\ follow-up can give a fundamental contribution in the identification of AGN and in the estimate of the nuclear properties through X-ray luminosity and spectroscopy: the WFI will follow-up the SASS1 sources in multiple systems and detect, even in the case of strong absorption, all nuclei at moderate angular separation ($>$5\arcsec); in addition, the X-IFU will be able to separate the non-thermal from the soft thermal (starburst) component, and provide an accurate determination of basic astrophysical parameters for both components (luminosity, temperature/spectral index, strength and ionization state of nuclear reprocessing features). 

 Identification of the OIR counterparts of sources detected in the few tens square degrees of the \athena\ WFI and \erosita\ surveys will rely on a large set of multi-wavelength data that will be available through e.g.  Subaru-HSC, SDSS V,  LSST (see Sect.~\ref{sssec:future_opticalspec}), and WFIRST. The last two facilities, however, will not be available at the time the \erosita\ survey is completed. Moreover, the comparatively poor angular resolution of \erosita\ adds further uncertainties to the counterpart identification with respect to, e.g., \xmm.
 
\subsection{Optical spectroscopy}
\label{sssec:future_opticalspec}

While the upcoming X-ray missions \erosita\ and \athena\ will provide a large number of X-ray sources (as discussed above), subsequent characterization of the AGN nature and precise redshift measurements are only possible with large spectroscopic surveys. New wide-field optical spectrographs such as WHT/WEAVE \citep{dalton2012}, VISTA/4MOST (see \citealt{merloni12_erosita} for an in-depth discussion of the natural synergies with \erosita) or ELT/HARMONI (see \citealt{Padovani2017} for a review) will provide the necessary numbers and depth to statistically evaluate the fraction of dual AGN among all galaxy pairs down to a certain separation and AGN luminosity. 

The characterization and identification of obscured AGN in optical spectroscopy still rely mainly on classical narrow emission-line diagnostics \citep[e.g.][]{Baldwin81,Veilleux1987}, which are able to distinguish AGN photoionization from other ionization mechansims due to the hardness of the AGN spectrum at UV wavelength. However, line ratios may be significantly affected by mixing AGN photoionization with ionization by star-forming regions \citep[e.g.][]{Davies2014,Trump2015} which makes a unique characterization difficult, in particular at high redshifts where the systematic decrease in gas-phase metallicity leads to rather similar line ratios for the different ionization processes \citep[e.g][]{Kewley2013,Trump2013,Steidel2014,Coil2015}. Integral-field unit (IFU) spectroscopy can migitate those effects to some degree as ionization sources can be spatially separated much better than in fiber-based or long-slit spectroscopy. Large IFU spectroscopic surveys like CALIFA \citep{Sanchez2012}, SAMI \citep{Croom2012} and MaNGA \citep{Bundy2015} allow to systematically identify AGN at low redshifts through kpc-scale AGN ionization even if the integrated or central galaxy spectra are dominated by ionization from star formation \citep{Wylezalek2018}. In addition, the detection of higher ionization lines, like the HeII $\lambda$4685 in the rest-frame optical wavelength range \citep{Baer2017} or the HeII $\lambda$1480 in the rest-frame far-UV wavelength range \citep{Feltre2016}, offer a promising alternative to characterize faint obscured AGN at low and high redshifts. The HeII $\lambda$1480 line can be observed at high redshift in the optical wavelength so that high-sensitive wide-field IFUs such as VLT/MUSE or Keck/KCWI offer the unique opportunity to detect faint type 2 obscured narrow-line AGN at small separation down to the seeing limit. 

Overdensities of multiple AGN separated by a few 10s of kpc have been detected this way within a few giant Ly$\alpha$ nebulae \citep{Cantalupo2014,Hennawi2015,Cai2017,Arrigoni-Battaia2018}. Hence, dual AGN can be resolved and studied in merging galaxy systems at high redshifts with separations of much less than 30\,kpc \citep[e.g.][]{Husemann2018}. Many more close obscured AGN multiple systems with kpc-scale separation will be likely been detected at $z\sim2-4$ using this technique in the future.
 
\subsection{Mid-IR imaging and spectroscopy}
\label{sssec:future_mir}

Motivated by the outstanding scientific results obtained by the {\it Hubble Space Telescope} (HST), its successor, the JWST, will push our view to longer wavelengths beyond the HST capabilities. With its larger collecting area (i.e. 25.4 m$^2$, 6.25 times bigger than HST), higher sensitivity, and finer angular resolution at wavelengths shorter than 28.8$\mu$m, JWST will outclass Hubble's imaging potential in the infrared and will provide a factor of 5-7 better spectral resolution than the Spitzer/IRS high-resolution instrument. As a result, JWST will provide both high angular resolution imaging and spectroscopy at mid-infrared wavelength to identify obscured dual AGN via the emission of the molecular torus in the mid-infrared.

Specifically, the NIRCam imaging and spectroscopic modes covering the wavelength range 0.6--6 $\mu$m (angular resolution of 0.07$^{\prime\prime}$, \citealt{2017JATIS...3c5001G}) and the Mid-Infrared Instrument (MIRI, 5--28 $\mu$m) will have a superior spatial resolution at $\lambda>1 \mu$m (i.e. 0.031$^{\prime\prime}$ at 0.6--2.3 $\mu$m against the 0.13$^{\prime\prime}$ of HST at 0.9--1.7 $\mu$m) and will provide a $\times$50 better sensitivity with respect to Spitzer/IRAC \citep{2015PASP..127..584R,2015PASP..127..665R}. The Infrared Field Unit Near InfraRed Spectrograph (NIRSpec), which operates over a wavelength range of 0.6 to 5.3 $\mu$m, in the highest spectral resolution mode is a factor of $\sim5$ better than the highest resolution spectrograph on Spitzer-IRS. 

With this exquisite spectral coverage, JWST offers us the opportunity to study the accretion disk emission reprocessed in the infrared by the dust surrounding the SMBH through accurate {\it torus} models, measure extended emission-line structures (i.e. ionization cones, outflows), and reconstruct the gas kinematics as never done before. 
Although JWST is not properly a survey instrument (because of the limited field-of-view), its imaging from NIR to MIR will cover significant areas on deep fields which will probe a significant volume at high redshift. 

Larger areas of the sky (10$\times$10 arcmin$^{2}$) will be covered at MIR wavelengths ($\sim17-34$~$\mu$m) in photometric/low-resolution spectroscopic mode by the {\it Space Infrared Telescope for Cosmology and Astrophysics, SPICA} with the SMI/CAM instrument. \spica\ \citep{roelfsema2018}, selected as a new-mission concept study by ESA for M5, will  also probe the physics of accreting systems using the mid-IR/far-IR emission lines thanks to the $\sim34-230$~$\mu$m coverage of the Safari instrument, thus potentially extending the studies of such systems up to very high redshifts. 

\section{The quest for hard SMBH binaries}
\label{ssec:future_SMBHB}

\subsection{AGN variability}
\label{sssec:future_agn_variability}

Time-domain surveys in the 2020s will provide powerful datasets for selecting SMBHB candidates whose nature can be tested with follow-up observations.  The Large Synoptic Survey Telescope (LSST), scheduled for first light in 2020, will be the flagship time-domain machine.  The well sampled, long-duration, high signal-to-noise light curves from LSST, in combination with the unprecedented sample size of about a million quasars, will be the best database for identifying candidate SMBHBs, based on (semi)periodic variability.

LSST is currently under construction on Cerro Pach{\'o}n in the Chilean Andes.  The telescope features an effective aperture of 6.7~m and will be equipped with a 9.2~deg$^2$ field-of-view survey camera and {\it ugrizy} filter set, spanning the range from the atmospheric cutoff in the UV to the limit of CCD sensitivity in the near-infrared.  A dedicated survey telescope, LSST will rapidly scan large areas of the sky to faint magnitudes, fulfilling its ``wide-fast-deep'' mantra.

LSST will spend 85-90 per cent of its time on a 10-year time-domain photometric survey covering 18,000~deg$^2$ of the sky.  The ``baseline cadence'' that allows LSST to meet its science goals is laid out by \cite{ivezic08} and described in the LSST observing strategy white paper\footnote{The LSST Observing Strategy white paper is a live document available \url{https://github.com/LSSTScienceCollaborations/ObservingStrategy/tree/master/whitepaper}.}.  According to this cadence, LSST would scan the observable sky every 3 nights in back-to-back pairs of 15~s exposures, called visits.  The survey would reach median single-visit depths of 23.14, 24.47, 24.16, 23.40, 22.23, 21.57 and have a median number of 62, 88, 199, 201, 180, 180 visits in {\it ugrizy}, respectively, from 2,293 overlapping fields.  However, the LSST cadence is {\it not yet set} and the Observing Strategy White Paper describes the ongoing work in the community to optimize it.  One alternative to the baseline cadence's spatially uniform annual tiling of the sky that is being explored is a ``rolling cadence'' that focuses on different parts of the survey area in different years.

LSST will spend 10-15 per cent of its time on specialized projects, including the Deep Drilling Fields (DDFs).  These will receive a higher cadence than the 10-year survey, with the possibility of doing AGN variability science for 10$^{4-5}$~AGN.  So far, four multi-wavelength fields have been selected: ELAIS-S1, XMM-LSS, Extended CDF-S, and COSMOS. The definition of additional DDFs and other specialized projects (mini-surveys) is extremely open, with decisions ongoing up to and beyond LSST first light. 
It is worth noting that the final decisions on the observing strategy will affect how efficiently LSST can detect SMBHB candidates. For instance, if the ``rolling cadence'' is selected, it may not allow LSST to build the long baselines that are necessary for the search of SMBHB. Additionally, even though LSST will have a nominal cadence of 3 days, given that the filters are successively alternated, the light curves in each band will contain a dozen 
datapoints every year. Techniques like the multi-band periodogram \citep{VanderPlas2015} have been developed to coherently take full advantage of the multiple time series. Nevertheless, it is unclear if they can be extended to quasar light curves, where the variability is colour-dependent. Another challenge is presented from the large available sample; this will pose significant statistical challenges in order to robustly filter false periodic detections. On the positive side, LSST is likely to dramatically improve our understanding of the underlying quasar variability, which eventually will facilitate the periodicity search. Additionally, the LSST data streams can be combined with existing time-domain data, from survey like CRTS, providing light curves with long baselines and high-quality data, which are necessary for this kind of analysis. 

In addition to candidates found from the growing body of all-sky photometric surveys including LSST, many more will come with the dawn of complementary ``panoptic spectroscopy''.  Pioneering this frontier, the Sloan Digital Sky Survey V (SDSS V, \citealt{2017arXiv171103234K}) will be the first homogeneous, wide spectral coverage all-sky multi-epoch spectroscopic survey.  

SDSS~V will last 5 years and science will be grouped into the Local Volume Mapper (LVM), Milky Way Mapper (MWM), and Black Hole Mapper (BHM) programs, with that last being the most relevant for finding gravitationally bound SMBHBs.  
The BHM program encompasses three scientific areas: \erosita\ spectroscopic follow-up, reverberation mapping, and an all-sky multi-epoch spectroscopic survey. \erosita\ will be a key player in searches for dual AGN via sensitive X-ray imaging surveys (Sect.~\ref{sssec:future_xray}) and BHM will provide spectroscopic identification and redshifts for $\sim$400,000 \erosita\ X-ray sources (primarily AGN at high Galactic latitude) in the first $\sim1.5$ years of the survey. BHM will also build on the SDSS-IV reverberation-mapping and time-domain spectroscopic survey (TDSS) programs to deliver new black hole mass measurements for $\sim1000-1500$ $0.1<z<4.5$ quasars/AGN and a few to a dozen epochs of spectroscopy per target for 25,000 quasars spanning temporal baselines of months to a decade.  For most of the sky, this dataset will not yield enough spectroscopic epochs to fully populate radial velocity curves for SMBHB candidates. However, it will provide a starting point for follow-up campaigns and enable a critical benchmark measurement of how normal quasars vary on a wide range of timescales.

In the high-energy domain, it is worth mentioning 
The planned future Chinese X-ray mission {\sl Einstein Probe}, a dedicated time-domain soft X-ray all-sky monitor aiming at detecting X-ray transients including TDEs in large numbers \citep{yuan2016}. 
It will provide well-covered lightcurves. These will allow us to search systematically for the characteristic phases of intermittency and recovery in the lightcurves from TDEs which happen in binary SMBHs (see Sect.~\ref{sssec:subpc_photometry}). This method will provide a census of the SMBHB fraction in quiescent galaxies, once enough lightcurves have been obtained. 

From an X-ray spectroscopic perspective, for relatively bright and local sources, \athena\ will allow to associate peculiarities in the iron line features (i.e., double-peaked iron K$\alpha$ emission-line profile, as reported in \citealt{Severgnini2018}; see Sect.~\ref{sssec:subpc_opt_xray}) with the presence of a sub-kpc scale dual AGN. Moreover, variability of the iron line, as described in Sect.~\ref{sssec:future_xray} (see \citealt{McKernan_ford2015}), will offer another viable, though challenging, possibility to disclose the presence of dual nuclei at close separation. 

\subsection{Theory and Simulations}
\label{sssec:future_sims}

We discuss here in detail the properties of SMBHBs embedded in circum-binary disks, expanding on the theoretical background summarized in Sect.~\ref{ssec:subpctheory}. Simulations have now followed binaries to merger in pseudo-Newtonian potentials \citep{Tang+2018} or very close to merger in full GR \citep{Bowen19} where space-time is violently changing (see also Sect.~\ref{sssec:LISA}). These, and previous simulations of SMBHBs embedded in circumbinary disks are converging on the following signatures.  

\bigskip
{\it Periodicities.}  The mass inflow rate across the cavity folows particular patterns, arising from modulation of the gas inflow on the binary's orbital period and half-orbital period, as well as on the factor of few longer time-scales corresponding to the orbital period at the cavity wall.  In particular, the variability structure of the mass accretion rates seen in
simulations can be roughly divided into four distinct categories,
based on the binary mass ratio $q\equiv M_2/M_1$. For $q\lsim0.05$,
the disk is steady and the BH accretion rate displays no strong
variability \citep{dorazio13,farris14,Dorazio+2016}. 
For $0.05 \lsim q \lsim 0.3$, the accretion rate
varies periodically on the timescale $t_{\rm bin}$, with additional
periodicity at $\approx 0.5t_{\rm bin}$. Binaries with $0.3 \lsim q
\lsim 0.8$ clear a lopsided central cavity in the disk, causing
variability on three timescales. The dominant period, $(3-8) t_{\rm
  bin}$ is that of an over-dense lump, orbiting at the ridge of the
cavity, with additional periodicities at $t_{\rm bin}$ and $\approx
0.5t_{\rm bin}$ \citep{macfadyen08,shi12,noble12,roedig12,farris14,Dorazio+2016}. 
The dominant period depends on the size of the cavity, and thus on disk parameters,
such as temperature and viscosity. Finally, equal-mass ($q=1$)
binaries display variability at the longer lump period and at $\approx
0.5t_{\rm bin}$.

\bigskip

{\it Enhanced brightness.} As emphasized by \citet{Farris+2014b}, the heating of the gas near the binary is dominated by shocks, rather than the usual viscous dissipation for a single-BH disk. As a result, the total luminosity of the binary can significantly exceed, even by 1-2 orders of magnitude, that of a Shakura-Sunyaev disk with the same mass and external large-scale accretion rate (see also \citealt{Lodato+2009, kocsis12a, kocsis12b}).  Interestingly, the additional power must come at the expense of the binary's binding energy, and is therefore directly tied to disk's contribution to the binary's inspiral rate (although this extra energy source does not exist if the disk torques are positive and cause an outspiral, rather than an inspiral; see \citealt{Tang+2017, Munoz+2019, Moody+2019}).   
Also interestingly, this point was noted in the context of pre-main sequence {\em stellar} binaries by \cite{TP2017}. 

\bigskip

{\it Unusual spectral shapes.} The shock-heating, together with the evacuation of the gas from the central cavity around the binary, also changes significant distortions in the spectral shape, compared to the usual Shakura-Sunyaev disk.  In particular, there can be a ``notch'' in the spectrum, at the frequencies where the emission, in the absence of a binary, would have been dominated by the gas missing from the cavity \citep{Sesana12,2012ApJ...761...90G,Roedig+2014}, although this notch can be partially filled in by the emission from the shock-heated mini-disks and accretion streams in the cavity \citep{Farris+2014b}.   On the other hand, strong shock-heating in the innermost regions (near the cavity's ridge, and inside the cavity) can give rise to unusually hard spectra at higher frequencies, making the spectrum extend to much harder photon energies than for a Shakura-Sunyaev disk, possibly also leading to unusually strong broad-line emission. \\ \\

As anticipated in Sect.~\ref{sssec:future_agn_variability}, variability with LSST will provide a powerful set of data for selecting good SMBHBs through their periodic signal.

More challenging will be the observations of brightest sources or AGN with a peculiar X-ray profile in the hardest energy bands (above 10~keV). In fact, a bump above 10~keV peaking at 30~keV is commonly observed in AGN and possibly associated to the well known Compton hump \citep{george&fabian91}. This component, along  with the fluorescence Fe line, is produced through the inverse Compton of photons from the central hot region (the so-called emitting corona) on the optically-thick accretion disk within the AGN. Nevertheless, the two models (single BH accretion disk and circumbinary disks) should produce flux variability in different timescales (order of few tens of seconds in single 10$^8$ M$_\odot$ BH accretion disks). In this direction, future time-resolved spectral analysis would help to break the degeneracy, and then disentangle the origin of the observed spectral signature.

\section{GW horizons}
\label{ssec:gw}

\begin{figure*}[!t]
\centering
\vspace{4.0pt}
\includegraphics[width=1.7\columnwidth,angle=0]{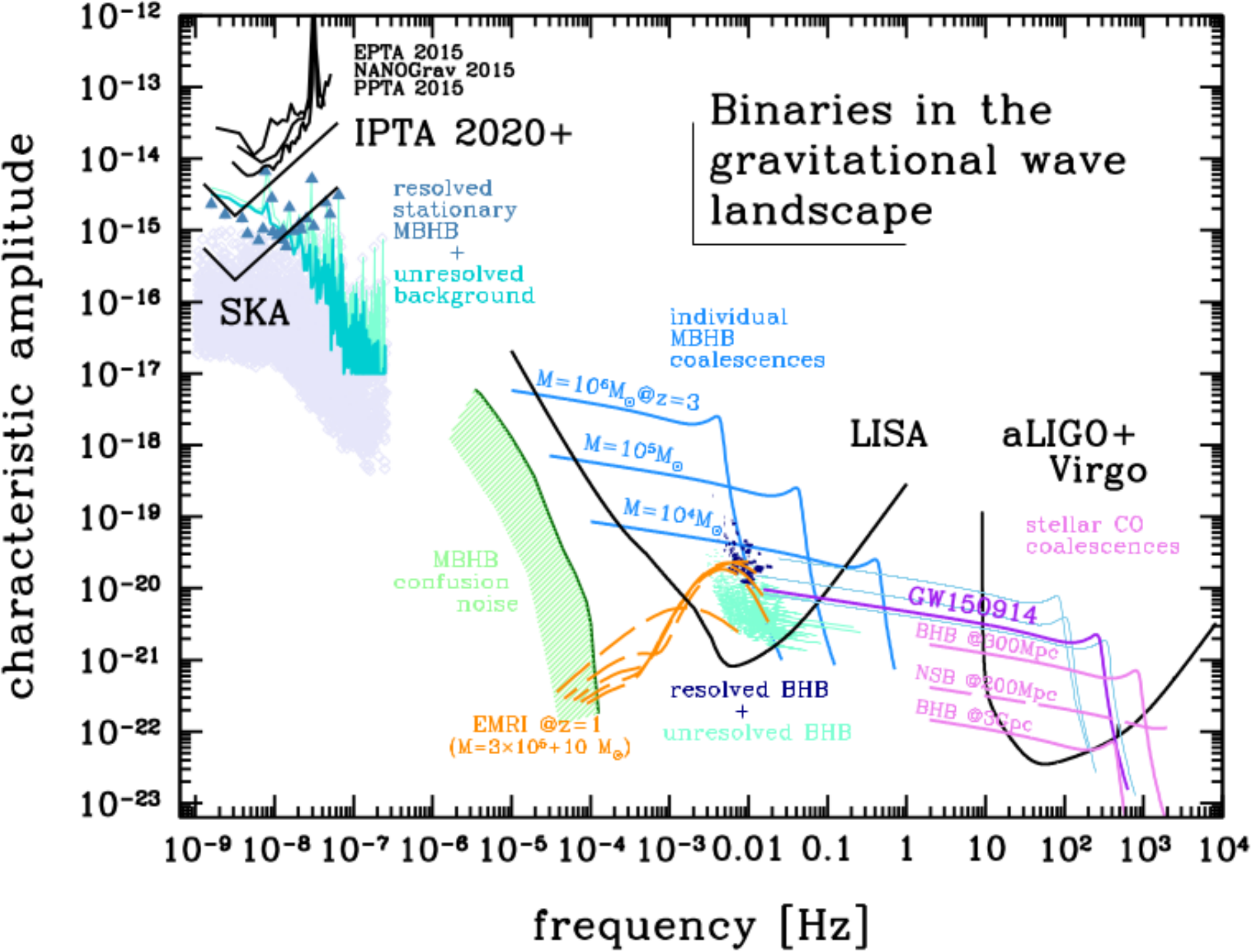}
\vspace{-0.0pt}
\caption[]{Characteristic strain amplitude versus frequency of binaries in the GW landscape. The black curves are the sensitivity of LISA, Advanced LIGO and Virgo, and various PTA experiments. In the PTA band, the GW signal (blue jagged line) is characterized by the incoherent superposition of quasi-monochromatic binaries (lavender diamonds). Blue triangles show a sample of individually resolvable SMBH binaries. In the LISA band, loudest sources will be merging SMBH binaries, shown here by the blue tracks. Other sources include extreme mass ratio inspirals (EMRIs, orange tracks) and stellar mass BHBs (pale and dark blue ticks). The LIGO/Virgo band is the realm of stellar mass compact objects, and typical tracks are shown in purple. From \cite{2017ogw..book...43C}.}
\label{fig:GW_all}
\end{figure*}

In the near future, observations of SMBHBs through their low-frequency GW emission (from nHz 
0.1Hz; see Fig.~\ref{fig:GW_all}) promise to be revolutionary in our understanding of the formation and growth of SMBHs.  

GWs are ripples in spacetime propagating at the speed of light, generated by accelerated masses with a time-varying non-zero mass quadrupole moment \citep{1987thyg.book..330T}.  Binary systems of two compact objects are therefore ideal sources of GWs.  In particular, SMBHBs are among the loudest sources of GWs when observed during the phases of late inspiral and coalescence \citep{LRRsathya09}. 

Four simple notions are worth of mention in this context.
First, circular binaries emit sinusoidal GWs at twice their orbital frequency, while eccentric binaries 
emit at frequencies that are multiples of the orbital frequency \citep[]{1963PhRv..131..435P}. As rule of thumb,  GWs circularize any initially eccentric orbit, but some residual eccentricity can remain at coalescence if, e.g., the binary hardened in a triple 
interaction \citep{Bonetti2019}, or is detect during its secualar adiabatic contraction (as for the case of PTA sources discussed below). 

Second, the strain amplitude  $h$ of a GW, which is the sum of the two polarization states of the wave, weighted through the antenna pattern of the detector, is a well understood function of the binary parameters and the main scaling of $h$ with key parameters to the source are highlighted in this formula  
\begin{equation}
  h\sim  {( G{\cal M}_z )^{5/3}\over c^4D_L}(\pi f)^{2/3},
\label{eq:gwstrain}    
\end{equation}
where ${\cal M}=(1+z)M_1M_2/(M_1+M_2)$ is the {\it redshifted} chirp mass of the binary, $D_L$ its luminosity distance and $f=f_r/(1+z)$ is the {\it observed} GW frequency (being $f_r$ the frequency in the source rest frame). 
Third, during the inspiral, the frequency $f_r$ sweeps to higher values as the binary contracts, in response to the energy loss by GWs. Its rate of change,
in the observer frame is 
\begin{equation}
 {\dot f}={96\over 5}\pi^{3/8}\left ({ G{\cal M}_z\over c^3}\right )^{5/3} f^{11/3}.
 % {\dot f}={96\over 5}\pi^{3/8}\left ({} G{\cal M}_z\over c^3}\right )^{5/3} f^{11/3}.
\label{eq:gwchirp}    
\end{equation}
The chirp refers  to the phase of inspiral when both $h$ and $f$ increase with time, until coalescence.
Measuring ${\dot f}$ provides an accurate determination of the chirp mass ${\cal M}_z,$ and an estimate of the time to coalescence.  Coalescence occurs at a frequency close to $\sim c^3/GM$, where $M=M_1+M_2$ is the total mass of the binary in the source frame. 

Fourth, the plus of GW observations is that they will not give ``candidates'' but rather ``secure detections''; providing that the analysis is not faulty, there is not much room left for alternative interpretation of clear GW signals \citep[see, e.g.,][]{2016PhRvX...6d1015A}. On the minus side, our ``secure detections'' of GWs events will have poor sky localization \citep{2016PhRvD..93b4003K,2018MNRAS.tmp..866G}, which will generally make coincident EM identification or follow-ups rather problematic (see detailed discussion in Sect.~\ref{sssec:LISA} and Sect.~\ref{sssec:PTA})\footnote{This was not the case for the merging neutron star binary GW170817 \citep{2017PhRvL.119p1101A}, which resulted in a spectacular multi-wavelength observational campaign \citep{2017ApJ...848L..12A} despite an initial GW sky localization of $\approx 30$deg$^2$. Although this is promising, one should bear in mind that low-frequency EM sources are expected to be observed at much further distances, and their emission is likely going to be Eddington limited, making them much fainter than short gamma-ray bursts.}.

% -----------
% -----------
In the following, we briefly discuss the two low-frequency observational windows that will soon become extremely relevant for SMBH binary science. The mHz frequency range will be probed by LISA  \citep{LISA17}, the third ESA L-class mission scheduled for launch in 2034. The nHz frequency window is currently probed by PTAs \citep{1990ApJ...361..300F}, and future radio instruments such as FAST \citep{2011IJMPD..20..989N} and SKA \citep{2009IEEEP..97.1482D} will provide a significant contribution.

\subsection{LISA: probing the SMBH assembly from their infancy}
\label{sssec:LISA}
LISA will detect merging SMBH binaries from few$\times 10^3~\msun$ to few$\times 10^7~\msun$ {\it everywhere in the Universe} \citep{Colpi2020}. Expected detection rates are uncertain, varying between several to few hundred over the planned 4-year mission lifetime \citep{2016PhRvD..93b4003K,Bonetti2019}. Note that, although the bulk of these events will involve SMBHBs with $M<10^5~\msun$ at $z>5$, EM observations will be greatly facilitated by more massive ($M>10^6~\msun$) and closer ($z<3$) systems \citep{2018arXiv181100050M}, which might be detected by LISA at a rate of few per year. 

One important question to address is whether the binary would remain bright {\em and} periodic all the way to merger. Naively, one would expect that this is not the case; as mentioned in Sect.~\ref{ssec:subpctheory}, at a binary separation of $\sim 100R_{\rm s}$ (Schwarzschild radii), the GW-driven inspiral timescale becomes shorter than the local viscous time and the binary ``decouples'' from the disk.
Past this stage, the runaway binary leaves the disk behind, which is unable to follow the rapidly shrinking binary~\citep{Liu+2003,MP2005}.  Encouragingly, it has now been demonstrated that this is not the case.   \citet{Farris+2015}, \citet{Tang+2018} and \citet{Bowen18, Bowen19} have performed simulations of GW-driven  nearly equal-mass binaries all the way to merger. The former simulations started binary evolution from an initial separation prior to the fiducial decoupling, and the last from a separation encompassing twelve orbits prior merging. 
These studies have shown that the gas is able to accrete onto the black holes, all the way to the merger, despite the rapid contraction of the binary orbit. The physical reason for this is that the angular momentum from the gas can be removed by gravitational torques and shocks caused by the binary, which dominate over viscosity, and operate on much shorter time-scales. 

3D MHD simulations in GR  \citep{Bowen18,Bowen19} have confirmed that strong gravity develops an $m=1$ azimuthal asymmetry (or lump) in the circumbinary disk, which quasi-periodically modifies the accretion flow into the central cavity and thus to the mini-disks.  Radial pressure gradients accelerate the inflow rate on the black holes well beyond that associated with stresses arising from MHD turbulence, dynamically coupling the mini-disks to the lump, directly. If the accretion rate makes the flow optically thick, soft X-ray emission comes from the inner rim of the circum-binary disk and harder radiation from the mini disks if also coronal emission is excited \citep[see Sect.~\ref{ssec:subpctheory}]{Tang+2018,Dascoli2018}. While an optical-UV chirp might be present in tandem with the GW emission in the very early phase of the inspiral, due to Doppler shift induced by the circular motion, near plunge an overall dimming and loss of periodicity might characterize the emission due to the erosion of each mini disk by the tidal field of the companion black hole, whose size shrinks  down to a few times the  innermost stable circular orbit \citep{Tang+2018,Bowen18}.
 Outside thermalized regions and in case of low accretion rates, coronal emission around the two SMBHs may give rise to hard X-ray emission. Its modulation might depend  on the orientation of the binary orbital plane relative to the line of sight, Doppler beaming and gravitational lensing.  At plunge and in the post-merger phase, numerical simulations demonstrated that an incipient relativistic jet is launched by the new spinning SMBH, which may spark gamma-ray emission and afterglow emission in its impact with the ISM \citep{gold14}.

Advance localization of the binary by {\it LISA} weeks to months prior to merger to a few square degrees on the sky \citep{KocsisHaimanMenou2008,LangHughes2008,McWilliams+2011} will enable a measurement of this EM chirp by wide-field X-ray (and possibly also optical) instruments.  
   
A comparison of the phases of the GW and EM chirp signals will help  break degeneracies between system parameters, and probe a fractional difference $\Delta v$ in the propagation speed  of photons and gravitons as low as $\Delta v/c \approx 10^{-17}$~\citep{Haiman2017}.

In order to explore to what extent LISA-EM synergies might be feasible, we consider here few selected cases, with masses ranging from $10^5~\msun$ to $10^7~\msun$, out to $z=7$, as listed in Table~\ref{tab:multimess}.

For the sake of the discussion, we make the following simple assumptions:
\begin{enumerate}
\item the merging binary emits either at its Eddington limit ($L_{\rm Edd}$) or at 10 per cent of this value ($0.1L_{\rm Edd}$)\footnote{Despite observed AGN have average luminosity of about $0.1L_{\rm Edd}$, we expect SMBHBs to form following major mergers, which are also known to trigger copious gas inflows in the remnant nucleus. Hydrodynamical simulations of SMBHB evolution in galaxy mergers show that the amount of gas bound to the binary can sustain Eddington-limited accretion throughout the merger process \citep[e.g.][]{2009MNRAS.396.1640D,Capelo_et_al_2015}, which justifies our choice of exploring also the $L_{\rm Edd}$ case (cf right panel of Fig.~\ref{fig:1to4_temporal_evolution}).}. For the two cases, the bolometric luminosity is $L=1.4\times 10^{44}M_6$ erg~s$^{-1}$ and $L=1.4\times 10^{43}M_6$~erg~s$^{-1}$, respectively, where $M_6=M/10^6~\msun$;  
\item at these values of luminosity, $\approx$10 per cent of the emission is in the $2-10$~keV band \citep{2012MNRAS.425..623L};
\item a generic 1.5 bolometric correction for observations in optical/infrared;
\item a radio luminosity based on the triggering of a powerful radio jet of $L_{\rm jet}\approx 10^{43}M_6$~erg~s$^{-1}$ at GHz frequencies (regardless of the assumed Eddington ratio of point 1).
\end{enumerate}
These assumptions translate in the observables detailed in Table~\ref{tab:multimess}.  Absorption is not considered in the X-ray fluxes estimates reported in Table~\ref{tab:multimess} while, as discussed above (see Sect.~\ref{sssec:xray}), galaxies in their advanced stage of merging are substantially obscured. However, we note that at high redshift ($z=3,5,7$) obscuration, with typical value of \nh=10$^{23}$ \cm2, will not affect the hard X-ray (2--10~keV) flux estimates, while at $z=1$ the estimated flux will decrease by about 10 per cent. 
We refer the reader to Sect.~5 of \cite{2016JCAP...04..002T} for further discussion about the modelling of the emission and the conversion into magnitudes and fluxes. We also stress that the emission models are very simplified and numbers in Table~\ref{tab:multimess} are indicative.

%%%%%%%%%%%%%%%%%%%%%%%%%%%%
\begin{table*}
\centering
\caption{Emission properties of selected GW sources observed by LISA. Besides the source mass and redshift, reported are the 1-$\sigma$ relative error in the luminosity distance measurement, $\Delta{D_L}/D_L$, and the sky localization region $\Delta\Omega$ at merger, defined such that the probability of finding the source within this region is $1-e^{-1}\approx 0.63$ \citep{2004PhRvD..69h2005B}. As for the EM emission, 10 per cent Eddington and Eddington-limited emission are assumed. The visual magnitude, $m_v$, has been obtained by applying a bolometric correction BC$=1.5$, the X-ray flux assumes 10 per cent of the bolometric luminosity in the $2-10$~keV band with a power-law distribution with slope $\Gamma=$1.7. The radio flux density is based on the jet emission model of \cite{2001ApJ...548L...9M}. Numbers should only be considered indicative as neither extinction nor absorption have been applied. See text in Sect.~\ref{sssec:LISA} for details
}
\label{tab:multimess}
\begin{tabular}{l|c|c|c|c|c|c|c|c} \hline  \hline
 \multicolumn{4}{c|}{}& \multicolumn{2}{|c|}{0.1$L_{\rm Edd}$}& \multicolumn{2}{|c|}{$L_{\rm Edd}$}&\\	
\hline

Mass & \multirow{2}{*}{redshift} & \multirow{2}{*}{$\Delta{D_L}/D_L$} & $\Delta\Omega$ &  \multirow{2}{*}{$m_v$} & X-ray flux  &  \multirow{2}{*}{$m_v$} & X-ray flux & Radio flux\\

[$\msun$] &  & & [deg$^2$] & & [erg s$^{-1}$cm$^{-2}$] & & [erg s$^{-1}$cm$^{-2}$]
& [$\mu$Jy]  \\

\hline
$10^7$ & 1 & 0.05 & 0.3 & 23.5 & $3.2\times 10^{-15}$ & 21   & $3.2\times 10^{-14}$ & $2\times 10^4$ \\
\hline
$10^6$ & 1 & 0.01 & 0.1 & 26   & $3.2\times 10^{-16}$ & 23.5 & $3.2\times 10^{-15}$ & $2\times 10^3$ \\
$10^6$ & 3 & 0.04 & 0.4 & 28.9 & $2.6\times 10^{-17}$ & 26.5 & $2.6\times 10^{-16}$ & 135 \\
$10^6$ & 5 & 0.07  & 2   & 30.3 & $8\times 10^{-18}$   & 27.8 & $8\times 10^{-17}$ & 38 \\
\hline
$10^5$ & 5 & 0.1 & 5   & 32.5 & $8\times 10^{-19}$   & 30   & $8\times 10^{-18}$ & 3.8 \\
$10^5$ & 7 & 0.1 & 10  & 33.5 & $4\times 10^{-19}$ & 31   & $4\times 10^{-18}$ & 1.7 \\
\hline 
\hline 
\end{tabular}
\end{table*}

For relatively massive sources at $z\leq 3$ (first three rows in Table~\ref{tab:multimess})  LISA might provide the sky location of the source with $\Delta\Omega < 0.4$ deg$^2$ {\it at merger}, a less than 10 per cent precision on the luminosity distance measurement (mostly limited by weak lensing), plus intrinsic source parameters such mass and mass ratio to better than 1 per cent, and the spin parameters of the two BHs to a $0.01-0.1$ absolute precision \citep{2016PhRvD..93b4003K}. Depending on the exact low-frequency performance of LISA, these systems might enter in band {\it months} before coalescence, allowing {\it pre-merger} identification and localisation (although with a much worse precision of several deg$^2$). For a LISA band entry frequency of $f<0.1$ mHz, the orbital period is several minutes. Note that both a $10^7\msun$ system emitting at $0.1 L_{\rm Edd}$ and an Eddington limited $10^6\msun$ binary at $z\leq 1$ have an apparent magnitude of $m_v\approx 23.5$, within the capabilities of LSST. The LISA error-box can therefore be covered periodically by LSST in the attempt to identify a varying source matching the orbital period \citep{Haiman2017}, as discussed in Sect.~\ref{sssec:subpc_photometry}. Note that at those small magnitudes a lot of transient optical signals are expected; matching a flux periodic variation to the (extremely well measured by LISA) frequency evolution of the binary might provide the key to discern the true GW source from other contaminants.

An arcsec sky localization precision during the late inspiral phase will allow to point any instrument in any band to obtain multi-wavelength coverage of the source, tracing its evolution as it proceeds to merger. This scenario would yield an unprecedented wealth of information about merging SMBHBs, allowing a time-dependent tracing of the emission at all wavelength, thus probing the interaction between the binary and the surrounding gas after dynamical decoupling \citep[see, e.g.,][]{Tang+2018,Bowen18}. Observations at multiple scales will probe the host of such system, providing unprecedented information about the physical environment of merging binary hosts at large. The emission properties of the system can be used to construct a consistent model that can be re-scaled to search for the much more abundant population of wider binaries \citep{haiman09} -- thus not emitting in the LISA band -- in, e.g., LSST data.   

As we move to higher redshifts, observations become more complicated. Beyond $z=1$, identifying a periodic {\it pre-merger} counterpart with LSST will be unfeasible, because even an Eddington-limited binary is simply too dim. Moreover, LISA parameters will not be so sharply determined, mostly because of the lower S/N and relative larger error in luminosity distance, which ramps up to 10 per cent, also because of weak lensing. Combined with the intrinsic low flux of such a high-redshift source, this will make EM identification problematic. Eddington-limited systems with $M\gsim 10^6\msun$ will still be within the nominal limiting flux of \athena\ up to $z \approx 5$, and to $z\approx 3$ the sky localization might still be good enough to fall within a single WFI FoV. Thus \athena\ can be pointed after merger, searching for a distinctive X-ray afterglow, maybe associated to the launch of a jet or to a post-merger re-brightening of the source. At even lower masses ($\approx 10^5\msun$) and higher redshifts, the best chance of finding an EM counterpart will be the identification of a $\mu$Jy transient with SKA, than might be associated with the emergence of a transient radio jet at merger \citep{2010Sci...329..927P,2011ApJ...734L..37K}. SKA will have a unique combination of sensitivity and large field of view (see Table~\ref{tabradiosrv} in Sect.~\ref{sssec:radiofuture}). Its sub-arcsec angular resolution will make follow up observation possible in all bands, characterizing the source and its environment. Detection of several such events will give invaluable insights on the high-redshift assembly of SMBHs in connection to their hosts \citep{2011PhRvD..83d4036S}.

\subsection{PTAs: giants in the low-redshift Universe}
\label{sssec:PTA}

A GW propagating into space affects the travel time  of photons, an effect that is measurable in very precise cosmic clocks like millisecond pulsars. The typical magnitude of the effect on the pulse time of arrival (ToAs) is of the order $\delta{t}\approx h/(2\pi f)$ (where $f$ is the GW frequency), which for an SMBH binary of $10^9\msun$ at 500~Mpc distance with an orbital period of a year gives $\approx 10$~ns \citep{2010PhRvD..81j4008S}. Several PTA projects around the world are monitoring {\cal O}(100) millisecond pulsars, some of them to better than 100~ns accuracy \citep{2016MNRAS.458.3341D,2016MNRAS.455.1751R,2016MNRAS.458.1267V,2018ApJS..235...37A}. The correlation of ToAs from multiple pulsars effectively increases the sensitivity of this technique, which is now starting to probe the interesting region of the astrophysical parameter space where SMBH binaries are expected to reside \citep{2018NatCo...9..573M}. In the long run, the SKA will likely boost current PTAs sensitivity by more than an order of magnitude \citep{2015aska.confE..37J}.

The primary target of PTA campaigns is the cosmic population of centi-pc separation SMBH binaries with $M>10^8\msun$ at $z<1$ \citep{2008MNRAS.390..192S}, emitting in the nHz frequency band (i.e. with orbital periods from decades to months). Crucially, this implies that {\it PTAs and time-domain surveys will be targeting the same sources}, opening synergies that can work in both directions. 

{\bf EM to PTA.} The first thing to notice is that sensitivity of EM probes is not a problem for typical PTA sources. A $M=10^9\msun$ system at  $z=1$ will be, for example, well within reach of single pointings with both LSST and \erosita. Assuming emission at $0.1 L_{\rm Edd}$, and neglecting obscuration, the flux in the 2--10~keV and 0.5-2~keV energy band would be 3 and 1.8 $\times 10^{-13}$~erg~s$^{-1}$cm$^{-2}$, respectively (adopting the same assumptions about the X-ray spectrum shape as in the previous section).
We know that the soft X-ray band is affected by absorption and AGN residing in galaxies in advanced merger stage are often substantially obscured (see Sect.~\ref{sssec:xray}). 
A better avenue would therefore be to observe in the 2--10~keV band, which is less affected by obscuration. 
In-fact, assuming an intrinsic column density $N_H=10^{23}$cm$^{-2}$ \citep{riccietal17,derosaetal2018},  the X-ray flux for the aforementioned binary decreases to 2.8 and 0.2 $\times 10^{-13}$~erg~s$^{-1}$cm$^{-2}$ in 2--10~keV and 0.5--2~keV, respectively.
This is still above the single-point flux limit of \erosita\ in both the hard and soft-X ray bands in 4 years all-sky survey (eRASS:4, see Sect.~\ref{sssec:future_xray}) and marginal within the average value expected at the end of the one-year all sky survey (eRASS:1).
In general, all sky time-domain surveys will generate a large amount of candidates that can be verified or dismissed by PTA observations. As already mentioned, this particular line of research has already been implemented on specific SMBH binary candidate samples \citep{Sesana2018}, casting doubts about their nature. Besides allowing verification of the individual brightest candidates, PTAs will also provide a strong consistency check of the overall population statistics. A high S/N detection of the stochastic GW background and of its spatial anisotropies (although on large scales only) can be cross-correlated with the population of SMBH binary candidates identified in EM surveys, to see whether they match or not. In turn, this can provide strong constrains on the fraction of accreting SMBH binaries at $z<1$, and thus on their typical environment. For example, a majority of the GW background unaccounted for by EM candidates will be an indication that massive low-redshift binaries preferentially evolve in gas-poor environments, being mostly driven by stellar dynamics.

{\bf PTA to EM.} Although the incoherent sum of all the sources will create a stochastic GW background, several individual sources will be eventually singled out from the signal, some with S/N$>10$. Those will mostly be systems with $M>10^9\msun$ at $z<0.5$ \citep{2015MNRAS.451.2417R,Kelley2019}, for which PTAs will provide a relatively poor sky localization within $\lesssim 100$ deg$^2$. Although this figure seems discouraging, two things should be borne in mind. First, those are extremely massive systems at relatively low redshift; if they are active they are bound to be extremely bright (unlike the case of LISA sources). Second, they will necessarily reside in very massive galaxies, limiting the number of likely hosts to $<1000$ even over such a large region of the sky \citep{goldstein2019}. Periodic coverage of the relevant sky region in optical, radio and X-rays, will allow to scrutinize all potential brightest hosts, as the GW emission is stationary or slowly chirping. Again, if no convincing variable AGN is seen, one can confidently say either that the system is severely sub-Eddington or that the AGN does not have a periodically variable signature at that given explored wavelength and timescales. 
On the other hand, a confident detection of a periodic source matching the GW period will allow deep follow-ups at all wavelengths. Identification of the host will also break the mass-distance degeneracy in equation (\ref{eq:gwstrain}), allowing a 10 per cent estimate of the source chirp mass. High resolution integral field spectroscopy will allow to probe the environment of the source from Mpc to sub-kpc scales, providing a wealth of information about the interplay of the binary with its host at different scales. Periodicity and variability at all wavelengths, spectroscopy of broad emission lines (including, possibly, K$\alpha$ lines) will permit a detailed characterization of the accretion flow onto the SMBH binary and the efficiency of the emission processes at play, that can be used to test our theoretical models of SMBH binary-disk interaction. 

\part{Concluding remarks}
\label{sec:concluding_remarks}

We reviewed the multi-faceted astrophysics of  dual and binary AGN systems, and discussed topics that emerged in the meeting held at the Lorentz Center in Leiden ``The Quest for Multiple Supermassive Black Holes: A Multi-Messenger View''. 
Dual and binary AGN are rare and often obscured sources. Their discovery and their modeling
is complex but their importance is overarching, as the astrophysics ruling the formation of dual AGN  encompasses many scales and different cosmic environments. This requires on the one hand costly, multiband and coordinated  follow-ups  in order to firmly establish the presence of multiple AGN systems on all scales and on the other hand multi-scale theoretical models, bridging the AU scale of accretion disks with the Mpc scales of cosmic structures.

 Within our broader community,  judging the reliability and/or the significance of an observational (or even a theoretical) result is often difficult because of the limited knowledge of the methodology used by groups working in different fields. 
The aim of this review is to provide a comprehensive summary of results and techniques adopted by different groups working in this field, with focus on  observations, theory, and numerical simulations.

While there is a rich literature on the topic, systematic searches of binary/dual SMBHs, or studies of their candidate hosts are rare. The future missions in different energy domains, from radio to X-rays, will increase  the number of candidate dual AGN systems by orders of magnitude due to enhanced resolution and sensitivity. Simultaneously, the combination of all surveys (e.g., optical/NIR or MIR) available in the near future will allow to filter out the ``real'' dual AGN from the false positives.

Current state-of-the-art cosmological simulations of galaxy evolution and simulations of isolated galaxy  mergers provide important tools for the interpretation of observations and allow us to detect the presence of dual, obscured AGN systems. In particular, cosmological simulations suggest that dual AGN activity is not a unique tracer of galaxy mergers, since many inactive SMBH pairs at kpc-scale separations exist.

Although isolated merger simulations typically reach a higher resolution compared to the coarser-grained cosmological simulations, dust obscuration from the torus in the vicinity of the SMBHs is not resolved yet. This might lead to an over-estimate of the  lifetime of dual AGN, if inferred from observations in the X-rays. Future higher resolution and improved subgrid recipes need to be developed in order to describe the behavior of these systems.

We then provided an overview on the current theoretical understanding of the orbital decay of SMBHBs  from the hundred pc to sub-pc scale. Several physical processes can cause the evolution of SMBH pairs to stall in this regime of separations. Such systems would be detectable as single/dual AGN, if fed individually (on the larger scale) or if surrounded by a circumbinary disk which feeds the two SMBHs through mini-disks. A quantity of interest for observations is the characteristic residency time which, on these scales, is of order of $\sim 10$ Myrs or less, yet sufficiently extended to enable detection.

At the time of writing, there are few hundred pc and sub-pc scale SMBHB {\it candidates} described in the literature but their nature as true binaries is inconclusive and remains to be tested through time-domain and continued multi-wavelength monitoring. The new photometric (e.g., ZTF, LSST and others) and spectroscopic searches (e.g., SDSS-V) promise to provide even longer baselines and larger datasets of AGN which can be used to search for binary signatures -- a crucial advancement given the low incidence of sub-pc SMBHBs expected from theory. Combined with a growing sophistication of simulations of gravitationally bound binaries and improved theoretical predictions for EM and GW signatures, this bodes well for future detections of SMBHBs. In the next decades, GW observatories, such as LISA and PTAs, will provide first direct evidence of binary and merging high-$z$ SMBHs, and of a GW foreground of inspiralling SMBHBs at $z\sim1$. These prospected observations will open a new chapter in studies of binary and dual SMBHBs.

%%%%%
\newpage
%%%%%

\section{Acknowledgments}
All authors acknowledge the hospitality of the Lorentz Center for international workshops (Leiden, The Netherlands), where the idea of this work was born, and acknowledge the support of the International Space Science Institute (ISSI Bern, Switzerland), where the collaboration was originated. We thank the reviewers for having provided valuable and constructive comments that improved the clarity of the manuscript, and J.E. Barnes, L. Blecha, M. Eracleous, B.D. Farris, H. Fu, X. Liu, R. Pfeifle, L.C. Popovic, C. Ricci, S. Rodriguez, S. Tang, G. B. Taylor for their kind permission to reuse figures from their publications. ADR, CV and SB acknowledge financial support from ASI under grant ASI-INAF I/037/12/0, and from the agreement ASI-INAF n. 2017-14-H.O. TB acknowledges support by the National Aeronautics and Space Administration under Grant No. NNX15AK84G and 80NSSC19K0319 issued through the Astrophysics Theory Program and by the Research Corporation for Science Advancement through a Cottrell Scholar Award. SF and K\'EG thank the Hungarian National Research, Development, and Innovation Office (OTKA NN110333) for support. K\'EG was supported by the J\'anos Bolyai Research Scholarship of the Hungarian Academy of Sciences and by the \'UNKP-19-4 New National Excellence Program of the Ministry for Innovation and Technology.
BH acknowledges financial support by the DFG grant GE625/17-1.
EL is supported by a European Union COFUND/Durham Junior Research Fellowship (under EU grant agreement no. 609412). DL acknowledges support from the European Research Council (ERC) under grant 647208 (PI Jonker). AS is supported by the ERC CoG grant 818691 (B Massive).
MGi is supported by the ``Programa de Atracci\'on de Talento'' of the Comunidad de Madrid grant 2018-T1/TIC-11733 for the project: ``Unveiling Black Hole Winds from Space'', and
by the Spanish State Research Agency (AEI) grant MDM-2017-0737 Unidad de Excelencia ``Mar\'ia de Maeztu'' - Centro de Astrobiolog\'ia (INTA-CSIC).
PRC and LM acknowledge support from the Swiss National
Science Foundation under the Grant 200020\_178949. 
MC acknowledges support from the National Science Foundation (NSF) NANOGrav Physics Frontier Center, award number 1430284.
NHR acknowledges support from the BMBF Verbundforschung under FKZ: 05A17PC1 and FKZ: 05A17PC2.
ZH acknowledges support from NASA grants NNX17AL82G and 80NSSC19K0149 and NSF grant 1715661. KI acknowledges support by the Spanish MINECO under grant AYA2016-76012-C3-1-P and MDM-2014-0369 of ICCUB (Unidad de Excelencia 'Mar\'ia de Maeztu').
MPT acknowledges financial support from the Spanish MCIU through
the ``Center of Excellence Severo Ochoa'' award for the Instituto de
Astrof\'isica de Andaluc\'ia (SEV-2017-0709) and through the MINECO grants
AYA2012-38491-C02-02 and AYA2015-63939-C2-1-P.

\section*{References}

\bibliography{references}

\begin{thebibliography}{581}
\expandafter\ifx\csname natexlab\endcsname\relax\def\natexlab#1{#1}\fi
\expandafter\ifx\csname url\endcsname\relax
  \def\url#1{\texttt{#1}}\fi
\expandafter\ifx\csname urlprefix\endcsname\relax\def\urlprefix{URL }\fi

\bibitem[{{Abazajian} et~al.(2009){Abazajian}, {Adelman-McCarthy},
  {Ag{\"u}eros}, {Allam}, {Allende Prieto}, {An}, {Anderson}, {Anderson},
  {Annis}, {Bahcall}, and et~al.}]{AbazajianAMC09}
{Abazajian}, K.~N., {Adelman-McCarthy}, J.~K., {Ag{\"u}eros}, M.~A., {Allam},
  S.~S., {Allende Prieto}, C., {An}, D., {Anderson}, K.~S.~J., {Anderson},
  S.~F., {Annis}, J., {Bahcall}, N.~A., et~al., Jun. 2009. {The Seventh Data
  Release of the Sloan Digital Sky Survey}. \apjs 182, 543--558.

\bibitem[{{Abbott} et~al.(2016){Abbott}, {Abbott}, {Abbott}, {Abernathy},
  {Acernese}, {Ackley}, {Adams}, {Adams}, {Addesso}, {Adhikari}, and
  et~al.}]{2016PhRvX...6d1015A}
{Abbott}, B.~P., {Abbott}, R., {Abbott}, T.~D., {Abernathy}, M.~R., {Acernese},
  F., {Ackley}, K., {Adams}, C., {Adams}, T., {Addesso}, P., {Adhikari}, R.~X.,
  et~al., Oct. 2016. {Binary Black Hole Mergers in the First Advanced LIGO
  Observing Run}. Physical Review X 6~(4), 041015.

\bibitem[{{Abbott} et~al.(2017{\natexlab{a}}){Abbott}, {Abbott}, {Abbott},
  {Acernese}, {Ackley}, {Adams}, {Adams}, {Addesso}, {Adhikari}, {Adya}, and
  et~al.}]{2017PhRvL.119p1101A}
{Abbott}, B.~P., {Abbott}, R., {Abbott}, T.~D., {Acernese}, F., {Ackley}, K.,
  {Adams}, C., {Adams}, T., {Addesso}, P., {Adhikari}, R.~X., {Adya}, V.~B.,
  et~al., Oct. 2017{\natexlab{a}}. {GW170817: Observation of Gravitational
  Waves from a Binary Neutron Star Inspiral}. Physical Review Letters 119~(16),
  161101.

\bibitem[{{Abbott} et~al.(2017{\natexlab{b}}){Abbott}, {Abbott}, {Abbott},
  {Acernese}, {Ackley}, {Adams}, {Adams}, {Addesso}, {Adhikari}, {Adya}, and
  et~al.}]{2017ApJ...848L..12A}
{Abbott}, B.~P., {Abbott}, R., {Abbott}, T.~D., {Acernese}, F., {Ackley}, K.,
  {Adams}, C., {Adams}, T., {Addesso}, P., {Adhikari}, R.~X., {Adya}, V.~B.,
  et~al., Oct. 2017{\natexlab{b}}. {Multi-messenger Observations of a Binary
  Neutron Star Merger}. \apjl 848, L12.

\bibitem[{{Ackermann} et~al.(2015){Ackermann}, {Ajello}, {Albert}, {Atwood},
  {Baldini}, {Ballet}, {Barbiellini}, {Bastieri}, {Becerra Gonzalez},
  {Bellazzini}, {Bissaldi}, {Blandford}, {Bloom}, {Bonino}, {Bottacini},
  {Bregeon}, {Bruel}, {Buehler}, {Buson}, {Caliandro}, {Cameron}, {Caputo},
  {Caragiulo}, {Caraveo}, {Cavazzuti}, {Cecchi}, {Chekhtman}, {Chiang},
  {Chiaro}, {Ciprini}, {Cohen-Tanugi}, {Conrad}, {Cutini}, {D'Ammando}, {de
  Angelis}, {de Palma}, {Desiante}, {Di Venere}, {Dom{\'{\i}}nguez}, {Drell},
  {Favuzzi}, {Fegan}, {Ferrara}, {Focke}, {Fuhrmann}, {Fukazawa}, {Fusco},
  {Gargano}, {Gasparrini}, {Giglietto}, {Giommi}, {Giordano}, {Giroletti},
  {Godfrey}, {Green}, {Grenier}, {Grove}, {Guiriec}, {Harding}, {Hays},
  {Hewitt}, {Hill}, {Horan}, {Jogler}, {J{\'o}hannesson}, {Johnson}, {Kamae},
  {Kuss}, {Larsson}, {Latronico}, {Li}, {Li}, {Longo}, {Loparco}, {Lott},
  {Lovellette}, {Lubrano}, {Magill}, {Maldera}, {Manfreda}, {Max-Moerbeck},
  {Mayer}, {Mazziotta}, {McEnery}, {Michelson}, {Mizuno}, {Monzani},
  {Morselli}, {Moskalenko}, {Murgia}, {Nuss}, {Ohno}, {Ohsugi}, {Ojha},
  {Omodei}, {Orlando}, {Ormes}, {Paneque}, {Pearson}, {Perkins}, {Perri},
  {Pesce-Rollins}, {Petrosian}, {Piron}, {Pivato}, {Porter}, {Rain{\`o}},
  {Rando}, {Razzano}, {Readhead}, {Reimer}, {Reimer}, {Schulz}, {Sgr{\`o}},
  {Siskind}, {Spada}, {Spandre}, {Spinelli}, {Suson}, {Takahashi}, {Thayer},
  {Thompson}, {Tibaldo}, {Torres}, {Tosti}, {Troja}, {Uchiyama}, {Vianello},
  {Wood}, {Wood}, {Zimmer}, {Berdyugin}, {Corbet}, {Hovatta}, {Lindfors},
  {Nilsson}, {Reinthal}, {Sillanp{\"a}{\"a}}, {Stamerra}, {Takalo}, and
  {Valtonen}}]{ackermannetal15}
{Ackermann}, M., {Ajello}, M., {Albert}, A., {Atwood}, W.~B., {Baldini}, L.,
  {Ballet}, J., {Barbiellini}, G., {Bastieri}, D., {Becerra Gonzalez}, J.,
  {Bellazzini}, R., {Bissaldi}, E., {Blandford}, R.~D., {Bloom}, E.~D.,
  {Bonino}, R., {Bottacini}, E., {Bregeon}, J., {Bruel}, P., {Buehler}, R.,
  {Buson}, S., {Caliandro}, G.~A., {Cameron}, R.~A., {Caputo}, R., {Caragiulo},
  M., {Caraveo}, P.~A., {Cavazzuti}, E., {Cecchi}, C., {Chekhtman}, A.,
  {Chiang}, J., {Chiaro}, G., {Ciprini}, S., {Cohen-Tanugi}, J., {Conrad}, J.,
  {Cutini}, S., {D'Ammando}, F., {de Angelis}, A., {de Palma}, F., {Desiante},
  R., {Di Venere}, L., {Dom{\'{\i}}nguez}, A., {Drell}, P.~S., {Favuzzi}, C.,
  {Fegan}, S.~J., {Ferrara}, E.~C., {Focke}, W.~B., {Fuhrmann}, L., {Fukazawa},
  Y., {Fusco}, P., {Gargano}, F., {Gasparrini}, D., {Giglietto}, N., {Giommi},
  P., {Giordano}, F., {Giroletti}, M., {Godfrey}, G., {Green}, D., {Grenier},
  I.~A., {Grove}, J.~E., {Guiriec}, S., {Harding}, A.~K., {Hays}, E., {Hewitt},
  J.~W., {Hill}, A.~B., {Horan}, D., {Jogler}, T., {J{\'o}hannesson}, G.,
  {Johnson}, A.~S., {Kamae}, T., {Kuss}, M., {Larsson}, S., {Latronico}, L.,
  {Li}, J., {Li}, L., {Longo}, F., {Loparco}, F., {Lott}, B., {Lovellette},
  M.~N., {Lubrano}, P., {Magill}, J., {Maldera}, S., {Manfreda}, A.,
  {Max-Moerbeck}, W., {Mayer}, M., {Mazziotta}, M.~N., {McEnery}, J.~E.,
  {Michelson}, P.~F., {Mizuno}, T., {Monzani}, M.~E., {Morselli}, A.,
  {Moskalenko}, I.~V., {Murgia}, S., {Nuss}, E., {Ohno}, M., {Ohsugi}, T.,
  {Ojha}, R., {Omodei}, N., {Orlando}, E., {Ormes}, J.~F., {Paneque}, D.,
  {Pearson}, T.~J., {Perkins}, J.~S., {Perri}, M., {Pesce-Rollins}, M.,
  {Petrosian}, V., {Piron}, F., {Pivato}, G., {Porter}, T.~A., {Rain{\`o}}, S.,
  {Rando}, R., {Razzano}, M., {Readhead}, A., {Reimer}, A., {Reimer}, O.,
  {Schulz}, A., {Sgr{\`o}}, C., {Siskind}, E.~J., {Spada}, F., {Spandre}, G.,
  {Spinelli}, P., {Suson}, D.~J., {Takahashi}, H., {Thayer}, J.~B., {Thompson},
  D.~J., {Tibaldo}, L., {Torres}, D.~F., {Tosti}, G., {Troja}, E., {Uchiyama},
  Y., {Vianello}, G., {Wood}, K.~S., {Wood}, M., {Zimmer}, S., {Berdyugin}, A.,
  {Corbet}, R.~H.~D., {Hovatta}, T., {Lindfors}, E., {Nilsson}, K., {Reinthal},
  R., {Sillanp{\"a}{\"a}}, A., {Stamerra}, A., {Takalo}, L.~O., {Valtonen},
  M.~J., Nov. 2015. {Multiwavelength Evidence for Quasi-periodic Modulation in
  the Gamma-Ray Blazar PG 1553+113}. \apjl 813, L41.

\bibitem[{{Agudo} et~al.(2012){Agudo}, {Marscher}, {Jorstad}, {G{\'o}mez},
  {Perucho}, {Piner}, {Rioja}, and {Dodson}}]{Agudo_et_al_2012}
{Agudo}, I., {Marscher}, A.~P., {Jorstad}, S.~G., {G{\'o}mez}, J.~L.,
  {Perucho}, M., {Piner}, B.~G., {Rioja}, M., {Dodson}, R., Mar. 2012. {Erratic
  Jet Wobbling in the BL Lacertae Object OJ287 Revealed by Sixteen Years of 7
  mm VLBA Observations}. \apj 747, 63.

\bibitem[{{Ahn} et~al.(2012){Ahn}, {Alexandroff}, {Allende Prieto}, {Anderson},
  {Anderton}, {Andrews}, {Aubourg}, {Bailey}, {Balbinot}, {Barnes}, and
  et~al.}]{2012ApJS..203...21A}
{Ahn}, C.~P., {Alexandroff}, R., {Allende Prieto}, C., {Anderson}, S.~F.,
  {Anderton}, T., {Andrews}, B.~H., {Aubourg}, {\'E}., {Bailey}, S.,
  {Balbinot}, E., {Barnes}, R., et~al., Dec. 2012. {The Ninth Data Release of
  the Sloan Digital Sky Survey: First Spectroscopic Data from the SDSS-III
  Baryon Oscillation Spectroscopic Survey}. \apjs 203, 21.

\bibitem[{{Aldering} et~al.(2002){Aldering}, {Adam}, {Antilogus}, {Astier},
  {Bacon}, {Bongard}, {Bonnaud}, {Copin}, {Hardin}, {Henault}, {Howell},
  {Lemonnier}, {Levy}, {Loken}, {Nugent}, {Pain}, {Pecontal}, {Pecontal},
  {Perlmutter}, {Quimby}, {Schahmaneche}, {Smadja}, and
  {Wood-Vasey}}]{AlderingAA02}
{Aldering}, G., {Adam}, G., {Antilogus}, P., {Astier}, P., {Bacon}, R.,
  {Bongard}, S., {Bonnaud}, C., {Copin}, Y., {Hardin}, D., {Henault}, F.,
  {Howell}, D.~A., {Lemonnier}, J.-P., {Levy}, J.-M., {Loken}, S.~C., {Nugent},
  P.~E., {Pain}, R., {Pecontal}, A., {Pecontal}, E., {Perlmutter}, S.,
  {Quimby}, R.~M., {Schahmaneche}, K., {Smadja}, G., {Wood-Vasey}, W.~M., Dec.
  2002. {Overview of the Nearby Supernova Factory}. In: {Tyson}, J.~A.,
  {Wolff}, S. (Eds.), Survey and Other Telescope Technologies and Discoveries.
  Vol. 4836 of \procspie. pp. 61--72.

\bibitem[{{Alexander} and {Hickox}(2012)}]{2012NewAR..56...93A}
{Alexander}, D.~M., {Hickox}, R.~C., Jun. 2012. {What drives the growth of
  black holes?} \nar 56, 93--121.

\bibitem[{{Amaro-Seoane} et~al.(2017){Amaro-Seoane}, {Audley}, {Babak},
  {Baker}, {Barausse}, {Bender}, {Berti}, {Binetruy}, {Born}, {Bortoluzzi},
  {Camp}, {Caprini}, {Cardoso}, {Colpi}, {Conklin}, {Cornish}, {Cutler},
  {Danzmann}, {Dolesi}, {Ferraioli}, {Ferroni}, {Fitzsimons}, {Gair}, {Gesa
  Bote}, {Giardini}, {Gibert}, {Grimani}, {Halloin}, {Heinzel}, {Hertog},
  {Hewitson}, {Holley-Bockelmann}, {Hollington}, {Hueller}, {Inchauspe},
  {Jetzer}, {Karnesis}, {Killow}, {Klein}, {Klipstein}, {Korsakova}, {Larson},
  {Livas}, {Lloro}, {Man}, {Mance}, {Martino}, {Mateos}, {McKenzie},
  {McWilliams}, {Miller}, {Mueller}, {Nardini}, {Nelemans}, {Nofrarias},
  {Petiteau}, {Pivato}, {Plagnol}, {Porter}, {Reiche}, {Robertson},
  {Robertson}, {Rossi}, {Russano}, {Schutz}, {Sesana}, {Shoemaker}, {Slutsky},
  {Sopuerta}, {Sumner}, {Tamanini}, {Thorpe}, {Troebs}, {Vallisneri},
  {Vecchio}, {Vetrugno}, {Vitale}, {Volonteri}, {Wanner}, {Ward}, {Wass},
  {Weber}, {Ziemer}, and {Zweifel}}]{LISA17}
{Amaro-Seoane}, P., {Audley}, H., {Babak}, S., {Baker}, J., {Barausse}, E.,
  {Bender}, P., {Berti}, E., {Binetruy}, P., {Born}, M., {Bortoluzzi}, D.,
  {Camp}, J., {Caprini}, C., {Cardoso}, V., {Colpi}, M., {Conklin}, J.,
  {Cornish}, N., {Cutler}, C., {Danzmann}, K., {Dolesi}, R., {Ferraioli}, L.,
  {Ferroni}, V., {Fitzsimons}, E., {Gair}, J., {Gesa Bote}, L., {Giardini}, D.,
  {Gibert}, F., {Grimani}, C., {Halloin}, H., {Heinzel}, G., {Hertog}, T.,
  {Hewitson}, M., {Holley-Bockelmann}, K., {Hollington}, D., {Hueller}, M.,
  {Inchauspe}, H., {Jetzer}, P., {Karnesis}, N., {Killow}, C., {Klein}, A.,
  {Klipstein}, B., {Korsakova}, N., {Larson}, S.~L., {Livas}, J., {Lloro}, I.,
  {Man}, N., {Mance}, D., {Martino}, J., {Mateos}, I., {McKenzie}, K.,
  {McWilliams}, S.~T., {Miller}, C., {Mueller}, G., {Nardini}, G., {Nelemans},
  G., {Nofrarias}, M., {Petiteau}, A., {Pivato}, P., {Plagnol}, E., {Porter},
  E., {Reiche}, J., {Robertson}, D., {Robertson}, N., {Rossi}, E., {Russano},
  G., {Schutz}, B., {Sesana}, A., {Shoemaker}, D., {Slutsky}, J., {Sopuerta},
  C.~F., {Sumner}, T., {Tamanini}, N., {Thorpe}, I., {Troebs}, M.,
  {Vallisneri}, M., {Vecchio}, A., {Vetrugno}, D., {Vitale}, S., {Volonteri},
  M., {Wanner}, G., {Ward}, H., {Wass}, P., {Weber}, W., {Ziemer}, J.,
  {Zweifel}, P., Feb. 2017. {Laser Interferometer Space Antenna}. arXiv
  e-prints.

\bibitem[{{An} et~al.(2018){An}, {Mohan}, and {Frey}}]{2018RaSc...53.1211A}
{An}, T., {Mohan}, P., {Frey}, S., Oct. 2018. {VLBI Studies of DAGN and SMBHB
  Hosting Galaxies}. Radio Science 53, 1211--1217.

\bibitem[{{Anninos} et~al.(2018){Anninos}, {Fragile}, {Olivier}, {Hoffman},
  {Mishra}, and {Camarda}}]{Anninos2018}
{Anninos}, P., {Fragile}, P.~C., {Olivier}, S.~S., {Hoffman}, R., {Mishra}, B.,
  {Camarda}, K., Sep. 2018. {Relativistic Tidal Disruption and Nuclear Ignition
  of White Dwarf Stars by Intermediate-mass Black Holes}. \apj 865, 3.

\bibitem[{{Armitage}(2007)}]{ArmitageReview}
{Armitage}, P.~J., Jan 2007. {Lecture notes on the formation and early
  evolution of planetary systems}. arXiv e-prints, astro--ph/0701485.

\bibitem[{{Armitage} and {Natarajan}(2002)}]{2002ApJ...567L...9A}
{Armitage}, P.~J., {Natarajan}, P., Mar 2002. {Accretion during the Merger of
  Supermassive Black Holes}. \apj 567~(1), L9--L12.

\bibitem[{{Armitage} and {Natarajan}(2005)}]{an05}
{Armitage}, P.~J., {Natarajan}, P., Dec. 2005. {Eccentricity of Supermassive
  Black Hole Binaries Coalescing from Gas-rich Mergers}. \apj 634, 921--927.

\bibitem[{{Armus} et~al.(2009){Armus}, {Mazzarella}, {Evans}, {Surace},
  {Sanders}, {Iwasawa}, {Frayer}, {Howell}, {Chan}, {Petric}, {Vavilkin},
  {Kim}, {Haan}, {Inami}, {Murphy}, {Appleton}, {Barnes}, {Bothun}, {Bridge},
  {Charmandaris}, {Jensen}, {Kewley}, {Lord}, {Madore}, {Marshall},
  {Melbourne}, {Rich}, {Satyapal}, {Schulz}, {Spoon}, {Sturm}, {U}, {Veilleux},
  and {Xu}}]{armus09}
{Armus}, L., {Mazzarella}, J.~M., {Evans}, A.~S., {Surace}, J.~A., {Sanders},
  D.~B., {Iwasawa}, K., {Frayer}, D.~T., {Howell}, J.~H., {Chan}, B., {Petric},
  A., {Vavilkin}, T., {Kim}, D.~C., {Haan}, S., {Inami}, H., {Murphy}, E.~J.,
  {Appleton}, P.~N., {Barnes}, J.~E., {Bothun}, G., {Bridge}, C.~R.,
  {Charmandaris}, V., {Jensen}, J.~B., {Kewley}, L.~J., {Lord}, S., {Madore},
  B.~F., {Marshall}, J.~A., {Melbourne}, J.~E., {Rich}, J., {Satyapal}, S.,
  {Schulz}, B., {Spoon}, H.~W.~W., {Sturm}, E., {U}, V., {Veilleux}, S., {Xu},
  K., Jun. 2009. {GOALS: The Great Observatories All-Sky LIRG Survey}. \pasp
  121, 559.

\bibitem[{{Arrigoni Battaia} et~al.(2019){Arrigoni Battaia}, {Hennawi},
  {Prochaska}, {O{\~n}orbe}, {Farina}, {Cantalupo}, and
  {Lusso}}]{Arrigoni-Battaia2019}
{Arrigoni Battaia}, F., {Hennawi}, J.~F., {Prochaska}, J.~X., {O{\~n}orbe}, J.,
  {Farina}, E.~P., {Cantalupo}, S., {Lusso}, E., Jan. 2019. {QSO MUSEUM I: a
  sample of 61 extended Ly $\alpha$-emission nebulae surrounding z $\sim$ 3
  quasars}. \mnras 482, 3162--3205.

\bibitem[{{Arrigoni Battaia} et~al.(2018){Arrigoni Battaia}, {Prochaska},
  {Hennawi}, {Obreja}, {Buck}, {Cantalupo}, {Dutton}, and
  {Macci{\`o}}}]{Arrigoni-Battaia2018}
{Arrigoni Battaia}, F., {Prochaska}, J.~X., {Hennawi}, J.~F., {Obreja}, A.,
  {Buck}, T., {Cantalupo}, S., {Dutton}, A.~A., {Macci{\`o}}, A.~V., Jan. 2018.
  {Inspiraling halo accretion mapped in Ly {$\alpha$} emission around a
  z$\sim$3 quasar}. \mnras 473, 3907--3940.

\bibitem[{{Artymowicz} and {Lubow}(1994)}]{artymowicz1994}
{Artymowicz}, P., {Lubow}, S.~H., Feb. 1994. {Dynamics of binary-disk
  interaction. 1: Resonances and disk gap sizes}. \apj 421, 651--667.

\bibitem[{{Arzoumanian} et~al.(2018){Arzoumanian}, {Brazier}, {Burke-Spolaor},
  {Chamberlin}, {Chatterjee}, {Christy}, {Cordes}, {Cornish}, {Crawford},
  {Thankful Cromartie}, {Crowter}, {DeCesar}, {Demorest}, {Dolch}, {Ellis},
  {Ferdman}, {Ferrara}, {Fonseca}, {Garver-Daniels}, {Gentile}, {Halmrast},
  {Huerta}, {Jenet}, {Jessup}, {Jones}, {Jones}, {Kaplan}, {Lam}, {Lazio},
  {Levin}, {Lommen}, {Lorimer}, {Luo}, {Lynch}, {Madison}, {Matthews},
  {McLaughlin}, {McWilliams}, {Mingarelli}, {Ng}, {Nice}, {Pennucci}, {Ransom},
  {Ray}, {Siemens}, {Simon}, {Spiewak}, {Stairs}, {Stinebring}, {Stovall},
  {Swiggum}, {Taylor}, {Vallisneri}, {van Haasteren}, {Vigeland}, {Zhu}, and
  {The NANOGrav Collaboration}}]{2018ApJS..235...37A}
{Arzoumanian}, Z., {Brazier}, A., {Burke-Spolaor}, S., {Chamberlin}, S.,
  {Chatterjee}, S., {Christy}, B., {Cordes}, J.~M., {Cornish}, N.~J.,
  {Crawford}, F., {Thankful Cromartie}, H., {Crowter}, K., {DeCesar}, M.~E.,
  {Demorest}, P.~B., {Dolch}, T., {Ellis}, J.~A., {Ferdman}, R.~D., {Ferrara},
  E.~C., {Fonseca}, E., {Garver-Daniels}, N., {Gentile}, P.~A., {Halmrast}, D.,
  {Huerta}, E.~A., {Jenet}, F.~A., {Jessup}, C., {Jones}, G., {Jones}, M.~L.,
  {Kaplan}, D.~L., {Lam}, M.~T., {Lazio}, T.~J.~W., {Levin}, L., {Lommen}, A.,
  {Lorimer}, D.~R., {Luo}, J., {Lynch}, R.~S., {Madison}, D., {Matthews},
  A.~M., {McLaughlin}, M.~A., {McWilliams}, S.~T., {Mingarelli}, C., {Ng}, C.,
  {Nice}, D.~J., {Pennucci}, T.~T., {Ransom}, S.~M., {Ray}, P.~S., {Siemens},
  X., {Simon}, J., {Spiewak}, R., {Stairs}, I.~H., {Stinebring}, D.~R.,
  {Stovall}, K., {Swiggum}, J.~K., {Taylor}, S.~R., {Vallisneri}, M., {van
  Haasteren}, R., {Vigeland}, S.~J., {Zhu}, W., {The NANOGrav Collaboration},
  Apr. 2018. {The NANOGrav 11-year Data Set: High-precision Timing of 45
  Millisecond Pulsars}. \apjs 235, 37.

\bibitem[{{Assef} et~al.(2013){Assef}, {Stern}, {Kochanek}, {Blain}, {Brodwin},
  {Brown}, {Donoso}, {Eisenhardt}, {Jannuzi}, {Jarrett}, {Stanford}, {Tsai},
  {Wu}, and {Yan}}]{Assef2013}
{Assef}, R.~J., {Stern}, D., {Kochanek}, C.~S., {Blain}, A.~W., {Brodwin}, M.,
  {Brown}, M.~J.~I., {Donoso}, E., {Eisenhardt}, P.~R.~M., {Jannuzi}, B.~T.,
  {Jarrett}, T.~H., {Stanford}, S.~A., {Tsai}, C.~W., {Wu}, J., {Yan}, L., Jul
  2013. {Mid-infrared Selection of Active Galactic Nuclei with the Wide-field
  Infrared Survey Explorer. II. Properties of WISE-selected Active Galactic
  Nuclei in the NDWFS Bo{\"o}tes Field}. \apj 772~(1), 26.

\bibitem[{{Baldassare} et~al.(2015){Baldassare}, {Reines}, {Gallo}, and
  {Greene}}]{Baldassare_et_al_2015}
{Baldassare}, V.~F., {Reines}, A.~E., {Gallo}, E., {Greene}, J.~E., Aug 2015.
  {A $\sim$50,000 M$_\odot$ Solar Mass Black Hole in the Nucleus of RGG 118}.
  \apjl 809~(1), L14.

\bibitem[{{Baldwin} et~al.(1995){Baldwin}, {Ferland}, {Korista}, and
  {Verner}}]{Baldwin95}
{Baldwin}, J., {Ferland}, G., {Korista}, K., {Verner}, D., Dec. 1995. {Locally
  Optimally Emitting Clouds and the Origin of Quasar Emission Lines}. \apjl
  455, L119.

\bibitem[{{Baldwin} et~al.(1981){Baldwin}, {Phillips}, and
  {Terlevich}}]{Baldwin81}
{Baldwin}, J.~A., {Phillips}, M.~M., {Terlevich}, R., Feb. 1981.
  {Classification parameters for the emission-line spectra of extragalactic
  objects}. \pasp 93, 5--19.

\bibitem[{{Ballo} et~al.(2004){Ballo}, {Braito}, {Della Ceca}, {Maraschi},
  {Tavecchio}, and {Dadina}}]{Ballo2004}
{Ballo}, L., {Braito}, V., {Della Ceca}, R., {Maraschi}, L., {Tavecchio}, F.,
  {Dadina}, M., Jan. 2004. {Arp 299: A Second Merging System with Two Active
  Nuclei?} \apj 600, 634--639.

\bibitem[{{Bansal} et~al.(2017){Bansal}, {Taylor}, {Peck}, {Zavala}, and
  {Romani}}]{bansal2017}
{Bansal}, K., {Taylor}, G.~B., {Peck}, A.~B., {Zavala}, R.~T., {Romani}, R.~W.,
  Jul. 2017. {Constraining the Orbit of the Supermassive Black Hole Binary
  0402+379}. \apj 843, 14.

\bibitem[{{B{\"a}r} et~al.(2017){B{\"a}r}, {Weigel}, {Sartori}, {Oh}, {Koss},
  and {Schawinski}}]{Baer2017}
{B{\"a}r}, R.~E., {Weigel}, A.~K., {Sartori}, L.~F., {Oh}, K., {Koss}, M.,
  {Schawinski}, K., Apr. 2017. {Active galactic nuclei from He II: a more
  complete census of AGN in SDSS galaxies yields a new population of
  low-luminosity AGN in highly star-forming galaxies}. \mnras 466, 2879--2887.

\bibitem[{{Barack} et~al.(2019){Barack}, {Cardoso}, {Nissanke}, {Sotiriou},
  {Askar}, {Belczynski}, {Bertone}, {Bon}, {Blas}, {Brito}, {Bulik}, {Burrage},
  {Byrnes}, {Caprini}, {Chernyakova}, {Chrusciel}, {Colpi}, {Ferrari},
  {Gaggero}, {Gair}, {Garcia-Bellido}, {Hassan}, {Heisenberg}, {Hendry},
  {Heng}, {Herdeiro}, {Hinderer}, {Horesh}, {Kavanagh}, {Kocsis}, {Kramer}, {Le
  Tiec}, {Mingarelli}, {Nardini}, {Nelemans}, {Palenzuela}, {Pani}, {Perego},
  {Porter}, {Rossi}, {Schmidt}, {Sesana}, {Sperhake}, {Stamerra}, {Stein},
  {Tamanini}, {Tauris}, {Urena-Lopez}, {Vincent}, {Volonteri}, {Wardell},
  {Wex}, {Yagi}, {Abdelsalhin}, {Aloy}, {Amaro-Seoane}, {Annulli},
  {Arca-Sedda}, {Bah}, {Barausse}, {Barakovic}, {Benkel}, {Bennett}, {Bernard},
  {Bernuzzi}, {Berry}, {Berti}, {Bezares}, {Juan Blanco-Pillado},
  {Blazquez-Salcedo}, {Bonetti}, {Boskovic}, {Bosnjak}, {Bricman}, {Bruegmann},
  {Capelo}, {Carloni}, {Cerda-Duran}, {Charmousis}, {Chaty}, {Clerici},
  {Coates}, {Colleoni}, {Collodel}, {Compere}, {Cook}, {Cordero-Carrion},
  {Correia}, {de la Cruz-Dombriz}, {Czinner}, {Destounis}, {Dialektopoulos},
  {Doneva}, {Dotti}, {Drew}, {Eckner}, {Edholm}, {Emparan}, {Erdem},
  {Ferreira}, {Ferreira}, {Finch}, {Font}, {Franchini}, {Fransen}, {Gal'tsov},
  {Ganguly}, {Gerosa}, {Glampedakis}, {Gomboc}, {Goobar}, {Gualtieri},
  {Guendelman}, {Haardt}, {Harmark}, {Hejda}, {Hertog}, {Hopper}, {Husa},
  {Ihanec}, {Ikeda}, {Jaodand}, {Jetzer Xisco Jimenez-Forteza}, {Kamionkowski},
  {Kaplan}, {Kazantzidis}, {Kimura}, {Kobayashi}, {Kokkotas}, {Krolik}, {Kunz},
  {Lammerzahl}, {Lasky}, {Lemos}, {Levi Said}, {Liberati}, {Lopes}, {Luna},
  {Ma}, {Maggio}, {Martinez Montero}, {Maselli}, {Mayer}, {Mazumdar},
  {Messenger}, {Menard}, {Minamitsuji}, {Moore}, {Mota}, {Nampalliwar},
  {Nerozzi}, {Nichols}, {Nissimov}, {Obergaulinger}, {Obers}, {Oliveri},
  {Pappas}, {Pasic}, {Peiris}, {Petrushevska}, {Pollney}, {Pratten}, {Rakic},
  {Racz}, {Radia}, {Ramazanouglu}, {Ramos-Buades}, {Raposo}, {Rosca-Mead},
  {Rogatko}, {Rosinska}, {Rosswog}, {Ruiz Morales}, {Sakellariadou},
  {Sanchis-Gual}, {Sharan Salafia}, {Samajdar}, {Sintes}, {Smole}, {Sopuerta},
  {Souza-Lima}, {Stalevski}, {Stergioulas}, {Stevens}, {Tamfal},
  {Torres-Forne}, {Tsygankov}, {Unluturk}, {Valiante}, {van de Meent},
  {Velhinho}, {Verbin}, {Vercnocke}, {Vernieri}, {Vicente}, {Vitagliano},
  {Weltman}, {Whiting}, {Williamson}, {Witek}, {Wojnar}, {Yakut}, {Yan},
  {Yazadjiev}, {Zaharijas}, and {Zilhao}}]{gwverse_paper}
{Barack}, L., {Cardoso}, V., {Nissanke}, S., {Sotiriou}, T.~P., {Askar}, A.,
  {Belczynski}, K., {Bertone}, G., {Bon}, E., {Blas}, D., {Brito}, R., {Bulik},
  T., {Burrage}, C., {Byrnes}, C.~T., {Caprini}, C., {Chernyakova}, M.,
  {Chrusciel}, P., {Colpi}, M., {Ferrari}, V., {Gaggero}, D., {Gair}, J.,
  {Garcia-Bellido}, J., {Hassan}, S.~F., {Heisenberg}, L., {Hendry}, M.,
  {Heng}, I.~S., {Herdeiro}, C., {Hinderer}, T., {Horesh}, A., {Kavanagh},
  B.~J., {Kocsis}, B., {Kramer}, M., {Le Tiec}, A., {Mingarelli}, C.,
  {Nardini}, G., {Nelemans}, G., {Palenzuela}, C., {Pani}, P., {Perego}, A.,
  {Porter}, E.~K., {Rossi}, E.~M., {Schmidt}, P., {Sesana}, A., {Sperhake}, U.,
  {Stamerra}, A., {Stein}, L.~C., {Tamanini}, N., {Tauris}, T.~M.,
  {Urena-Lopez}, L.~A., {Vincent}, F., {Volonteri}, M., {Wardell}, B., {Wex},
  N., {Yagi}, K., {Abdelsalhin}, T., {Aloy}, M.~A., {Amaro-Seoane}, P.,
  {Annulli}, L., {Arca-Sedda}, M., {Bah}, I., {Barausse}, E., {Barakovic}, E.,
  {Benkel}, R., {Bennett}, C.~L., {Bernard}, L., {Bernuzzi}, S., {Berry},
  C.~P.~L., {Berti}, E., {Bezares}, M., {Juan Blanco-Pillado}, J.,
  {Blazquez-Salcedo}, J.~L., {Bonetti}, M., {Boskovic}, M., {Bosnjak}, Z.,
  {Bricman}, K., {Bruegmann}, B., {Capelo}, P.~R., {Carloni}, S.,
  {Cerda-Duran}, P., {Charmousis}, C., {Chaty}, S., {Clerici}, A., {Coates},
  A., {Colleoni}, M., {Collodel}, L.~G., {Compere}, G., {Cook}, W.,
  {Cordero-Carrion}, I., {Correia}, M., {de la Cruz-Dombriz}, A., {Czinner},
  V.~G., {Destounis}, K., {Dialektopoulos}, K., {Doneva}, D., {Dotti}, M.,
  {Drew}, A., {Eckner}, C., {Edholm}, J., {Emparan}, R., {Erdem}, R.,
  {Ferreira}, M., {Ferreira}, P.~G., {Finch}, A., {Font}, J.~A., {Franchini},
  N., {Fransen}, K., {Gal'tsov}, D., {Ganguly}, A., {Gerosa}, D.,
  {Glampedakis}, K., {Gomboc}, A., {Goobar}, A., {Gualtieri}, L., {Guendelman},
  E., {Haardt}, F., {Harmark}, T., {Hejda}, F., {Hertog}, T., {Hopper}, S.,
  {Husa}, S., {Ihanec}, N., {Ikeda}, T., {Jaodand}, A., {Jetzer Xisco
  Jimenez-Forteza}, P., {Kamionkowski}, M., {Kaplan}, D.~E., {Kazantzidis}, S.,
  {Kimura}, M., {Kobayashi}, S., {Kokkotas}, K., {Krolik}, J., {Kunz}, J.,
  {Lammerzahl}, C., {Lasky}, P., {Lemos}, J.~P.~S., {Levi Said}, J.,
  {Liberati}, S., {Lopes}, J., {Luna}, R., {Ma}, Y.-Z., {Maggio}, E., {Martinez
  Montero}, M., {Maselli}, A., {Mayer}, L., {Mazumdar}, A., {Messenger}, C.,
  {Menard}, B., {Minamitsuji}, M., {Moore}, C.~J., {Mota}, D., {Nampalliwar},
  S., {Nerozzi}, A., {Nichols}, D., {Nissimov}, E., {Obergaulinger}, M.,
  {Obers}, N.~A., {Oliveri}, R., {Pappas}, G., {Pasic}, V., {Peiris}, H.,
  {Petrushevska}, T., {Pollney}, D., {Pratten}, G., {Rakic}, N., {Racz}, I.,
  {Radia}, M., {Ramazanouglu}, F.~M., {Ramos-Buades}, A., {Raposo}, G.,
  {Rosca-Mead}, R., {Rogatko}, M., {Rosinska}, D., {Rosswog}, S., {Ruiz
  Morales}, E., {Sakellariadou}, M., {Sanchis-Gual}, N., {Sharan Salafia}, O.,
  {Samajdar}, A., {Sintes}, A., {Smole}, M., {Sopuerta}, C., {Souza-Lima}, R.,
  {Stalevski}, M., {Stergioulas}, N., {Stevens}, C., {Tamfal}, T.,
  {Torres-Forne}, A., {Tsygankov}, S., {Unluturk}, K., {Valiante}, R., {van de
  Meent}, M., {Velhinho}, J., {Verbin}, Y., {Vercnocke}, B., {Vernieri}, D.,
  {Vicente}, R., {Vitagliano}, V., {Weltman}, A., {Whiting}, B., {Williamson},
  A., {Witek}, H., {Wojnar}, A., {Yakut}, K., {Yan}, H., {Yazadjiev}, S.,
  {Zaharijas}, G., {Zilhao}, M., Jul 2019. {Black holes, gravitational waves
  and fundamental physics: a roadmap}. Classical and Quantum Gravity 36~(14),
  143001.

\bibitem[{{Barack} and {Cutler}(2004)}]{2004PhRvD..69h2005B}
{Barack}, L., {Cutler}, C., Apr. 2004. {LISA capture sources: Approximate
  waveforms, signal-to-noise ratios, and parameter estimation accuracy}. \prd
  69~(8), 082005.

\bibitem[{{Barnes}(1992)}]{Barnes_1992}
{Barnes}, J.~E., Jul. 1992. {Transformations of galaxies. I - Mergers of
  equal-mass stellar disks}. \apj 393, 484--507.

\bibitem[{{Barnes}(2002)}]{Barnes_2002}
{Barnes}, J.~E., Jul. 2002. {Formation of gas discs in merging galaxies}.
  \mnras 333, 481--494.

\bibitem[{{Barnes} and {Hernquist}(1996)}]{Barnes_Hernquist_1996}
{Barnes}, J.~E., {Hernquist}, L., Nov. 1996. {Transformations of Galaxies. II.
  Gasdynamics in Merging Disk Galaxies}. \apj 471, 115.

\bibitem[{{Barnes} and {Hernquist}(1991)}]{Barnes_Hernquist_1991}
{Barnes}, J.~E., {Hernquist}, L.~E., Apr. 1991. {Fueling starburst galaxies
  with gas-rich mergers}. \apjl 370, L65--L68.

\bibitem[{{Barrows} et~al.(2013){Barrows}, {Sandberg Lacy}, {Kennefick},
  {Comerford}, {Kennefick}, and {Berrier}}]{Barrows:2013}
{Barrows}, R.~S., {Sandberg Lacy}, C.~H., {Kennefick}, J., {Comerford}, J.~M.,
  {Kennefick}, D., {Berrier}, J.~C., Jun. 2013. {Identification of Outflows and
  Candidate Dual Active Galactic Nuclei in SDSS Quasars at z = 0.8-1.6}. \apj
  769, 95.

\bibitem[{{Barrows} et~al.(2012){Barrows}, {Stern}, {Madsen}, {Harrison},
  {Assef}, {Comerford}, {Cushing}, {Fassnacht}, {Gonzalez}, {Griffith},
  {Hickox}, {Kirkpatrick}, and {Lagattuta}}]{Barrows:2012}
{Barrows}, R.~S., {Stern}, D., {Madsen}, K., {Harrison}, F., {Assef}, R.~J.,
  {Comerford}, J.~M., {Cushing}, M.~C., {Fassnacht}, C.~D., {Gonzalez}, A.~H.,
  {Griffith}, R., {Hickox}, R., {Kirkpatrick}, J.~D., {Lagattuta}, D.~J., Jan.
  2012. {A Candidate Dual Active Galactic Nucleus at z = 1.175}. \apj 744, 7.

\bibitem[{{Barth} et~al.(2015){Barth}, {Bennert}, {Canalizo}, {Filippenko},
  {Gates}, {Greene}, {Li}, {Malkan}, {Pancoast}, {Sand}, {Stern}, {Treu},
  {Woo}, {Assef}, {Bae}, {Brewer}, {Cenko}, {Clubb}, {Cooper},
  {Diamond-Stanic}, {Hiner}, {H{\"o}nig}, {Hsiao}, {Kandrashoff}, {Lazarova},
  {Nierenberg}, {Rex}, {Silverman}, {Tollerud}, and {Walsh}}]{barth15}
{Barth}, A.~J., {Bennert}, V.~N., {Canalizo}, G., {Filippenko}, A.~V., {Gates},
  E.~L., {Greene}, J.~E., {Li}, W., {Malkan}, M.~A., {Pancoast}, A., {Sand},
  D.~J., {Stern}, D., {Treu}, T., {Woo}, J.-H., {Assef}, R.~J., {Bae}, H.-J.,
  {Brewer}, B.~J., {Cenko}, S.~B., {Clubb}, K.~I., {Cooper}, M.~C.,
  {Diamond-Stanic}, A.~M., {Hiner}, K.~D., {H{\"o}nig}, S.~F., {Hsiao}, E.,
  {Kandrashoff}, M.~T., {Lazarova}, M.~S., {Nierenberg}, A.~M., {Rex}, J.,
  {Silverman}, J.~M., {Tollerud}, E.~J., {Walsh}, J.~L., Apr. 2015. {The Lick
  AGN Monitoring Project 2011: Spectroscopic Campaign and Emission-line Light
  Curves}. \apjs 217, 26.

\bibitem[{{Barth} and {Stern}(2018)}]{Barth2018}
{Barth}, A.~J., {Stern}, D., May 2018. {No Evidence of Periodic Variability in
  the Light Curve of Active Galaxy J0045+41}. \apj 859~(1), 10.

\bibitem[{{Beckmann} et~al.(2018){Beckmann}, {Slyz}, and
  {Devriendt}}]{2018MNRAS.478..995B}
{Beckmann}, R.~S., {Slyz}, A., {Devriendt}, J., Jul 2018. {Bondi or not Bondi:
  the impact of resolution on accretion and drag force modelling for
  supermassive black holes}. \mnras 478~(1), 995--1016.

\bibitem[{{Begelman} et~al.(1980){Begelman}, {Blandford}, and
  {Rees}}]{begelman1980}
{Begelman}, M.~C., {Blandford}, R.~D., {Rees}, M.~J., Sep. 1980. {Massive black
  hole binaries in active galactic nuclei}. \nat 287, 307--309.

\bibitem[{{Begelman} et~al.(2006){Begelman}, {Volonteri}, and
  {Rees}}]{Begelman2006}
{Begelman}, M.~C., {Volonteri}, M., {Rees}, M.~J., Jul 2006. {Formation of
  supermassive black holes by direct collapse in pre-galactic haloes}. \mnras
  370~(1), 289--298.

\bibitem[{{Bellovary} et~al.(2019){Bellovary}, {Cleary}, {Munshi}, {Tremmel},
  {Christensen}, {Brooks}, and {Quinn}}]{Bellovary_et_al_2018}
{Bellovary}, J.~M., {Cleary}, C.~E., {Munshi}, F., {Tremmel}, M.,
  {Christensen}, C.~R., {Brooks}, A., {Quinn}, T.~R., Jan. 2019.
  {Multimessenger signatures of massive black holes in dwarf galaxies}. \mnras
  482, 2913--2923.

\bibitem[{{Bellovary} et~al.(2010){Bellovary}, {Governato}, {Quinn}, {Wadsley},
  {Shen}, and {Volonteri}}]{Bellovary_et_al_2010}
{Bellovary}, J.~M., {Governato}, F., {Quinn}, T.~R., {Wadsley}, J., {Shen}, S.,
  {Volonteri}, M., Oct. 2010. {Wandering Black Holes in Bright Disk Galaxy
  Halos}. \apjl 721, L148--L152.

\bibitem[{{Ben{\'{\i}}tez} et~al.(2013){Ben{\'{\i}}tez}, {M{\'e}ndez-Abreu},
  {Fuentes-Carrera}, {Cruz-Gonz{\'a}lez}, {Mart{\'{\i}}nez},
  {L{\'o}pez-Martin}, {Jim{\'e}nez-Bail{\'o}n}, {Le{\'o}n-Tavares}, and
  {Chavushyan}}]{Benitez:2013}
{Ben{\'{\i}}tez}, E., {M{\'e}ndez-Abreu}, J., {Fuentes-Carrera}, I.,
  {Cruz-Gonz{\'a}lez}, I., {Mart{\'{\i}}nez}, B., {L{\'o}pez-Martin}, L.,
  {Jim{\'e}nez-Bail{\'o}n}, E., {Le{\'o}n-Tavares}, J., {Chavushyan}, V.~H.,
  Jan. 2013. {Characterization of a Sample of Intermediate-type AGNs. I.
  Spectroscopic Properties and Serendipitous Discovery of New Dual AGNs}. \apj
  763, 36.

\bibitem[{{Bentz} et~al.(2009){Bentz}, {Peterson}, {Netzer}, {Pogge}, and
  {Vestergaard}}]{bentz09}
{Bentz}, M.~C., {Peterson}, B.~M., {Netzer}, H., {Pogge}, R.~W., {Vestergaard},
  M., May 2009. {The Radius-Luminosity Relationship for Active Galactic Nuclei:
  The Effect of Host-Galaxy Starlight on Luminosity Measurements. II. The Full
  Sample of Reverberation-Mapped AGNs}. \apj 697, 160--181.

\bibitem[{{Berczik} et~al.(2006){Berczik}, {Merritt}, {Spurzem}, and
  {Bischof}}]{berczik06}
{Berczik}, P., {Merritt}, D., {Spurzem}, R., {Bischof}, H.-P., May 2006.
  {Efficient Merger of Binary Supermassive Black Holes in Nonaxisymmetric
  Galaxies}. \apjl 642, L21--L24.

\bibitem[{{Bianchi} et~al.(2008){Bianchi}, {Chiaberge}, {Piconcelli},
  {Guainazzi}, and {Matt}}]{Bianchi2008}
{Bianchi}, S., {Chiaberge}, M., {Piconcelli}, E., {Guainazzi}, M., {Matt}, G.,
  May 2008. {Chandra unveils a binary active galactic nucleus in Mrk 463}.
  \mnras 386, 105--110.

\bibitem[{{Biava} et~al.(2019){Biava}, {Colpi}, {Capelo}, {Bonetti},
  {Volonteri}, {Tamfal}, {Mayer}, and {Sesana}}]{Biava_et_al_2019}
{Biava}, N., {Colpi}, M., {Capelo}, P.~R., {Bonetti}, M., {Volonteri}, M.,
  {Tamfal}, T., {Mayer}, L., {Sesana}, A., Aug 2019. {The lifetime of binary
  black holes in S{\'e}rsic galaxy models}. \mnras 487~(4), 4985--4994.

\bibitem[{{Biernacki} et~al.(2017){Biernacki}, {Teyssier}, and
  {Bleuler}}]{Biernacki_et_al_2017}
{Biernacki}, P., {Teyssier}, R., {Bleuler}, A., Jul. 2017. {On the dynamics of
  supermassive black holes in gas-rich, star-forming galaxies: the case for
  nuclear star cluster co-evolution}. \mnras 469, 295--313.

\bibitem[{{Blaes} et~al.(2002){Blaes}, {Lee}, and {Socrates}}]{Blaes2002}
{Blaes}, O., {Lee}, M.~H., {Socrates}, A., Oct 2002. {The Kozai Mechanism and
  the Evolution of Binary Supermassive Black Holes}. \apj 578~(2), 775--786.

\bibitem[{{Blanton} et~al.(2003){Blanton}, {Lin}, {Lupton}, {Maley}, {Young},
  {Zehavi}, and {Loveday}}]{2003AJ....125.2276B}
{Blanton}, M.~R., {Lin}, H., {Lupton}, R.~H., {Maley}, F.~M., {Young}, N.,
  {Zehavi}, I., {Loveday}, J., Apr. 2003. {An Efficient Targeting Strategy for
  Multiobject Spectrograph Surveys: the Sloan Digital Sky Survey ``Tiling''
  Algorithm}. \aj 125, 2276--2286.

\bibitem[{{Blecha} et~al.(2011){Blecha}, {Cox}, {Loeb}, and
  {Hernquist}}]{Blecha2011}
{Blecha}, L., {Cox}, T.~J., {Loeb}, A., {Hernquist}, L., Apr 2011. {Recoiling
  black holes in merging galaxies: relationship to active galactic nucleus
  lifetimes, starbursts and the M$_{BH}$-{\ensuremath{\sigma}}$_{*}$ relation}.
  \mnras 412~(4), 2154--2182.

\bibitem[{{Blecha} et~al.(2013){Blecha}, {Loeb}, and
  {Narayan}}]{Blecha_et_al_2013}
{Blecha}, L., {Loeb}, A., {Narayan}, R., Mar. 2013. {Double-peaked narrow-line
  signatures of dual supermassive black holes in galaxy merger simulations}.
  MNRAS 429, 2594--2616.

\bibitem[{{Blecha} et~al.(2016){Blecha}, {Sijacki}, {Kelley}, {Torrey},
  {Vogelsberger}, {Nelson}, {Springel}, {Snyder}, and
  {Hernquist}}]{Blecha_et_al_2016}
{Blecha}, L., {Sijacki}, D., {Kelley}, L.~Z., {Torrey}, P., {Vogelsberger}, M.,
  {Nelson}, D., {Springel}, V., {Snyder}, G., {Hernquist}, L., Feb. 2016.
  {Recoiling black holes: prospects for detection and implications of spin
  alignment}. \mnras 456, 961--989.

\bibitem[{{Blecha} et~al.(2018){Blecha}, {Snyder}, {Satyapal}, and
  {Ellison}}]{Blecha_et_al_2017}
{Blecha}, L., {Snyder}, G.~F., {Satyapal}, S., {Ellison}, S.~L., Aug. 2018.
  {The power of infrared AGN selection in mergers: a theoretical study}. \mnras
  478, 3056--3071.

\bibitem[{{Blumenthal} and {Barnes}(2018)}]{Blumenthal_Barnes_2018}
{Blumenthal}, K.~A., {Barnes}, J.~E., Sep. 2018. {Go with the Flow:
  Understanding inflow mechanisms in galaxy collisions}. \mnras 479,
  3952--3965.

\bibitem[{{Bogdanovi{\'c}}(2015)}]{bogdanovic15}
{Bogdanovi{\'c}}, T., 2015. {Supermassive Black Hole Binaries: The Search
  Continues}. In: {Sopuerta}, C.~F. (Ed.), Gravitational Wave Astrophysics.
  Vol.~40 of Astrophysics and Space Science Proceedings. p. 103.

\bibitem[{{Bogdanovi{\'c}} et~al.(2011){Bogdanovi{\'c}}, {Bode}, {Haas},
  {Laguna}, and {Shoemaker}}]{bogdanovic11}
{Bogdanovi{\'c}}, T., {Bode}, T., {Haas}, R., {Laguna}, P., {Shoemaker}, D.,
  May 2011. {Properties of accretion flows around coalescing supermassive black
  holes}. Classical and Quantum Gravity 28~(9), 094020.

\bibitem[{{Bogdanovi{\'c}} et~al.(2008){Bogdanovi{\'c}}, {Smith}, {Sigurdsson},
  and {Eracleous}}]{bogdanovic08}
{Bogdanovi{\'c}}, T., {Smith}, B.~D., {Sigurdsson}, S., {Eracleous}, M., Feb.
  2008. {Modeling of Emission Signatures of Massive Black Hole Binaries. I.
  Methods}. \apjs 174, 455--480.

\bibitem[{{Boller} et~al.(2016){Boller}, {Freyberg}, {Tr{\"u}mper}, {Haberl},
  {Voges}, and {Nandra}}]{Boller2016}
{Boller}, T., {Freyberg}, M.~J., {Tr{\"u}mper}, J., {Haberl}, F., {Voges}, W.,
  {Nandra}, K., Apr. 2016. {Second ROSAT all-sky survey (2RXS) source
  catalogue}. \aap 588, A103.

\bibitem[{{Bon} et~al.(2016){Bon}, {Zucker}, {Netzer}, {Marziani}, {Bon},
  {Jovanovi{\'c}}, {Shapovalova}, {Komossa}, {Gaskell}, {Popovi{\'c}},
  {Britzen}, {Chavushyan}, {Burenkov}, {Sergeev}, {La Mura}, {Vald{\'e}s}, and
  {Stalevski}}]{Bon2016}
{Bon}, E., {Zucker}, S., {Netzer}, H., {Marziani}, P., {Bon}, N.,
  {Jovanovi{\'c}}, P., {Shapovalova}, A.~I., {Komossa}, S., {Gaskell}, C.~M.,
  {Popovi{\'c}}, L.~{\v C}., {Britzen}, S., {Chavushyan}, V.~H., {Burenkov},
  A.~N., {Sergeev}, S., {La Mura}, G., {Vald{\'e}s}, J.~R., {Stalevski}, M.,
  Aug. 2016. {Evidence for Periodicity in 43 year-long Monitoring of NGC 5548}.
  \apjs 225, 29.

\bibitem[{{Bondi} and {P{\'e}rez-Torres}(2010)}]{bondi2010}
{Bondi}, M., {P{\'e}rez-Torres}, M.-A., May 2010. {VLBI Detection of an Active
  Galactic Nucleus Pair in the Binary Black Hole Candidate SDSS J1536+0441}.
  Astrophysical Journal Letters 714, L271--L274.

\bibitem[{{Bondi} et~al.(2016){Bondi}, {P{\'e}rez-Torres}, {Piconcelli}, and
  {Fu}}]{2016A&A...588A.102B}
{Bondi}, M., {P{\'e}rez-Torres}, M.~A., {Piconcelli}, E., {Fu}, H., Apr. 2016.
  {Unveiling the radio counterparts of two binary AGN candidates: J1108+0659
  and J1131-0204}. \aap 588, A102.

\bibitem[{{Bonetti} et~al.(2018){Bonetti}, {Sesana}, {Barausse}, and
  {Haardt}}]{Bonetti+2018}
{Bonetti}, M., {Sesana}, A., {Barausse}, E., {Haardt}, F., Jun. 2018.
  {Post-Newtonian evolution of massive black hole triplets in galactic nuclei -
  III. A robust lower limit to the nHz stochastic background of gravitational
  waves}. \mnras 477, 2599--2612.

\bibitem[{{Bonetti} et~al.(2019){Bonetti}, {Sesana}, {Haardt}, {Barausse}, and
  {Colpi}}]{Bonetti2019}
{Bonetti}, M., {Sesana}, A., {Haardt}, F., {Barausse}, E., {Colpi}, M., Jul
  2019. {Post-Newtonian evolution of massive black hole triplets in galactic
  nuclei - IV. Implications for LISA}. \mnras 486~(3), 4044--4060.

\bibitem[{{Borisova} et~al.(2016){Borisova}, {Cantalupo}, {Lilly}, {Marino},
  {Gallego}, {Bacon}, {Blaizot}, {Bouch{\'e}}, {Brinchmann}, {Carollo},
  {Caruana}, {Finley}, {Herenz}, {Richard}, {Schaye}, {Straka}, {Turner},
  {Urrutia}, {Verhamme}, and {Wisotzki}}]{Borisova2016}
{Borisova}, E., {Cantalupo}, S., {Lilly}, S.~J., {Marino}, R.~A., {Gallego},
  S.~G., {Bacon}, R., {Blaizot}, J., {Bouch{\'e}}, N., {Brinchmann}, J.,
  {Carollo}, C.~M., {Caruana}, J., {Finley}, H., {Herenz}, E.~C., {Richard},
  J., {Schaye}, J., {Straka}, L.~A., {Turner}, M.~L., {Urrutia}, T.,
  {Verhamme}, A., {Wisotzki}, L., Nov. 2016. {Ubiquitous Giant Ly{$\alpha$}
  Nebulae around the Brightest Quasars at z$\sim$3.5 Revealed with MUSE}. \apj
  831, 39.

\bibitem[{{Boroson} and {Lauer}(2009)}]{boroson2009}
{Boroson}, T.~A., {Lauer}, T.~R., Mar. 2009. {A candidate sub-parsec
  supermassive binary black hole system}. Nature 458, 53--55.

\bibitem[{{Bovy} et~al.(2011){Bovy}, {Hennawi}, {Hogg}, {Myers}, {Kirkpatrick},
  {Schlegel}, {Ross}, {Sheldon}, {McGreer}, {Schneider}, and
  {Weaver}}]{2011ApJ...729..141B}
{Bovy}, J., {Hennawi}, J.~F., {Hogg}, D.~W., {Myers}, A.~D., {Kirkpatrick},
  J.~A., {Schlegel}, D.~J., {Ross}, N.~P., {Sheldon}, E.~S., {McGreer}, I.~D.,
  {Schneider}, D.~P., {Weaver}, B.~A., Mar. 2011. {Think Outside the Color Box:
  Probabilistic Target Selection and the SDSS-XDQSO Quasar Targeting Catalog}.
  \apj 729, 141.

\bibitem[{{Bovy} et~al.(2012){Bovy}, {Myers}, {Hennawi}, {Hogg}, {McMahon},
  {Schiminovich}, {Sheldon}, {Brinkmann}, {Schneider}, and
  {Weaver}}]{2012ApJ...749...41B}
{Bovy}, J., {Myers}, A.~D., {Hennawi}, J.~F., {Hogg}, D.~W., {McMahon}, R.~G.,
  {Schiminovich}, D., {Sheldon}, E.~S., {Brinkmann}, J., {Schneider}, D.~P.,
  {Weaver}, B.~A., Apr. 2012. {Photometric Redshifts and Quasar Probabilities
  from a Single, Data-driven Generative Model}. \apj 749, 41.

\bibitem[{{Bowen} et~al.(2018){Bowen}, {Mewes}, {Campanelli}, {Noble},
  {Krolik}, and {Zilh{\~a}o}}]{Bowen18}
{Bowen}, D.~B., {Mewes}, V., {Campanelli}, M., {Noble}, S.~C., {Krolik}, J.~H.,
  {Zilh{\~a}o}, M., Jan. 2018. {Quasi-periodic Behavior of Mini-disks in Binary
  Black Holes Approaching Merger}. \apjl 853, L17.

\bibitem[{{Bowen} et~al.(2019){Bowen}, {Mewes}, {Noble}, {Avara}, {Campanelli},
  and {Krolik}}]{Bowen19}
{Bowen}, D.~B., {Mewes}, V., {Noble}, S.~C., {Avara}, M., {Campanelli}, M.,
  {Krolik}, J.~H., Jul 2019. {Quasi-periodicity of Supermassive Binary Black
  Hole Accretion Approaching Merger}. \apj 879~(2), 76.

\bibitem[{{Britzen} et~al.(2018){Britzen}, {Fendt}, {Witzel}, {Qian},
  {Pashchenko}, {Kurtanidze}, {Zajacek}, {Martinez}, {Karas}, {Aller}, {Aller},
  {Eckart}, {Nilsson}, {Ar{\'e}valo}, {Cuadra}, {Subroweit}, and
  {Witzel}}]{Britzen_et_al_2018}
{Britzen}, S., {Fendt}, C., {Witzel}, G., {Qian}, S.-J., {Pashchenko}, I.~N.,
  {Kurtanidze}, O., {Zajacek}, M., {Martinez}, G., {Karas}, V., {Aller}, M.,
  {Aller}, H., {Eckart}, A., {Nilsson}, K., {Ar{\'e}valo}, P., {Cuadra}, J.,
  {Subroweit}, M., {Witzel}, A., Aug. 2018. {OJ287: deciphering the `Rosetta
  stone of blazars}. \mnras 478, 3199--3219.

\bibitem[{{Bundy} et~al.(2015){Bundy}, {Bershady}, {Law}, {Yan}, {Drory},
  {MacDonald}, {Wake}, {Cherinka}, {S{\'a}nchez-Gallego}, {Weijmans}, {Thomas},
  {Tremonti}, {Masters}, {Coccato}, {Diamond-Stanic}, {Arag{\'o}n-Salamanca},
  {Avila-Reese}, {Badenes}, {Falc{\'o}n-Barroso}, {Belfiore}, {Bizyaev},
  {Blanc}, {Bland-Hawthorn}, {Blanton}, {Brownstein}, {Byler}, {Cappellari},
  {Conroy}, {Dutton}, {Emsellem}, {Etherington}, {Frinchaboy}, {Fu}, {Gunn},
  {Harding}, {Johnston}, {Kauffmann}, {Kinemuchi}, {Klaene}, {Knapen},
  {Leauthaud}, {Li}, {Lin}, {Maiolino}, {Malanushenko}, {Malanushenko}, {Mao},
  {Maraston}, {McDermid}, {Merrifield}, {Nichol}, {Oravetz}, {Pan}, {Parejko},
  {Sanchez}, {Schlegel}, {Simmons}, {Steele}, {Steinmetz}, {Thanjavur},
  {Thompson}, {Tinker}, {van den Bosch}, {Westfall}, {Wilkinson}, {Wright},
  {Xiao}, and {Zhang}}]{Bundy2015}
{Bundy}, K., {Bershady}, M.~A., {Law}, D.~R., {Yan}, R., {Drory}, N.,
  {MacDonald}, N., {Wake}, D.~A., {Cherinka}, B., {S{\'a}nchez-Gallego}, J.~R.,
  {Weijmans}, A.-M., {Thomas}, D., {Tremonti}, C., {Masters}, K., {Coccato},
  L., {Diamond-Stanic}, A.~M., {Arag{\'o}n-Salamanca}, A., {Avila-Reese}, V.,
  {Badenes}, C., {Falc{\'o}n-Barroso}, J., {Belfiore}, F., {Bizyaev}, D.,
  {Blanc}, G.~A., {Bland-Hawthorn}, J., {Blanton}, M.~R., {Brownstein}, J.~R.,
  {Byler}, N., {Cappellari}, M., {Conroy}, C., {Dutton}, A.~A., {Emsellem}, E.,
  {Etherington}, J., {Frinchaboy}, P.~M., {Fu}, H., {Gunn}, J.~E., {Harding},
  P., {Johnston}, E.~J., {Kauffmann}, G., {Kinemuchi}, K., {Klaene}, M.~A.,
  {Knapen}, J.~H., {Leauthaud}, A., {Li}, C., {Lin}, L., {Maiolino}, R.,
  {Malanushenko}, V., {Malanushenko}, E., {Mao}, S., {Maraston}, C.,
  {McDermid}, R.~M., {Merrifield}, M.~R., {Nichol}, R.~C., {Oravetz}, D.,
  {Pan}, K., {Parejko}, J.~K., {Sanchez}, S.~F., {Schlegel}, D., {Simmons}, A.,
  {Steele}, O., {Steinmetz}, M., {Thanjavur}, K., {Thompson}, B.~A., {Tinker},
  J.~L., {van den Bosch}, R.~C.~E., {Westfall}, K.~B., {Wilkinson}, D.,
  {Wright}, S., {Xiao}, T., {Zhang}, K., Jan. 2015. {Overview of the SDSS-IV
  MaNGA Survey: Mapping nearby Galaxies at Apache Point Observatory}. \apj 798,
  7.

\bibitem[{{Burke-Spolaor}(2011)}]{burkespolaor2011}
{Burke-Spolaor}, S., Feb. 2011. {A radio Census of binary supermassive black
  holes}. Monthly Notices of the Royal Astronomical Society 410, 2113--2122.

\bibitem[{{Burke-Spolaor} et~al.(2018){Burke-Spolaor}, {Blecha}, {Bogdanovic},
  {Comerford}, {Lazio}, {Liu}, {Maccarone}, {Pesce}, {Shen}, and
  {Taylor}}]{Burke-Spolaor2018}
{Burke-Spolaor}, S., {Blecha}, L., {Bogdanovic}, T., {Comerford}, J.~M.,
  {Lazio}, T.~J.~W., {Liu}, X., {Maccarone}, T.~J., {Pesce}, D., {Shen}, Y.,
  {Taylor}, G., Aug. 2018. {The Next-Generation Very Large Array: Supermassive
  Black Hole Pairs and Binaries}. arXiv e-prints.

\bibitem[{{Burke-Spolaor} et~al.(2014){Burke-Spolaor}, {Brazier}, {Chatterjee},
  {Comerford}, {Cordes}, {Lazio}, {Liu}, and {Shen}}]{Burke-spolaor2014}
{Burke-Spolaor}, S., {Brazier}, A., {Chatterjee}, S., {Comerford}, J.,
  {Cordes}, J., {Lazio}, T.~J.~W., {Liu}, X., {Shen}, Y., Feb. 2014. {A hunt
  for dual radio active galactic nuclei in the VLASS}. arXiv e-prints.

\bibitem[{{Cai} et~al.(2017){Cai}, {Fan}, {Yang}, {Bian}, {Prochaska},
  {Zabludoff}, {McGreer}, {Zheng}, {Green}, {Cantalupo}, {Frye}, {Hamden},
  {Jiang}, {Kashikawa}, and {Wang}}]{Cai2017}
{Cai}, Z., {Fan}, X., {Yang}, Y., {Bian}, F., {Prochaska}, J.~X., {Zabludoff},
  A., {McGreer}, I., {Zheng}, Z.-Y., {Green}, R., {Cantalupo}, S., {Frye}, B.,
  {Hamden}, E., {Jiang}, L., {Kashikawa}, N., {Wang}, R., Mar. 2017. {Discovery
  of an Enormous Ly{$\alpha$} Nebula in a Massive Galaxy Overdensity at z =
  2.3}. \apj 837, 71.

\bibitem[{{Callegari} et~al.(2011){Callegari}, {Kazantzidis}, {Mayer}, {Colpi},
  {Bellovary}, {Quinn}, and {Wadsley}}]{Callegari_et_al_2011}
{Callegari}, S., {Kazantzidis}, S., {Mayer}, L., {Colpi}, M., {Bellovary},
  J.~M., {Quinn}, T., {Wadsley}, J., Mar. 2011. {Growing Massive Black Hole
  Pairs in Minor Mergers of Disk Galaxies}. \apj 729, 85.

\bibitem[{{Callegari} et~al.(2009){Callegari}, {Mayer}, {Kazantzidis}, {Colpi},
  {Governato}, {Quinn}, and {Wadsley}}]{Callegari_et_al_2009}
{Callegari}, S., {Mayer}, L., {Kazantzidis}, S., {Colpi}, M., {Governato}, F.,
  {Quinn}, T., {Wadsley}, J., May 2009. {Pairing of Supermassive Black Holes in
  Unequal-Mass Galaxy Mergers}. \apjl 696, L89--L92.

\bibitem[{{Campanelli} et~al.(2007){Campanelli}, {Lousto}, {Zlochower}, and
  {Merritt}}]{Campanelli2007}
{Campanelli}, M., {Lousto}, C.~O., {Zlochower}, Y., {Merritt}, D., Jun 2007.
  {Maximum Gravitational Recoil}. \prl 98~(23), 231102.

\bibitem[{{Canalizo} et~al.(2003){Canalizo}, {Max}, {Whysong}, {Antonucci}, and
  {Dahm}}]{Canalizo2003}
{Canalizo}, G., {Max}, C., {Whysong}, D., {Antonucci}, R., {Dahm}, S.~E., Nov.
  2003. {Adaptive Optics Imaging and Spectroscopy of Cygnus A. I. Evidence for
  a Minor Merger}. \apj 597, 823--831.

\bibitem[{{Cantalupo} et~al.(2014){Cantalupo}, {Arrigoni-Battaia}, {Prochaska},
  {Hennawi}, and {Madau}}]{Cantalupo2014}
{Cantalupo}, S., {Arrigoni-Battaia}, F., {Prochaska}, J.~X., {Hennawi}, J.~F.,
  {Madau}, P., Feb. 2014. {A cosmic web filament revealed in Lyman-{$\alpha$}
  emission around a luminous high-redshift quasar}. \nat 506, 63--66.

\bibitem[{{Capelo} and {Dotti}(2017)}]{Capelo_Dotti_2017}
{Capelo}, P.~R., {Dotti}, M., Mar. 2017. {Shocks and angular momentum flips: a
  different path to feeding the nuclear regions of merging galaxies}. \mnras
  465, 2643--2653.

\bibitem[{{Capelo} et~al.(2017){Capelo}, {Dotti}, {Volonteri}, {Mayer},
  {Bellovary}, and {Shen}}]{Capelo_et_al_2017}
{Capelo}, P.~R., {Dotti}, M., {Volonteri}, M., {Mayer}, L., {Bellovary}, J.~M.,
  {Shen}, S., Aug. 2017. {A survey of dual active galactic nuclei in
  simulations of galaxy mergers: frequency and properties}. \mnras 469,
  4437--4454.

\bibitem[{{Capelo} et~al.(2015){Capelo}, {Volonteri}, {Dotti}, {Bellovary},
  {Mayer}, and {Governato}}]{Capelo_et_al_2015}
{Capelo}, P.~R., {Volonteri}, M., {Dotti}, M., {Bellovary}, J.~M., {Mayer}, L.,
  {Governato}, F., Mar. 2015. {Growth and activity of black holes in galaxy
  mergers with varying mass ratios}. \mnras 447, 2123--2143.

\bibitem[{{Cavaliere} et~al.(2017){Cavaliere}, {Tavani}, and
  {Vittorini}}]{cavaliereatal17}
{Cavaliere}, A., {Tavani}, M., {Vittorini}, V., Feb. 2017. {Blazar Jets
  Perturbed by Magneto-gravitational Stresses in Supermassive Binaries}. \apj
  836, 220.

\bibitem[{{Chandrasekhar}(1943)}]{Chandrasekhar_1943}
{Chandrasekhar}, S., Mar. 1943. {Dynamical Friction. I. General Considerations:
  the Coefficient of Dynamical Friction.} \apj 97, 255.

\bibitem[{{Charisi} et~al.(2016){Charisi}, {Bartos}, {Haiman}, {Price-Whelan},
  {Graham}, {Bellm}, {Laher}, and {M{\'a}rka}}]{Charisi2016}
{Charisi}, M., {Bartos}, I., {Haiman}, Z., {Price-Whelan}, A.~M., {Graham},
  M.~J., {Bellm}, E.~C., {Laher}, R.~R., {M{\'a}rka}, S., Dec. 2016. {A
  population of short-period variable quasars from PTF as supermassive black
  hole binary candidates}. \mnras 463, 2145--2171.

\bibitem[{{Charisi} et~al.(2015){Charisi}, {Bartos}, {Haiman}, {Price-Whelan},
  and {M{\'a}rka}}]{Charisi2015}
{Charisi}, M., {Bartos}, I., {Haiman}, Z., {Price-Whelan}, A.~M., {M{\'a}rka},
  S., Nov. 2015. {Multiple periods in the variability of the supermassive black
  hole binary candidate quasar PG1302-102?} \mnras 454, L21--L25.

\bibitem[{{Charisi} et~al.(2018){Charisi}, {Haiman}, {Schiminovich}, and
  {D'Orazio}}]{Charisi2018}
{Charisi}, M., {Haiman}, Z., {Schiminovich}, D., {D'Orazio}, D.~J., Jun. 2018.
  {Testing the relativistic Doppler boost hypothesis for supermassive black
  hole binary candidates}. \mnras 476, 4617--4628.

\bibitem[{{Chen} et~al.(1989){Chen}, {Halpern}, and {Filippenko}}]{Chen1989}
{Chen}, K., {Halpern}, J.~P., {Filippenko}, A.~V., Apr. 1989. {Kinematic
  evidence for a relativistic Keplerian disk - ARP 102B}. \apj 339, 742--751.

\bibitem[{{Cheung}(2007)}]{2007AJ....133.2097C}
{Cheung}, C.~C., May 2007. {FIRST ``Winged'' and X-Shaped Radio Source
  Candidates}. \aj 133~(5), 2097--2121.

\bibitem[{{Chi} et~al.(2013){Chi}, {Barthel}, and {Garrett}}]{chi2013}
{Chi}, S., {Barthel}, P.~D., {Garrett}, M.~A., Feb. 2013. {Deep, wide-field,
  global VLBI observations of the Hubble deep field north (HDF-N) and flanking
  fields (HFF)}. Astronomy \& Astrophysics 550, A68.

\bibitem[{{Cisternas} et~al.(2011){Cisternas}, {Jahnke}, {Inskip},
  {Kartaltepe}, {Koekemoer}, {Lisker}, {Robaina}, {Scodeggio}, {Sheth},
  {Trump}, {Andrae}, and {Miyaji}}]{Cisternas:2011}
{Cisternas}, M., {Jahnke}, K., {Inskip}, K.~J., {Kartaltepe}, J., {Koekemoer},
  A.~M., {Lisker}, T., {Robaina}, A.~R., {Scodeggio}, M., {Sheth}, K., {Trump},
  J.~R., {Andrae}, R., {Miyaji}, T., Jan. 2011. {The Bulk of the Black Hole
  Growth Since z \~{} 1 Occurs in a Secular Universe: No Major Merger-AGN
  Connection}. \apj 726, 57--+.

\bibitem[{{Civano} et~al.(2012){Civano}, {Elvis}, {Lanzuisi}, {Aldcroft},
  {Trichas}, {Bongiorno}, {Brusa}, {Blecha}, {Comastri}, {Loeb}, {Salvato},
  {Fruscione}, {Koekemoer}, {Komossa}, {Gilli}, {Mainieri}, {Piconcelli}, and
  {Vignali}}]{Civano2012}
{Civano}, F., {Elvis}, M., {Lanzuisi}, G., {Aldcroft}, T., {Trichas}, M.,
  {Bongiorno}, A., {Brusa}, M., {Blecha}, L., {Comastri}, A., {Loeb}, A.,
  {Salvato}, M., {Fruscione}, A., {Koekemoer}, A., {Komossa}, S., {Gilli}, R.,
  {Mainieri}, V., {Piconcelli}, E., {Vignali}, C., Jun. 2012. {Chandra
  High-resolution observations of CID-42, a Candidate Recoiling Supermassive
  Black Hole}. \apj 752, 49.

\bibitem[{{Civano} et~al.(2010){Civano}, {Elvis}, {Lanzuisi}, {Jahnke},
  {Zamorani}, {Blecha}, {Bongiorno}, {Brusa}, {Comastri}, {Hao}, {Leauthaud},
  {Loeb}, {Mainieri}, {Piconcelli}, {Salvato}, {Scoville}, {Trump}, {Vignali},
  {Aldcroft}, {Bolzonella}, {Bressert}, {Finoguenov}, {Fruscione}, {Koekemoer},
  {Cappelluti}, {Fiore}, {Giodini}, {Gilli}, {Impey}, {Lilly}, {Lusso},
  {Puccetti}, {Silverman}, {Aussel}, {Capak}, {Frayer}, {Le Floch},
  {McCracken}, {Sanders}, {Schiminovich}, and {Taniguchi}}]{Civano2010}
{Civano}, F., {Elvis}, M., {Lanzuisi}, G., {Jahnke}, K., {Zamorani}, G.,
  {Blecha}, L., {Bongiorno}, A., {Brusa}, M., {Comastri}, A., {Hao}, H.,
  {Leauthaud}, A., {Loeb}, A., {Mainieri}, V., {Piconcelli}, E., {Salvato}, M.,
  {Scoville}, N., {Trump}, J., {Vignali}, C., {Aldcroft}, T., {Bolzonella}, M.,
  {Bressert}, E., {Finoguenov}, A., {Fruscione}, A., {Koekemoer}, A.~M.,
  {Cappelluti}, N., {Fiore}, F., {Giodini}, S., {Gilli}, R., {Impey}, C.~D.,
  {Lilly}, S.~J., {Lusso}, E., {Puccetti}, S., {Silverman}, J.~D., {Aussel},
  H., {Capak}, P., {Frayer}, D., {Le Floch}, E., {McCracken}, H.~J., {Sanders},
  D.~B., {Schiminovich}, D., {Taniguchi}, Y., Jul. 2010. {A Runaway Black Hole
  in COSMOS: Gravitational Wave or Slingshot Recoil?} \apj 717, 209--222.

\bibitem[{{Coil} et~al.(2015){Coil}, {Aird}, {Reddy}, {Shapley}, {Kriek},
  {Siana}, {Mobasher}, {Freeman}, {Price}, and {Shivaei}}]{Coil2015}
{Coil}, A.~L., {Aird}, J., {Reddy}, N., {Shapley}, A.~E., {Kriek}, M., {Siana},
  B., {Mobasher}, B., {Freeman}, W.~R., {Price}, S.~H., {Shivaei}, I., Mar.
  2015. {The MOSDEF Survey: Optical Active Galactic Nucleus Diagnostics at
  z$\sim$2.3}. \apj 801, 35.

\bibitem[{{Cole} et~al.(2012){Cole}, {Dehnen}, {Read}, and
  {Wilkinson}}]{Cole_et_al_2012}
{Cole}, D.~R., {Dehnen}, W., {Read}, J.~I., {Wilkinson}, M.~I., Oct. 2012. {The
  mass distribution of the Fornax dSph: constraints from its globular cluster
  distribution}. \mnras 426, 601--613.

\bibitem[{{Colpi}(2014)}]{Colpi2014}
{Colpi}, M., Sep. 2014. {Massive Binary Black Holes in Galactic Nuclei and
  Their Path to Coalescence}. \ssr 183, 189--221.

\bibitem[{{Colpi} et~al.(2019){Colpi}, {Holley-Bockelmann}, {Bogdanovic},
  {Natarajan}, {Bellovary}, {Sesana}, {Tremmel}, {Schnittman}, {Comerford},
  {Barausse}, {Berti}, {Volonteri}, {Khan}, {McWilliams}, {Burke-Spolaor},
  {Hazboun}, {Conklin}, {Mueller}, and {Larson}}]{Colpi2020}
{Colpi}, M., {Holley-Bockelmann}, K., {Bogdanovic}, T., {Natarajan}, P.,
  {Bellovary}, J., {Sesana}, A., {Tremmel}, M., {Schnittman}, J., {Comerford},
  J., {Barausse}, E., {Berti}, E., {Volonteri}, M., {Khan}, F., {McWilliams},
  S., {Burke-Spolaor}, S., {Hazboun}, J., {Conklin}, J., {Mueller}, G.,
  {Larson}, S., Mar. 2019. {Astro2020 science white paper: The gravitational
  wave view of massive black holes}. arXiv e-prints.

\bibitem[{{Colpi} and {Sesana}(2017)}]{2017ogw..book...43C}
{Colpi}, M., {Sesana}, A., 2017. {Gravitational Wave Sources in the Era of
  Multi-Band Gravitational Wave Astronomy}. World Scientific Publishing Co, pp.
  43--140.

\bibitem[{{Comerford} et~al.(2009){Comerford}, {Griffith}, {Gerke}, {Cooper},
  {Newman}, {Davis}, and {Stern}}]{Comerford2009}
{Comerford}, J.~M., {Griffith}, R.~L., {Gerke}, B.~F., {Cooper}, M.~C.,
  {Newman}, J.~A., {Davis}, M., {Stern}, D., Sep. 2009. {1.75 h $^{-1}$ kpc
  Separation Dual Active Galactic Nuclei at z = 0.36 in the Cosmos Field}.
  \apjl 702, L82--L86.

\bibitem[{{Comerford} et~al.(2015){Comerford}, {Pooley}, {Barrows}, {Greene},
  {Zakamska}, {Madejski}, and {Cooper}}]{Comerford2015}
{Comerford}, J.~M., {Pooley}, D., {Barrows}, R.~S., {Greene}, J.~E.,
  {Zakamska}, N.~L., {Madejski}, G.~M., {Cooper}, M.~C., Jun 2015.
  {Merger-driven Fueling of Active Galactic Nuclei: Six Dual and Offset AGNs
  Discovered with Chandra and Hubble Space Telescope Observations}. \apj
  806~(2), 219.

\bibitem[{{Comerford} et~al.(2013){Comerford}, {Schluns}, {Greene}, and
  {Cool}}]{comerford2013}
{Comerford}, J.~M., {Schluns}, K., {Greene}, J.~E., {Cool}, R.~J., Nov 2013.
  {Dual Supermassive Black Hole Candidates in the AGN and Galaxy Evolution
  Survey}. \apj 777~(1), 64.

\bibitem[{{Condon}(1992)}]{condon1992}
{Condon}, J.~J., 1992. {Radio emission from normal galaxies}. Annual Review of
  Astronomy \& Astrophysics 30, 575--611.

\bibitem[{{Cotini} et~al.(2013){Cotini}, {Ripamonti}, {Caccianiga}, {Colpi},
  {Della Ceca}, {Mapelli}, {Severgnini}, and {Segreto}}]{Cotini2013}
{Cotini}, S., {Ripamonti}, E., {Caccianiga}, A., {Colpi}, M., {Della Ceca}, R.,
  {Mapelli}, M., {Severgnini}, P., {Segreto}, A., May 2013. {The merger
  fraction of active and inactive galaxies in the local Universe through an
  improved non-parametric classification}. \mnras 431, 2661--2672.

\bibitem[{{Cox} et~al.(2008){Cox}, {Jonsson}, {Somerville}, {Primack}, and
  {Dekel}}]{Cox_et_al_2008}
{Cox}, T.~J., {Jonsson}, P., {Somerville}, R.~S., {Primack}, J.~R., {Dekel},
  A., Feb. 2008. {The effect of galaxy mass ratio on merger-driven starbursts}.
  \mnras 384, 386--409.

\bibitem[{{Croom} et~al.(2012){Croom}, {Lawrence}, {Bland-Hawthorn}, {Bryant},
  {Fogarty}, {Richards}, {Goodwin}, {Farrell}, {Miziarski}, {Heald}, {Jones},
  {Lee}, {Colless}, {Brough}, {Hopkins}, {Bauer}, {Birchall}, {Ellis},
  {Horton}, {Leon-Saval}, {Lewis}, {L{\'o}pez-S{\'a}nchez}, {Min}, {Trinh}, and
  {Trowland}}]{Croom2012}
{Croom}, S.~M., {Lawrence}, J.~S., {Bland-Hawthorn}, J., {Bryant}, J.~J.,
  {Fogarty}, L., {Richards}, S., {Goodwin}, M., {Farrell}, T., {Miziarski}, S.,
  {Heald}, R., {Jones}, D.~H., {Lee}, S., {Colless}, M., {Brough}, S.,
  {Hopkins}, A.~M., {Bauer}, A.~E., {Birchall}, M.~N., {Ellis}, S., {Horton},
  A., {Leon-Saval}, S., {Lewis}, G., {L{\'o}pez-S{\'a}nchez}, {\'A}.~R., {Min},
  S.-S., {Trinh}, C., {Trowland}, H., Mar. 2012. {The Sydney-AAO Multi-object
  Integral field spectrograph}. \mnras 421, 872--893.

\bibitem[{{Cuadra} et~al.(2009{\natexlab{a}}){Cuadra}, {Armitage}, {Alexander},
  and {Begelman}}]{cuadra09}
{Cuadra}, J., {Armitage}, P.~J., {Alexander}, R.~D., {Begelman}, M.~C., Mar.
  2009{\natexlab{a}}. {Massive black hole binary mergers within subparsec scale
  gas discs}. \mnras 393, 1423--1432.

\bibitem[{{Cuadra} et~al.(2009{\natexlab{b}}){Cuadra}, {Armitage}, {Alexander},
  and {Begelman}}]{Cuadra+2009}
{Cuadra}, J., {Armitage}, P.~J., {Alexander}, R.~D., {Begelman}, M.~C., Mar.
  2009{\natexlab{b}}. {Massive black hole binary mergers within subparsec scale
  gas discs}. \mnras 393, 1423--1432.

\bibitem[{{Dalton} et~al.(2012){Dalton}, {Trager}, {Abrams}, {Carter},
  {Bonifacio}, {Aguerri}, {MacIntosh}, {Evans}, {Lewis}, {Navarro}, {Agocs},
  {Dee}, {Rousset}, {Tosh}, {Middleton}, {Pragt}, {Terrett}, {Brock}, {Benn},
  {Verheijen}, {Cano Infantes}, {Bevil}, {Steele}, {Mottram}, {Bates},
  {Gribbin}, {Rey}, {Rodriguez}, {Delgado}, {Guinouard}, {Walton}, {Irwin},
  {Jagourel}, {Stuik}, {Gerlofsma}, {Roelfsma}, {Skillen}, {Ridings},
  {Balcells}, {Daban}, {Gouvret}, {Venema}, and {Girard}}]{dalton2012}
{Dalton}, G., {Trager}, S.~C., {Abrams}, D.~C., {Carter}, D., {Bonifacio}, P.,
  {Aguerri}, J.~A.~L., {MacIntosh}, M., {Evans}, C., {Lewis}, I., {Navarro},
  R., {Agocs}, T., {Dee}, K., {Rousset}, S., {Tosh}, I., {Middleton}, K.,
  {Pragt}, J., {Terrett}, D., {Brock}, M., {Benn}, C., {Verheijen}, M., {Cano
  Infantes}, D., {Bevil}, C., {Steele}, I., {Mottram}, C., {Bates}, S.,
  {Gribbin}, F.~J., {Rey}, J., {Rodriguez}, L.~F., {Delgado}, J.~M.,
  {Guinouard}, I., {Walton}, N., {Irwin}, M.~J., {Jagourel}, P., {Stuik}, R.,
  {Gerlofsma}, G., {Roelfsma}, R., {Skillen}, I., {Ridings}, A., {Balcells},
  M., {Daban}, J.-B., {Gouvret}, C., {Venema}, L., {Girard}, P., Sep. 2012.
  {WEAVE: the next generation wide-field spectroscopy facility for the William
  Herschel Telescope}. In: Ground-based and Airborne Instrumentation for
  Astronomy IV. Vol. 8446 of \procspie. p. 84460P.

\bibitem[{{d'Ascoli} et~al.(2018){d'Ascoli}, {Noble}, {Bowen}, {Campanelli},
  {Krolik}, and {Mewes}}]{Dascoli2018}
{d'Ascoli}, S., {Noble}, S.~C., {Bowen}, D.~B., {Campanelli}, M., {Krolik},
  J.~H., {Mewes}, V., Oct 2018. {Electromagnetic Emission from Supermassive
  Binary Black Holes Approaching Merger}. \apj 865~(2), 140.

\bibitem[{{Dav{\'e}} et~al.(2016){Dav{\'e}}, {Thompson}, and
  {Hopkins}}]{Dave_et_al_2016}
{Dav{\'e}}, R., {Thompson}, R., {Hopkins}, P.~F., Nov. 2016. {MUFASA: galaxy
  formation simulations with meshless hydrodynamics}. \mnras 462, 3265--3284.

\bibitem[{{Davies} et~al.(2014){Davies}, {Kewley}, {Ho}, and
  {Dopita}}]{Davies2014}
{Davies}, R.~L., {Kewley}, L.~J., {Ho}, I.-T., {Dopita}, M.~A., Nov. 2014.
  {Starburst-AGN mixing - II. Optically selected active galaxies}. \mnras 444,
  3961--3974.

\bibitem[{{De Rosa} et~al.(2018){De Rosa}, {Vignali}, {Husemann}, {Bianchi},
  {Bogdanovi{\'c}}, {Guainazzi}, {Herrero-Illana}, {Komossa}, {Kun}, {Loiseau},
  {Paragi}, {Perez-Torres}, and {Piconcelli}}]{derosaetal2018}
{De Rosa}, A., {Vignali}, C., {Husemann}, B., {Bianchi}, S., {Bogdanovi{\'c}},
  T., {Guainazzi}, M., {Herrero-Illana}, R., {Komossa}, S., {Kun}, E.,
  {Loiseau}, N., {Paragi}, Z., {Perez-Torres}, M., {Piconcelli}, E., Oct. 2018.
  {Disclosing the properties of low-redshift dual AGN through XMM-Newton and
  SDSS spectroscopy}. \mnras 480, 1639--1655.

\bibitem[{{Deane} et~al.(2014){Deane}, {Paragi}, {Jarvis}, {Coriat},
  {Bernardi}, {Fender}, {Frey}, {Heywood}, {Kl{\"o}ckner}, {Grainge}, and
  {Rumsey}}]{deane2014}
{Deane}, R.~P., {Paragi}, Z., {Jarvis}, M.~J., {Coriat}, M., {Bernardi}, G.,
  {Fender}, R.~P., {Frey}, S., {Heywood}, I., {Kl{\"o}ckner}, H.-R., {Grainge},
  K., {Rumsey}, C., Jul. 2014. {A close-pair binary in a distant triple
  supermassive black hole system}. \nat 511, 57--60.

\bibitem[{{Debuhr} et~al.(2010){Debuhr}, {Quataert}, {Ma}, and
  {Hopkins}}]{Debuhr_et_al_2010}
{Debuhr}, J., {Quataert}, E., {Ma}, C.-P., {Hopkins}, P., Jul. 2010.
  {Self-regulated black hole growth via momentum deposition in galaxy merger
  simulations}. \mnras 406, L55--L59.

\bibitem[{{Decarli} et~al.(2013){Decarli}, {Dotti}, {Fumagalli}, {Tsalmantza},
  {Montuori}, {Lusso}, {Hogg}, and {Prochaska}}]{decarli13}
{Decarli}, R., {Dotti}, M., {Fumagalli}, M., {Tsalmantza}, P., {Montuori}, C.,
  {Lusso}, E., {Hogg}, D.~W., {Prochaska}, J.~X., Aug. 2013. {The nature of
  massive black hole binary candidates - I. Spectral properties and evolution}.
  \mnras 433, 1492--1504.

\bibitem[{{Decarli} et~al.(2010){Decarli}, {Dotti}, {Montuori}, {Liimets}, and
  {Ederoclite}}]{Decarli10}
{Decarli}, R., {Dotti}, M., {Montuori}, C., {Liimets}, T., {Ederoclite}, A.,
  Sep. 2010. {The Peculiar Optical Spectrum of 4C+22.25: Imprint of a Massive
  Black Hole Binary?} \apjl 720, L93--L96.

\bibitem[{{del Valle} et~al.(2015){del Valle}, {Escala}, {Maureira-Fredes},
  {Molina}, {Cuadra}, and {Amaro-Seoane}}]{delValle_et_al_2015}
{del Valle}, L., {Escala}, A., {Maureira-Fredes}, C., {Molina}, J., {Cuadra},
  J., {Amaro-Seoane}, P., Sep. 2015. {Supermassive Black Holes in a
  Star-forming Gaseous Circumnuclear Disk}. \apj 811, 59.

\bibitem[{{Denney} et~al.(2010){Denney}, {Peterson}, {Pogge}, {Adair}, {Atlee},
  {Au-Yong}, {Bentz}, {Bird}, {Brokofsky}, {Chisholm}, {Comins}, {Dietrich},
  {Doroshenko}, {Eastman}, {Efimov}, {Ewald}, {Ferbey}, {Gaskell}, {Hedrick},
  {Jackson}, {Klimanov}, {Klimek}, {Kruse}, {Lad{\'e}route}, {Lamb}, {Leighly},
  {Minezaki}, {Nazarov}, {Onken}, {Petersen}, {Peterson}, {Poindexter},
  {Sakata}, {Schlesinger}, {Sergeev}, {Skolski}, {Stieglitz}, {Tobin},
  {Unterborn}, {Vestergaard}, {Watkins}, {Watson}, and {Yoshii}}]{Denney2010}
{Denney}, K.~D., {Peterson}, B.~M., {Pogge}, R.~W., {Adair}, A., {Atlee},
  D.~W., {Au-Yong}, K., {Bentz}, M.~C., {Bird}, J.~C., {Brokofsky}, D.~J.,
  {Chisholm}, E., {Comins}, M.~L., {Dietrich}, M., {Doroshenko}, V.~T.,
  {Eastman}, J.~D., {Efimov}, Y.~S., {Ewald}, S., {Ferbey}, S., {Gaskell},
  C.~M., {Hedrick}, C.~H., {Jackson}, K., {Klimanov}, S.~A., {Klimek}, E.~S.,
  {Kruse}, A.~K., {Lad{\'e}route}, A., {Lamb}, J.~B., {Leighly}, K.,
  {Minezaki}, T., {Nazarov}, S.~V., {Onken}, C.~A., {Petersen}, E.~A.,
  {Peterson}, P., {Poindexter}, S., {Sakata}, Y., {Schlesinger}, K.~J.,
  {Sergeev}, S.~G., {Skolski}, N., {Stieglitz}, L., {Tobin}, J.~J.,
  {Unterborn}, C., {Vestergaard}, M., {Watkins}, A.~E., {Watson}, L.~C.,
  {Yoshii}, Y., Sep. 2010. {Reverberation Mapping Measurements of Black Hole
  Masses in Six Local Seyfert Galaxies}. \apj 721, 715--737.

\bibitem[{{Desvignes} et~al.(2016){Desvignes}, {Caballero}, {Lentati},
  {Verbiest}, {Champion}, {Stappers}, {Janssen}, {Lazarus}, {Os{\l}owski},
  {Babak}, {Bassa}, {Brem}, {Burgay}, {Cognard}, {Gair}, {Graikou},
  {Guillemot}, {Hessels}, {Jessner}, {Jordan}, {Karuppusamy}, {Kramer},
  {Lassus}, {Lazaridis}, {Lee}, {Liu}, {Lyne}, {McKee}, {Mingarelli},
  {Perrodin}, {Petiteau}, {Possenti}, {Purver}, {Rosado}, {Sanidas}, {Sesana},
  {Shaifullah}, {Smits}, {Taylor}, {Theureau}, {Tiburzi}, {van Haasteren}, and
  {Vecchio}}]{2016MNRAS.458.3341D}
{Desvignes}, G., {Caballero}, R.~N., {Lentati}, L., {Verbiest}, J.~P.~W.,
  {Champion}, D.~J., {Stappers}, B.~W., {Janssen}, G.~H., {Lazarus}, P.,
  {Os{\l}owski}, S., {Babak}, S., {Bassa}, C.~G., {Brem}, P., {Burgay}, M.,
  {Cognard}, I., {Gair}, J.~R., {Graikou}, E., {Guillemot}, L., {Hessels},
  J.~W.~T., {Jessner}, A., {Jordan}, C., {Karuppusamy}, R., {Kramer}, M.,
  {Lassus}, A., {Lazaridis}, K., {Lee}, K.~J., {Liu}, K., {Lyne}, A.~G.,
  {McKee}, J., {Mingarelli}, C.~M.~F., {Perrodin}, D., {Petiteau}, A.,
  {Possenti}, A., {Purver}, M.~B., {Rosado}, P.~A., {Sanidas}, S., {Sesana},
  A., {Shaifullah}, G., {Smits}, R., {Taylor}, S.~R., {Theureau}, G.,
  {Tiburzi}, C., {van Haasteren}, R., {Vecchio}, A., May 2016. {High-precision
  timing of 42 millisecond pulsars with the European Pulsar Timing Array}.
  \mnras 458, 3341--3380.

\bibitem[{{Dewdney} et~al.(2009){Dewdney}, {Hall}, {Schilizzi}, and
  {Lazio}}]{2009IEEEP..97.1482D}
{Dewdney}, P.~E., {Hall}, P.~J., {Schilizzi}, R.~T., {Lazio}, T.~J.~L.~W., Aug.
  2009. {The Square Kilometre Array}. IEEE Proceedings 97, 1482--1496.

\bibitem[{{Dey} et~al.(2019){Dey}, {Gopakumar}, {Valtonen}, {Zola},
  {Susobhanan}, {Hudec}, {Pihajoki}, {Pursimo}, {Berdyugin}, {Piirola},
  {Ciprini}, {Nilsson}, {Jermak}, {Kidger}, and {Komossa}}]{dey19}
{Dey}, L., {Gopakumar}, A., {Valtonen}, M., {Zola}, S., {Susobhanan}, A.,
  {Hudec}, R., {Pihajoki}, P., {Pursimo}, T., {Berdyugin}, A., {Piirola}, V.,
  {Ciprini}, S., {Nilsson}, K., {Jermak}, H., {Kidger}, M., {Komossa}, S., May
  2019. {The Unique Blazar OJ 287 and Its Massive Binary Black Hole Central
  Engine}. Universe 5~(5), 108.

\bibitem[{{Di Cintio} et~al.(2017){Di Cintio}, {Tremmel}, {Governato},
  {Pontzen}, {Zavala}, {Bastidas Fry}, {Brooks}, and
  {Vogelsberger}}]{DiCintio_et_al_2017}
{Di Cintio}, A., {Tremmel}, M., {Governato}, F., {Pontzen}, A., {Zavala}, J.,
  {Bastidas Fry}, A., {Brooks}, A., {Vogelsberger}, M., Aug. 2017. {A rumble in
  the dark: signatures of self-interacting dark matter in supermassive black
  hole dynamics and galaxy density profiles}. \mnras 469, 2845--2854.

\bibitem[{{Di Matteo} et~al.(2005){Di Matteo}, {Springel}, and
  {Hernquist}}]{DiMatteo_et_al_2005}
{Di Matteo}, T., {Springel}, V., {Hernquist}, L., Feb. 2005. {Energy input from
  quasars regulates the growth and activity of black holes and their host
  galaxies}. \nat 433, 604--607.

\bibitem[{{Dietrich} et~al.(1998){Dietrich}, {Peterson}, {Albrecht}, {Altmann},
  {Barth}, {Bennie}, {Bertram}, {Bochkarev}, {Bock}, {Braun}, {Burenkov},
  {Collier}, {Fang}, {Francis}, {Filippenko}, {Foltz}, {G{\"a}ssler},
  {Gaskell}, {Geffert}, {Ghosh}, {Hilditch}, {Honeycutt}, {Horne}, {Huchra},
  {Kaspi}, {K{\"u}mmel}, {Leighly}, {Leonard}, {Malkov}, {Mikhailov}, {Miller},
  {Morrill}, {Noble}, {O'Brien}, {Oswalt}, {Pebley}, {Pfeiffer}, {Pronik},
  {Qian}, {Robertson}, {Robinson}, {Rumstay}, {Schmoll}, {Sergeev}, {Sergeeva},
  {Shapovalova}, {Skillman}, {Snedden}, {Soundararajaperumal}, {Stirpe}, {Tao},
  {Turner}, {Wagner}, {Wagner}, {Wei}, {Wu}, {Zheng}, and {Zou}}]{Dietrich1998}
{Dietrich}, M., {Peterson}, B.~M., {Albrecht}, P., {Altmann}, M., {Barth},
  A.~J., {Bennie}, P.~J., {Bertram}, R., {Bochkarev}, N.~G., {Bock}, H.,
  {Braun}, J.~M., {Burenkov}, A., {Collier}, S., {Fang}, L.-Z., {Francis},
  O.~P., {Filippenko}, A.~V., {Foltz}, C.~B., {G{\"a}ssler}, W., {Gaskell},
  C.~M., {Geffert}, M., {Ghosh}, K.~K., {Hilditch}, R.~W., {Honeycutt}, R.~K.,
  {Horne}, K., {Huchra}, J.~P., {Kaspi}, S., {K{\"u}mmel}, M., {Leighly},
  K.~M., {Leonard}, D.~C., {Malkov}, Y.~F., {Mikhailov}, V., {Miller}, H.~R.,
  {Morrill}, A.~C., {Noble}, J., {O'Brien}, P.~T., {Oswalt}, T.~D., {Pebley},
  S.~P., {Pfeiffer}, M., {Pronik}, V.~I., {Qian}, B.-C., {Robertson}, J.~W.,
  {Robinson}, A., {Rumstay}, K.~S., {Schmoll}, J., {Sergeev}, S.~G.,
  {Sergeeva}, E.~A., {Shapovalova}, A.~I., {Skillman}, D.~R., {Snedden}, S.~A.,
  {Soundararajaperumal}, S., {Stirpe}, G.~M., {Tao}, J., {Turner}, G.~W.,
  {Wagner}, R.~M., {Wagner}, S.~J., {Wei}, J.~Y., {Wu}, H., {Zheng}, W., {Zou},
  Z.~L., Apr. 1998. {Steps toward Determination of the Size and Structure of
  the Broad-Line Region in Active Galactic Nuclei. XII. Ground-based Monitoring
  of 3C 390.3}. \apjs 115, 185--202.

\bibitem[{{DiPompeo} et~al.(2015){DiPompeo}, {Bovy}, {Myers}, and
  {Lang}}]{2015MNRAS.452.3124D}
{DiPompeo}, M.~A., {Bovy}, J., {Myers}, A.~D., {Lang}, D., Sep. 2015. {Quasar
  probabilities and redshifts from WISE mid-IR through GALEX UV photometry}.
  \mnras 452, 3124--3138.

\bibitem[{{Djorgovski} et~al.(2007){Djorgovski}, {Courbin}, {Meylan}, {Sluse},
  {Thompson}, {Mahabal}, and {Glikman}}]{Djorgovski2007}
{Djorgovski}, S.~G., {Courbin}, F., {Meylan}, G., {Sluse}, D., {Thompson}, D.,
  {Mahabal}, A., {Glikman}, E., Jun. 2007. {Discovery of a Probable Physical
  Triple Quasar}. \apjl 662, L1--L5.

\bibitem[{{Dolag} et~al.(2017){Dolag}, {Mevius}, and
  {Remus}}]{Dolag_et_al_2017}
{Dolag}, K., {Mevius}, E., {Remus}, R.-S., Aug. 2017. {Distribution and
  Evolution of Metals in the Magneticum Simulations}. Galaxies 5, 35.

\bibitem[{{Donley} et~al.(2007){Donley}, {Rieke}, {P{\'e}rez-Gonz{\'a}lez},
  {Rigby}, and {Alonso-Herrero}}]{Donley2007}
{Donley}, J.~L., {Rieke}, G.~H., {P{\'e}rez-Gonz{\'a}lez}, P.~G., {Rigby},
  J.~R., {Alonso-Herrero}, A., May 2007. {Spitzer Power-Law Active Galactic
  Nucleus Candidates in the Chandra Deep Field-North}. \apj 660~(1), 167--190.

\bibitem[{{D'Orazio} and {Di Stefano}(2018)}]{DorazioDiStefano2018}
{D'Orazio}, D.~J., {Di Stefano}, R., Mar. 2018. {Periodic self-lensing from
  accreting massive black hole binaries}. \mnras 474, 2975--2986.

\bibitem[{{D'Orazio} and {Haiman}(2017)}]{Dorazio2017}
{D'Orazio}, D.~J., {Haiman}, Z., Sep 2017. {Lighthouse in the dust: infrared
  echoes of periodic emission from massive black hole binaries}. \mnras
  470~(1), 1198--1217.

\bibitem[{{D'Orazio} et~al.(2016){D'Orazio}, {Haiman}, {Duffell}, {MacFadyen},
  and {Farris}}]{Dorazio+2016}
{D'Orazio}, D.~J., {Haiman}, Z., {Duffell}, P., {MacFadyen}, A., {Farris}, B.,
  Jul. 2016. {A transition in circumbinary accretion discs at a binary mass
  ratio of 1:25}. \mnras 459, 2379--2393.

\bibitem[{{D'Orazio} et~al.(2013){D'Orazio}, {Haiman}, and
  {MacFadyen}}]{dorazio13}
{D'Orazio}, D.~J., {Haiman}, Z., {MacFadyen}, A., Dec. 2013. {Accretion into
  the central cavity of a circumbinary disc}. \mnras 436, 2997--3020.

\bibitem[{{D'Orazio} et~al.(2015){D'Orazio}, {Haiman}, and
  {Schiminovich}}]{Dorazio+2015}
{D'Orazio}, D.~J., {Haiman}, Z., {Schiminovich}, D., Sep. 2015. {Relativistic
  boost as the cause of periodicity in a massive black-hole binary candidate}.
  \nat 525, 351--353.

\bibitem[{{Dorn-Wallenstein} et~al.(2017){Dorn-Wallenstein}, {Levesque}, and
  {Ruan}}]{Dorn_Wallenstein2017}
{Dorn-Wallenstein}, T., {Levesque}, E.~M., {Ruan}, J.~J., Nov. 2017. {A Mote in
  Andromeda's Disk: A Misidentified Periodic AGN behind M31}. \apj 850, 86.

\bibitem[{{Dotti} et~al.(2006){Dotti}, {Colpi}, and
  {Haardt}}]{Dotti_et_al_2006}
{Dotti}, M., {Colpi}, M., {Haardt}, F., Mar. 2006. {Laser Interferometer Space
  Antenna double black holes: dynamics in gaseous nuclear discs}. \mnras 367,
  103--112.

\bibitem[{{Dotti} et~al.(2007){Dotti}, {Colpi}, {Haardt}, and
  {Mayer}}]{dotti07}
{Dotti}, M., {Colpi}, M., {Haardt}, F., {Mayer}, L., Aug. 2007. {Supermassive
  black hole binaries in gaseous and stellar circumnuclear discs: orbital
  dynamics and gas accretion}. \mnras 379, 956--962.

\bibitem[{{Dotti} et~al.(2009){Dotti}, {Ruszkowski}, {Paredi}, {Colpi},
  {Volonteri}, and {Haardt}}]{2009MNRAS.396.1640D}
{Dotti}, M., {Ruszkowski}, M., {Paredi}, L., {Colpi}, M., {Volonteri}, M.,
  {Haardt}, F., Jul. 2009. {Dual black holes in merger remnants - I. Linking
  accretion to dynamics}. \mnras 396, 1640--1646.

\bibitem[{{Downes} and {Solomon}(1998)}]{downes_solomon1998}
{Downes}, D., {Solomon}, P.~M., Nov 1998. {Rotating Nuclear Rings and Extreme
  Starbursts in Ultraluminous Galaxies}. \apj 507~(2), 615--654.

\bibitem[{{Dubois} et~al.(2012){Dubois}, {Devriendt}, {Slyz}, and
  {Teyssier}}]{2012MNRAS.420.2662D}
{Dubois}, Y., {Devriendt}, J., {Slyz}, A., {Teyssier}, R., Mar 2012.
  {Self-regulated growth of supermassive black holes by a dual jet-heating
  active galactic nucleus feedback mechanism: methods, tests and implications
  for cosmological simulations}. \mnras 420~(3), 2662--2683.

\bibitem[{{Dubois} et~al.(2014){Dubois}, {Pichon}, {Welker}, {Le Borgne},
  {Devriendt}, {Laigle}, {Codis}, {Pogosyan}, {Arnouts}, {Benabed}, {Bertin},
  {Blaizot}, {Bouchet}, {Cardoso}, {Colombi}, {de Lapparent}, {Desjacques},
  {Gavazzi}, {Kassin}, {Kimm}, {McCracken}, {Milliard}, {Peirani}, {Prunet},
  {Rouberol}, {Silk}, {Slyz}, {Sousbie}, {Teyssier}, {Tresse}, {Treyer},
  {Vibert}, and {Volonteri}}]{2014MNRAS.444.1453D}
{Dubois}, Y., {Pichon}, C., {Welker}, C., {Le Borgne}, D., {Devriendt}, J.,
  {Laigle}, C., {Codis}, S., {Pogosyan}, D., {Arnouts}, S., {Benabed}, K.,
  {Bertin}, E., {Blaizot}, J., {Bouchet}, F., {Cardoso}, J.~F., {Colombi}, S.,
  {de Lapparent}, V., {Desjacques}, V., {Gavazzi}, R., {Kassin}, S., {Kimm},
  T., {McCracken}, H., {Milliard}, B., {Peirani}, S., {Prunet}, S., {Rouberol},
  S., {Silk}, J., {Slyz}, A., {Sousbie}, T., {Teyssier}, R., {Tresse}, L.,
  {Treyer}, M., {Vibert}, D., {Volonteri}, M., Oct 2014. {Dancing in the dark:
  galactic properties trace spin swings along the cosmic web}. \mnras 444~(2),
  1453--1468.

\bibitem[{{Duffell} et~al.(2014){Duffell}, {Haiman}, {MacFadyen}, {D'Orazio},
  and {Farris}}]{Duffell+2014}
{Duffell}, P.~C., {Haiman}, Z., {MacFadyen}, A.~I., {D'Orazio}, D.~J.,
  {Farris}, B.~D., Sep. 2014. {The Migration of Gap-opening Planets is Not
  Locked to Viscous Disk Evolution}. \apjl 792, L10.

\bibitem[{{D{\"u}rmann} and {Kley}(2017)}]{DK2017}
{D{\"u}rmann}, C., {Kley}, W., Feb. 2017. {The accretion of migrating giant
  planets}. \aap 598, A80.

\bibitem[{{Eftekharzadeh} et~al.(2017){Eftekharzadeh}, {Myers}, {Hennawi},
  {Djorgovski}, {Richards}, {Mahabal}, and {Graham}}]{2017MNRAS.468...77E}
{Eftekharzadeh}, S., {Myers}, A.~D., {Hennawi}, J.~F., {Djorgovski}, S.~G.,
  {Richards}, G.~T., {Mahabal}, A.~A., {Graham}, M.~J., Jun. 2017. {Clustering
  on very small scales from a large sample of confirmed quasar pairs: does
  quasar clustering track from Mpc to kpc scales?} \mnras 468, 77--90.

\bibitem[{{Eftekharzadeh} et~al.(2019){Eftekharzadeh}, {Myers}, and
  {Kourkchi}}]{Eftekharzadeh2019}
{Eftekharzadeh}, S., {Myers}, A.~D., {Kourkchi}, E., Jun 2019. {A Halo
  Occupation Interpretation of Quasars at z $\sim$ 1.5 Using Very Small-Scale
  Clustering Information}. \mnras 486~(1), 274--282.

\bibitem[{{Ellison} et~al.(2015){Ellison}, {Patton}, and
  {Hickox}}]{Ellison2015}
{Ellison}, S.~L., {Patton}, D.~R., {Hickox}, R.~C., Jul 2015. {Galaxy pairs in
  the Sloan Digital Sky Survey - XII. The fuelling mechanism of low-excitation
  radio-loud AGN.} \mnras 451, L35--L39.

\bibitem[{{Ellison} et~al.(2011){Ellison}, {Patton}, {Mendel}, and
  {Scudder}}]{Ellison2011}
{Ellison}, S.~L., {Patton}, D.~R., {Mendel}, J.~T., {Scudder}, J.~M., Dec.
  2011. {Galaxy pairs in the Sloan Digital Sky Survey - IV. Interactions
  trigger active galactic nuclei}. \mnras 418, 2043--2053.

\bibitem[{{Ellison} et~al.(2017){Ellison}, {Secrest}, {Mendel}, {Satyapal}, and
  {Simard}}]{EllisonSM17}
{Ellison}, S.~L., {Secrest}, N.~J., {Mendel}, J.~T., {Satyapal}, S., {Simard},
  L., Sep. 2017. {Discovery of a dual active galactic nucleus with 8 kpc
  separation}. \mnras 470, L49--L53.

\bibitem[{{Eracleous} et~al.(2012){Eracleous}, {Boroson}, {Halpern}, and
  {Liu}}]{Eracleous2012}
{Eracleous}, M., {Boroson}, T.~A., {Halpern}, J.~P., {Liu}, J., Aug. 2012. {A
  Large Systematic Search for Close Supermassive Binary and Rapidly Recoiling
  Black Holes}. \apjs 201, 23.

\bibitem[{{Eracleous} and {Halpern}(1994)}]{Eracleous1994}
{Eracleous}, M., {Halpern}, J.~P., Jan. 1994. {Doubled-peaked emission lines in
  active galactic nuclei}. \apjs 90, 1--30.

\bibitem[{{Eracleous} et~al.(1997){Eracleous}, {Halpern}, {M.~Gilbert},
  {Newman}, and {Filippenko}}]{Eracleous1997}
{Eracleous}, M., {Halpern}, J.~P., {M.~Gilbert}, A., {Newman}, J.~A.,
  {Filippenko}, A.~V., Nov. 1997. {Rejection of the Binary Broad-Line Region
  Interpretation of Double-peaked Emission Lines in Three Active Galactic
  Nuclei}. \apj 490, 216--226.

\bibitem[{{Eracleous} et~al.(2009){Eracleous}, {Lewis}, and
  {Flohic}}]{Eracleous2009}
{Eracleous}, M., {Lewis}, K.~T., {Flohic}, H.~M.~L.~G., Jul. 2009.
  {Double-peaked emission lines as a probe of the broad-line regions of active
  galactic nuclei}. \nar 53, 133--139.

\bibitem[{{Escala} et~al.(2005){Escala}, {Larson}, {Coppi}, and
  {Mardones}}]{escala05}
{Escala}, A., {Larson}, R.~B., {Coppi}, P.~S., {Mardones}, D., Sep. 2005. {The
  Role of Gas in the Merging of Massive Black Holes in Galactic Nuclei. II.
  Black Hole Merging in a Nuclear Gas Disk}. \apj 630, 152--166.

\bibitem[{{Fabbiano} et~al.(2011){Fabbiano}, {Wang}, {Elvis}, and
  {Risaliti}}]{Fabbiano2011}
{Fabbiano}, G., {Wang}, J., {Elvis}, M., {Risaliti}, G., Sep. 2011. {A close
  nuclear black-hole pair in the spiral galaxy NGC3393}. \nat 477, 431--434.

\bibitem[{{Fabjan} et~al.(2010){Fabjan}, {Borgani}, {Tornatore}, {Saro},
  {Murante}, and {Dolag}}]{Fabjan_et_al_2010}
{Fabjan}, D., {Borgani}, S., {Tornatore}, L., {Saro}, A., {Murante}, G.,
  {Dolag}, K., Jan. 2010. {Simulating the effect of active galactic nuclei
  feedback on the metal enrichment of galaxy clusters}. \mnras 401, 1670--1690.

\bibitem[{{Fan} et~al.(2016){Fan}, {Han}, {Nikutta}, {Drouart}, and
  {Knudsen}}]{fan_etal2016}
{Fan}, L., {Han}, Y., {Nikutta}, R., {Drouart}, G., {Knudsen}, K.~K., Jun 2016.
  {Infrared Spectral Energy Distribution Decomposition of WISE-selected,
  Hyperluminous Hot Dust-obscured Galaxies}. \apj 823~(2), 107.

\bibitem[{{Farina} et~al.(2013){Farina}, {Montuori}, {Decarli}, and
  {Fumagalli}}]{2013MNRAS.431.1019F}
{Farina}, E.~P., {Montuori}, C., {Decarli}, R., {Fumagalli}, M., May 2013.
  {Caught in the act: discovery of a physical quasar triplet}. \mnras 431,
  1019--1025.

\bibitem[{{Farris} et~al.(2014){Farris}, {Duffell}, {MacFadyen}, and
  {Haiman}}]{farris14}
{Farris}, B.~D., {Duffell}, P., {MacFadyen}, A.~I., {Haiman}, Z., Mar. 2014.
  {Binary Black Hole Accretion from a Circumbinary Disk: Gas Dynamics inside
  the Central Cavity}. \apj 783, 134.

\bibitem[{{Farris} et~al.(2015{\natexlab{a}}){Farris}, {Duffell}, {MacFadyen},
  and {Haiman}}]{Farris+2015}
{Farris}, B.~D., {Duffell}, P., {MacFadyen}, A.~I., {Haiman}, Z., Feb.
  2015{\natexlab{a}}. {Binary black hole accretion during inspiral and merger}.
  \mnras 447, L80--L84.

\bibitem[{{Farris} et~al.(2015{\natexlab{b}}){Farris}, {Duffell}, {MacFadyen},
  and {Haiman}}]{Farris+2014b}
{Farris}, B.~D., {Duffell}, P., {MacFadyen}, A.~I., {Haiman}, Z., Jan.
  2015{\natexlab{b}}. {Characteristic signatures in the thermal emission from
  accreting binary black holes}. \mnras 446, L36--L40.

\bibitem[{{Farris} et~al.(2012){Farris}, {Gold}, {Paschalidis}, {Etienne}, and
  {Shapiro}}]{farris12}
{Farris}, B.~D., {Gold}, R., {Paschalidis}, V., {Etienne}, Z.~B., {Shapiro},
  S.~L., Nov. 2012. {Binary Black-Hole Mergers in Magnetized Disks: Simulations
  in Full General Relativity}. Physical Review Letters 109~(22), 221102.

\bibitem[{{Feltre} et~al.(2016){Feltre}, {Charlot}, and {Gutkin}}]{Feltre2016}
{Feltre}, A., {Charlot}, S., {Gutkin}, J., Mar. 2016. {Nuclear activity versus
  star formation: emission-line diagnostics at ultraviolet and optical
  wavelengths}. \mnras 456, 3354--3374.

\bibitem[{{Feng} et~al.(2016){Feng}, {Di-Matteo}, {Croft}, {Bird}, {Battaglia},
  and {Wilkins}}]{Feng_et_al_2016}
{Feng}, Y., {Di-Matteo}, T., {Croft}, R.~A., {Bird}, S., {Battaglia}, N.,
  {Wilkins}, S., Jan. 2016. {The BlueTides simulation: first galaxies and
  reionization}. \mnras 455, 2778--2791.

\bibitem[{{Ferland} et~al.(1998){Ferland}, {Korista}, {Verner}, {Ferguson},
  {Kingdon}, and {Verner}}]{Ferland98}
{Ferland}, G.~J., {Korista}, K.~T., {Verner}, D.~A., {Ferguson}, J.~W.,
  {Kingdon}, J.~B., {Verner}, E.~M., Jul. 1998. {CLOUDY 90: Numerical
  Simulation of Plasmas and Their Spectra}. \pasp 110, 761--778.

\bibitem[{{Ferrarese} and {Merritt}(2000)}]{ferrarese&merritt00}
{Ferrarese}, L., {Merritt}, D., Aug. 2000. {A Fundamental Relation between
  Supermassive Black Holes and Their Host Galaxies}. \apjl 539, L9--L12.

\bibitem[{{Fey} et~al.(2015){Fey}, {Gordon}, {Jacobs}, {Ma}, {Gaume}, {Arias},
  {Bianco}, {Boboltz}, {B{\"o}ckmann}, {Bolotin}, {Charlot}, {Collioud},
  {Engelhardt}, {Gipson}, {Gontier}, {Heinkelmann}, {Kurdubov}, {Lambert},
  {Lytvyn}, {MacMillan}, {Malkin}, {Nothnagel}, {Ojha}, {Skurikhina},
  {Sokolova}, {Souchay}, {Sovers}, {Tesmer}, {Titov}, {Wang}, and
  {Zharov}}]{Fey2015}
{Fey}, A.~L., {Gordon}, D., {Jacobs}, C.~S., {Ma}, C., {Gaume}, R.~A., {Arias},
  E.~F., {Bianco}, G., {Boboltz}, D.~A., {B{\"o}ckmann}, S., {Bolotin}, S.,
  {Charlot}, P., {Collioud}, A., {Engelhardt}, G., {Gipson}, J., {Gontier},
  A.-M., {Heinkelmann}, R., {Kurdubov}, S., {Lambert}, S., {Lytvyn}, S.,
  {MacMillan}, D.~S., {Malkin}, Z., {Nothnagel}, A., {Ojha}, R., {Skurikhina},
  E., {Sokolova}, J., {Souchay}, J., {Sovers}, O.~J., {Tesmer}, V., {Titov},
  O., {Wang}, G., {Zharov}, V., Aug. 2015. {The Second Realization of the
  International Celestial Reference Frame by Very Long Baseline
  Interferometry}. \aj 150, 58.

\bibitem[{{Fiacconi} et~al.(2013){Fiacconi}, {Mayer}, {Ro{\v s}kar}, and
  {Colpi}}]{Fiacconi_et_al_2013}
{Fiacconi}, D., {Mayer}, L., {Ro{\v s}kar}, R., {Colpi}, M., Nov. 2013.
  {Massive Black Hole Pairs in Clumpy, Self-gravitating Circumnuclear Disks:
  Stochastic Orbital Decay}. \apjl 777, L14.

\bibitem[{{Findlay} et~al.(2018){Findlay}, {Prochaska}, {Hennawi}, {Fumagalli},
  {Myers}, {Bartle}, {Chehade}, {DiPompeo}, {Shanks}, {Lau}, and
  {Rubin}}]{findlay2018}
{Findlay}, J.~R., {Prochaska}, J.~X., {Hennawi}, J.~F., {Fumagalli}, M.,
  {Myers}, A.~D., {Bartle}, S., {Chehade}, B., {DiPompeo}, M.~A., {Shanks}, T.,
  {Lau}, M.~W., {Rubin}, K. H.~R., Jun 2018. {Quasars Probing Quasars. X. The
  Quasar Pair Spectral Database}. \apjs 236~(2), 44.

\bibitem[{{Foord} et~al.(2019){Foord}, {G{\"u}ltekin}, {Reynolds},
  {Hodges-Kluck}, {Cackett}, {Comerford}, {King}, {Miller}, and
  {Runnoe}}]{foord19}
{Foord}, A., {G{\"u}ltekin}, K., {Reynolds}, M.~T., {Hodges-Kluck}, E.,
  {Cackett}, E.~M., {Comerford}, J.~M., {King}, A.~L., {Miller}, J.~M.,
  {Runnoe}, J.~C., May 2019. {A Bayesian Analysis of SDSS J0914+0853, a
  Low-mass Dual AGN Candidate}. \apj 877~(1), 17.

\bibitem[{{Foreman} et~al.(2009){Foreman}, {Volonteri}, and
  {Dotti}}]{2009ApJ...693.1554F}
{Foreman}, G., {Volonteri}, M., {Dotti}, M., Mar. 2009. {Double Quasars: Probes
  of Black Hole Scaling Relationships and Merger Scenarios}. \apj 693,
  1554--1562.

\bibitem[{{F{\"o}rster Schreiber} et~al.(2011){F{\"o}rster Schreiber},
  {Shapley}, {Genzel}, {Bouch{\'e}}, {Cresci}, {Davies}, {Erb}, {Genel},
  {Lutz}, {Newman}, {Shapiro}, {Steidel}, {Sternberg}, and
  {Tacconi}}]{ForsterSchreiber_et_al_2011}
{F{\"o}rster Schreiber}, N.~M., {Shapley}, A.~E., {Genzel}, R., {Bouch{\'e}},
  N., {Cresci}, G., {Davies}, R., {Erb}, D.~K., {Genel}, S., {Lutz}, D.,
  {Newman}, S., {Shapiro}, K.~L., {Steidel}, C.~C., {Sternberg}, A., {Tacconi},
  L.~J., Sep. 2011. {Constraints on the Assembly and Dynamics of Galaxies. II.
  Properties of Kiloparsec-scale Clumps in Rest-frame Optical Emission of z
  \~{} 2 Star-forming Galaxies}. \apj 739, 45.

\bibitem[{{Foster} and {Backer}(1990)}]{1990ApJ...361..300F}
{Foster}, R.~S., {Backer}, D.~C., Sep. 1990. {Constructing a pulsar timing
  array}. \apj 361, 300--308.

\bibitem[{{Fragos} et~al.(2013){Fragos}, {Lehmer}, {Tremmel}, {Tzanavaris},
  {Basu-Zych}, {Belczynski}, {Hornschemeier}, {Jenkins}, {Kalogera}, and
  {Ptak}}]{fragosetal2013}
{Fragos}, T., {Lehmer}, B., {Tremmel}, M., {Tzanavaris}, P., {Basu-Zych}, A.,
  {Belczynski}, K., {Hornschemeier}, A., {Jenkins}, L., {Kalogera}, V., {Ptak},
  A., Feb 2013. {X-Ray Binary Evolution Across Cosmic Time}. \apj 764~(1), 41.

\bibitem[{{French} et~al.(2016){French}, {Arcavi}, and
  {Zabludoff}}]{French2016}
{French}, K.~D., {Arcavi}, I., {Zabludoff}, A., Feb. 2016. {Tidal Disruption
  Events Prefer Unusual Host Galaxies}. \apjl 818, L21.

\bibitem[{{Frey} et~al.(2012){Frey}, {Paragi}, {An}, and
  {Gab{\'a}nyi}}]{frey2012}
{Frey}, S., {Paragi}, Z., {An}, T., {Gab{\'a}nyi}, K.~{\'E}., Sep. 2012. {Two
  in one? A possible dual radio-emitting nucleus in the quasar SDSS
  J1425+3231}. Monthly Notices of the Royal Astronomical Society 425,
  1185--1191.

\bibitem[{{Fu} et~al.(2012){Fu}, {Yan}, {Myers}, {Stockton}, {Djorgovski},
  {Aldering}, and {Rich}}]{FuYM12}
{Fu}, H., {Yan}, L., {Myers}, A.~D., {Stockton}, A., {Djorgovski}, S.~G.,
  {Aldering}, G., {Rich}, J.~A., Jan. 2012. {The Nature of Double-peaked [O
  III] Active Galactic Nuclei}. \apj 745, 67.

\bibitem[{{Fu} et~al.(2011){Fu}, {Zhang}, {Assef}, {Stockton}, {Myers}, {Yan},
  {Djorgovski}, {Wrobel}, and {Riechers}}]{fu2011}
{Fu}, H., {Zhang}, Z.-Y., {Assef}, R.~J., {Stockton}, A., {Myers}, A.~D.,
  {Yan}, L., {Djorgovski}, S.~G., {Wrobel}, J.~M., {Riechers}, D.~A., Oct.
  2011. {A Kiloparsec-scale Binary Active Galactic Nucleus Confirmed by the
  Expanded Very Large Array}. Astrophysical Journal Letters 740, L44.

\bibitem[{{Fumagalli} et~al.(2017){Fumagalli}, {Mackenzie}, {Trayford},
  {Theuns}, {Cantalupo}, {Christensen}, {Fynbo}, {M{\o}ller}, {O'Meara},
  {Prochaska}, {Rafelski}, and {Shanks}}]{Fumagalli2017}
{Fumagalli}, M., {Mackenzie}, R., {Trayford}, J., {Theuns}, T., {Cantalupo},
  S., {Christensen}, L., {Fynbo}, J.~P.~U., {M{\o}ller}, P., {O'Meara}, J.,
  {Prochaska}, J.~X., {Rafelski}, M., {Shanks}, T., Nov. 2017. {Witnessing
  galaxy assembly in an extended z$\approx$3 structure}. \mnras 471,
  3686--3698.

\bibitem[{{Gab{\'a}nyi} et~al.(2016){Gab{\'a}nyi}, {An}, {Frey}, {Komossa},
  {Paragi}, {Hong}, and {Shen}}]{gabanyi2016}
{Gab{\'a}nyi}, K.~{\'E}., {An}, T., {Frey}, S., {Komossa}, S., {Paragi}, Z.,
  {Hong}, X.-Y., {Shen}, Z.-Q., Aug. 2016. {Four Dual AGN Candidates Observed
  with the VLBA}. Astrophysical Journal 826, 106.

\bibitem[{{Gab{\'a}nyi} et~al.(2017){Gab{\'a}nyi}, {Frey}, {Paragi}, {An}, and
  {Komossa}}]{gabanyi2017}
{Gab{\'a}nyi}, K.~{\'E}., {Frey}, S., {Paragi}, Z., {An}, T., {Komossa}, S.,
  2017. {Searching for a pair of accreting supermassive black holes in
  J1425+3231}. In: {Gomboc}, A. (Ed.), New Frontiers in Black Hole
  Astrophysics. Vol. 324 of IAU Symposium. pp. 223--226.

\bibitem[{{Gabor} et~al.(2016){Gabor}, {Capelo}, {Volonteri}, {Bournaud},
  {Bellovary}, {Governato}, and {Quinn}}]{Gabor_et_al_2016}
{Gabor}, J.~M., {Capelo}, P.~R., {Volonteri}, M., {Bournaud}, F., {Bellovary},
  J., {Governato}, F., {Quinn}, T., Jul. 2016. {Comparison of black hole growth
  in galaxy mergers with gasoline and ramses}. \aap 592, A62.

\bibitem[{{Gaskell}(1983)}]{Gaskell1983}
{Gaskell}, C.~M., Jun. 1983. {Quasars as supermassive binaries}. In: {Swings},
  J.-P. (Ed.), Liege International Astrophysical Colloquia. Vol.~24 of Liege
  International Astrophysical Colloquia. pp. 473--477.

\bibitem[{{Gaskell}(1984)}]{Gaskell1984}
{Gaskell}, C.~M., 1984. {Quasars as supermassive binaries.} Annals of the New
  York Academy of Sciences 422, 349--349.

\bibitem[{{Gaskell}(1988)}]{Gaskell1988}
{Gaskell}, C.~M., 1988. {Double peaked broad line profiles -- edge on accretion
  disks or double quasar nuclei?} In: {Miller}, H.~R., {Wiita}, P.~J. (Eds.),
  Active Galactic Nuclei. Vol. 307 of Lecture Notes in Physics, Berlin Springer
  Verlag. p.~61.

\bibitem[{{Ge} et~al.(2012){Ge}, {Hu}, {Wang}, {Bai}, and {Zhang}}]{Ge:2012}
{Ge}, J.-Q., {Hu}, C., {Wang}, J.-M., {Bai}, J.-M., {Zhang}, S., Aug. 2012.
  {Double-peaked Narrow Emission-line Galaxies from the Sloan Digital Sky
  Survey. I. Sample and Basic Properties}. \apjs 201, 31.

\bibitem[{{Gebhardt} et~al.(2000){Gebhardt}, {Bender}, {Bower}, {Dressler},
  {Faber}, {Filippenko}, {Green}, {Grillmair}, {Ho}, {Kormendy}, {Lauer},
  {Magorrian}, {Pinkney}, {Richstone}, and {Tremaine}}]{gebhardtetal2000}
{Gebhardt}, K., {Bender}, R., {Bower}, G., {Dressler}, A., {Faber}, S.~M.,
  {Filippenko}, A.~V., {Green}, R., {Grillmair}, C., {Ho}, L.~C., {Kormendy},
  J., {Lauer}, T.~R., {Magorrian}, J., {Pinkney}, J., {Richstone}, D.,
  {Tremaine}, S., Aug. 2000. {A Relationship between Nuclear Black Hole Mass
  and Galaxy Velocity Dispersion}. \apjl 539, L13--L16.

\bibitem[{{George} and {Fabian}(1991)}]{george&fabian91}
{George}, I.~M., {Fabian}, A.~C., Mar. 1991. {X-ray reflection from cold matter
  in active galactic nuclei and X-ray binaries}. \mnras 249, 352--367.

\bibitem[{{Gerke} et~al.(2007){Gerke}, {Newman}, {Lotz}, {Yan}, {Barmby},
  {Coil}, {Conselice}, {Ivison}, {Lin}, {Koo}, {Nandra}, {Salim}, {Small},
  {Weiner}, {Cooper}, {Davis}, {Faber}, and {Guhathakurta}}]{Gerke:2007}
{Gerke}, B.~F., {Newman}, J.~A., {Lotz}, J., {Yan}, R., {Barmby}, P., {Coil},
  A.~L., {Conselice}, C.~J., {Ivison}, R.~J., {Lin}, L., {Koo}, D.~C.,
  {Nandra}, K., {Salim}, S., {Small}, T., {Weiner}, B.~J., {Cooper}, M.~C.,
  {Davis}, M., {Faber}, S.~M., {Guhathakurta}, P., May 2007. {The DEEP2 Galaxy
  Redshift Survey: AEGIS Observations of a Dual AGN at z = 0.7}. \apjl 660,
  L23--L26.

\bibitem[{{Gezari} et~al.(2007){Gezari}, {Halpern}, and
  {Eracleous}}]{Gezari2007}
{Gezari}, S., {Halpern}, J.~P., {Eracleous}, M., Apr. 2007. {Long-Term Profile
  Variability of Double-peaked Emission Lines in Active Galactic Nuclei}. \apjs
  169, 167--212.

\bibitem[{{Giacomazzo} et~al.(2012){Giacomazzo}, {Baker}, {Miller}, {Reynolds},
  and {van Meter}}]{giaco12}
{Giacomazzo}, B., {Baker}, J.~G., {Miller}, M.~C., {Reynolds}, C.~S., {van
  Meter}, J.~R., Jun. 2012. {General Relativistic Simulations of Magnetized
  Plasmas around Merging Supermassive Black Holes}. \apjl 752, L15.

\bibitem[{{Glikman} et~al.(2015){Glikman}, {Simmons}, {Mailly}, {Schawinski},
  {Urry}, and {Lacy}}]{Glikman2015}
{Glikman}, E., {Simmons}, B., {Mailly}, M., {Schawinski}, K., {Urry}, C.~M.,
  {Lacy}, M., Jun 2015. {Major Mergers Host the Most-luminous Red Quasars at z
  \raisebox{-0.5ex}\textasciitilde 2: A Hubble Space Telescope WFC3/IR Study}.
  \apj 806~(2), 218.

\bibitem[{{Gold} et~al.(2014){Gold}, {Paschalidis}, {Etienne}, {Shapiro}, and
  {Pfeiffer}}]{gold14}
{Gold}, R., {Paschalidis}, V., {Etienne}, Z.~B., {Shapiro}, S.~L., {Pfeiffer},
  H.~P., Mar. 2014. {Accretion disks around binary black holes of unequal mass:
  General relativistic magnetohydrodynamic simulations near decoupling}. \prd
  89~(6), 064060.

\bibitem[{{Goldreich} and {Tremaine}(1980)}]{GT80}
{Goldreich}, P., {Tremaine}, S., Oct. 1980. {Disk-satellite interactions}. \apj
  241, 425--441.

\bibitem[{{Goldstein} et~al.(2018){Goldstein}, {Veitch}, {Sesana}, and
  {Vecchio}}]{2018MNRAS.tmp..866G}
{Goldstein}, J., {Veitch}, J., {Sesana}, A., {Vecchio}, A., Apr. 2018. {Null
  stream analysis of Pulsar Timing Array data: localisation of resolvable
  gravitational wave sources}. \mnras.

\bibitem[{{Goldstein} et~al.(2019){Goldstein}, {Sesana}, {Holgado}, and
  {Veitch}}]{goldstein2019}
{Goldstein}, J.~M., {Sesana}, A., {Holgado}, A.~M., {Veitch}, J., May 2019.
  {Associating host galaxy candidates to massive black hole binaries resolved
  by pulsar timing arrays}. \mnras 485~(1), 248--259.

\bibitem[{{Goulding} et~al.(2018){Goulding}, {Greene}, {Bezanson}, {Greco},
  {Johnson}, {Leauthaud}, {Matsuoka}, {Medezinski}, and
  {Price-Whelan}}]{goulding+2018}
{Goulding}, A.~D., {Greene}, J.~E., {Bezanson}, R., {Greco}, J., {Johnson}, S.,
  {Leauthaud}, A., {Matsuoka}, Y., {Medezinski}, E., {Price-Whelan}, A.~M.,
  Jan. 2018. {Galaxy interactions trigger rapid black hole growth: An
  unprecedented view from the Hyper Suprime-Cam survey}. \pasj 70, S37.

\bibitem[{{Gower} et~al.(1982){Gower}, {Gregory}, {Unruh}, and
  {Hutchings}}]{gower1982}
{Gower}, A.~C., {Gregory}, P.~C., {Unruh}, W.~G., {Hutchings}, J.~B., Nov.
  1982. {Relativistic precessing jets in quasars and radio galaxies - Models to
  fit high resolution data}. \apj 262, 478--496.

\bibitem[{{Graham} et~al.(2015){Graham}, {Djorgovski}, {Stern}, {Drake},
  {Mahabal}, {Donalek}, {Glikman}, {Larson}, and {Christensen}}]{Graham2015}
{Graham}, M.~J., {Djorgovski}, S.~G., {Stern}, D., {Drake}, A.~J., {Mahabal},
  A.~A., {Donalek}, C., {Glikman}, E., {Larson}, S., {Christensen}, E., Oct.
  2015. {A systematic search for close supermassive black hole binaries in the
  Catalina Real-time Transient Survey}. \mnras 453, 1562--1576.

\bibitem[{{Green} et~al.(2011){Green}, {Myers}, {Barkhouse}, {Aldcroft},
  {Trichas}, {Richards}, {Ruiz}, and {Hopkins}}]{Green2011}
{Green}, P.~J., {Myers}, A.~D., {Barkhouse}, W.~A., {Aldcroft}, T.~L.,
  {Trichas}, M., {Richards}, G.~T., {Ruiz}, {\'A}., {Hopkins}, P.~F., Dec.
  2011. {A Multiwavelength Study of Binary Quasars and Their Environments}.
  \apj 743, 81.

\bibitem[{{Green} et~al.(2010){Green}, {Myers}, {Barkhouse}, {Mulchaey},
  {Bennert}, {Cox}, and {Aldcroft}}]{Green2010}
{Green}, P.~J., {Myers}, A.~D., {Barkhouse}, W.~A., {Mulchaey}, J.~S.,
  {Bennert}, V.~N., {Cox}, T.~J., {Aldcroft}, T.~L., Feb. 2010. {SDSS
  J1254+0846: A Binary Quasar Caught in the Act of Merging}. \apj 710,
  1578--1588.

\bibitem[{{Greene} and {Ho}(2004)}]{Greene_Ho_2004}
{Greene}, J.~E., {Ho}, L.~C., Aug 2004. {Active Galactic Nuclei with Candidate
  Intermediate-Mass Black Holes}. \apj 610~(2), 722--736.

\bibitem[{{Greene} and {Ho}(2007)}]{Greene_Ho_2007}
{Greene}, J.~E., {Ho}, L.~C., Nov 2007. {A New Sample of Low-Mass Black Holes
  in Active Galaxies}. \apj 670~(1), 92--104.

\bibitem[{{Greene} et~al.(2017){Greene}, {Kelly}, {Stansberry}, {Leisenring},
  {Egami}, {Schlawin}, {Chu}, {Hodapp}, and {Rieke}}]{2017JATIS...3c5001G}
{Greene}, T.~P., {Kelly}, D.~M., {Stansberry}, J., {Leisenring}, J., {Egami},
  E., {Schlawin}, E., {Chu}, L., {Hodapp}, K.~W., {Rieke}, M., Jul. 2017.
  {{$\lambda$} = 2.4 to 5 {$\mu$}m spectroscopy with the James Webb Space
  Telescope NIRCam instrument}. Journal of Astronomical Telescopes,
  Instruments, and Systems 3~(3), 035001.

\bibitem[{{Grier} et~al.(2013){Grier}, {Peterson}, {Horne}, {Bentz}, {Pogge},
  {Denney}, {De Rosa}, {Martini}, {Kochanek}, {Zu}, {Shappee}, {Siverd},
  {Beatty}, {Sergeev}, {Kaspi}, {Araya Salvo}, {Bird}, {Bord}, {Borman}, {Che},
  {Chen}, {Cohen}, {Dietrich}, {Doroshenko}, {Efimov}, {Free}, {Ginsburg},
  {Henderson}, {King}, {Mogren}, {Molina}, {Mosquera}, {Nazarov}, {Okhmat},
  {Pejcha}, {Rafter}, {Shields}, {Skowron}, {Szczygiel}, {Valluri}, and {van
  Saders}}]{Grier2013}
{Grier}, C.~J., {Peterson}, B.~M., {Horne}, K., {Bentz}, M.~C., {Pogge}, R.~W.,
  {Denney}, K.~D., {De Rosa}, G., {Martini}, P., {Kochanek}, C.~S., {Zu}, Y.,
  {Shappee}, B., {Siverd}, R., {Beatty}, T.~G., {Sergeev}, S.~G., {Kaspi}, S.,
  {Araya Salvo}, C., {Bird}, J.~C., {Bord}, D.~J., {Borman}, G.~A., {Che}, X.,
  {Chen}, C., {Cohen}, S.~A., {Dietrich}, M., {Doroshenko}, V.~T., {Efimov},
  Y.~S., {Free}, N., {Ginsburg}, I., {Henderson}, C.~B., {King}, A.~L.,
  {Mogren}, K., {Molina}, M., {Mosquera}, A.~M., {Nazarov}, S.~V., {Okhmat},
  D.~N., {Pejcha}, O., {Rafter}, S., {Shields}, J.~C., {Skowron}, J.,
  {Szczygiel}, D.~M., {Valluri}, M., {van Saders}, J.~L., Feb. 2013. {The
  Structure of the Broad-line Region in Active Galactic Nuclei. I.
  Reconstructed Velocity-delay Maps}. \apj 764, 47.

\bibitem[{{Gross} et~al.(2019){Gross}, {Fu}, {Myers}, {Wrobel}, and
  {Djorgovski}}]{gross2019}
{Gross}, A.~C., {Fu}, H., {Myers}, A.~D., {Wrobel}, J.~M., {Djorgovski}, S.~G.,
  Sep 2019. {X-Ray Properties of Radio-selected Dual Active Galactic Nuclei}.
  \apj 883~(1), 50.

\bibitem[{{Guainazzi} et~al.(2005){Guainazzi}, {Piconcelli},
  {Jim{\'e}nez-Bail{\'o}n}, and {Matt}}]{Guainazzi2005}
{Guainazzi}, M., {Piconcelli}, E., {Jim{\'e}nez-Bail{\'o}n}, E., {Matt}, G.,
  Jan. 2005. {The early stage of a cosmic collision? XMM-Newton unveils two
  obscured AGN in the galaxy pair ESO509-IG066}. \aap 429, L9--L12.

\bibitem[{{Guedes} et~al.(2011){Guedes}, {Callegari}, {Madau}, and
  {Mayer}}]{Guedes_et_al_2011}
{Guedes}, J., {Callegari}, S., {Madau}, P., {Mayer}, L., Dec. 2011. {Forming
  Realistic Late-type Spirals in a {$\Lambda$}CDM Universe: The Eris
  Simulation}. \apj 742, 76.

\bibitem[{{G{\"u}ltekin} and {Miller}(2012)}]{2012ApJ...761...90G}
{G{\"u}ltekin}, K., {Miller}, J.~M., Dec 2012. {Observable Consequences of
  Merger-driven Gaps and Holes in Black Hole Accretion Disks}. \apj 761~(2),
  90.

\bibitem[{{G{\"u}nther} and {Kley}(2002)}]{gunther02}
{G{\"u}nther}, R., {Kley}, W., May 2002. {Circumbinary disk evolution}. \aap
  387, 550--559.

\bibitem[{{Guo} et~al.(2019){Guo}, {Liu}, {Shen}, {Loeb}, {Monroe}, and
  {Prochaska}}]{Guo2019}
{Guo}, H., {Liu}, X., {Shen}, Y., {Loeb}, A., {Monroe}, T., {Prochaska}, J.~X.,
  Jan. 2019. {Constraining sub-parsec binary supermassive black holes in
  quasars with multi-epoch spectroscopy - III. Candidates from continued radial
  velocity tests}. \mnras 482, 3288--3307.

\bibitem[{{Haiman}(2017)}]{Haiman2017}
{Haiman}, Z., Jul. 2017. {Electromagnetic chirp of a compact binary black hole:
  A phase template for the gravitational wave inspiral}. \prd 96~(2), 023004.

\bibitem[{{Haiman} et~al.(2009){Haiman}, {Kocsis}, and {Menou}}]{haiman09}
{Haiman}, Z., {Kocsis}, B., {Menou}, K., Aug. 2009. {The Population of
  Viscosity- and Gravitational Wave-driven Supermassive Black Hole Binaries
  Among Luminous Active Galactic Nuclei}. \apj 700, 1952--1969.

\bibitem[{{Hainline} et~al.(2016){Hainline}, {Reines}, {Greene}, and
  {Stern}}]{2016ApJ...832..119H}
{Hainline}, K.~N., {Reines}, A.~E., {Greene}, J.~E., {Stern}, D., Dec 2016.
  {Mid-infrared Colors of Dwarf Galaxies: Young Starbursts Mimicking Active
  Galactic Nuclei}. \apj 832~(2), 119.

\bibitem[{{Halpern} and {Eracleous}(2000)}]{Halpern2000}
{Halpern}, J.~P., {Eracleous}, M., Mar. 2000. {The End of the Lines for OX 169:
  No Binary Broad-Line Region}. \apj 531, 647--653.

\bibitem[{{Hardee} et~al.(1994){Hardee}, {Cooper}, and {Clarke}}]{hardee1994}
{Hardee}, P.~E., {Cooper}, M.~A., {Clarke}, D.~A., Mar. 1994. {On jet response
  to a driving frequency and the jets in 3C 449}. \apj 424, 126--137.

\bibitem[{{Hayasaki}(2009)}]{hayasaki09}
{Hayasaki}, K., Feb. 2009. {A New Mechanism for Massive Binary Black-Hole
  Evolution}. \pasj 61, 65--.

\bibitem[{{Hayasaki} et~al.(2007{\natexlab{a}}){Hayasaki}, {Mineshige}, and
  {Sudou}}]{hayasaki07}
{Hayasaki}, K., {Mineshige}, S., {Sudou}, H., Apr. 2007{\natexlab{a}}. {Binary
  Black Hole Accretion Flows in Merged Galactic Nuclei}. \pasj 59, 427--441.

\bibitem[{{Hayasaki} et~al.(2007{\natexlab{b}}){Hayasaki}, {Mineshige}, and
  {Sudou}}]{Hayasaki2007}
{Hayasaki}, K., {Mineshige}, S., {Sudou}, H., Apr. 2007{\natexlab{b}}. {Binary
  Black Hole Accretion Flows in Merged Galactic Nuclei}. \pasj 59, 427--441.

\bibitem[{{Hayward} et~al.(2014){Hayward}, {Torrey}, {Springel}, {Hernquist},
  and {Vogelsberger}}]{Hayward_et_al_2014}
{Hayward}, C.~C., {Torrey}, P., {Springel}, V., {Hernquist}, L.,
  {Vogelsberger}, M., Aug. 2014. {Galaxy mergers on a moving mesh: a comparison
  with smoothed particle hydrodynamics}. \mnras 442, 1992--2016.

\bibitem[{{Hennawi} et~al.(2010){Hennawi}, {Myers}, {Shen}, {Strauss},
  {Djorgovski}, {Fan}, {Glikman}, {Mahabal}, {Martin}, {Richards}, {Schneider},
  and {Shankar}}]{hennawi2010}
{Hennawi}, J.~F., {Myers}, A.~D., {Shen}, Y., {Strauss}, M.~A., {Djorgovski},
  S.~G., {Fan}, X., {Glikman}, E., {Mahabal}, A., {Martin}, C.~L., {Richards},
  G.~T., {Schneider}, D.~P., {Shankar}, F., Aug. 2010. {Binary Quasars at High
  Redshift. I. 24 New Quasar Pairs at z \~{} 3-4}. \apj 719, 1672--1692.

\bibitem[{{Hennawi} and {Prochaska}(2013)}]{Hennawi2013}
{Hennawi}, J.~F., {Prochaska}, J.~X., Mar. 2013. {Quasars Probing Quasars. IV.
  Joint Constraints on the Circumgalactic Medium from Absorption and Emission}.
  \apj 766, 58.

\bibitem[{{Hennawi} et~al.(2015){Hennawi}, {Prochaska}, {Cantalupo}, and
  {Arrigoni-Battaia}}]{Hennawi2015}
{Hennawi}, J.~F., {Prochaska}, J.~X., {Cantalupo}, S., {Arrigoni-Battaia}, F.,
  May 2015. {Quasar quartet embedded in giant nebula reveals rare massive
  structure in distant universe}. Science 348, 779--783.

\bibitem[{{Hennawi} et~al.(2006){Hennawi}, {Strauss}, {Oguri}, {Inada},
  {Richards}, {Pindor}, {Schneider}, {Becker}, {Gregg}, {Hall}, {Johnston},
  {Fan}, {Burles}, {Schlegel}, {Gunn}, {Lupton}, {Bahcall}, {Brunner}, and
  {Brinkmann}}]{hennawi2006}
{Hennawi}, J.~F., {Strauss}, M.~A., {Oguri}, M., {Inada}, N., {Richards},
  G.~T., {Pindor}, B., {Schneider}, D.~P., {Becker}, R.~H., {Gregg}, M.~D.,
  {Hall}, P.~B., {Johnston}, D.~E., {Fan}, X., {Burles}, S., {Schlegel}, D.~J.,
  {Gunn}, J.~E., {Lupton}, R.~H., {Bahcall}, N.~A., {Brunner}, R.~J.,
  {Brinkmann}, J., Jan. 2006. {Binary Quasars in the Sloan Digital Sky Survey:
  Evidence for Excess Clustering on Small Scales}. \aj 131, 1--23.

\bibitem[{{Hernquist}(1989)}]{Hernquist_1989}
{Hernquist}, L., Aug. 1989. {Tidal triggering of starbursts and nuclear
  activity in galaxies}. \nat 340, 687--691.

\bibitem[{{Hernquist}(1990)}]{Hernquist_1990}
{Hernquist}, L., Jun. 1990. {An analytical model for spherical galaxies and
  bulges}. \apj 356, 359--364.

\bibitem[{{Herrera Ruiz} et~al.(2017){Herrera Ruiz}, {Middelberg}, {Deller},
  {Norris}, {Best}, {Brisken}, {Schinnerer}, {Smol{\v c}i{\'c}}, {Delvecchio},
  {Momjian}, {Bomans}, {Scoville}, and {Carilli}}]{herreraruiz2017}
{Herrera Ruiz}, N., {Middelberg}, E., {Deller}, A., {Norris}, R.~P., {Best},
  P.~N., {Brisken}, W., {Schinnerer}, E., {Smol{\v c}i{\'c}}, V., {Delvecchio},
  I., {Momjian}, E., {Bomans}, D., {Scoville}, N.~Z., {Carilli}, C., Nov. 2017.
  {The faint radio sky: VLBA observations of the COSMOS field}. Astronomy \&
  Astrophysics 607, A132.

\bibitem[{{Herrera Ruiz} et~al.(2016){Herrera Ruiz}, {Middelberg}, {Norris},
  and {Maini}}]{herreraruiz2016}
{Herrera Ruiz}, N., {Middelberg}, E., {Norris}, R.~P., {Maini}, A., May 2016.
  {Unveiling the origin of the radio emission in radio-quiet quasars}.
  Astronomy \& Astrophysics 589, L2.

\bibitem[{{Hirschmann} et~al.(2014){Hirschmann}, {Dolag}, {Saro}, {Bachmann},
  {Borgani}, and {Burkert}}]{Hirschmann_et_al_2014}
{Hirschmann}, M., {Dolag}, K., {Saro}, A., {Bachmann}, L., {Borgani}, S.,
  {Burkert}, A., Aug. 2014. {Cosmological simulations of black hole growth: AGN
  luminosities and downsizing}. \mnras 442, 2304--2324.

\bibitem[{{Hoffman} and {Loeb}(2007)}]{Hoffman2007}
{Hoffman}, L., {Loeb}, A., May 2007. {Dynamics of triple black hole systems in
  hierarchically merging massive galaxies}. \mnras 377~(3), 957--976.

\bibitem[{{Holley-Bockelmann} and {Khan}(2015)}]{holley-bockelmann15}
{Holley-Bockelmann}, K., {Khan}, F.~M., Sep. 2015. {Galaxy Rotation and Rapid
  Supermassive Binary Coalescence}. \apj 810, 139.

\bibitem[{{Holmberg}(1941)}]{Holmberg_1941}
{Holmberg}, E., Nov. 1941. {On the Clustering Tendencies among the Nebulae. II.
  a Study of Encounters Between Laboratory Models of Stellar Systems by a New
  Integration Procedure.} \apj 94, 385.

\bibitem[{{Hopkins} et~al.(2006){Hopkins}, {Hernquist}, {Cox}, {Di Matteo},
  {Robertson}, and {Springel}}]{Hopkins_et_al_2006}
{Hopkins}, P.~F., {Hernquist}, L., {Cox}, T.~J., {Di Matteo}, T., {Robertson},
  B., {Springel}, V., Mar. 2006. {A Unified, Merger-driven Model of the Origin
  of Starbursts, Quasars, the Cosmic X-Ray Background, Supermassive Black
  Holes, and Galaxy Spheroids}. \apjs 163, 1--49.

\bibitem[{{Hopkins} and {Quataert}(2010)}]{Hopkins_Quataert_2010}
{Hopkins}, P.~F., {Quataert}, E., Sep. 2010. {How do massive black holes get
  their gas?} \mnras 407, 1529--1564.

\bibitem[{{Hopkins} and {Quataert}(2011)}]{Hopkins_Quataert_2011}
{Hopkins}, P.~F., {Quataert}, E., Aug. 2011. {An analytic model of angular
  momentum transport by gravitational torques: from galaxies to massive black
  holes}. \mnras 415, 1027--1050.

\bibitem[{{Hopkins} et~al.(2007){Hopkins}, {Richards}, and
  {Hernquist}}]{Hopkins_et_al_2007}
{Hopkins}, P.~F., {Richards}, G.~T., {Hernquist}, L., Jan. 2007. {An
  Observational Determination of the Bolometric Quasar Luminosity Function}.
  \apj 654, 731--753.

\bibitem[{{Hou} et~al.(2019){Hou}, {Liu}, {Guo}, {Li}, {Shen}, and
  {Green}}]{hou2019}
{Hou}, M., {Liu}, X., {Guo}, H., {Li}, Z., {Shen}, Y., {Green}, P.~J., Sep
  2019. {Active Galactic Nucleus Pairs from the Sloan Digital Sky Survey. III.
  Chandra X-Ray Observations Unveil Obscured Double Nuclei}. \apj 882~(1), 41.

\bibitem[{{Hudson} et~al.(2006){Hudson}, {Reiprich}, {Clarke}, and
  {Sarazin}}]{hudson2006}
{Hudson}, D.~S., {Reiprich}, T.~H., {Clarke}, T.~E., {Sarazin}, C.~L., Jul.
  2006. {X-ray detection of the proto supermassive binary black hole at the
  centre of Abell 400}. Astronomy \& Astrophysics 453, 433--446.

\bibitem[{{Husemann} et~al.(2018){Husemann}, {Worseck}, {Arrigoni Battaia}, and
  {Shanks}}]{Husemann2018}
{Husemann}, B., {Worseck}, G., {Arrigoni Battaia}, F., {Shanks}, T., Feb. 2018.
  {Discovery of a dual AGN at z {$\sim$} 3.3 with 20 kpc separation}. \aap 610,
  L7.

\bibitem[{{Imanishi} and {Saito}(2014)}]{Imanishi_Saito2014}
{Imanishi}, M., {Saito}, Y., Jan. 2014. {Subaru Adaptive-optics
  High-spatial-resolution Infrared K- and L'-band Imaging Search for Deeply
  Buried Dual AGNs in Merging Galaxies}. \apj 780, 106.

\bibitem[{Ivanov et~al.(1999)Ivanov, Papaloizou, and Polnarev}]{Ivanov99}
Ivanov, P.~B., Papaloizou, J. C.~B., Polnarev, A.~G., Jul 1999. The evolution
  of a supermassive binary caused by an accretion disc. MNRAS 307, 79.

\bibitem[{{Ivezi{\'c}} et~al.(2008){Ivezi{\'c}}, {Kahn}, {Tyson}, {Abel},
  {Acosta}, {Allsman}, {Alonso}, {AlSayyad}, {Anderson}, {Andrew}, and
  et~al.}]{ivezic08}
{Ivezi{\'c}}, {\v Z}., {Kahn}, S.~M., {Tyson}, J.~A., {Abel}, B., {Acosta}, E.,
  {Allsman}, R., {Alonso}, D., {AlSayyad}, Y., {Anderson}, S.~F., {Andrew}, J.,
  et~al., May 2008. {LSST: from Science Drivers to Reference Design and
  Anticipated Data Products}. ArXiv e-prints.

\bibitem[{{Ivezi{\'c}} et~al.(2002){Ivezi{\'c}}, {Menou}, {Knapp}, {Strauss},
  {Lupton}, {Vanden Berk}, {Richards}, {Tremonti}, {Weinstein}, {Anderson},
  {Bahcall}, {Becker}, {Bernardi}, {Blanton}, {Eisenstein}, {Fan},
  {Finkbeiner}, {Finlator}, {Frieman}, {Gunn}, {Hall}, {Kim}, {Kinkhabwala},
  {Narayanan}, {Rockosi}, {Schlegel}, {Schneider}, {Strateva}, {SubbaRao},
  {Thakar}, {Voges}, {White}, {Yanny}, {Brinkmann}, {Doi}, {Fukugita},
  {Hennessy}, {Munn}, {Nichol}, and {York}}]{ivezic2002}
{Ivezi{\'c}}, {\v Z}., {Menou}, K., {Knapp}, G.~R., {Strauss}, M.~A., {Lupton},
  R.~H., {Vanden Berk}, D.~E., {Richards}, G.~T., {Tremonti}, C., {Weinstein},
  M.~A., {Anderson}, S., {Bahcall}, N.~A., {Becker}, R.~H., {Bernardi}, M.,
  {Blanton}, M., {Eisenstein}, D., {Fan}, X., {Finkbeiner}, D., {Finlator}, K.,
  {Frieman}, J., {Gunn}, J.~E., {Hall}, P.~B., {Kim}, R.~S.~J., {Kinkhabwala},
  A., {Narayanan}, V.~K., {Rockosi}, C.~M., {Schlegel}, D., {Schneider}, D.~P.,
  {Strateva}, I., {SubbaRao}, M., {Thakar}, A.~R., {Voges}, W., {White}, R.~L.,
  {Yanny}, B., {Brinkmann}, J., {Doi}, M., {Fukugita}, M., {Hennessy}, G.~S.,
  {Munn}, J.~A., {Nichol}, R.~C., {York}, D.~G., Nov. 2002. {Optical and Radio
  Properties of Extragalactic Sources Observed by the FIRST Survey and the
  Sloan Digital Sky Survey}. Astronomical Journal 124, 2364--2400.

\bibitem[{{Iwasawa} et~al.(2011){Iwasawa}, {Sanders}, {Teng}, {U}, {Armus},
  {Evans}, {Howell}, {Komossa}, {Mazzarella}, {Petric}, {Surace}, {Vavilkin},
  {Veilleux}, and {Trentham}}]{iwasawa11}
{Iwasawa}, K., {Sanders}, D.~B., {Teng}, S.~H., {U}, V., {Armus}, L., {Evans},
  A.~S., {Howell}, J.~H., {Komossa}, S., {Mazzarella}, J.~M., {Petric}, A.~O.,
  {Surace}, J.~A., {Vavilkin}, T., {Veilleux}, S., {Trentham}, N., May 2011.
  {C-GOALS: Chandra observations of a complete sample of luminous infrared
  galaxies from the IRAS Revised Bright Galaxy Survey}. \aap 529, A106.

\bibitem[{{Iwasawa} et~al.(2018){Iwasawa}, {U}, {Mazzarella}, {Medling},
  {Sanders}, and {Evans}}]{Iwasawa2018}
{Iwasawa}, K., {U}, V., {Mazzarella}, J.~M., {Medling}, A.~M., {Sanders},
  D.~B., {Evans}, A.~S., Apr. 2018. {Testing a double AGN hypothesis for Mrk
  273}. \aap 611, A71.

\bibitem[{{Izumi} et~al.(2016){Izumi}, {Kawakatu}, and
  {Kohno}}]{Izumi_et_al_2016}
{Izumi}, T., {Kawakatu}, N., {Kohno}, K., Aug. 2016. {Do Circumnuclear Dense
  Gas Disks Drive Mass Accretion onto Supermassive Black Holes?} \apj 827, 81.

\bibitem[{{Janssen} et~al.(2015){Janssen}, {Hobbs}, {McLaughlin}, {Bassa},
  {Deller}, {Kramer}, {Lee}, {Mingarelli}, {Rosado}, {Sanidas}, {Sesana},
  {Shao}, {Stairs}, {Stappers}, and {Verbiest}}]{2015aska.confE..37J}
{Janssen}, G., {Hobbs}, G., {McLaughlin}, M., {Bassa}, C., {Deller}, A.,
  {Kramer}, M., {Lee}, K., {Mingarelli}, C., {Rosado}, P., {Sanidas}, S.,
  {Sesana}, A., {Shao}, L., {Stairs}, I., {Stappers}, B., {Verbiest}, J.~P.~W.,
  Apr. 2015. {Gravitational Wave Astronomy with the SKA}. Advancing
  Astrophysics with the Square Kilometre Array (AASKA14), 37.

\bibitem[{{Jarrett} et~al.(2011){Jarrett}, {Cohen}, {Masci}, {Wright}, {Stern},
  {Benford}, {Blain}, {Carey}, {Cutri}, {Eisenhardt}, {Lonsdale}, {Mainzer},
  {Marsh}, {Padgett}, {Petty}, {Ressler}, {Skrutskie}, {Stanford}, {Surace},
  {Tsai}, {Wheelock}, and {Yan}}]{Jarrett_et_al_2011}
{Jarrett}, T.~H., {Cohen}, M., {Masci}, F., {Wright}, E., {Stern}, D.,
  {Benford}, D., {Blain}, A., {Carey}, S., {Cutri}, R.~M., {Eisenhardt}, P.,
  {Lonsdale}, C., {Mainzer}, A., {Marsh}, K., {Padgett}, D., {Petty}, S.,
  {Ressler}, M., {Skrutskie}, M., {Stanford}, S., {Surace}, J., {Tsai}, C.~W.,
  {Wheelock}, S., {Yan}, D.~L., Jul. 2011. {The Spitzer-WISE Survey of the
  Ecliptic Poles}. \apj 735, 112.

\bibitem[{{Johansson} et~al.(2009){Johansson}, {Burkert}, and
  {Naab}}]{Johansson_et_al_2009}
{Johansson}, P.~H., {Burkert}, A., {Naab}, T., Dec. 2009. {The Evolution of
  Black Hole Scaling Relations in Galaxy Mergers}. \apjl 707, L184--L189.

\bibitem[{{Johnston} et~al.(2007){Johnston}, {Bailes}, {Bartel}, {Baugh},
  {Bietenholz}, {Blake}, {Braun}, {Brown}, {Chatterjee}, {Darling}, {Deller},
  {Dodson}, {Edwards}, {Ekers}, {Ellingsen}, {Feain}, {Gaensler}, {Haverkorn},
  {Hobbs}, {Hopkins}, {Jackson}, {James}, {Joncas}, {Kaspi}, {Kilborn},
  {Koribalski}, {Kothes}, {Landecker}, {Lenc}, {Lovell}, {Macquart},
  {Manchester}, {Matthews}, {McClure-Griffiths}, {Norris}, {Pen}, {Phillips},
  {Power}, {Protheroe}, {Sadler}, {Schmidt}, {Stairs}, {Staveley-Smith},
  {Stil}, {Taylor}, {Tingay}, {Tzioumis}, {Walker}, {Wall}, and
  {Wolleben}}]{Johnston2007}
{Johnston}, S., {Bailes}, M., {Bartel}, N., {Baugh}, C., {Bietenholz}, M.,
  {Blake}, C., {Braun}, R., {Brown}, J., {Chatterjee}, S., {Darling}, J.,
  {Deller}, A., {Dodson}, R., {Edwards}, P.~G., {Ekers}, R., {Ellingsen}, S.,
  {Feain}, I., {Gaensler}, B.~M., {Haverkorn}, M., {Hobbs}, G., {Hopkins}, A.,
  {Jackson}, C., {James}, C., {Joncas}, G., {Kaspi}, V., {Kilborn}, V.,
  {Koribalski}, B., {Kothes}, R., {Landecker}, T.~L., {Lenc}, E., {Lovell}, J.,
  {Macquart}, J.-P., {Manchester}, R., {Matthews}, D., {McClure-Griffiths},
  N.~M., {Norris}, R., {Pen}, U.-L., {Phillips}, C., {Power}, C., {Protheroe},
  R., {Sadler}, E., {Schmidt}, B., {Stairs}, I., {Staveley-Smith}, L., {Stil},
  J., {Taylor}, R., {Tingay}, S., {Tzioumis}, A., {Walker}, M., {Wall}, J.,
  {Wolleben}, M., Dec. 2007. {Science with the Australian Square Kilometre
  Array Pathfinder}. \pasa 24, 174--188.

\bibitem[{{Ju} et~al.(2013){Ju}, {Greene}, {Rafikov}, {Bickerton}, and
  {Badenes}}]{ju13}
{Ju}, W., {Greene}, J.~E., {Rafikov}, R.~R., {Bickerton}, S.~J., {Badenes}, C.,
  Nov. 2013. {Search for Supermassive Black Hole Binaries in the Sloan Digital
  Sky Survey Spectroscopic Sample}. \apj 777, 44.

\bibitem[{{Jun} et~al.(2015){Jun}, {Stern}, {Graham}, {Djorgovski}, {Mainzer},
  {Cutri}, {Drake}, and {Mahabal}}]{Jun2015}
{Jun}, H.~D., {Stern}, D., {Graham}, M.~J., {Djorgovski}, S.~G., {Mainzer}, A.,
  {Cutri}, R.~M., {Drake}, A.~J., {Mahabal}, A.~A., Nov. 2015. {Infrared Time
  Lags for the Periodic Quasar PG 1302-102}. \apjl 814, L12.

\bibitem[{{Kaastra} and {Roos}(1992)}]{kaastra1992}
{Kaastra}, J.~S., {Roos}, N., Feb. 1992. {Massive Binary Black-Holes and
  Wiggling Jets}. \aap 254, 96.

\bibitem[{{Kaplan} et~al.(2011){Kaplan}, {O'Shaughnessy}, {Sesana}, and
  {Volonteri}}]{2011ApJ...734L..37K}
{Kaplan}, D.~L., {O'Shaughnessy}, R., {Sesana}, A., {Volonteri}, M., Jun. 2011.
  {Blindly Detecting Merging Supermassive Black Holes with Radio Surveys}.
  \apjl 734, L37.

\bibitem[{{Kauffmann} and {Haehnelt}(2000)}]{KauffmanHaehnelt2000}
{Kauffmann}, G., {Haehnelt}, M., Jan. 2000. {A unified model for the evolution
  of galaxies and quasars}. \mnras 311, 576--588.

\bibitem[{{Kayo} and {Oguri}(2012)}]{2012MNRAS.424.1363K}
{Kayo}, I., {Oguri}, M., Aug 2012. {Very small scale clustering of quasars from
  a complete quasar lens survey}. \mnras 424~(2), 1363--1371.

\bibitem[{{Kelley} et~al.(2019{\natexlab{a}}){Kelley}, {Charisi},
  {Burke-Spolaor}, {Simon}, {Blecha}, {Bogdanovic}, {Colpi}, {Comerford},
  {D'Orazio}, {Dotti}, {Eracleous}, {Graham}, {Greene}, {Haiman},
  {Holley-Bockelmann}, {Kara}, {Kelly}, {Komossa}, {Larson}, {Liu}, {Ma},
  {Noble}, {Paschalidis}, {Rafikov}, {Ravi}, {Runnoe}, {Sesana}, {Stern},
  {Strauss}, {U}, {Volonteri}, and {The Nanograv Collaboration}}]{Kelley2020}
{Kelley}, L., {Charisi}, M., {Burke-Spolaor}, S., {Simon}, J., {Blecha}, L.,
  {Bogdanovic}, T., {Colpi}, M., {Comerford}, J., {D'Orazio}, D., {Dotti}, M.,
  {Eracleous}, M., {Graham}, M., {Greene}, J., {Haiman}, Z.,
  {Holley-Bockelmann}, K., {Kara}, E., {Kelly}, B., {Komossa}, S., {Larson},
  S., {Liu}, X., {Ma}, C.~P., {Noble}, S., {Paschalidis}, V., {Rafikov}, R.,
  {Ravi}, V., {Runnoe}, J., {Sesana}, A., {Stern}, D., {Strauss}, M.~A., {U},
  V., {Volonteri}, M., {The Nanograv Collaboration}, May 2019{\natexlab{a}}.
  {Multi-Messenger Astrophysics With Pulsar Timing Arrays}. \baas 51~(3), 490.

\bibitem[{{Kelley} et~al.(2017){Kelley}, {Blecha}, and
  {Hernquist}}]{2017MNRAS.464.3131K}
{Kelley}, L.~Z., {Blecha}, L., {Hernquist}, L., Jan 2017. {Massive black hole
  binary mergers in dynamical galactic environments}. \mnras 464~(3),
  3131--3157.

\bibitem[{{Kelley} et~al.(2019{\natexlab{b}}){Kelley}, {Haiman}, {Sesana}, and
  {Hernquist}}]{2019MNRAS.485.1579K}
{Kelley}, L.~Z., {Haiman}, Z., {Sesana}, A., {Hernquist}, L., May
  2019{\natexlab{b}}. {Massive BH binaries as periodically variable AGN}.
  \mnras 485~(2), 1579--1594.

\bibitem[{{Kelley} et~al.(2019{\natexlab{c}}){Kelley}, {Haiman}, {Sesana}, and
  {Hernquist}}]{Kelley2019}
{Kelley}, L.~Z., {Haiman}, Z., {Sesana}, A., {Hernquist}, L., May
  2019{\natexlab{c}}. {Massive BH binaries as periodically variable AGN}.
  \mnras 485~(2), 1579--1594.

\bibitem[{{Kelly} et~al.(2017){Kelly}, {Baker}, {Etienne}, {Giacomazzo}, and
  {Schnittman}}]{Kelly2017}
{Kelly}, B.~J., {Baker}, J.~G., {Etienne}, Z.~B., {Giacomazzo}, B.,
  {Schnittman}, J., Dec 2017. {Prompt electromagnetic transients from binary
  black hole mergers}. \prd 96~(12), 123003.

\bibitem[{{Kewley} et~al.(2013){Kewley}, {Maier}, {Yabe}, {Ohta}, {Akiyama},
  {Dopita}, and {Yuan}}]{Kewley2013}
{Kewley}, L.~J., {Maier}, C., {Yabe}, K., {Ohta}, K., {Akiyama}, M., {Dopita},
  M.~A., {Yuan}, T., Sep. 2013. {The Cosmic BPT Diagram: Confronting Theory
  with Observations}. \apjl 774, L10.

\bibitem[{{Khan} et~al.(2012{\natexlab{a}}){Khan}, {Berentzen}, {Berczik},
  {Just}, {Mayer}, {Nitadori}, and {Callegari}}]{khan12}
{Khan}, F.~M., {Berentzen}, I., {Berczik}, P., {Just}, A., {Mayer}, L.,
  {Nitadori}, K., {Callegari}, S., Sep. 2012{\natexlab{a}}. {Formation and
  Hardening of Supermassive Black Hole Binaries in Minor Mergers of Disk
  Galaxies}. \apj 756, 30.

\bibitem[{{Khan} et~al.(2018){Khan}, {Capelo}, {Mayer}, and
  {Berczik}}]{Khan_et_al_2018}
{Khan}, F.~M., {Capelo}, P.~R., {Mayer}, L., {Berczik}, P., Dec 2018.
  {Dynamical Evolution and Merger Timescales of LISA Massive Black Hole
  Binaries in Disk Galaxy Mergers}. \apj 868~(2), 97.

\bibitem[{{Khan} et~al.(2013){Khan}, {Holley-Bockelmann}, {Berczik}, and
  {Just}}]{khan13}
{Khan}, F.~M., {Holley-Bockelmann}, K., {Berczik}, P., {Just}, A., Aug. 2013.
  {Supermassive Black Hole Binary Evolution in Axisymmetric Galaxies: The Final
  Parsec Problem is Not a Problem}. \apj 773, 100.

\bibitem[{{Khan} et~al.(2011){Khan}, {Just}, and {Merritt}}]{khan11}
{Khan}, F.~M., {Just}, A., {Merritt}, D., May 2011. {Efficient Merger of Binary
  Supermassive Black Holes in Merging Galaxies}. \apj 732, 89.

\bibitem[{{Khan} et~al.(2012{\natexlab{b}}){Khan}, {Preto}, {Berczik},
  {Berentzen}, {Just}, and {Spurzem}}]{khan12a}
{Khan}, F.~M., {Preto}, M., {Berczik}, P., {Berentzen}, I., {Just}, A.,
  {Spurzem}, R., Apr. 2012{\natexlab{b}}. {Mergers of Unequal-mass Galaxies:
  Supermassive Black Hole Binary Evolution and Structure of Merger Remnants}.
  \apj 749, 147.

\bibitem[{{Khandai} et~al.(2015){Khandai}, {Di Matteo}, {Croft}, {Wilkins},
  {Feng}, {Tucker}, {DeGraf}, and {Liu}}]{Khandai_et_al_2015}
{Khandai}, N., {Di Matteo}, T., {Croft}, R., {Wilkins}, S., {Feng}, Y.,
  {Tucker}, E., {DeGraf}, C., {Liu}, M.-S., Jun. 2015. {The MassiveBlack-II
  simulation: the evolution of haloes and galaxies to $z \tilde 0$}. \mnras
  450, 1349--1374.

\bibitem[{{Kharb} et~al.(2017){Kharb}, {Lal}, and {Merritt}}]{kharb2017}
{Kharb}, P., {Lal}, D.~V., {Merritt}, D., Oct. 2017. {A candidate sub-parsec
  binary black hole in the Seyfert galaxy NGC 7674}. Nature Astronomy 1,
  727--733.

\bibitem[{{Kim} et~al.(2013){Kim}, {Evans}, {Vavilkin}, {Armus}, {Mazzarella},
  {Sheth}, {Surace}, {Haan}, {Howell}, {D{\'{\i}}az-Santos}, {Petric},
  {Iwasawa}, {Privon}, and {Sanders}}]{2013ApJ...768..102K}
{Kim}, D.-C., {Evans}, A.~S., {Vavilkin}, T., {Armus}, L., {Mazzarella}, J.~M.,
  {Sheth}, K., {Surace}, J.~A., {Haan}, S., {Howell}, J.~H.,
  {D{\'{\i}}az-Santos}, T., {Petric}, A., {Iwasawa}, K., {Privon}, G.~C.,
  {Sanders}, D.~B., May 2013. {Hubble Space Telescope ACS Imaging of the GOALS
  Sample: Quantitative Structural Properties of Nearby Luminous Infrared
  Galaxies with L$_{IR}$ $>$ 10$^{11.4}$ L$_{\odot}$}. \apj 768, 102.

\bibitem[{{King} and {Pounds}(2015)}]{2015ARA&A..53..115K}
{King}, A., {Pounds}, K., Aug. 2015. {Powerful Outflows and Feedback from
  Active Galactic Nuclei}. \araa 53, 115--154.

\bibitem[{{Klein} et~al.(2016){Klein}, {Barausse}, {Sesana}, {Petiteau},
  {Berti}, {Babak}, {Gair}, {Aoudia}, {Hinder}, {Ohme}, and
  {Wardell}}]{2016PhRvD..93b4003K}
{Klein}, A., {Barausse}, E., {Sesana}, A., {Petiteau}, A., {Berti}, E.,
  {Babak}, S., {Gair}, J., {Aoudia}, S., {Hinder}, I., {Ohme}, F., {Wardell},
  B., Jan. 2016. {Science with the space-based interferometer eLISA:
  Supermassive black hole binaries}. \prd 93~(2), 024003.

\bibitem[{{Kocevski} et~al.(2015){Kocevski}, {Brightman}, {Nandra},
  {Koekemoer}, {Salvato}, {Aird}, {Bell}, {Hsu}, {Kartaltepe}, {Koo}, {Lotz},
  {McIntosh}, {Mozena}, {Rosario}, and {Trump}}]{kocevskietal15}
{Kocevski}, D.~D., {Brightman}, M., {Nandra}, K., {Koekemoer}, A.~M.,
  {Salvato}, M., {Aird}, J., {Bell}, E.~F., {Hsu}, L.-T., {Kartaltepe}, J.~S.,
  {Koo}, D.~C., {Lotz}, J.~M., {McIntosh}, D.~H., {Mozena}, M., {Rosario}, D.,
  {Trump}, J.~R., Dec. 2015. {Are Compton-thick AGNs the Missing Link between
  Mergers and Black Hole Growth?} \apj 814, 104.

\bibitem[{{Kochanek} et~al.(2017){Kochanek}, {Shappee}, {Stanek}, {Holoien},
  {Thompson}, {Prieto}, {Dong}, {Shields}, {Will}, {Britt}, {Perzanowski}, and
  {Pojma{\'n}ski}}]{ASASSN_2}
{Kochanek}, C.~S., {Shappee}, B.~J., {Stanek}, K.~Z., {Holoien}, T.~W.-S.,
  {Thompson}, T.~A., {Prieto}, J.~L., {Dong}, S., {Shields}, J.~V., {Will}, D.,
  {Britt}, C., {Perzanowski}, D., {Pojma{\'n}ski}, G., Oct. 2017. {The All-Sky
  Automated Survey for Supernovae (ASAS-SN) Light Curve Server v1.0}. \pasp
  129~(10), 104502.

\bibitem[{{Kocsis} et~al.(2012{\natexlab{a}}){Kocsis}, {Haiman}, and
  {Loeb}}]{kocsis12b}
{Kocsis}, B., {Haiman}, Z., {Loeb}, A., Dec. 2012{\natexlab{a}}. {Gas pile-up,
  gap overflow and Type 1.5 migration in circumbinary discs: application to
  supermassive black hole binaries}. \mnras 427, 2680--2700.

\bibitem[{{Kocsis} et~al.(2012{\natexlab{b}}){Kocsis}, {Haiman}, and
  {Loeb}}]{kocsis12a}
{Kocsis}, B., {Haiman}, Z., {Loeb}, A., Dec. 2012{\natexlab{b}}. {Gas pile-up,
  gap overflow and Type 1.5 migration in circumbinary discs: general theory}.
  \mnras 427, 2660--2679.

\bibitem[{{Kocsis} et~al.(2008){Kocsis}, {Haiman}, and
  {Menou}}]{KocsisHaimanMenou2008}
{Kocsis}, B., {Haiman}, Z., {Menou}, K., Sep. 2008. {Premerger Localization of
  Gravitational Wave Standard Sirens with LISA: Triggered Search for an
  Electromagnetic Counterpart}. \apj 684, 870--887.

\bibitem[{{Kollatschny} et~al.(2019){Kollatschny}, {Weilbacher}, {Ochmann},
  {Chelouche}, {Monreal-Ibero}, {Bacon}, and {Contini}}]{Kollatschny19}
{Kollatschny}, W., {Weilbacher}, P.~M., {Ochmann}, M.~W., {Chelouche}, D.,
  {Monreal-Ibero}, A., {Bacon}, R., {Contini}, T., Oct 2019. {NGC6240: A triple
  nucleus system in the advanced/final state of merging}. arXiv e-prints,
  arXiv:1910.12813.

\bibitem[{{Kollmeier} et~al.(2017){Kollmeier}, {Zasowski}, {Rix}, {Johns},
  {Anderson}, {Drory}, {Johnson}, {Pogge}, {Bird}, {Blanc}, {Brownstein},
  {Crane}, {De Lee}, {Klaene}, {Kreckel}, {MacDonald}, {Merloni}, {Ness},
  {O'Brien}, {Sanchez-Gallego}, {Sayres}, {Shen}, {Thakar}, {Tkachenko},
  {Aerts}, {Blanton}, {Eisenstein}, {Holtzman}, {Maoz}, {Nandra}, {Rockosi},
  {Weinberg}, {Bovy}, {Casey}, {Chaname}, {Clerc}, {Conroy}, {Eracleous},
  {G{\"a}nsicke}, {Hekker}, {Horne}, {Kauffmann}, {McQuinn}, {Pellegrini},
  {Schinnerer}, {Schlafly}, {Schwope}, {Seibert}, {Teske}, and {van
  Saders}}]{2017arXiv171103234K}
{Kollmeier}, J.~A., {Zasowski}, G., {Rix}, H.-W., {Johns}, M., {Anderson},
  S.~F., {Drory}, N., {Johnson}, J.~A., {Pogge}, R.~W., {Bird}, J.~C., {Blanc},
  G.~A., {Brownstein}, J.~R., {Crane}, J.~D., {De Lee}, N.~M., {Klaene}, M.~A.,
  {Kreckel}, K., {MacDonald}, N., {Merloni}, A., {Ness}, M.~K., {O'Brien}, T.,
  {Sanchez-Gallego}, J.~R., {Sayres}, C.~C., {Shen}, Y., {Thakar}, A.~R.,
  {Tkachenko}, A., {Aerts}, C., {Blanton}, M.~R., {Eisenstein}, D.~J.,
  {Holtzman}, J.~A., {Maoz}, D., {Nandra}, K., {Rockosi}, C., {Weinberg},
  D.~H., {Bovy}, J., {Casey}, A.~R., {Chaname}, J., {Clerc}, N., {Conroy}, C.,
  {Eracleous}, M., {G{\"a}nsicke}, B.~T., {Hekker}, S., {Horne}, K.,
  {Kauffmann}, J., {McQuinn}, K. B.~W., {Pellegrini}, E.~W., {Schinnerer}, E.,
  {Schlafly}, E.~F., {Schwope}, A.~D., {Seibert}, M., {Teske}, J.~K., {van
  Saders}, J.~L., Nov 2017. {SDSS-V: Pioneering Panoptic Spectroscopy}. arXiv
  e-prints, arXiv:1711.03234.

\bibitem[{{Komberg}(1968)}]{Komberg1968}
{Komberg}, B.~V., Feb. 1968. {A Binary System as a Quasar Model.} \sovast 11,
  727.

\bibitem[{{Komossa} et~al.(2003){Komossa}, {Burwitz}, {Hasinger}, {Predehl},
  {Kaastra}, and {Ikebe}}]{Komossa2003}
{Komossa}, S., {Burwitz}, V., {Hasinger}, G., {Predehl}, P., {Kaastra}, J.~S.,
  {Ikebe}, Y., Jan. 2003. {Discovery of a Binary Active Galactic Nucleus in the
  Ultraluminous Infrared Galaxy NGC 6240 Using Chandra}. \apjl 582, L15--L19.

\bibitem[{{Komossa} and {Zensus}(2016)}]{komossa_zensus2016}
{Komossa}, S., {Zensus}, J.~A., Feb. 2016. {Compact object mergers:
  observations of supermassive binary black holes and stellar tidal disruption
  events}. In: {Meiron}, Y., {Li}, S., {Liu}, F.-K., {Spurzem}, R. (Eds.), Star
  Clusters and Black Holes in Galaxies across Cosmic Time. Vol. 312 of IAU
  Symposium. pp. 13--25.

\bibitem[{{Koratkar} and {Gaskell}(1991)}]{Koratkar1991}
{Koratkar}, A.~P., {Gaskell}, C.~M., Mar. 1991. {Structure and kinematics of
  the broad-line regions in active galaxies from IUE variability data}. \apjs
  75, 719--750.

\bibitem[{{Kormendy} and {Ho}(2013)}]{Kormendy13}
{Kormendy}, J., {Ho}, L.~C., Aug. 2013. {Coevolution (Or Not) of Supermassive
  Black Holes and Host Galaxies}. \araa 51, 511--653.

\bibitem[{Kosec et~al.(2017)Kosec, Brightman, Stern, M{\"u}ller-S{\'a}nchez,
  Koss, Oh, Assef, Gandhi, Harrison, Jun, and et~al.}]{KosecBS17}
Kosec, P., Brightman, M., Stern, D., M{\"u}ller-S{\'a}nchez, F., Koss, M., Oh,
  K., Assef, R.~J., Gandhi, P., Harrison, F.~A., Jun, H., et~al., Nov 2017.
  Investigating the evolution of the dual agn system eso 509-ig066. The
  Astrophysical Journal 850~(2), 168.
\newline\urlprefix\url{http://dx.doi.org/10.3847/1538-4357/aa932e}

\bibitem[{{Koss} et~al.(2011){Koss}, {Mushotzky}, {Treister}, {Veilleux},
  {Vasudevan}, {Miller}, {Sanders}, {Schawinski}, and {Trippe}}]{Koss2011}
{Koss}, M., {Mushotzky}, R., {Treister}, E., {Veilleux}, S., {Vasudevan}, R.,
  {Miller}, N., {Sanders}, D.~B., {Schawinski}, K., {Trippe}, M., Jul. 2011.
  {Chandra Discovery of a Binary Active Galactic Nucleus in Mrk 739}. \apjl
  735, L42.

\bibitem[{{Koss} et~al.(2012){Koss}, {Mushotzky}, {Treister}, {Veilleux},
  {Vasudevan}, and {Trippe}}]{Koss_et_al_2012}
{Koss}, M., {Mushotzky}, R., {Treister}, E., {Veilleux}, S., {Vasudevan}, R.,
  {Trippe}, M., Feb. 2012. {Understanding Dual Active Galactic Nucleus
  Activation in the nearby Universe}. \apjl 746, L22.

\bibitem[{{Koss} et~al.(2010){Koss}, {Mushotzky}, {Veilleux}, and
  {Winter}}]{Koss_et_al_2010}
{Koss}, M., {Mushotzky}, R., {Veilleux}, S., {Winter}, L., Jun. 2010. {Merging
  and Clustering of the Swift BAT AGN Sample}. \apjl 716, L125--L130.

\bibitem[{{Koss} et~al.(2018){Koss}, {Blecha}, {Bernhard}, {Hung}, {Lu},
  {Trakthenbrot}, {Treister}, {Weigel}, {Sartori}, {Mushotzky}, {Schawinski},
  {Ricci}, {Veilleux}, and {Sanders}}]{kossetal2018}
{Koss}, M.~J., {Blecha}, L., {Bernhard}, P., {Hung}, C.-L., {Lu}, J.~R.,
  {Trakthenbrot}, B., {Treister}, E., {Weigel}, A., {Sartori}, L.~F.,
  {Mushotzky}, R., {Schawinski}, K., {Ricci}, C., {Veilleux}, S., {Sanders},
  D.~B., Nov. 2018. {A population of luminous accreting black holes with hidden
  mergers}. \nat 563, 214--216.

\bibitem[{{Koss} et~al.(2015){Koss}, {Romero-Ca{\~n}izales}, {Baronchelli},
  {Teng}, {Balokovi{\'c}}, {Puccetti}, {Bauer}, {Ar{\'e}valo}, {Assef},
  {Ballantyne}, {Brandt}, {Brightman}, {Comastri}, {Gandhi}, {Harrison}, {Luo},
  {Schawinski}, {Stern}, and {Treister}}]{Koss2015}
{Koss}, M.~J., {Romero-Ca{\~n}izales}, C., {Baronchelli}, L., {Teng}, S.~H.,
  {Balokovi{\'c}}, M., {Puccetti}, S., {Bauer}, F.~E., {Ar{\'e}valo}, P.,
  {Assef}, R., {Ballantyne}, D.~R., {Brandt}, W.~N., {Brightman}, M.,
  {Comastri}, A., {Gandhi}, P., {Harrison}, F.~A., {Luo}, B., {Schawinski}, K.,
  {Stern}, D., {Treister}, E., Jul. 2015. {Broadband Observations of the
  Compton-thick Nucleus of NGC 3393}. \apj 807, 149.

\bibitem[{{Kovalev} et~al.(2017){Kovalev}, {Petrov}, and
  {Plavin}}]{Kovalev2017}
{Kovalev}, Y.~Y., {Petrov}, L., {Plavin}, A.~V., Feb. 2017. {VLBI-Gaia offsets
  favor parsec-scale jet direction in active galactic nuclei}. \aap 598, L1.

\bibitem[{{Krolik} et~al.(2019){Krolik}, {Volonteri}, {Dubois}, and
  {Devriendt}}]{krolik2019}
{Krolik}, J.~H., {Volonteri}, M., {Dubois}, Y., {Devriendt}, J., Jul 2019.
  {Population Estimates for Electromagnetically Distinguishable Supermassive
  Binary Black Holes}. \apj 879~(2), 110.

\bibitem[{{Kun} et~al.(2015){Kun}, {Frey}, {Gab{\'a}nyi}, {Britzen}, {Cseh},
  and {Gergely}}]{kun2015}
{Kun}, E., {Frey}, S., {Gab{\'a}nyi}, K.~{\'E}., {Britzen}, S., {Cseh}, D.,
  {Gergely}, L.~{\'A}., Dec. 2015. {Constraining the parameters of the putative
  supermassive binary black hole in PG 1302-102 from its radio structure}.
  \mnras 454, 1290--1296.

\bibitem[{{Kun} et~al.(2014){Kun}, {Gab{\'a}nyi}, {Karouzos}, {Britzen}, and
  {Gergely}}]{kun2014}
{Kun}, E., {Gab{\'a}nyi}, K.~{\'E}., {Karouzos}, M., {Britzen}, S., {Gergely},
  L.~{\'A}., Dec. 2014. {A spinning supermassive black hole binary model
  consistent with VLBI observations of the S5 1928+738 jet}. \mnras 445,
  1370--1382.

\bibitem[{{Lacy} et~al.(2015){Lacy}, {Ridgway}, {Sajina}, {Petric}, {Gates},
  {Urrutia}, and {Storrie-Lombardi}}]{Lacy_et_al_2015}
{Lacy}, M., {Ridgway}, S.~E., {Sajina}, A., {Petric}, A.~O., {Gates}, E.~L.,
  {Urrutia}, T., {Storrie-Lombardi}, L.~J., Apr. 2015. {The Spitzer
  Mid-infrared AGN Survey. II. The Demographics and Cosmic Evolution of the AGN
  Population}. \apj 802, 102.

\bibitem[{{Lang} and {Hughes}(2008)}]{LangHughes2008}
{Lang}, R.~N., {Hughes}, S.~A., Apr. 2008. {Localizing Coalescing Massive Black
  Hole Binaries with Gravitational Waves}. \apj 677, 1184--1200.

\bibitem[{{Lantz} et~al.(2004){Lantz}, {Aldering}, {Antilogus}, {Bonnaud},
  {Capoani}, {Castera}, {Copin}, {Dubet}, {Gangler}, {Henault}, {Lemonnier},
  {Pain}, {Pecontal}, {Pecontal}, and {Smadja}}]{LantzAA04}
{Lantz}, B., {Aldering}, G., {Antilogus}, P., {Bonnaud}, C., {Capoani}, L.,
  {Castera}, A., {Copin}, Y., {Dubet}, D., {Gangler}, E., {Henault}, F.,
  {Lemonnier}, J.-P., {Pain}, R., {Pecontal}, A., {Pecontal}, E., {Smadja}, G.,
  Feb. 2004. {SNIFS: a wideband integral field spectrograph with microlens
  arrays}. In: {Mazuray}, L., {Rogers}, P.~J., {Wartmann}, R. (Eds.), Optical
  Design and Engineering. Vol. 5249 of \procspie. pp. 146--155.

\bibitem[{{Lanzuisi} et~al.(2018){Lanzuisi}, {Civano}, {Marchesi}, {Comastri},
  {Brusa}, {Gilli}, {Vignali}, {Zamorani}, {Brightman}, {Griffiths}, and
  {Koekemoer}}]{Lanzuisi2018}
{Lanzuisi}, G., {Civano}, F., {Marchesi}, S., {Comastri}, A., {Brusa}, M.,
  {Gilli}, R., {Vignali}, C., {Zamorani}, G., {Brightman}, M., {Griffiths},
  R.~E., {Koekemoer}, A.~M., Oct. 2018. {The Chandra COSMOS Legacy Survey:
  Compton thick AGN at high redshift}. \mnras 480, 2578--2592.

\bibitem[{{Lanzuisi} et~al.(2013){Lanzuisi}, {Civano}, {Marchesi}, {Comastri},
  {Costantini}, {Elvis}, {Mainieri}, {Hickox}, {Jahnke}, {Komossa},
  {Piconcelli}, {Vignali}, {Brusa}, {Cappelluti}, and
  {Fruscione}}]{Lanzuisi2013}
{Lanzuisi}, G., {Civano}, F., {Marchesi}, S., {Comastri}, A., {Costantini}, E.,
  {Elvis}, M., {Mainieri}, V., {Hickox}, R., {Jahnke}, K., {Komossa}, S.,
  {Piconcelli}, E., {Vignali}, C., {Brusa}, M., {Cappelluti}, N., {Fruscione},
  A., Nov. 2013. {The XMM-Newton Spectrum of a Candidate Recoiling Supermassive
  Black Hole: An Elusive Inverted P-Cygni Profile}. \apj 778, 62.

\bibitem[{{Larkin} et~al.(2006){Larkin}, {Barczys}, {Krabbe}, {Adkins},
  {Aliado}, {Amico}, {Brims}, {Campbell}, {Canfield}, {Gasaway}, {Honey},
  {Iserlohe}, {Johnson}, {Kress}, {LaFreniere}, {Lyke}, {Magnone}, {Magnone},
  {McElwain}, {Moon}, {Quirrenbach}, {Skulason}, {Song}, {Spencer}, {Weiss},
  and {Wright}}]{LarkinBK06}
{Larkin}, J., {Barczys}, M., {Krabbe}, A., {Adkins}, S., {Aliado}, T., {Amico},
  P., {Brims}, G., {Campbell}, R., {Canfield}, J., {Gasaway}, T., {Honey}, A.,
  {Iserlohe}, C., {Johnson}, C., {Kress}, E., {LaFreniere}, D., {Lyke}, J.,
  {Magnone}, K., {Magnone}, N., {McElwain}, M., {Moon}, J., {Quirrenbach}, A.,
  {Skulason}, G., {Song}, I., {Spencer}, M., {Weiss}, J., {Wright}, S., Jun.
  2006. {OSIRIS: a diffraction limited integral field spectrograph for Keck}.
  In: Society of Photo-Optical Instrumentation Engineers (SPIE) Conference
  Series. Vol. 6269 of \procspie. p. 62691A.

\bibitem[{{Leahy} and {Parma}(1992)}]{Leahy_Parma_1992}
{Leahy}, J.~P., {Parma}, P., 1992. {Multiple outbursts in radio galaxies.} In:
  {Roland}, J., {Sol}, H., {Pelletier}, G. (Eds.), Extragalactic Radio Sources.
  From Beams to Jets. pp. 307--308.

\bibitem[{{Lehmer} et~al.(2016){Lehmer}, {Basu-Zych}, {Mineo}, {Brand t},
  {Eufrasio}, {Fragos}, {Hornschemeier}, {Luo}, {Xue}, and
  {Bauer}}]{lehmeretal2016}
{Lehmer}, B.~D., {Basu-Zych}, A.~R., {Mineo}, S., {Brand t}, W.~N., {Eufrasio},
  R.~T., {Fragos}, T., {Hornschemeier}, A.~E., {Luo}, B., {Xue}, Y.~Q.,
  {Bauer}, F.~E., Jul 2016. {The Evolution of Normal Galaxy X-Ray Emission
  through Cosmic History: Constraints from the 6 MS Chandra Deep Field-South}.
  \apj 825~(1), 7.

\bibitem[{{Lemons} et~al.(2015){Lemons}, {Reines}, {Plotkin}, {Gallo}, and
  {Greene}}]{Lemons_et_al_2015}
{Lemons}, S.~M., {Reines}, A.~E., {Plotkin}, R.~M., {Gallo}, E., {Greene},
  J.~E., May 2015. {An X-Ray Selected Sample of Candidate Black Holes in Dwarf
  Galaxies}. \apj 805~(1), 12.

\bibitem[{{Lena}(2015)}]{Lena15}
{Lena}, D., 2015. {Aspects of Supermassive Black Hole Growth in Nearby Active
  Galactic Nuclei}. Ph.D. thesis, Rochester Institute of Technology.

\bibitem[{{Lena} et~al.(2018){Lena}, {Panizo-Espinar}, {Jonker}, {Torres}, and
  {Heida}}]{LenaPE18}
{Lena}, D., {Panizo-Espinar}, G., {Jonker}, P.~G., {Torres}, M., {Heida}, M.,
  May 2018. {Characterisation of a candidate dual AGN}. \mnras.

\bibitem[{{Lewis} et~al.(2010){Lewis}, {Eracleous}, and
  {Storchi-Bergmann}}]{Lewis2010}
{Lewis}, K.~T., {Eracleous}, M., {Storchi-Bergmann}, T., Apr. 2010. {Long-term
  Profile Variability in Active Galactic Nucleus with Double-peaked Balmer
  Emission Lines}. \apjs 187, 416--446.

\bibitem[{{Li} et~al.(2017){Li}, {Liu}, {Berczik}, and {Spurzem}}]{Lietal2017}
{Li}, S., {Liu}, F.~K., {Berczik}, P., {Spurzem}, R., Jan 2017. {Boosted Tidal
  Disruption by Massive Black Hole Binaries During Galaxy Mergers from the View
  of N-Body Simulation}. \apj 834~(2), 195.

\bibitem[{{Li} et~al.(2016){Li}, {Wang}, {Ho}, {Lu}, {Qiu}, {Du}, {Hu},
  {Huang}, {Zhang}, {Wang}, and {Bai}}]{Li2016}
{Li}, Y.-R., {Wang}, J.-M., {Ho}, L.~C., {Lu}, K.-X., {Qiu}, J., {Du}, P.,
  {Hu}, C., {Huang}, Y.-K., {Zhang}, Z.-X., {Wang}, K., {Bai}, J.-M., May 2016.
  {Spectroscopic Indication of a Centi-parsec Supermassive Black Hole Binary in
  the Galactic Center of NGC 5548}. \apj 822, 4.

\bibitem[{{Lien} et~al.(2011){Lien}, {Chakraborty}, {Fields}, and
  {Kemball}}]{lien2011}
{Lien}, A., {Chakraborty}, N., {Fields}, B.~D., {Kemball}, A., Oct. 2011.
  {Radio Supernovae in the Great Survey Era}. \apj 740, 23.

\bibitem[{{Lin} and {Papaloizou}(1986)}]{LinPapa1986}
{Lin}, D.~N.~C., {Papaloizou}, J., Oct. 1986. {On the tidal interaction between
  protoplanets and the protoplanetary disk. III - Orbital migration of
  protoplanets}. \apj 309, 846--857.

\bibitem[{{Liska} et~al.(2018){Liska}, {Hesp}, {Tchekhovskoy}, {Ingram}, {van
  der Klis}, and {Markoff}}]{liska2018}
{Liska}, M., {Hesp}, C., {Tchekhovskoy}, A., {Ingram}, A., {van der Klis}, M.,
  {Markoff}, S., Feb. 2018. {Formation of precessing jets by tilted black hole
  discs in 3D general relativistic MHD simulations}. \mnras 474, L81--L85.

\bibitem[{{Liu} et~al.(2009){Liu}, {Li}, and {Chen}}]{liu2009}
{Liu}, F.~K., {Li}, S., {Chen}, X., Nov 2009. {Interruption of Tidal-Disruption
  Flares by Supermassive Black Hole Binaries}. \apjl 706~(1), L133--L137.

\bibitem[{{Liu} et~al.(2014{\natexlab{a}}){Liu}, {Li}, and {Komossa}}]{liu2014}
{Liu}, F.~K., {Li}, S., {Komossa}, S., May 2014{\natexlab{a}}. {A Milliparsec
  Supermassive Black Hole Binary Candidate in the Galaxy SDSS
  J120136.02+300305.5}. \apj 786~(2), 103.

\bibitem[{{Liu} et~al.(2003){Liu}, {Wu}, and {Cao}}]{Liu+2003}
{Liu}, F.~K., {Wu}, X., {Cao}, S.~L., Apr. 2003. {Double-double radio galaxies:
  remnants of merged supermassive binary black holes}. \mnras 340, 411--416.

\bibitem[{{Liu} et~al.(2016{\natexlab{a}}){Liu}, {Eracleous}, and
  {Halpern}}]{JLiu2016}
{Liu}, J., {Eracleous}, M., {Halpern}, J.~P., Jan. 2016{\natexlab{a}}. {A
  Radial Velocity Test for Supermassive Black Hole Binaries as an Explanation
  for Broad, Double-peaked Emission Lines in Active Galactic Nuclei}. \apj 817,
  42.

\bibitem[{{Liu} et~al.(2016{\natexlab{b}}){Liu}, {Gezari}, {Burgett},
  {Chambers}, {Draper}, {Hodapp}, {Huber}, {Kudritzki}, {Magnier}, {Metcalfe},
  {Tonry}, {Wainscoat}, and {Waters}}]{Liu2016}
{Liu}, T., {Gezari}, S., {Burgett}, W., {Chambers}, K., {Draper}, P., {Hodapp},
  K., {Huber}, M., {Kudritzki}, R.-P., {Magnier}, E., {Metcalfe}, N., {Tonry},
  J., {Wainscoat}, R., {Waters}, C., Dec. 2016{\natexlab{b}}. {A Systematic
  Search for Periodically Varying Quasars in Pan-STARRS1: An Extended Baseline
  Test in Medium Deep Survey Field MD09}. \apj 833, 6.

\bibitem[{{Liu} et~al.(2015){Liu}, {Gezari}, {Heinis}, {Magnier}, {Burgett},
  {Chambers}, {Flewelling}, {Huber}, {Hodapp}, {Kaiser}, {Kudritzki}, {Tonry},
  {Wainscoat}, and {Waters}}]{Liu2015}
{Liu}, T., {Gezari}, S., {Heinis}, S., {Magnier}, E.~A., {Burgett}, W.~S.,
  {Chambers}, K., {Flewelling}, H., {Huber}, M., {Hodapp}, K.~W., {Kaiser}, N.,
  {Kudritzki}, R.-P., {Tonry}, J.~L., {Wainscoat}, R.~J., {Waters}, C., Apr.
  2015. {A Periodically Varying Luminous Quasar at z = 2 from the Pan-STARRS1
  Medium Deep Survey: A Candidate Supermassive Black Hole Binary in the
  Gravitational Wave-driven Regime}. \apjl 803, L16.

\bibitem[{{Liu} et~al.(2018{\natexlab{a}}){Liu}, {Gezari}, and
  {Miller}}]{Liu2018}
{Liu}, T., {Gezari}, S., {Miller}, M.~C., May 2018{\natexlab{a}}. {Did ASAS-SN
  Kill the Supermassive Black Hole Binary Candidate PG1302-102?} \apjl 859~(1),
  L12.

\bibitem[{{Liu} et~al.(2013){Liu}, {Civano}, {Shen}, {Green}, {Greene}, and
  {Strauss}}]{LiuCS13}
{Liu}, X., {Civano}, F., {Shen}, Y., {Green}, P., {Greene}, J.~E., {Strauss},
  M.~A., Jan. 2013. {Chandra X-Ray and Hubble Space Telescope Imaging of
  Optically Selected Kiloparsec-scale Binary Active Galactic Nuclei. I. Nature
  of the Nuclear Ionizing Sources}. \apj 762, 110.

\bibitem[{{Liu} et~al.(2010{\natexlab{a}}){Liu}, {Greene}, {Shen}, and
  {Strauss}}]{2010ApJ...715L..30L}
{Liu}, X., {Greene}, J.~E., {Shen}, Y., {Strauss}, M.~A., May
  2010{\natexlab{a}}. {Discovery of Four kpc-scale Binary Active Galactic
  Nuclei}. \apjl 715, L30--L34.

\bibitem[{{Liu} et~al.(2018{\natexlab{b}}){Liu}, {Guo}, {Shen}, {Greene}, and
  {Strauss}}]{liux2018}
{Liu}, X., {Guo}, H., {Shen}, Y., {Greene}, J.~E., {Strauss}, M.~A., Jul
  2018{\natexlab{b}}. {Hubble Space Telescope Wide Field Camera 3 Identifies an
  r $_{ p }$ = 1 Kpc Dual Active Galactic Nucleus in the Minor Galaxy Merger
  SDSS J0924+0510 at z = 0.1495}. \apj 862~(1), 29.

\bibitem[{{Liu} et~al.(2014{\natexlab{b}}){Liu}, {Shen}, {Bian}, {Loeb}, and
  {Tremaine}}]{liu14}
{Liu}, X., {Shen}, Y., {Bian}, F., {Loeb}, A., {Tremaine}, S., Jul.
  2014{\natexlab{b}}. {Constraining Sub-parsec Binary Supermassive Black Holes
  in Quasars with Multi-epoch Spectroscopy. II. The Population with
  Kinematically Offset Broad Balmer Emission Lines}. \apj 789, 140.

\bibitem[{{Liu} et~al.(2011){Liu}, {Shen}, and {Strauss}}]{2011ApJ...736L...7L}
{Liu}, X., {Shen}, Y., {Strauss}, M.~A., Jul. 2011. {Cosmic Train Wreck by
  Massive Black Holes: Discovery of a Kiloparsec-scale Triple Active Galactic
  Nucleus}. \apjl 736, L7.

\bibitem[{{Liu} et~al.(2010{\natexlab{b}}){Liu}, {Shen}, {Strauss}, and
  {Greene}}]{Liu:2010}
{Liu}, X., {Shen}, Y., {Strauss}, M.~A., {Greene}, J.~E., Jan.
  2010{\natexlab{b}}. {Type 2 Active Galactic Nuclei with Double-Peaked [O III]
  Lines: Narrow-Line Region Kinematics or Merging Supermassive Black Hole
  Pairs?} \apj 708, 427--434.

\bibitem[{{Lobanov} and {Roland}(2005)}]{lobanov2005}
{Lobanov}, A.~P., {Roland}, J., Mar. 2005. {A supermassive binary black hole in
  the quasar 3C 345}. \aap 431, 831--846.

\bibitem[{{Lodato} et~al.(2009){Lodato}, {Nayakshin}, {King}, and
  {Pringle}}]{Lodato+2009}
{Lodato}, G., {Nayakshin}, S., {King}, A.~R., {Pringle}, J.~E., Sep. 2009.
  {Black hole mergers: can gas discs solve the `final parsec' problem?} \mnras
  398, 1392--1402.

\bibitem[{{Luo} et~al.(2017){Luo}, {Brandt}, {Xue}, {Lehmer}, {Alexander},
  {Bauer}, {Vito}, {Yang}, {Basu-Zych}, {Comastri}, {Gilli}, {Gu},
  {Hornschemeier}, {Koekemoer}, {Liu}, {Mainieri}, {Paolillo}, {Ranalli},
  {Rosati}, {Schneider}, {Shemmer}, {Smail}, {Sun}, {Tozzi}, {Vignali}, and
  {Wang}}]{Luo2017}
{Luo}, B., {Brandt}, W.~N., {Xue}, Y.~Q., {Lehmer}, B., {Alexander}, D.~M.,
  {Bauer}, F.~E., {Vito}, F., {Yang}, G., {Basu-Zych}, A.~R., {Comastri}, A.,
  {Gilli}, R., {Gu}, Q.-S., {Hornschemeier}, A.~E., {Koekemoer}, A., {Liu}, T.,
  {Mainieri}, V., {Paolillo}, M., {Ranalli}, P., {Rosati}, P., {Schneider},
  D.~P., {Shemmer}, O., {Smail}, I., {Sun}, M., {Tozzi}, P., {Vignali}, C.,
  {Wang}, J.-X., Jan. 2017. {The Chandra Deep Field-South Survey: 7 Ms Source
  Catalogs}. \apjs 228, 2.

\bibitem[{{Lupi} et~al.(2015{\natexlab{a}}){Lupi}, {Haardt}, and
  {Dotti}}]{Lupi_et_al_2015a}
{Lupi}, A., {Haardt}, F., {Dotti}, M., Jan. 2015{\natexlab{a}}. {Massive black
  hole and gas dynamics in galaxy nuclei mergers - I. Numerical
  implementation}. \mnras 446, 1765--1774.

\bibitem[{{Lupi} et~al.(2015{\natexlab{b}}){Lupi}, {Haardt}, {Dotti}, and
  {Colpi}}]{Lupi_et_al_2015b}
{Lupi}, A., {Haardt}, F., {Dotti}, M., {Colpi}, M., Nov. 2015{\natexlab{b}}.
  {Massive black hole and gas dynamics in mergers of galaxy nuclei - II. Black
  hole sinking in star-forming nuclear discs}. \mnras 453, 3437--3446.

\bibitem[{{Lusso} et~al.(2018){Lusso}, {Fumagalli}, {Rafelski}, {Neeleman},
  {Prochaska}, {Hennawi}, {O'Meara}, and {Theuns}}]{lusso2018}
{Lusso}, E., {Fumagalli}, M., {Rafelski}, M., {Neeleman}, M., {Prochaska},
  J.~X., {Hennawi}, J.~F., {O'Meara}, J.~M., {Theuns}, T., Jun. 2018. {The
  Spectral and Environment Properties of z$\sim$2.0-2.5 Quasar Pairs}. \apj
  860, 41.

\bibitem[{{Lusso} et~al.(2012)}]{2012MNRAS.425..623L}
{Lusso}, E., et~al., Sep. 2012. {Bolometric luminosities and Eddington ratios
  of X-ray selected active galactic nuclei in the XMM-COSMOS survey}. \mnras
  425, 623--640.

\bibitem[{{Lyu} and {Liu}(2016)}]{Lyu:2016}
{Lyu}, Y., {Liu}, X., Nov. 2016. {A high fraction of double-peaked narrow
  emission lines in powerful active galactic nuclei}. \mnras 463, 24--36.

\bibitem[{{MacFadyen} and {Milosavljevi{\'c}}(2008)}]{macfadyen08}
{MacFadyen}, A.~I., {Milosavljevi{\'c}}, M., Jan. 2008. {An Eccentric
  Circumbinary Accretion Disk and the Detection of Binary Massive Black Holes}.
  \apj 672, 83--93.

\bibitem[{{Marchesi} et~al.(2016){Marchesi}, {Civano}, {Elvis}, {Salvato},
  {Brusa}, {Comastri}, {Gilli}, {Hasinger}, {Lanzuisi}, {Miyaji}, {Treister},
  {Urry}, {Vignali}, {Zamorani}, {Allevato}, {Cappelluti}, {Cardamone},
  {Finoguenov}, {Griffiths}, {Karim}, {Laigle}, {LaMassa}, {Jahnke}, {Ranalli},
  {Schawinski}, {Schinnerer}, {Silverman}, {Smolcic}, {Suh}, and
  {Trakhtenbrot}}]{Marchesi2016}
{Marchesi}, S., {Civano}, F., {Elvis}, M., {Salvato}, M., {Brusa}, M.,
  {Comastri}, A., {Gilli}, R., {Hasinger}, G., {Lanzuisi}, G., {Miyaji}, T.,
  {Treister}, E., {Urry}, C.~M., {Vignali}, C., {Zamorani}, G., {Allevato}, V.,
  {Cappelluti}, N., {Cardamone}, C., {Finoguenov}, A., {Griffiths}, R.~E.,
  {Karim}, A., {Laigle}, C., {LaMassa}, S.~M., {Jahnke}, K., {Ranalli}, P.,
  {Schawinski}, K., {Schinnerer}, E., {Silverman}, J.~D., {Smolcic}, V., {Suh},
  H., {Trakhtenbrot}, B., Jan. 2016. {The Chandra COSMOS Legacy survey:
  optical/IR identifications}. \apj 817, 34.

\bibitem[{{Marconi} et~al.(2004){Marconi}, {Risaliti}, {Gilli}, {Hunt},
  {Maiolino}, and {Salvati}}]{marconietal04}
{Marconi}, A., {Risaliti}, G., {Gilli}, R., {Hunt}, L.~K., {Maiolino}, R.,
  {Salvati}, M., Jun. 2004. {Local supermassive black holes, relics of active
  galactic nuclei and the X-ray background}. \mnras 351, 169--185.

\bibitem[{{Martin} et~al.(2018){Martin}, {Kaviraj}, {Volonteri}, {Simmons},
  {Devriendt}, {Lintott}, {Smethurst}, {Dubois}, and
  {Pichon}}]{2018MNRAS.476.2801M}
{Martin}, G., {Kaviraj}, S., {Volonteri}, M., {Simmons}, B.~D., {Devriendt},
  J.~E.~G., {Lintott}, C.~J., {Smethurst}, R.~J., {Dubois}, Y., {Pichon}, C.,
  May 2018. {Normal black holes in bulge-less galaxies: the largely quiescent,
  merger-free growth of black holes over cosmic time}. \mnras 476~(2),
  2801--2812.

\bibitem[{{Mateos} et~al.(2012){Mateos}, {Alonso-Herrero}, {Carrera}, {Blain},
  {Watson}, {Barcons}, {Braito}, {Severgnini}, {Donley}, and
  {Stern}}]{mateos2012}
{Mateos}, S., {Alonso-Herrero}, A., {Carrera}, F.~J., {Blain}, A., {Watson},
  M.~G., {Barcons}, X., {Braito}, V., {Severgnini}, P., {Donley}, J.~L.,
  {Stern}, D., Nov. 2012. {Using the Bright Ultrahard XMM-Newton survey to
  define an IR selection of luminous AGN based on WISE colours}. \mnras 426,
  3271--3281.

\bibitem[{{Mateos} et~al.(2017){Mateos}, {Carrera}, {Barcons}, {Alonso-
  Herrero}, {Hern{\'a}n-Caballero}, {Page}, {Ramos Almeida}, {Caccianiga},
  {Miyaji}, and {Blain}}]{mateos17}
{Mateos}, S., {Carrera}, F.~J., {Barcons}, X., {Alonso- Herrero}, A.,
  {Hern{\'a}n-Caballero}, A., {Page}, M., {Ramos Almeida}, C., {Caccianiga},
  A., {Miyaji}, T., {Blain}, A., Jun. 2017. {Survival of the Obscuring Torus in
  the Most Powerful Active Galactic Nuclei}. \apj 841, L18.

\bibitem[{{Mayer}(2013)}]{Mayer:2013:MBHBGasRev}
{Mayer}, L., Dec. 2013. {Massive black hole binaries in gas-rich galaxy
  mergers; multiple regimes of orbital decay and interplay with gas inflows}.
  Classical and Quantum Gravity 30~(24), 244008.

\bibitem[{{Mayer} and {Bonoli}(2019)}]{Mayer19}
{Mayer}, L., {Bonoli}, S., Jan. 2019. {The route to massive black hole
  formation via merger-driven direct collapse: a review}. Reports on Progress
  in Physics 82~(1), 016901.

\bibitem[{{Mayer} et~al.(2010){Mayer}, {Kazantzidis}, {Escala}, and
  {Callegari}}]{Mayer2010}
{Mayer}, L., {Kazantzidis}, S., {Escala}, A., {Callegari}, S., Aug 2010.
  {Direct formation of supermassive black holes via multi-scale gas inflows in
  galaxy mergers}. \nat 466~(7310), 1082--1084.

\bibitem[{{Mayer} et~al.(2007){Mayer}, {Kazantzidis}, {Madau}, {Colpi},
  {Quinn}, and {Wadsley}}]{mayer07}
{Mayer}, L., {Kazantzidis}, S., {Madau}, P., {Colpi}, M., {Quinn}, T.,
  {Wadsley}, J., Jun. 2007. {Rapid Formation of Supermassive Black Hole
  Binaries in Galaxy Mergers with Gas}. Science 316, 1874--.

\bibitem[{{Mazzarella} et~al.(2012){Mazzarella}, {Iwasawa}, {Vavilkin},
  {Armus}, {Kim}, {Bothun}, {Evans}, {Spoon}, {Haan}, {Howell}, {Lord},
  {Marshall}, {Ishida}, {Xu}, {Petric}, {Sanders}, {Surace}, {Appleton},
  {Chan}, {Frayer}, {Inami}, {Khachikian}, {Madore}, {Privon}, {Sturm}, {U},
  and {Veilleux}}]{MazzarellaIV12}
{Mazzarella}, J.~M., {Iwasawa}, K., {Vavilkin}, T., {Armus}, L., {Kim}, D.-C.,
  {Bothun}, G., {Evans}, A.~S., {Spoon}, H.~W.~W., {Haan}, S., {Howell}, J.~H.,
  {Lord}, S., {Marshall}, J.~A., {Ishida}, C.~M., {Xu}, C.~K., {Petric}, A.,
  {Sanders}, D.~B., {Surace}, J.~A., {Appleton}, P., {Chan}, B.~H.~P.,
  {Frayer}, D.~T., {Inami}, H., {Khachikian}, E.~Y., {Madore}, B.~F., {Privon},
  G.~C., {Sturm}, E., {U}, V., {Veilleux}, S., Nov. 2012. {Investigation of
  Dual Active Nuclei, Outflows, Shock-heated Gas, and Young Star Clusters in
  Markarian 266}. \aj 144, 125.

\bibitem[{{McAlpine} et~al.(2018){McAlpine}, {Bower}, {Rosario}, {Crain},
  {Schaye}, and {Theuns}}]{2018MNRAS.481.3118M}
{McAlpine}, S., {Bower}, R.~G., {Rosario}, D.~J., {Crain}, R.~A., {Schaye}, J.,
  {Theuns}, T., Dec 2018. {The rapid growth phase of supermassive black holes}.
  \mnras 481~(3), 3118--3128.

\bibitem[{{McGee} et~al.(2018){McGee}, {Sesana}, and
  {Vecchio}}]{2018arXiv181100050M}
{McGee}, S., {Sesana}, A., {Vecchio}, A., Oct. 2018. {The assembly of cosmic
  structure from baryons to black holes with joint gravitational-wave and X-ray
  observations}. arXiv e-prints.

\bibitem[{McGurk et~al.(2015)McGurk, Max, Medling, Shields, and
  Comerford}]{McGurk_2015}
McGurk, R.~C., Max, C.~E., Medling, A.~M., Shields, G.~A., Comerford, J.~M.,
  Sep 2015. Spatially resolved imaging and spectroscopy of candidate dual
  active galactic nuclei. The Astrophysical Journal 811~(1), 14.
\newline\urlprefix\url{http://dx.doi.org/10.1088/0004-637X/811/1/14}

\bibitem[{{McGurk} et~al.(2011){McGurk}, {Max}, {Rosario}, {Shields}, {Smith},
  and {Wright}}]{McGurkMR11}
{McGurk}, R.~C., {Max}, C.~E., {Rosario}, D.~J., {Shields}, G.~A., {Smith},
  K.~L., {Wright}, S.~A., Sep. 2011. {Spatially Resolved Spectroscopy of SDSS
  J0952+2552: A Confirmed Dual Active Galactic Nucleus}. \apjl 738, L2.

\bibitem[{{McKernan} and {Ford}(2015)}]{McKernan_ford2015}
{McKernan}, B., {Ford}, K.~E.~S., Sep 2015. {Detection of radial velocity
  shifts due to black hole binaries near merger.} \mnras 452, L1--L5.

\bibitem[{{McWilliams} et~al.(2011){McWilliams}, {Lang}, {Baker}, and
  {Thorpe}}]{McWilliams+2011}
{McWilliams}, S.~T., {Lang}, R.~N., {Baker}, J.~G., {Thorpe}, J.~I., Sep. 2011.
  {Sky localization of complete inspiral-merger-ringdown signals for
  nonspinning massive black hole binaries}. \prd 84~(6), 064003.

\bibitem[{{Mechtley} et~al.(2016){Mechtley}, {Jahnke}, {Windhorst}, {Andrae},
  {Cisternas}, {Cohen}, {Hewlett}, {Koekemoer}, {Schramm}, {Schulze},
  {Silverman}, {Villforth}, {van der Wel}, and {Wisotzki}}]{Mechtley:2016}
{Mechtley}, M., {Jahnke}, K., {Windhorst}, R.~A., {Andrae}, R., {Cisternas},
  M., {Cohen}, S.~H., {Hewlett}, T., {Koekemoer}, A.~M., {Schramm}, M.,
  {Schulze}, A., {Silverman}, J.~D., {Villforth}, C., {van der Wel}, A.,
  {Wisotzki}, L., Oct. 2016. {Do the Most Massive Black Holes at z = 2 Grow via
  Major Mergers?} \apj 830, 156.

\bibitem[{{Medling} et~al.(2014){Medling}, {U}, {Guedes}, {Max}, {Mayer},
  {Armus}, {Holden}, {Ro{\v s}kar}, and {Sanders}}]{Medling_et_al_2014}
{Medling}, A.~M., {U}, V., {Guedes}, J., {Max}, C.~E., {Mayer}, L., {Armus},
  L., {Holden}, B., {Ro{\v s}kar}, R., {Sanders}, D., Mar. 2014. {Stellar and
  Gaseous Nuclear Disks Observed in Nearby (U)LIRGs}. \apj 784, 70.

\bibitem[{{Medling} et~al.(2015){Medling}, {U}, {Max}, {Sanders}, {Armus},
  {Holden}, {Mieda}, {Wright}, and {Larkin}}]{Medling_et_al_2015}
{Medling}, A.~M., {U}, V., {Max}, C.~E., {Sanders}, D.~B., {Armus}, L.,
  {Holden}, B., {Mieda}, E., {Wright}, S.~A., {Larkin}, J.~E., Apr. 2015.
  {Following Black Hole Scaling Relations through Gas-rich Mergers}. \apj 803,
  61.

\bibitem[{{Meier}(2001)}]{2001ApJ...548L...9M}
{Meier}, D.~L., Feb. 2001. {The Association of Jet Production with
  Geometrically Thick Accretion Flows and Black Hole Rotation}. \apjl 548,
  L9--L12.

\bibitem[{{Merloni} et~al.(2012){Merloni}, {Predehl}, {Becker},
  {B{\"o}hringer}, {Boller}, {Brunner}, {Brusa}, {Dennerl}, {Freyberg},
  {Friedrich}, {Georgakakis}, {Haberl}, {Hasinger}, {Meidinger}, {Mohr},
  {Nandra}, {Rau}, {Reiprich}, {Robrade}, {Salvato}, {Santangelo}, {Sasaki},
  {Schwope}, {Wilms}, and {German eROSITA Consortium}}]{merloni12_erosita}
{Merloni}, A., {Predehl}, P., {Becker}, W., {B{\"o}hringer}, H., {Boller}, T.,
  {Brunner}, H., {Brusa}, M., {Dennerl}, K., {Freyberg}, M., {Friedrich}, P.,
  {Georgakakis}, A., {Haberl}, F., {Hasinger}, G., {Meidinger}, N., {Mohr}, J.,
  {Nandra}, K., {Rau}, A., {Reiprich}, T.~H., {Robrade}, J., {Salvato}, M.,
  {Santangelo}, A., {Sasaki}, M., {Schwope}, A., {Wilms}, J., {German eROSITA
  Consortium}, t., Sep. 2012. {eROSITA Science Book: Mapping the Structure of
  the Energetic Universe}. arXiv e-prints.

\bibitem[{{Merritt} and {Ekers}(2002)}]{Merritt_Ekers_2002}
{Merritt}, D., {Ekers}, R.~D., Aug. 2002. {Tracing Black Hole Mergers Through
  Radio Lobe Morphology}. Science 297, 1310--1313.

\bibitem[{{Merritt} and {Poon}(2004)}]{merritt04}
{Merritt}, D., {Poon}, M.~Y., May 2004. {Chaotic Loss Cones and Black Hole
  Fueling}. \apj 606, 788--798.

\bibitem[{{Mezcua} et~al.(2018){Mezcua}, {Civano}, {Marchesi}, {Suh},
  {Fabbiano}, and {Volonteri}}]{Mezcua_et_al_2018}
{Mezcua}, M., {Civano}, F., {Marchesi}, S., {Suh}, H., {Fabbiano}, G.,
  {Volonteri}, M., Aug. 2018. {Intermediate-mass black holes in dwarf galaxies
  out to redshift $\sim$ 2.4 in the Chandra COSMOS Legacy Survey}. \mnras 478,
  2576--2591.

\bibitem[{{Middelberg} et~al.(2013){Middelberg}, {Deller}, {Norris},
  {Fotopoulou}, {Salvato}, {Morgan}, {Brisken}, {Lutz}, and
  {Rovilos}}]{middelberg2013}
{Middelberg}, E., {Deller}, A.~T., {Norris}, R.~P., {Fotopoulou}, S.,
  {Salvato}, M., {Morgan}, J.~S., {Brisken}, W., {Lutz}, D., {Rovilos}, E.,
  Mar. 2013. {Mosaiced wide-field VLBI observations of the Lockman Hole/XMM}.
  Astronomy \& Astrophysics 551, A97.

\bibitem[{{Middleton} et~al.(2018){Middleton}, {Chen}, {Del Pozzo}, {Sesana},
  and {Vecchio}}]{2018NatCo...9..573M}
{Middleton}, H., {Chen}, S., {Del Pozzo}, W., {Sesana}, A., {Vecchio}, A., Mar.
  2018. {No tension between assembly models of super massive black hole
  binaries and pulsar observations}. Nature Communications 9, 573.

\bibitem[{{Mihos} and {Hernquist}(1996)}]{Mihos_Hernquist_1996}
{Mihos}, J.~C., {Hernquist}, L., Jun. 1996. {Gasdynamics and Starbursts in
  Major Mergers}. \apj 464, 641.

\bibitem[{{Milosavljevi{\'c}} and {Merritt}(2001)}]{mm01}
{Milosavljevi{\'c}}, M., {Merritt}, D., Dec. 2001. {Formation of Galactic
  Nuclei}. \apj 563, 34--62.

\bibitem[{{Milosavljevi{\'c}} and {Phinney}(2005)}]{MP2005}
{Milosavljevi{\'c}}, M., {Phinney}, E.~S., Apr. 2005. {The Afterglow of Massive
  Black Hole Coalescence}. \apjl 622, L93--L96.

\bibitem[{{Miranda} et~al.(2017){Miranda}, {Mu{\~n}oz}, and
  {Lai}}]{Miranda+2017}
{Miranda}, R., {Mu{\~n}oz}, D.~J., {Lai}, D., Apr. 2017. {Viscous hydrodynamics
  simulations of circumbinary accretion discs: variability, quasi-steady state
  and angular momentum transfer}. \mnras 466, 1170--1191.

\bibitem[{{Mirza} et~al.(2017){Mirza}, {Tahir}, {Khan}, {Holley-Bockelmann},
  {Baig}, {Berczik}, and {Chishtie}}]{Mirza2017}
{Mirza}, M.~A., {Tahir}, A., {Khan}, F.~M., {Holley-Bockelmann}, H., {Baig},
  A.~M., {Berczik}, P., {Chishtie}, F., Sep 2017. {Galaxy rotation and
  supermassive black hole binary evolution}. \mnras 470~(1), 940--947.

\bibitem[{{Mohan} et~al.(2016){Mohan}, {An}, {Frey}, {Mangalam}, {Gab{\'a}nyi},
  and {Kun}}]{mohan2016}
{Mohan}, P., {An}, T., {Frey}, S., {Mangalam}, A., {Gab{\'a}nyi}, K.~{\'E}.,
  {Kun}, E., Dec. 2016. {Parsec-scale jet properties of the quasar PG
  1302-102}. \mnras 463, 1812--1821.

\bibitem[{{Montuori} et~al.(2011){Montuori}, {Dotti}, {Colpi}, {Decarli}, and
  {Haardt}}]{Montuori11}
{Montuori}, C., {Dotti}, M., {Colpi}, M., {Decarli}, R., {Haardt}, F., Mar.
  2011. {Search for sub-parsec massive binary black holes through line
  diagnosis}. \mnras 412, 26--32.

\bibitem[{{Montuori} et~al.(2012){Montuori}, {Dotti}, {Haardt}, {Colpi}, and
  {Decarli}}]{Montuori12}
{Montuori}, C., {Dotti}, M., {Haardt}, F., {Colpi}, M., {Decarli}, R., Sep.
  2012. {Search for sub-parsec massive binary black holes through line
  diagnosis - II}. \mnras 425, 1633--1639.

\bibitem[{{Moody} et~al.(2019){Moody}, {Shi}, and {Stone}}]{Moody+2019}
{Moody}, M. S.~L., {Shi}, J.-M., {Stone}, J.~M., Apr 2019. {Hydrodynamic
  Torques in Circumbinary Accretion Disks}. \apj 875~(1), 66.

\bibitem[{{Mooley} et~al.(2018){Mooley}, {Wrobel}, {Anderson}, and
  {Hallinan}}]{mooley2018}
{Mooley}, K.~P., {Wrobel}, J.~M., {Anderson}, M.~M., {Hallinan}, G., Jan. 2018.
  {The twisted radio structure of PSO J334.2028+01.4075, still a supermassive
  binary black hole candidate}. \mnras 473, 1388--1393.

\bibitem[{{Moran} et~al.(2014){Moran}, {Shahinyan}, {Sugarman}, {V{\'e}lez},
  and {Eracleous}}]{Moran_et_al_2014}
{Moran}, E.~C., {Shahinyan}, K., {Sugarman}, H.~R., {V{\'e}lez}, D.~O.,
  {Eracleous}, M., Dec 2014. {Black Holes At the Centers of Nearby Dwarf
  Galaxies}. \aj 148~(6), 136.

\bibitem[{{Mu{\~n}oz} et~al.(2019){Mu{\~n}oz}, {Miranda}, and
  {Lai}}]{Munoz+2019}
{Mu{\~n}oz}, D.~J., {Miranda}, R., {Lai}, D., Jan. 2019. {Hydrodynamics of
  Circumbinary Accretion: Angular Momentum Transfer and Binary Orbital
  Evolution}. \apj 871, 84.

\bibitem[{{M{\"u}ller-S{\'a}nchez} et~al.(2015){M{\"u}ller-S{\'a}nchez},
  {Comerford}, {Nevin}, {Barrows}, {Cooper}, and {Greene}}]{mullersanchez2015}
{M{\"u}ller-S{\'a}nchez}, F., {Comerford}, J.~M., {Nevin}, R., {Barrows},
  R.~S., {Cooper}, M.~C., {Greene}, J.~E., Nov. 2015. {The Origin of
  Double-peaked Narrow Lines in Active Galactic Nuclei. I. Very Large Array
  Detections of Dual AGNs and AGN Outflows}. Astrophysical Journal 813, 103.

\bibitem[{{Munshi} et~al.(2013){Munshi}, {Governato}, {Brooks}, {Christensen},
  {Shen}, {Loebman}, {Moster}, {Quinn}, and {Wadsley}}]{Munshi_et_al_2013}
{Munshi}, F., {Governato}, F., {Brooks}, A.~M., {Christensen}, C., {Shen}, S.,
  {Loebman}, S., {Moster}, B., {Quinn}, T., {Wadsley}, J., Mar. 2013.
  {Reproducing the Stellar Mass/Halo Mass Relation in Simulated {$\Lambda$}CDM
  Galaxies: Theory versus Observational Estimates}. \apj 766, 56.

\bibitem[{{Murphy}(2018)}]{Murphy2018}
{Murphy}, E.~J., Aug. 2018. {A next-generation Very Large Array}. In: {Tarchi},
  A., {Reid}, M.~J., {Castangia}, P. (Eds.), Astrophysical Masers: Unlocking
  the Mysteries of the Universe. Vol. 336 of IAU Symposium. pp. 426--432.

\bibitem[{{Mushotzky}(2018)}]{mushotzky2018}
{Mushotzky}, R., Jul 2018. {AXIS: a probe class next generation high angular
  resolution x-ray imaging satellite}. In: \procspie. Vol. 10699 of Society of
  Photo-Optical Instrumentation Engineers (SPIE) Conference Series. p. 1069929.

\bibitem[{{Myers} et~al.(2008){Myers}, {Richards}, {Brunner}, {Schneider},
  {Strand}, {Hall}, {Blomquist}, and {York}}]{2008ApJ...678..635M}
{Myers}, A.~D., {Richards}, G.~T., {Brunner}, R.~J., {Schneider}, D.~P.,
  {Strand}, N.~E., {Hall}, P.~B., {Blomquist}, J.~A., {York}, D.~G., May 2008.
  {Quasar Clustering at 25 h$^{-1}$ kpc from a Complete Sample of Binaries}.
  \apj 678, 635--646.

\bibitem[{{Nan} et~al.(2011){Nan}, {Li}, {Jin}, {Wang}, {Zhu}, {Zhu}, {Zhang},
  {Yue}, and {Qian}}]{2011IJMPD..20..989N}
{Nan}, R., {Li}, D., {Jin}, C., {Wang}, Q., {Zhu}, L., {Zhu}, W., {Zhang}, H.,
  {Yue}, Y., {Qian}, L., 2011. {The Five-Hundred Aperture Spherical Radio
  Telescope (fast) Project}. International Journal of Modern Physics D 20,
  989--1024.

\bibitem[{{Nandra} et~al.(2013){Nandra}, {Barret}, {Barcons}, {Fabian}, {den
  Herder}, {Piro}, {Watson}, {Adami}, {Aird}, {Afonso}, and
  et~al.}]{Nandra2013}
{Nandra}, K., {Barret}, D., {Barcons}, X., {Fabian}, A., {den Herder}, J.-W.,
  {Piro}, L., {Watson}, M., {Adami}, C., {Aird}, J., {Afonso}, J.~M., et~al.,
  Jun. 2013. {The Hot and Energetic Universe: A White Paper presenting the
  science theme motivating the Athena+ mission}. arXiv e-prints.

\bibitem[{{Nardini}(2017)}]{Nardini2017}
{Nardini}, E., Nov. 2017. {Nuclear absorption and emission in the AGN merger
  NGC 6240 : the hard X-ray view}. \mnras 471, 3483--3493.

\bibitem[{{Navarro} et~al.(1996){Navarro}, {Frenk}, and {White}}]{NFW96}
{Navarro}, J.~F., {Frenk}, C.~S., {White}, S.~D.~M., May 1996. {The Structure
  of Cold Dark Matter Halos}. \apj 462, 563.

\bibitem[{{Newton} and {Kay}(2013)}]{Newton_Kay_2013}
{Newton}, R.~D.~A., {Kay}, S.~T., Oct. 2013. {A study of AGN and supernova
  feedback in simulations of isolated and merging disc galaxies}. \mnras 434,
  3606--3627.

\bibitem[{{Nguyen} and {Bogdanovi{\'c}}(2016)}]{nguyen16}
{Nguyen}, K., {Bogdanovi{\'c}}, T., Sep. 2016. {Emission Signatures from
  Sub-parsec Binary Supermassive Black Holes. I. Diagnostic Power of Broad
  Emission Lines}. \apj 828, 68.

\bibitem[{{Nguyen} et~al.(2019{\natexlab{a}}){Nguyen}, {Bogdanovi{\'c}},
  {Runnoe}, {Eracleous}, {Sigurdsson}, and {Boroson}}]{nguyen19}
{Nguyen}, K., {Bogdanovi{\'c}}, T., {Runnoe}, J.~C., {Eracleous}, M.,
  {Sigurdsson}, S., {Boroson}, T., Jan. 2019{\natexlab{a}}. {Emission
  Signatures from Sub-parsec Binary Supermassive Black Holes. II. Effect of
  Accretion Disk Wind on Broad Emission Lines}. \apj 870, 16.

\bibitem[{{Nguyen} et~al.(2019{\natexlab{b}}){Nguyen}, {Bogdanovic}, {Runnoe},
  {Eracleous}, {Sigurdsson}, and {Boroson}}]{nguyen19b}
{Nguyen}, K., {Bogdanovic}, T., {Runnoe}, J.~C., {Eracleous}, M., {Sigurdsson},
  S., {Boroson}, T., Aug 2019{\natexlab{b}}. {Emission Signatures from
  Sub-parsec Binary Supermassive Black Holes III: Comparison of Models with
  Observations}. arXiv e-prints, arXiv:1908.01799.

\bibitem[{{Noble} et~al.(2012){Noble}, {Mundim}, {Nakano}, {Krolik},
  {Campanelli}, {Zlochower}, and {Yunes}}]{noble12}
{Noble}, S.~C., {Mundim}, B.~C., {Nakano}, H., {Krolik}, J.~H., {Campanelli},
  M., {Zlochower}, Y., {Yunes}, N., Aug. 2012. {Circumbinary
  Magnetohydrodynamic Accretion into Inspiraling Binary Black Holes}. \apj 755,
  51.

\bibitem[{{Norris} et~al.(2015){Norris}, {Basu}, {Brown}, {Carretti},
  {Kapinska}, {Prandoni}, {Rudnick}, and {Seymour}}]{Norris2015}
{Norris}, R., {Basu}, K., {Brown}, M., {Carretti}, E., {Kapinska}, A.~D.,
  {Prandoni}, I., {Rudnick}, L., {Seymour}, N., Apr. 2015. {The SKA
  Mid-frequency All-sky Continuum Survey: Discovering the unexpected and
  transforming radio-astronomy}. Advancing Astrophysics with the Square
  Kilometre Array (AASKA14), 86.

\bibitem[{{Norris} et~al.(2011){Norris}, {Hopkins}, {Afonso}, {Brown},
  {Condon}, {Dunne}, {Feain}, {Hollow}, {Jarvis}, {Johnston-Hollitt}, {Lenc},
  {Middelberg}, {Padovani}, {Prandoni}, {Rudnick}, {Seymour}, {Umana},
  {Andernach}, {Alexander}, {Appleton}, {Bacon}, {Banfield}, {Becker}, {Brown},
  {Ciliegi}, {Jackson}, {Eales}, {Edge}, {Gaensler}, {Giovannini}, {Hales},
  {Hancock}, {Huynh}, {Ibar}, {Ivison}, {Kennicutt}, {Kimball}, {Koekemoer},
  {Koribalski}, {L{\'o}pez-S{\'a}nchez}, {Mao}, {Murphy}, {Messias},
  {Pimbblet}, {Raccanelli}, {Randall}, {Reiprich}, {Roseboom},
  {R{\"o}ttgering}, {Saikia}, {Sharp}, {Slee}, {Smail}, {Thompson}, {Urquhart},
  {Wall}, and {Zhao}}]{Norris2011}
{Norris}, R.~P., {Hopkins}, A.~M., {Afonso}, J., {Brown}, S., {Condon}, J.~J.,
  {Dunne}, L., {Feain}, I., {Hollow}, R., {Jarvis}, M., {Johnston-Hollitt}, M.,
  {Lenc}, E., {Middelberg}, E., {Padovani}, P., {Prandoni}, I., {Rudnick}, L.,
  {Seymour}, N., {Umana}, G., {Andernach}, H., {Alexander}, D.~M., {Appleton},
  P.~N., {Bacon}, D., {Banfield}, J., {Becker}, W., {Brown}, M.~J.~I.,
  {Ciliegi}, P., {Jackson}, C., {Eales}, S., {Edge}, A.~C., {Gaensler}, B.~M.,
  {Giovannini}, G., {Hales}, C.~A., {Hancock}, P., {Huynh}, M.~T., {Ibar}, E.,
  {Ivison}, R.~J., {Kennicutt}, R., {Kimball}, A.~E., {Koekemoer}, A.~M.,
  {Koribalski}, B.~S., {L{\'o}pez-S{\'a}nchez}, {\'A}.~R., {Mao}, M.~Y.,
  {Murphy}, T., {Messias}, H., {Pimbblet}, K.~A., {Raccanelli}, A., {Randall},
  K.~E., {Reiprich}, T.~H., {Roseboom}, I.~G., {R{\"o}ttgering}, H., {Saikia},
  D.~J., {Sharp}, R.~G., {Slee}, O.~B., {Smail}, I., {Thompson}, M.~A.,
  {Urquhart}, J.~S., {Wall}, J.~V., {Zhao}, G.-B., Aug. 2011. {EMU:
  Evolutionary Map of the Universe}. \pasa 28, 215--248.

\bibitem[{{Novak} et~al.(2015){Novak}, {Smol{\v{c}}i{\'c}}, {Civano}, {Bondi},
  {Ciliegi}, {Wang}, {Loeb}, {Banfield}, {Bourke}, {Elvis}, {Hallinan},
  {Intema}, {Kl{\"o}ckner}, {Mooley}, and {Navarrete}}]{Novak2015}
{Novak}, M., {Smol{\v{c}}i{\'c}}, V., {Civano}, F., {Bondi}, M., {Ciliegi}, P.,
  {Wang}, X., {Loeb}, A., {Banfield}, J., {Bourke}, S., {Elvis}, M.,
  {Hallinan}, G., {Intema}, H.~T., {Kl{\"o}ckner}, H.-R., {Mooley}, K.,
  {Navarrete}, F., Feb 2015. {New insights from deep VLA data on the
  potentially recoiling black hole CID-42 in the COSMOS field}. \mnras 447~(2),
  1282--1288.

\bibitem[{{Nyland} et~al.(2018){Nyland}, {Harwood}, {Mukherjee}, {Jagannathan},
  {Rujopakarn}, {Emonts}, {Alatalo}, {Bicknell}, {Davis}, {Greene}, {Kimball},
  {Lacy}, {Lonsdale}, {Lonsdale}, {Maksym}, {Moln{\'a}r}, {Morabito}, {Murphy},
  {Patil}, {Prandoni}, {Sargent}, and {Vlahakis}}]{Nyland2018}
{Nyland}, K., {Harwood}, J.~J., {Mukherjee}, D., {Jagannathan}, P.,
  {Rujopakarn}, W., {Emonts}, B., {Alatalo}, K., {Bicknell}, G.~V., {Davis},
  T.~A., {Greene}, J.~E., {Kimball}, A., {Lacy}, M., {Lonsdale}, C.,
  {Lonsdale}, C., {Maksym}, W.~P., {Moln{\'a}r}, D.~C., {Morabito}, L.,
  {Murphy}, E.~J., {Patil}, P., {Prandoni}, I., {Sargent}, M., {Vlahakis}, C.,
  May 2018. {Revolutionizing Our Understanding of AGN Feedback and its
  Importance to Galaxy Evolution in the Era of the Next Generation Very Large
  Array}. \apj 859, 23.

\bibitem[{{O'Brien} et~al.(1998){O'Brien}, {Dietrich}, {Leighly}, {Alloin},
  {Clavel}, {Crenshaw}, {Horne}, {Kriss}, {Krolik}, {Malkan}, {Netzer},
  {Peterson}, {Reichert}, {Rodr{\'{\i}}guez-Pascual}, {Wamsteker}, {Anderson},
  {Bochkarev}, {Cheng}, {Filippenko}, {Gaskell}, {George}, {Goad}, {Ho},
  {Kaspi}, {Kollatschny}, {Korista}, {MacAlpine}, {Marlow}, {Martin}, {Morris},
  {Pogge}, {Qian}, {Recondo-Gonzalez}, {Espinosa}, {Santos-Lle{\'o}},
  {Shapovalova}, {Shull}, {Stirpe}, {Sun}, {Turner}, {Vio}, {Wagner},
  {Wanders}, {Wills}, {Wu}, {Xue}, and {Zou}}]{OBrien1998}
{O'Brien}, P.~T., {Dietrich}, M., {Leighly}, K., {Alloin}, D., {Clavel}, J.,
  {Crenshaw}, D.~M., {Horne}, K., {Kriss}, G.~A., {Krolik}, J.~H., {Malkan},
  M.~A., {Netzer}, H., {Peterson}, B.~M., {Reichert}, G.~A.,
  {Rodr{\'{\i}}guez-Pascual}, P.~M., {Wamsteker}, W., {Anderson}, K.~S.~J.,
  {Bochkarev}, N.~G., {Cheng}, F.-Z., {Filippenko}, A.~V., {Gaskell}, C.~M.,
  {George}, I.~M., {Goad}, M.~R., {Ho}, L.~C., {Kaspi}, S., {Kollatschny}, W.,
  {Korista}, K.~T., {MacAlpine}, G., {Marlow}, D., {Martin}, P.~G., {Morris},
  S.~L., {Pogge}, R.~W., {Qian}, B.-C., {Recondo-Gonzalez}, M.~C., {Espinosa},
  J.~M.~R., {Santos-Lle{\'o}}, M., {Shapovalova}, A.~I., {Shull}, J.~M.,
  {Stirpe}, G.~M., {Sun}, W.-H., {Turner}, T.~J., {Vio}, R., {Wagner}, S.,
  {Wanders}, I., {Wills}, K.~A., {Wu}, H., {Xue}, S.-J., {Zou}, Z.-L., Dec.
  1998. {Steps toward Determination of the Size and Structure of the Broad-Line
  Region in Active Galactic Nuclei. XIII. Ultraviolet Observations of the
  Broad-Line Radio Galaxy 3C 390.3}. \apj 509, 163--176.

\bibitem[{{O'Dea}(1998)}]{1998PASP..110..493O}
{O'Dea}, C.~P., May 1998. {The Compact Steep-Spectrum and Gigahertz
  Peaked-Spectrum Radio Sources}. \pasp 110, 493--532.

\bibitem[{{Okamoto} et~al.(2008){Okamoto}, {Nemmen}, and
  {Bower}}]{Okamoto_et_al_2008}
{Okamoto}, T., {Nemmen}, R.~S., {Bower}, R.~G., Mar. 2008. {The impact of radio
  feedback from active galactic nuclei in cosmological simulations: formation
  of disc galaxies}. \mnras 385, 161--180.

\bibitem[{{Orosz} and {Frey}(2013)}]{Orosz2013}
{Orosz}, G., {Frey}, S., May 2013. {Optical-radio positional offsets for active
  galactic nuclei}. \aap 553, A13.

\bibitem[{{Owen} et~al.(1985){Owen}, {O'Dea}, {Inoue}, and {Eilek}}]{owen1985}
{Owen}, F.~N., {O'Dea}, C.~P., {Inoue}, M., {Eilek}, J.~A., Jul. 1985. {VLA
  observations of the multiple jet galaxy 3C 75}. Astrophysical Journal Letters
  294, L85--L88.

\bibitem[{{Paardekooper} and {Mellema}(2006)}]{PM2006}
{Paardekooper}, S.-J., {Mellema}, G., Nov. 2006. {Halting type I planet
  migration in non-isothermal disks}. \aap 459, L17--L20.

\bibitem[{{Padovani} et~al.(2017){Padovani}, {Combes}, {Diaz Trigo}, {Ettori},
  {Hatziminaoglou}, {Jonker}, {Salvato}, {Viti}, {Adami}, {Aird}, {Alexander},
  {Casella}, {Ceccarelli}, {Churazov}, {Cirasuolo}, {Daddi}, {Edge},
  {Feruglio}, {Mainieri}, {Markoff}, {Merloni}, {Nicastro}, {O'Brien},
  {Oskinova}, {Panessa}, {Pointecouteau}, {Rau}, {Robrade}, {Schaye}, {Stoehr},
  {Testi}, and {Tombesi}}]{Padovani2017}
{Padovani}, P., {Combes}, F., {Diaz Trigo}, M., {Ettori}, S., {Hatziminaoglou},
  E., {Jonker}, P., {Salvato}, M., {Viti}, S., {Adami}, C., {Aird}, J.,
  {Alexander}, D., {Casella}, P., {Ceccarelli}, C., {Churazov}, E.,
  {Cirasuolo}, M., {Daddi}, E., {Edge}, A., {Feruglio}, C., {Mainieri}, V.,
  {Markoff}, S., {Merloni}, A., {Nicastro}, F., {O'Brien}, P., {Oskinova}, L.,
  {Panessa}, F., {Pointecouteau}, E., {Rau}, A., {Robrade}, J., {Schaye}, J.,
  {Stoehr}, F., {Testi}, L., {Tombesi}, F., May 2017. {ESO-Athena Synergy White
  Paper}. arXiv e-prints.

\bibitem[{{Pakmor} et~al.(2011){Pakmor}, {Bauer}, and
  {Springel}}]{Pakmor_et_al_2011}
{Pakmor}, R., {Bauer}, A., {Springel}, V., Dec. 2011. {Magnetohydrodynamics on
  an unstructured moving grid}. \mnras 418, 1392--1401.

\bibitem[{{Pakmor} and {Springel}(2013)}]{Pakmor_Springel_2013}
{Pakmor}, R., {Springel}, V., Jun. 2013. {Simulations of magnetic fields in
  isolated disc galaxies}. \mnras 432, 176--193.

\bibitem[{{Palenzuela} et~al.(2010){Palenzuela}, {Lehner}, and
  {Liebling}}]{2010Sci...329..927P}
{Palenzuela}, C., {Lehner}, L., {Liebling}, S.~L., Aug. 2010. {Dual Jets from
  Binary Black Holes}. Science 329, 927--930.

\bibitem[{{Paragi} et~al.(2015){Paragi}, {Godfrey}, {Reynolds}, {Rioja},
  {Deller}, {Zhang}, {Gurvits}, {Bietenholz}, {Szomoru}, {Bignall}, {Boven},
  {Charlot}, {Dodson}, {Frey}, {Garrett}, {Imai}, {Lobanov}, {Reid}, {Ros},
  {van Langevelde}, {Zensus}, {Zheng}, {Alberdi}, {Agudo}, {An}, {Argo},
  {Beswick}, {Biggs}, {Brunthaler}, {Campbell}, {Cimo}, {Colomer}, {Corbel},
  {Conway}, {Cseh}, {Deane}, {Falcke}, {Gawronski}, {Gaylard}, {Giovannini},
  {Giroletti}, {Goddi}, {Goedhart}, {G{\'o}mez}, {Gunn}, {Kharb}, {Kloeckner},
  {Koerding}, {Kovalev}, {Kunert-Bajraszewska}, {Lindqvist}, {Lister},
  {Mantovani}, {Marti-Vidal}, {Mezcua}, {McKean}, {Middelberg}, {Miller-Jones},
  {Moldon}, {Muxlow}, {O'Brien}, {Perez-Torres}, {Pogrebenko}, {Quick},
  {Rushton}, {Schilizzi}, {Smirnov}, {Sohn}, {Surcis}, {Taylor}, {Tingay},
  {Tudose}, {van der Horst}, {van Leeuwen}, {Venturi}, {Vermeulen},
  {Vlemmings}, {de Witt}, {Wucknitz}, {Yang}, {Gab{\"a}nyi}, and
  {Jung}}]{Paragi2015}
{Paragi}, Z., {Godfrey}, L., {Reynolds}, C., {Rioja}, M.~J., {Deller}, A.,
  {Zhang}, B., {Gurvits}, L., {Bietenholz}, M., {Szomoru}, A., {Bignall},
  H.~E., {Boven}, P., {Charlot}, P., {Dodson}, R., {Frey}, S., {Garrett},
  M.~A., {Imai}, H., {Lobanov}, A., {Reid}, M.~J., {Ros}, E., {van Langevelde},
  H.~J., {Zensus}, A.~J., {Zheng}, X.~W., {Alberdi}, A., {Agudo}, I., {An}, T.,
  {Argo}, M., {Beswick}, R., {Biggs}, A., {Brunthaler}, A., {Campbell}, B.,
  {Cimo}, G., {Colomer}, F., {Corbel}, S., {Conway}, J.~E., {Cseh}, D.,
  {Deane}, R., {Falcke}, H.~D.~E., {Gawronski}, M., {Gaylard}, M.,
  {Giovannini}, G., {Giroletti}, M., {Goddi}, C., {Goedhart}, S., {G{\'o}mez},
  J.~L., {Gunn}, A., {Kharb}, P., {Kloeckner}, H.~R., {Koerding}, E.,
  {Kovalev}, Y., {Kunert-Bajraszewska}, M., {Lindqvist}, M., {Lister}, M.,
  {Mantovani}, F., {Marti-Vidal}, I., {Mezcua}, M., {McKean}, J., {Middelberg},
  E., {Miller-Jones}, J.~C.~A., {Moldon}, J., {Muxlow}, T., {O'Brien}, T.,
  {Perez-Torres}, M., {Pogrebenko}, S.~V., {Quick}, J., {Rushton}, A.,
  {Schilizzi}, R., {Smirnov}, O., {Sohn}, B.~W., {Surcis}, G., {Taylor}, G.~B.,
  {Tingay}, S., {Tudose}, V.~M., {van der Horst}, A., {van Leeuwen}, J.,
  {Venturi}, T., {Vermeulen}, R., {Vlemmings}, W.~H.~T., {de Witt}, A.,
  {Wucknitz}, O., {Yang}, J., {Gab{\"a}nyi}, K., {Jung}, T., Apr. 2015. {Very
  Long Baseline Interferometry with the SKA}. Advancing Astrophysics with the
  Square Kilometre Array (AASKA14), 143.

\bibitem[{{Park} and {Bogdanovi{\'c}}(2017)}]{Park_Bogdanovic_2017}
{Park}, K., {Bogdanovi{\'c}}, T., Apr. 2017. {Gaseous Dynamical Friction in
  Presence of Black Hole Radiative Feedback}. \apj 838, 103.

\bibitem[{{Peng} et~al.(2011){Peng}, {Chen}, {Gu}, and {Hu}}]{peng2011}
{Peng}, Z.-X., {Chen}, Y.-M., {Gu}, Q.-S., {Hu}, C., Apr. 2011. {Quasar SDSS
  J142507.32+323137.4: dual AGNs?} Research in Astronomy and Astrophysics 11,
  411--418.

\bibitem[{{Peres}(1962)}]{Peres1962}
{Peres}, A., Dec 1962. {Classical Radiation Recoil}. Physical Review 128~(5),
  2471--2475.

\bibitem[{{Perley} et~al.(2017){Perley}, {Perley}, {Dhawan}, and
  {Carilli}}]{Perley2017}
{Perley}, D.~A., {Perley}, R.~A., {Dhawan}, V., {Carilli}, C.~L., Jun. 2017.
  {Discovery of a Luminous Radio Transient 460 pc from the Central Supermassive
  Black Hole in Cygnus A}. \apj 841, 117.

\bibitem[{{Peters} and {Mathews}(1963)}]{1963PhRv..131..435P}
{Peters}, P.~C., {Mathews}, J., Jul. 1963. {Gravitational Radiation from Point
  Masses in a Keplerian Orbit}. Physical Review 131, 435--440.

\bibitem[{{Peterson} and {Wandel}(2000)}]{Peterson2000}
{Peterson}, B.~M., {Wandel}, A., Sep. 2000. {Evidence for Supermassive Black
  Holes in Active Galactic Nuclei from Emission-Line Reverberation}. \apjl 540,
  L13--L16.

\bibitem[{{Petric} et~al.(2011){Petric}, {Armus}, {Howell}, {Chan},
  {Mazzarella}, {Evans}, {Surace}, {Sanders}, {Appleton}, {Charmandaris},
  {D{\'{\i}}az-Santos}, {Frayer}, {Haan}, {Inami}, {Iwasawa}, {Kim}, {Madore},
  {Marshall}, {Spoon}, {Stierwalt}, {Sturm}, {U}, {Vavilkin}, and
  {Veilleux}}]{petricetal11}
{Petric}, A.~O., {Armus}, L., {Howell}, J., {Chan}, B., {Mazzarella}, J.~M.,
  {Evans}, A.~S., {Surace}, J.~A., {Sanders}, D., {Appleton}, P.,
  {Charmandaris}, V., {D{\'{\i}}az-Santos}, T., {Frayer}, D., {Haan}, S.,
  {Inami}, H., {Iwasawa}, K., {Kim}, D., {Madore}, B., {Marshall}, J., {Spoon},
  H., {Stierwalt}, S., {Sturm}, E., {U}, V., {Vavilkin}, T., {Veilleux}, S.,
  Mar. 2011. {Mid-Infrared Spectral Diagnostics of Luminous Infrared Galaxies}.
  \apj 730, 28.

\bibitem[{{Petts} et~al.(2015){Petts}, {Gualandris}, and
  {Read}}]{Petts_et_al_2015}
{Petts}, J.~A., {Gualandris}, A., {Read}, J.~I., Dec. 2015. {A semi-analytic
  dynamical friction model that reproduces core stalling}. \mnras 454,
  3778--3791.

\bibitem[{{Petts} et~al.(2016){Petts}, {Read}, and
  {Gualandris}}]{Petts_et_al_2016}
{Petts}, J.~A., {Read}, J.~I., {Gualandris}, A., Nov. 2016. {A semi-analytic
  dynamical friction model for cored galaxies}. \mnras 463, 858--869.

\bibitem[{{Pfeifle} et~al.(2019{\natexlab{a}}){Pfeifle}, {Satyapal},
  {Manzano-King}, {Cann}, {Sexton}, {Rothberg}, {Canalizo}, {Ricci}, {Blecha},
  {Ellison}, {Gliozzi}, {Secrest}, {Constantin}, and {Harvey}}]{Pfeifle2019b}
{Pfeifle}, R.~W., {Satyapal}, S., {Manzano-King}, C., {Cann}, J., {Sexton},
  R.~O., {Rothberg}, B., {Canalizo}, G., {Ricci}, C., {Blecha}, L., {Ellison},
  S.~L., {Gliozzi}, M., {Secrest}, N.~J., {Constantin}, A., {Harvey}, J.~B.,
  Oct 2019{\natexlab{a}}. {A Triple AGN in a Mid-infrared Selected Late-stage
  Galaxy Merger}. \apj 883~(2), 167.

\bibitem[{{Pfeifle} et~al.(2019{\natexlab{b}}){Pfeifle}, {Satyapal}, {Secrest},
  {Gliozzi}, {Ricci}, {Ellison}, {Rothberg}, {Cann}, {Blecha}, {Williams}, and
  {Constantin}}]{pfeifle19}
{Pfeifle}, R.~W., {Satyapal}, S., {Secrest}, N.~J., {Gliozzi}, M., {Ricci}, C.,
  {Ellison}, S.~L., {Rothberg}, B., {Cann}, J., {Blecha}, L., {Williams},
  J.~K., {Constantin}, A., Apr. 2019{\natexlab{b}}. {Buried Black Hole Growth
  in IR-selected Mergers: New Results from Chandra}. \apj 875, 117.

\bibitem[{{Pfister} et~al.(2019{\natexlab{a}}){Pfister}, {Bar-Or}, {Volonteri},
  {Dubois}, and {Capelo}}]{Pfister_et_al_2019}
{Pfister}, H., {Bar-Or}, B., {Volonteri}, M., {Dubois}, Y., {Capelo}, P.~R.,
  Sep 2019{\natexlab{a}}. {Tidal disruption event rates in galaxy merger
  remnants}. \mnras 488~(1), L29--L34.

\bibitem[{{Pfister} et~al.(2017){Pfister}, {Lupi}, {Capelo}, {Volonteri},
  {Bellovary}, and {Dotti}}]{Pfister_et_al_2017}
{Pfister}, H., {Lupi}, A., {Capelo}, P.~R., {Volonteri}, M., {Bellovary},
  J.~M., {Dotti}, M., Nov. 2017. {The birth of a supermassive black hole
  binary}. \mnras 471, 3646--3656.

\bibitem[{{Pfister} et~al.(2019{\natexlab{b}}){Pfister}, {Volonteri}, {Dubois},
  {Dotti}, and {Colpi}}]{2019MNRAS.486..101P}
{Pfister}, H., {Volonteri}, M., {Dubois}, Y., {Dotti}, M., {Colpi}, M., Jun
  2019{\natexlab{b}}. {The erratic dynamical life of black hole seeds in
  high-redshift galaxies}. \mnras 486~(1), 101--111.

\bibitem[{{Pflueger} et~al.(2018){Pflueger}, {Nguyen}, {Bogdanovi{\'c}},
  {Eracleous}, {Runnoe}, {Sigurdsson}, and {Boroson}}]{Pflueger2018}
{Pflueger}, B.~J., {Nguyen}, K., {Bogdanovi{\'c}}, T., {Eracleous}, M.,
  {Runnoe}, J.~C., {Sigurdsson}, S., {Boroson}, T., Jul. 2018. {Likelihood for
  Detection of Subparsec Supermassive Black Hole Binaries in Spectroscopic
  Surveys}. \apj 861, 59.

\bibitem[{{Piconcelli} et~al.(2010){Piconcelli}, {Vignali}, {Bianchi},
  {Mathur}, {Fiore}, {Guainazzi}, {Lanzuisi}, {Maiolino}, and
  {Nicastro}}]{Piconcelli2010}
{Piconcelli}, E., {Vignali}, C., {Bianchi}, S., {Mathur}, S., {Fiore}, F.,
  {Guainazzi}, M., {Lanzuisi}, G., {Maiolino}, R., {Nicastro}, F., Oct. 2010.
  {Witnessing the Key Early Phase of Quasar Evolution: An Obscured Active
  Galactic Nucleus Pair in the Interacting Galaxy IRAS 20210+1121}. \apjl 722,
  L147--L151.

\bibitem[{{Pillepich} et~al.(2018){Pillepich}, {Springel}, {Nelson}, {Genel},
  {Naiman}, {Pakmor}, {Hernquist}, {Torrey}, {Vogelsberger}, {Weinberger}, and
  {Marinacci}}]{Pillepich_et_al_2018}
{Pillepich}, A., {Springel}, V., {Nelson}, D., {Genel}, S., {Naiman}, J.,
  {Pakmor}, R., {Hernquist}, L., {Torrey}, P., {Vogelsberger}, M.,
  {Weinberger}, R., {Marinacci}, F., Jan. 2018. {Simulating galaxy formation
  with the IllustrisTNG model}. \mnras 473, 4077--4106.

\bibitem[{{Pilyugin} et~al.(2012){Pilyugin}, {Zinchenko}, {Cedr{\'e}s}, {Cepa},
  {Bongiovanni}, {Mattsson}, and {V{\'{\i}}lchez}}]{Pilyugin:2012}
{Pilyugin}, L.~S., {Zinchenko}, I.~A., {Cedr{\'e}s}, B., {Cepa}, J.,
  {Bongiovanni}, A., {Mattsson}, L., {V{\'{\i}}lchez}, J.~M., Jan. 2012. {SDSS
  galaxies with double-peaked emission lines: double starbursts or active
  galactic nuclei?} \mnras 419, 490--502.

\bibitem[{{Plavin} et~al.(2019){Plavin}, {Kovalev}, and {Petrov}}]{Plavin2019}
{Plavin}, A.~V., {Kovalev}, Y.~Y., {Petrov}, L.~Y., Feb. 2019. {Dissecting the
  AGN Disk--Jet System with Joint VLBI-Gaia Analysis}. \apj 871, 143.

\bibitem[{{Popovi{\'c}}(2012)}]{Popovic2012}
{Popovi{\'c}}, L.~{\v{C}}., Feb 2012. {Super-massive binary black holes and
  emission lines in active galactic nuclei}. \nar 56~(2-3), 74--91.

\bibitem[{{Prandoni} and {Seymour}(2015)}]{Prandoni_Seymour2015}
{Prandoni}, I., {Seymour}, N., Apr. 2015. {Revealing the Physics and Evolution
  of Galaxies and Galaxy Clusters with SKA Continuum Surveys}. Advancing
  Astrophysics with the Square Kilometre Array (AASKA14), 67.

\bibitem[{{Preto} et~al.(2011){Preto}, {Berentzen}, {Berczik}, and
  {Spurzem}}]{preto11}
{Preto}, M., {Berentzen}, I., {Berczik}, P., {Spurzem}, R., May 2011. {Fast
  Coalescence of Massive Black Hole Binaries from Mergers of Galactic Nuclei:
  Implications for Low-frequency Gravitational-wave Astrophysics}. \apjl 732,
  L26.

\bibitem[{{Rafikov}(2016)}]{rafikov16}
{Rafikov}, R.~R., Aug 2016. {Accretion and Orbital Inspiral in Gas-assisted
  Supermassive Black Hole Binary Mergers}. \apj 827~(2), 111.

\bibitem[{{Ragusa} et~al.(2016){Ragusa}, {Lodato}, and {Price}}]{Ragusa+2016}
{Ragusa}, E., {Lodato}, G., {Price}, D.~J., Aug. 2016. {Suppression of the
  accretion rate in thin discs around binary black holes}. \mnras 460,
  1243--1253.

\bibitem[{{Ranalli} et~al.(2013){Ranalli}, {Comastri}, {Vignali}, {Carrera},
  {Cappelluti}, {Gilli}, {Puccetti}, {Brand t}, {Brunner}, {Brusa},
  {Georgantopoulos}, {Iwasawa}, and {Mainieri}}]{Ranalli2013}
{Ranalli}, P., {Comastri}, A., {Vignali}, C., {Carrera}, F.~J., {Cappelluti},
  N., {Gilli}, R., {Puccetti}, S., {Brand t}, W.~N., {Brunner}, H., {Brusa},
  M., {Georgantopoulos}, I., {Iwasawa}, K., {Mainieri}, V., Jul 2013. {The XMM
  deep survey in the CDF-S. III. Point source catalogue and number counts in
  the hard X-rays}. \aap 555, A42.

\bibitem[{{Reardon} et~al.(2016){Reardon}, {Hobbs}, {Coles}, {Levin}, {Keith},
  {Bailes}, {Bhat}, {Burke-Spolaor}, {Dai}, {Kerr}, {Lasky}, {Manchester},
  {Os{\l}owski}, {Ravi}, {Shannon}, {van Straten}, {Toomey}, {Wang}, {Wen},
  {You}, and {Zhu}}]{2016MNRAS.455.1751R}
{Reardon}, D.~J., {Hobbs}, G., {Coles}, W., {Levin}, Y., {Keith}, M.~J.,
  {Bailes}, M., {Bhat}, N.~D.~R., {Burke-Spolaor}, S., {Dai}, S., {Kerr}, M.,
  {Lasky}, P.~D., {Manchester}, R.~N., {Os{\l}owski}, S., {Ravi}, V.,
  {Shannon}, R.~M., {van Straten}, W., {Toomey}, L., {Wang}, J., {Wen}, L.,
  {You}, X.~P., {Zhu}, X.-J., Jan. 2016. {Timing analysis for 20 millisecond
  pulsars in the Parkes Pulsar Timing Array}. \mnras 455, 1751--1769.

\bibitem[{{Reines} et~al.(2013){Reines}, {Greene}, and
  {Geha}}]{Reines_et_al_2013}
{Reines}, A.~E., {Greene}, J.~E., {Geha}, M., Oct. 2013. {Dwarf Galaxies with
  Optical Signatures of Active Massive Black Holes}. \apj 775, 116.

\bibitem[{{Remus} et~al.(2017){Remus}, {Dolag}, {Naab}, {Burkert},
  {Hirschmann}, {Hoffmann}, and {Johansson}}]{Remus_et_al_2017}
{Remus}, R.-S., {Dolag}, K., {Naab}, T., {Burkert}, A., {Hirschmann}, M.,
  {Hoffmann}, T.~L., {Johansson}, P.~H., Jan. 2017. {The co-evolution of total
  density profiles and central dark matter fractions in simulated early-type
  galaxies}. \mnras 464, 3742--3756.

\bibitem[{{Ricarte} et~al.(2019){Ricarte}, {Tremmel}, {Natarajan}, and
  {Quinn}}]{Ricarte2019}
{Ricarte}, A., {Tremmel}, M., {Natarajan}, P., {Quinn}, T., Oct 2019. {Tracing
  black hole and galaxy co-evolution in the ROMULUS simulations}. \mnras
  489~(1), 802--819.

\bibitem[{{Ricci} et~al.(2017){Ricci}, {Bauer}, {Treister}, {Schawinski},
  {Privon}, {Blecha}, {Arevalo}, {Armus}, {Harrison}, {Ho}, {Iwasawa},
  {Sanders}, and {Stern}}]{riccietal17}
{Ricci}, C., {Bauer}, F.~E., {Treister}, E., {Schawinski}, K., {Privon}, G.~C.,
  {Blecha}, L., {Arevalo}, P., {Armus}, L., {Harrison}, F., {Ho}, L.~C.,
  {Iwasawa}, K., {Sanders}, D.~B., {Stern}, D., Jun. 2017. {Growing
  supermassive black holes in the late stages of galaxy mergers are heavily
  obscured}. \mnras 468, 1273--1299.

\bibitem[{{Ricci} et~al.(2015){Ricci}, {Ueda}, {Koss}, {Trakhtenbrot}, {Bauer},
  and {Gandhi}}]{riccietal15}
{Ricci}, C., {Ueda}, Y., {Koss}, M.~J., {Trakhtenbrot}, B., {Bauer}, F.~E.,
  {Gandhi}, P., Dec. 2015. {Compton-thick Accretion in the Local Universe}.
  \apjl 815, L13.

\bibitem[{{Rieke} et~al.(2015{\natexlab{a}}){Rieke}, {Ressler}, {Morrison},
  {Bergeron}, {Bouchet}, {Garc{\'{\i}}a-Mar{\'{\i}}n}, {Greene}, {Regan},
  {Sukhatme}, and {Walker}}]{2015PASP..127..665R}
{Rieke}, G.~H., {Ressler}, M.~E., {Morrison}, J.~E., {Bergeron}, L., {Bouchet},
  P., {Garc{\'{\i}}a-Mar{\'{\i}}n}, M., {Greene}, T.~P., {Regan}, M.~W.,
  {Sukhatme}, K.~G., {Walker}, H., Jul. 2015{\natexlab{a}}. {The Mid-Infrared
  Instrument for the James Webb Space Telescope, VII: The MIRI Detectors}.
  \pasp 127, 665.

\bibitem[{{Rieke} et~al.(2015{\natexlab{b}}){Rieke}, {Wright}, {B{\"o}ker},
  {Bouwman}, {Colina}, {Glasse}, {Gordon}, {Greene}, {G{\"u}del}, {Henning},
  {Justtanont}, {Lagage}, {Meixner}, {N{\o}rgaard-Nielsen}, {Ray}, {Ressler},
  {van Dishoeck}, and {Waelkens}}]{2015PASP..127..584R}
{Rieke}, G.~H., {Wright}, G.~S., {B{\"o}ker}, T., {Bouwman}, J., {Colina}, L.,
  {Glasse}, A., {Gordon}, K.~D., {Greene}, T.~P., {G{\"u}del}, M., {Henning},
  T., {Justtanont}, K., {Lagage}, P.-O., {Meixner}, M.~E.,
  {N{\o}rgaard-Nielsen}, H.-U., {Ray}, T.~P., {Ressler}, M.~E., {van Dishoeck},
  E.~F., {Waelkens}, C., Jul. 2015{\natexlab{b}}. {The Mid-Infrared Instrument
  for the James Webb Space Telescope, I: Introduction}. \pasp 127, 584.

\bibitem[{{Roberts} et~al.(2015{\natexlab{a}}){Roberts}, {Cohen}, {Lu},
  {Saripalli}, and {Subrahmanyan}}]{2015ApJS..220....7R}
{Roberts}, D.~H., {Cohen}, J.~P., {Lu}, J., {Saripalli}, L., {Subrahmanyan},
  R., Sep 2015{\natexlab{a}}. {The Abundance of X-shaped Radio Sources. I. VLA
  Survey of 52 Sources with Off-axis Distortions}. \apjs 220~(1), 7.

\bibitem[{{Roberts} et~al.(2015{\natexlab{b}}){Roberts}, {Saripalli}, and
  {Subrahmanyan}}]{Roberts_Saripalli_2015}
{Roberts}, D.~H., {Saripalli}, L., {Subrahmanyan}, R., Sep. 2015{\natexlab{b}}.
  {The Abundance of X-shaped Radio Sources: Implications for the Gravitational
  Wave Background}. \apjl 810, L6.

\bibitem[{{Rodriguez} et~al.(2006){Rodriguez}, {Taylor}, {Zavala}, {Peck},
  {Pollack}, and {Romani}}]{rodriguez2006}
{Rodriguez}, C., {Taylor}, G.~B., {Zavala}, R.~T., {Peck}, A.~B., {Pollack},
  L.~K., {Romani}, R.~W., Jul. 2006. {A Compact Supermassive Binary Black Hole
  System}. \apj 646, 49--60.

\bibitem[{{Roedig} et~al.(2011){Roedig}, {Dotti}, {Sesana}, {Cuadra}, and
  {Colpi}}]{Roedig11}
{Roedig}, C., {Dotti}, M., {Sesana}, A., {Cuadra}, J., {Colpi}, M., Aug. 2011.
  {Limiting eccentricity of subparsec massive black hole binaries surrounded by
  self-gravitating gas discs}. \mnras 415, 3033--3041.

\bibitem[{{Roedig} et~al.(2014){Roedig}, {Krolik}, and {Miller}}]{Roedig+2014}
{Roedig}, C., {Krolik}, J.~H., {Miller}, M.~C., Apr. 2014. {Observational
  Signatures of Binary Supermassive Black Holes}. \apj 785, 115.

\bibitem[{{Roedig} and {Sesana}(2014)}]{roedig_sesana2014}
{Roedig}, C., {Sesana}, A., Apr 2014. {Migration of massive black hole binaries
  in self-gravitating discs: retrograde versus prograde}. \mnras 439~(4),
  3476--3489.

\bibitem[{{Roedig} et~al.(2012){Roedig}, {Sesana}, {Dotti}, {Cuadra},
  {Amaro-Seoane}, and {Haardt}}]{roedig12}
{Roedig}, C., {Sesana}, A., {Dotti}, M., {Cuadra}, J., {Amaro-Seoane}, P.,
  {Haardt}, F., Sep. 2012. {Evolution of binary black holes in self gravitating
  discs. Dissecting the torques}. \aap 545, A127.

\bibitem[{{Roelfsema} et~al.(2018){Roelfsema}, {Shibai}, {Armus}, {Arrazola},
  {Audard}, {Audley}, {Bradford}, {Charles}, {Dieleman}, {Doi}, {Duband},
  {Eggens}, {Evers}, {Funaki}, {Gao}, {Giard}, {di Giorgio}, {Gonz{\'a}lez
  Fern{\'a}ndez}, {Griffin}, {Helmich}, {Hijmering}, {Huisman}, {Ishihara},
  {Isobe}, {Jackson}, {Jacobs}, {Jellema}, {Kamp}, {Kaneda}, {Kawada},
  {Kemper}, {Kerschbaum}, {Khosropanah}, {Kohno}, {Kooijman}, {Krause}, {van
  der Kuur}, {Kwon}, {Laauwen}, {de Lange}, {Larsson}, {van Loon}, {Madden},
  {Matsuhara}, {Najarro}, {Nakagawa}, {Naylor}, {Ogawa}, {Onaka}, {Oyabu},
  {Poglitsch}, {Reveret}, {Rodriguez}, {Spinoglio}, {Sakon}, {Sato},
  {Shinozaki}, {Shipman}, {Sugita}, {Suzuki}, {van der Tak}, {Torres Redondo},
  {Wada}, {Wang}, {Wafelbakker}, {van Weers}, {Withington}, {Vandenbussche},
  {Yamada}, and {Yamamura}}]{roelfsema2018}
{Roelfsema}, P.~R., {Shibai}, H., {Armus}, L., {Arrazola}, D., {Audard}, M.,
  {Audley}, M.~D., {Bradford}, C.~M., {Charles}, I., {Dieleman}, P., {Doi}, Y.,
  {Duband}, L., {Eggens}, M., {Evers}, J., {Funaki}, I., {Gao}, J.~R., {Giard},
  M., {di Giorgio}, A., {Gonz{\'a}lez Fern{\'a}ndez}, L.~M., {Griffin}, M.,
  {Helmich}, F.~P., {Hijmering}, R., {Huisman}, R., {Ishihara}, D., {Isobe},
  N., {Jackson}, B., {Jacobs}, H., {Jellema}, W., {Kamp}, I., {Kaneda}, H.,
  {Kawada}, M., {Kemper}, F., {Kerschbaum}, F., {Khosropanah}, P., {Kohno}, K.,
  {Kooijman}, P.~P., {Krause}, O., {van der Kuur}, J., {Kwon}, J., {Laauwen},
  W.~M., {de Lange}, G., {Larsson}, B., {van Loon}, D., {Madden}, S.~C.,
  {Matsuhara}, H., {Najarro}, F., {Nakagawa}, T., {Naylor}, D., {Ogawa}, H.,
  {Onaka}, T., {Oyabu}, S., {Poglitsch}, A., {Reveret}, V., {Rodriguez}, L.,
  {Spinoglio}, L., {Sakon}, I., {Sato}, Y., {Shinozaki}, K., {Shipman}, R.,
  {Sugita}, H., {Suzuki}, T., {van der Tak}, F.~F.~S., {Torres Redondo}, J.,
  {Wada}, T., {Wang}, S.~Y., {Wafelbakker}, C.~K., {van Weers}, H.,
  {Withington}, S., {Vandenbussche}, B., {Yamada}, T., {Yamamura}, I., Aug
  2018. {SPICA-A Large Cryogenic Infrared Space Telescope: Unveiling the
  Obscured Universe}. \pasa 35, e030.

\bibitem[{{Roos}(1988)}]{roos1988}
{Roos}, N., Nov. 1988. {Jet precession in active galaxies}. \apj 334, 95--103.

\bibitem[{{Roos} et~al.(1993){Roos}, {Kaastra}, and {Hummel}}]{roos1993}
{Roos}, N., {Kaastra}, J.~S., {Hummel}, C.~A., May 1993. {A massive binary
  black hole in 1928 + 738?} \apj 409, 130--133.

\bibitem[{{Rorai} et~al.(2017){Rorai}, {Hennawi}, {O{\~n}orbe}, {White},
  {Prochaska}, {Kulkarni}, {Walther}, {Luki{\'c}}, and
  {Lee}}]{2017Sci...356..418R}
{Rorai}, A., {Hennawi}, J.~F., {O{\~n}orbe}, J., {White}, M., {Prochaska},
  J.~X., {Kulkarni}, G., {Walther}, M., {Luki{\'c}}, Z., {Lee}, K.-G., Apr.
  2017. {Measurement of the small-scale structure of the intergalactic medium
  using close quasar pairs}. Science 356, 418--422.

\bibitem[{{Rosado} et~al.(2015){Rosado}, {Sesana}, and
  {Gair}}]{2015MNRAS.451.2417R}
{Rosado}, P.~A., {Sesana}, A., {Gair}, J., Aug. 2015. {Expected properties of
  the first gravitational wave signal detected with pulsar timing arrays}.
  \mnras 451, 2417--2433.

\bibitem[{{Rosario} et~al.(2011){Rosario}, {McGurk}, {Max}, {Shields}, {Smith},
  and {Ammons}}]{rosario2011}
{Rosario}, D.~J., {McGurk}, R.~C., {Max}, C.~E., {Shields}, G.~A., {Smith},
  K.~L., {Ammons}, S.~M., Sep 2011. {Adaptive Optics Imaging of Quasi-stellar
  Objects with Double-peaked Narrow Lines: Are They Dual Active Galactic
  Nuclei?} \apj 739~(1), 44.

\bibitem[{{Rosario} et~al.(2010){Rosario}, {Shields}, {Taylor}, {Salviand er},
  and {Smith}}]{rosario2010}
{Rosario}, D.~J., {Shields}, G.~A., {Taylor}, G.~B., {Salviand er}, S.,
  {Smith}, K.~L., Jun 2010. {The Jet-driven Outflow in the Radio Galaxy SDSS
  J1517+3353: Implications for Double-peaked Narrow-line Active Galactic
  Nucleus}. \apj 716~(1), 131--143.

\bibitem[{{Rosas-Guevara} et~al.(2019){Rosas-Guevara}, {Bower}, {McAlpine},
  {Bonoli}, and {Tissera}}]{Rosas_Guevara_et_al_2019}
{Rosas-Guevara}, Y.~M., {Bower}, R.~G., {McAlpine}, S., {Bonoli}, S.,
  {Tissera}, P.~B., Feb 2019. {The abundances and properties of Dual AGN and
  their host galaxies in the EAGLE simulations}. \mnras 483~(2), 2712--2720.

\bibitem[{{Rosas-Guevara} et~al.(2015){Rosas-Guevara}, {Bower}, {Schaye},
  {Furlong}, {Frenk}, {Booth}, {Crain}, {Dalla Vecchia}, {Schaller}, and
  {Theuns}}]{Rosas_Guevara_et_al_2015}
{Rosas-Guevara}, Y.~M., {Bower}, R.~G., {Schaye}, J., {Furlong}, M., {Frenk},
  C.~S., {Booth}, C.~M., {Crain}, R.~A., {Dalla Vecchia}, C., {Schaller}, M.,
  {Theuns}, T., Nov. 2015. {The impact of angular momentum on black hole
  accretion rates in simulations of galaxy formation}. \mnras 454, 1038--1057.

\bibitem[{{Rosswog} et~al.(2009){Rosswog}, {Ramirez-Ruiz}, and
  {Hix}}]{Rosswog2009}
{Rosswog}, S., {Ramirez-Ruiz}, E., {Hix}, W.~R., Apr. 2009. {Tidal Disruption
  and Ignition of White Dwarfs by Moderately Massive Black Holes}. \apj 695,
  404--419.

\bibitem[{{Ro{\v s}kar} et~al.(2015){Ro{\v s}kar}, {Fiacconi}, {Mayer},
  {Kazantzidis}, {Quinn}, and {Wadsley}}]{Roskar_et_al_2015}
{Ro{\v s}kar}, R., {Fiacconi}, D., {Mayer}, L., {Kazantzidis}, S., {Quinn},
  T.~R., {Wadsley}, J., May 2015. {Orbital decay of supermassive black hole
  binaries in clumpy multiphase merger remnants}. \mnras 449, 494--505.

\bibitem[{{Rovilos} et~al.(2014){Rovilos}, {Georgantopoulos}, {Akylas}, {Aird},
  {Alexander}, {Comastri}, {Del Moro}, {Gandhi}, {Georgakakis}, {Harrison}, and
  {Mullaney}}]{Rovilos2014}
{Rovilos}, E., {Georgantopoulos}, I., {Akylas}, A., {Aird}, J., {Alexander},
  D.~M., {Comastri}, A., {Del Moro}, A., {Gandhi}, P., {Georgakakis}, A.,
  {Harrison}, C.~M., {Mullaney}, J.~R., Feb. 2014. {A wide search for obscured
  active galactic nuclei using XMM-Newton and WISE}. \mnras 438, 494--512.

\bibitem[{{Runnoe} et~al.(2015){Runnoe}, {Eracleous}, {Mathes}, {Pennell},
  {Boroson}, {Sigurdsson}, {Bogdanovi{\'c}}, {Halpern}, and {Liu}}]{Runnoe2015}
{Runnoe}, J.~C., {Eracleous}, M., {Mathes}, G., {Pennell}, A., {Boroson}, T.,
  {Sigurdsson}, S., {Bogdanovi{\'c}}, T., {Halpern}, J.~P., {Liu}, J., Nov.
  2015. {A Large Systematic Search for Close Supermassive Binary and Rapidly
  Recoiling Black Holes. II. Continued Spectroscopic Monitoring and Optical
  Flux Variability}. \apjs 221, 7.

\bibitem[{{Runnoe} et~al.(2017){Runnoe}, {Eracleous}, {Pennell}, {Mathes},
  {Boroson}, {Sigurdsson}, {Bogdanovi{\'c}}, {Halpern}, {Liu}, and
  {Brown}}]{runnoe17}
{Runnoe}, J.~C., {Eracleous}, M., {Pennell}, A., {Mathes}, G., {Boroson}, T.,
  {Sigurdsson}, S., {Bogdanovi{\'c}}, T., {Halpern}, J.~P., {Liu}, J., {Brown},
  S., Jun. 2017. {A large systematic search for close supermassive binary and
  rapidly recoiling black holes - III. Radial velocity variations}. \mnras 468,
  1683--1702.

\bibitem[{{Ryu} et~al.(2018){Ryu}, {Perna}, {Haiman}, {Ostriker}, and
  {Stone}}]{Ryu+2018}
{Ryu}, T., {Perna}, R., {Haiman}, Z., {Ostriker}, J.~P., {Stone}, N.~C., Jan.
  2018. {Interactions between multiple supermassive black holes in galactic
  nuclei: a solution to the final parsec problem}. \mnras 473, 3410--3433.

\bibitem[{{Salcido} et~al.(2016){Salcido}, {Bower}, {Theuns}, {McAlpine},
  {Schaller}, {Crain}, {Schaye}, and {Regan}}]{Salcido_et_al_2016}
{Salcido}, J., {Bower}, R.~G., {Theuns}, T., {McAlpine}, S., {Schaller}, M.,
  {Crain}, R.~A., {Schaye}, J., {Regan}, J., Nov. 2016. {Music from the heavens
  - gravitational waves from supermassive black hole mergers in the EAGLE
  simulations}. \mnras 463, 870--885.

\bibitem[{{S{\'a}nchez} et~al.(2012){S{\'a}nchez}, {Kennicutt}, {Gil de Paz},
  {van de Ven}, {V{\'{\i}}lchez}, {Wisotzki}, {Walcher}, {Mast}, {Aguerri},
  {Albiol-P{\'e}rez}, {Alonso-Herrero}, {Alves}, {Bakos}, {Bart{\'a}kov{\'a}},
  {Bland-Hawthorn}, {Boselli}, {Bomans}, {Castillo-Morales}, {Cortijo-Ferrero},
  {de Lorenzo-C{\'a}ceres}, {Del Olmo}, {Dettmar}, {D{\'{\i}}az}, {Ellis},
  {Falc{\'o}n-Barroso}, {Flores}, {Gallazzi}, {Garc{\'{\i}}a-Lorenzo},
  {Gonz{\'a}lez Delgado}, {Gruel}, {Haines}, {Hao}, {Husemann},
  {Igl{\'e}sias-P{\'a}ramo}, {Jahnke}, {Johnson}, {Jungwiert}, {Kalinova},
  {Kehrig}, {Kupko}, {L{\'o}pez-S{\'a}nchez}, {Lyubenova}, {Marino},
  {M{\'a}rmol-Queralt{\'o}}, {M{\'a}rquez}, {Masegosa}, {Meidt},
  {Mendez-Abreu}, {Monreal-Ibero}, {Montijo}, {Mour{\~a}o}, {Palacios-Navarro},
  {Papaderos}, {Pasquali}, {Peletier}, {P{\'e}rez}, {P{\'e}rez}, {Quirrenbach},
  {Rela{\~n}o}, {Rosales-Ortega}, {Roth}, {Ruiz-Lara},
  {S{\'a}nchez-Bl{\'a}zquez}, {Sengupta}, {Singh}, {Stanishev}, {Trager},
  {Vazdekis}, {Viironen}, {Wild}, {Zibetti}, and {Ziegler}}]{Sanchez2012}
{S{\'a}nchez}, S.~F., {Kennicutt}, R.~C., {Gil de Paz}, A., {van de Ven}, G.,
  {V{\'{\i}}lchez}, J.~M., {Wisotzki}, L., {Walcher}, C.~J., {Mast}, D.,
  {Aguerri}, J.~A.~L., {Albiol-P{\'e}rez}, S., {Alonso-Herrero}, A., {Alves},
  J., {Bakos}, J., {Bart{\'a}kov{\'a}}, T., {Bland-Hawthorn}, J., {Boselli},
  A., {Bomans}, D.~J., {Castillo-Morales}, A., {Cortijo-Ferrero}, C., {de
  Lorenzo-C{\'a}ceres}, A., {Del Olmo}, A., {Dettmar}, R.-J., {D{\'{\i}}az},
  A., {Ellis}, S., {Falc{\'o}n-Barroso}, J., {Flores}, H., {Gallazzi}, A.,
  {Garc{\'{\i}}a-Lorenzo}, B., {Gonz{\'a}lez Delgado}, R., {Gruel}, N.,
  {Haines}, T., {Hao}, C., {Husemann}, B., {Igl{\'e}sias-P{\'a}ramo}, J.,
  {Jahnke}, K., {Johnson}, B., {Jungwiert}, B., {Kalinova}, V., {Kehrig}, C.,
  {Kupko}, D., {L{\'o}pez-S{\'a}nchez}, {\'A}.~R., {Lyubenova}, M., {Marino},
  R.~A., {M{\'a}rmol-Queralt{\'o}}, E., {M{\'a}rquez}, I., {Masegosa}, J.,
  {Meidt}, S., {Mendez-Abreu}, J., {Monreal-Ibero}, A., {Montijo}, C.,
  {Mour{\~a}o}, A.~M., {Palacios-Navarro}, G., {Papaderos}, P., {Pasquali}, A.,
  {Peletier}, R., {P{\'e}rez}, E., {P{\'e}rez}, I., {Quirrenbach}, A.,
  {Rela{\~n}o}, M., {Rosales-Ortega}, F.~F., {Roth}, M.~M., {Ruiz-Lara}, T.,
  {S{\'a}nchez-Bl{\'a}zquez}, P., {Sengupta}, C., {Singh}, R., {Stanishev}, V.,
  {Trager}, S.~C., {Vazdekis}, A., {Viironen}, K., {Wild}, V., {Zibetti}, S.,
  {Ziegler}, B., Feb. 2012. {CALIFA, the Calar Alto Legacy Integral Field Area
  survey. I. Survey presentation}. \aap 538, A8.

\bibitem[{{Sanders} and {Mirabel}(1996)}]{1996ARA&A..34..749S}
{Sanders}, D.~B., {Mirabel}, I.~F., Jan. 1996. {Luminous Infrared Galaxies}.
  Annual Review of Astronomy and Astrophysics 34, 749.

\bibitem[{{Sanders} et~al.(1988){Sanders}, {Soifer}, {Elias}, {Madore},
  {Matthews}, {Neugebauer}, and {Scoville}}]{Sanders1988}
{Sanders}, D.~B., {Soifer}, B.~T., {Elias}, J.~H., {Madore}, B.~F., {Matthews},
  K., {Neugebauer}, G., {Scoville}, N.~Z., Feb 1988. {Ultraluminous Infrared
  Galaxies and the Origin of Quasars}. \apj 325, 74.

\bibitem[{{Sandrinelli} et~al.(2016){Sandrinelli}, {Covino}, {Dotti}, and
  {Treves}}]{Sandrinelli_et_al_2016}
{Sandrinelli}, A., {Covino}, S., {Dotti}, M., {Treves}, A., Mar. 2016.
  {Quasi-periodicities at Year-like Timescales in Blazars}. \aj 151, 54.

\bibitem[{{Sandrinelli} et~al.(2018{\natexlab{a}}){Sandrinelli}, {Covino},
  {Treves}, {Holgado}, {Sesana}, {Lindfors}, and
  {Ramazani}}]{Sandrinelli_et_al_2018}
{Sandrinelli}, A., {Covino}, S., {Treves}, A., {Holgado}, A.~M., {Sesana}, A.,
  {Lindfors}, E., {Ramazani}, V.~F., Jul 2018{\natexlab{a}}.
  {Quasi-periodicities of BL Lacertae objects}. \aap 615, A118.

\bibitem[{{Sandrinelli} et~al.(2014){Sandrinelli}, {Falomo}, {Treves},
  {Farina}, and {Uslenghi}}]{2014MNRAS.444.1835S}
{Sandrinelli}, A., {Falomo}, R., {Treves}, A., {Farina}, E.~P., {Uslenghi}, M.,
  Oct. 2014. {The environment of low-redshift quasar pairs}. \mnras 444,
  1835--1841.

\bibitem[{{Sandrinelli} et~al.(2018{\natexlab{b}}){Sandrinelli}, {Falomo},
  {Treves}, {Scarpa}, and {Uslenghi}}]{2018MNRAS.474.4925S}
{Sandrinelli}, A., {Falomo}, R., {Treves}, A., {Scarpa}, R., {Uslenghi}, M.,
  Mar. 2018{\natexlab{b}}. {Overdensity of galaxies in the environment of
  quasar pairs}. \mnras 474, 4925--4936.

\bibitem[{{Saripalli} and {Roberts}(2018)}]{Saripalli_Roberts_2018}
{Saripalli}, L., {Roberts}, D.~H., Jan. 2018. {What Are X-shaped Radio Sources
  Telling Us? II. Properties of a Sample of 87}. \apj 852, 48.

\bibitem[{{Sathyaprakash} and {Schutz}(2009)}]{LRRsathya09}
{Sathyaprakash}, B.~S., {Schutz}, B.~F., Mar 2009. {Physics, Astrophysics and
  Cosmology with Gravitational Waves}. Living Reviews in Relativity 12~(1), 2.

\bibitem[{{Satyapal} et~al.(2018){Satyapal}, {Abel}, and
  {Secrest}}]{satyapal2018}
{Satyapal}, S., {Abel}, N.~P., {Secrest}, N.~J., May 2018. {Star-forming
  Galaxies as AGN Imposters? A Theoretical Investigation of the Mid-infrared
  Colors of AGNs and Extreme Starbursts}. \apj 858, 38.

\bibitem[{{Satyapal} et~al.(2014){Satyapal}, {Ellison}, {McAlpine}, {Hickox},
  {Patton}, and {Mendel}}]{satyapaletal14}
{Satyapal}, S., {Ellison}, S.~L., {McAlpine}, W., {Hickox}, R.~C., {Patton},
  D.~R., {Mendel}, J.~T., Jun. 2014. {Galaxy pairs in the Sloan Digital Sky
  Survey - IX. Merger-induced AGN activity as traced by the Wide-field Infrared
  Survey Explorer}. \mnras 441, 1297--1304.

\bibitem[{{Satyapal} et~al.(2017){Satyapal}, {Secrest}, {Ricci}, {Ellison},
  {Rothberg}, {Blecha}, {Constantin}, {Gliozzi}, {McNulty}, and
  {Ferguson}}]{satyapaletal17}
{Satyapal}, S., {Secrest}, N.~J., {Ricci}, C., {Ellison}, S.~L., {Rothberg},
  B., {Blecha}, L., {Constantin}, A., {Gliozzi}, M., {McNulty}, P., {Ferguson},
  J., Oct. 2017. {Buried AGNs in Advanced Mergers: Mid-infrared Color Selection
  as a Dual AGN Candidate Finder}. \apj 848, 126.

\bibitem[{{Saxton} et~al.(2012){Saxton}, {Read}, {Esquej}, {Komossa},
  {Dougherty}, {Rodriguez-Pascual}, and {Barrado}}]{saxton2012}
{Saxton}, R.~D., {Read}, A.~M., {Esquej}, P., {Komossa}, S., {Dougherty}, S.,
  {Rodriguez-Pascual}, P., {Barrado}, D., May 2012. {A tidal disruption-like
  X-ray flare from the quiescent galaxy SDSS J120136.02+300305.5}. \aap 541,
  A106.

\bibitem[{{Schaye} et~al.(2015){Schaye}, {Crain}, {Bower}, {Furlong},
  {Schaller}, {Theuns}, {Dalla Vecchia}, {Frenk}, {McCarthy}, {Helly},
  {Jenkins}, {Rosas-Guevara}, {White}, {Baes}, {Booth}, {Camps}, {Navarro},
  {Qu}, {Rahmati}, {Sawala}, {Thomas}, and {Trayford}}]{Schaye_et_al_2015}
{Schaye}, J., {Crain}, R.~A., {Bower}, R.~G., {Furlong}, M., {Schaller}, M.,
  {Theuns}, T., {Dalla Vecchia}, C., {Frenk}, C.~S., {McCarthy}, I.~G.,
  {Helly}, J.~C., {Jenkins}, A., {Rosas-Guevara}, Y.~M., {White}, S.~D.~M.,
  {Baes}, M., {Booth}, C.~M., {Camps}, P., {Navarro}, J.~F., {Qu}, Y.,
  {Rahmati}, A., {Sawala}, T., {Thomas}, P.~A., {Trayford}, J., Jan. 2015. {The
  EAGLE project: simulating the evolution and assembly of galaxies and their
  environments}. \mnras 446, 521--554.

\bibitem[{{Schaye} and {Dalla Vecchia}(2008)}]{Schaye_DallaVecchia_2008}
{Schaye}, J., {Dalla Vecchia}, C., Jan. 2008. {On the relation between the
  Schmidt and Kennicutt-Schmidt star formation laws and its implications for
  numerical simulations}. \mnras 383, 1210--1222.

\bibitem[{{Sesana} et~al.(2011{\natexlab{a}}){Sesana}, {Gair}, {Berti}, and
  {Volonteri}}]{2011PhRvD..83d4036S}
{Sesana}, A., {Gair}, J., {Berti}, E., {Volonteri}, M., Feb.
  2011{\natexlab{a}}. {Reconstructing the massive black hole cosmic history
  through gravitational waves}. \prd 83~(4), 044036.

\bibitem[{{Sesana} et~al.(2011{\natexlab{b}}){Sesana}, {Gualandris}, and
  {Dotti}}]{sesana11}
{Sesana}, A., {Gualandris}, A., {Dotti}, M., Jul. 2011{\natexlab{b}}. {Massive
  black hole binary eccentricity in rotating stellar systems}. \mnras 415,
  L35--L39.

\bibitem[{{Sesana} et~al.(2018){Sesana}, {Haiman}, {Kocsis}, and
  {Kelley}}]{Sesana2018}
{Sesana}, A., {Haiman}, Z., {Kocsis}, B., {Kelley}, L.~Z., Mar. 2018. {Testing
  the Binary Hypothesis: Pulsar Timing Constraints on Supermassive Black Hole
  Binary Candidates}. \apj 856, 42.

\bibitem[{{Sesana} et~al.(2012{\natexlab{a}}){Sesana}, {Roedig}, {Reynolds},
  and {Dotti}}]{2012MNRAS.420..860S}
{Sesana}, A., {Roedig}, C., {Reynolds}, M.~T., {Dotti}, M., Feb
  2012{\natexlab{a}}. {Multimessenger astronomy with pulsar timing and X-ray
  observations of massive black hole binaries}. \mnras 420~(1), 860--877.

\bibitem[{{Sesana} et~al.(2012{\natexlab{b}}){Sesana}, {Roedig}, {Reynolds},
  and {Dotti}}]{Sesana12}
{Sesana}, A., {Roedig}, C., {Reynolds}, M.~T., {Dotti}, M., Feb.
  2012{\natexlab{b}}. {Multimessenger astronomy with pulsar timing and X-ray
  observations of massive black hole binaries}. \mnras 420, 860--877.

\bibitem[{{Sesana} and {Vecchio}(2010)}]{2010PhRvD..81j4008S}
{Sesana}, A., {Vecchio}, A., May 2010. {Measuring the parameters of massive
  black hole binary systems with pulsar timing array observations of
  gravitational waves}. \prd 81~(10), 104008.

\bibitem[{{Sesana} et~al.(2008){Sesana}, {Vecchio}, and
  {Colacino}}]{2008MNRAS.390..192S}
{Sesana}, A., {Vecchio}, A., {Colacino}, C.~N., Oct. 2008. {The stochastic
  gravitational-wave background from massive black hole binary systems:
  implications for observations with Pulsar Timing Arrays}. \mnras 390,
  192--209.

\bibitem[{{Severgnini} et~al.(2018){Severgnini}, {Cicone}, {Della Ceca},
  {Braito}, {Caccianiga}, {Ballo}, {Campana}, {Moretti}, {La Parola},
  {Vignali}, {Zaino}, {Matzeu}, and {Landoni}}]{Severgnini2018}
{Severgnini}, P., {Cicone}, C., {Della Ceca}, R., {Braito}, V., {Caccianiga},
  A., {Ballo}, L., {Campana}, S., {Moretti}, A., {La Parola}, V., {Vignali},
  C., {Zaino}, A., {Matzeu}, G.~A., {Landoni}, M., Sep. 2018. {Swift data hint
  at a binary supermassive black hole candidate at sub-parsec separation}.
  \mnras 479, 3804--3813.

\bibitem[{{Shakura} and {Sunyaev}(1973)}]{SS1973}
{Shakura}, N.~I., {Sunyaev}, R.~A., 1973. {Black holes in binary systems.
  Observational appearance.} \aap 24, 337--355.

\bibitem[{{Shappee} et~al.(2014){Shappee}, {Prieto}, {Grupe}, {Kochanek},
  {Stanek}, {De Rosa}, {Mathur}, {Zu}, {Peterson}, {Pogge}, {Komossa}, {Im},
  {Jencson}, {Holoien}, {Basu}, {Beacom}, {Szczygie{\l}}, {Brimacombe},
  {Adams}, {Campillay}, {Choi}, {Contreras}, {Dietrich}, {Dubberley},
  {Elphick}, {Foale}, {Giustini}, {Gonzalez}, {Hawkins}, {Howell}, {Hsiao},
  {Koss}, {Leighly}, {Morrell}, {Mudd}, {Mullins}, {Nugent}, {Parrent},
  {Phillips}, {Pojmanski}, {Rosing}, {Ross}, {Sand}, {Terndrup}, {Valenti},
  {Walker}, and {Yoon}}]{ASASSN_1}
{Shappee}, B.~J., {Prieto}, J.~L., {Grupe}, D., {Kochanek}, C.~S., {Stanek},
  K.~Z., {De Rosa}, G., {Mathur}, S., {Zu}, Y., {Peterson}, B.~M., {Pogge},
  R.~W., {Komossa}, S., {Im}, M., {Jencson}, J., {Holoien}, T.~W.-S., {Basu},
  U., {Beacom}, J.~F., {Szczygie{\l}}, D.~M., {Brimacombe}, J., {Adams}, S.,
  {Campillay}, A., {Choi}, C., {Contreras}, C., {Dietrich}, M., {Dubberley},
  M., {Elphick}, M., {Foale}, S., {Giustini}, M., {Gonzalez}, C., {Hawkins},
  E., {Howell}, D.~A., {Hsiao}, E.~Y., {Koss}, M., {Leighly}, K.~M., {Morrell},
  N., {Mudd}, D., {Mullins}, D., {Nugent}, J.~M., {Parrent}, J., {Phillips},
  M.~M., {Pojmanski}, G., {Rosing}, W., {Ross}, R., {Sand}, D., {Terndrup},
  D.~M., {Valenti}, S., {Walker}, Z., {Yoon}, Y., Jun. 2014. {The Man behind
  the Curtain: X-Rays Drive the UV through NIR Variability in the 2013 Active
  Galactic Nucleus Outburst in NGC 2617}. \apj 788, 48.

\bibitem[{{Shen} et~al.(2013){Shen}, {Liu}, {Loeb}, and {Tremaine}}]{shen13}
{Shen}, Y., {Liu}, X., {Loeb}, A., {Tremaine}, S., Sep. 2013. {Constraining
  Sub-parsec Binary Supermassive Black Holes in Quasars with Multi-epoch
  Spectroscopy. I. The General Quasar Population}. \apj 775, 49.

\bibitem[{{Shen} and {Loeb}(2010)}]{Shen2010}
{Shen}, Y., {Loeb}, A., Dec. 2010. {Identifying Supermassive Black Hole
  Binaries with Broad Emission Line Diagnosis}. \apj 725, 249--260.

\bibitem[{{Shi} and {Krolik}(2015)}]{ShiKrolik2015}
{Shi}, J.-M., {Krolik}, J.~H., Jul. 2015. {Three-dimensional MHD Simulation of
  Circumbinary Accretion Disks. II. Net Accretion Rate}. \apj 807, 131.

\bibitem[{{Shi} et~al.(2012){Shi}, {Krolik}, {Lubow}, and {Hawley}}]{shi12}
{Shi}, J.-M., {Krolik}, J.~H., {Lubow}, S.~H., {Hawley}, J.~F., Apr. 2012.
  {Three-dimensional Magnetohydrodynamic Simulations of Circumbinary Accretion
  Disks: Disk Structures and Angular Momentum Transport}. \apj 749, 118.

\bibitem[{{Shi} et~al.(2014){Shi}, {Luo}, {Comte}, {Chen}, {Wei}, {Zhao}, {Wu},
  {Zhang}, {Shen}, {Yang}, {Wu}, {Wu}, {Zhang}, {Lei}, {Zhang}, {Wang}, {Jin},
  and {Zhang}}]{Shi:2014}
{Shi}, Z.-X., {Luo}, A.-L., {Comte}, G., {Chen}, X.-Y., {Wei}, P., {Zhao},
  Y.-H., {Wu}, F.-C., {Zhang}, Y.-X., {Shen}, S.-Y., {Yang}, M., {Wu}, H.,
  {Wu}, X.-B., {Zhang}, H.-T., {Lei}, Y.-J., {Zhang}, J.-N., {Wang}, T.-G.,
  {Jin}, G., {Zhang}, Y., Oct. 2014. {A search for double-peaked narrow
  emission line galaxies and AGNs in the LAMOST DR1}. Research in Astronomy and
  Astrophysics 14, 1234--1250.

\bibitem[{{Sijacki} et~al.(2015){Sijacki}, {Vogelsberger}, {Genel}, {Springel},
  {Torrey}, {Snyder}, {Nelson}, and {Hernquist}}]{Sijacki_et_al_2015}
{Sijacki}, D., {Vogelsberger}, M., {Genel}, S., {Springel}, V., {Torrey}, P.,
  {Snyder}, G.~F., {Nelson}, D., {Hernquist}, L., Sep 2015. {The Illustris
  simulation: the evolving population of black holes across cosmic time}.
  \mnras 452~(1), 575--596.

\bibitem[{{Silk} and {Rees}(1998)}]{silk&rees98}
{Silk}, J., {Rees}, M.~J., Mar. 1998. {Quasars and galaxy formation}. \aap 331,
  L1--L4.

\bibitem[{{Sillanpaa} et~al.(1988){Sillanpaa}, {Haarala}, {Valtonen},
  {Sundelius}, and {Byrd}}]{sillanpaa1988}
{Sillanpaa}, A., {Haarala}, S., {Valtonen}, M.~J., {Sundelius}, B., {Byrd},
  G.~G., Feb. 1988. {OJ 287 - Binary pair of supermassive black holes}. \apj
  325, 628--634.

\bibitem[{{Silverman} et~al.(2011){Silverman}, {Kampczyk}, {Jahnke}, {Andrae},
  {Lilly}, {Elvis}, {Civano}, {Mainieri}, {Vignali}, {Zamorani}, {Nair}, {Le
  F{\`e}vre}, {de Ravel}, {Bardelli}, {Bongiorno}, {Bolzonella}, {Cappi},
  {Caputi}, {Carollo}, {Contini}, {Coppa}, {Cucciati}, {de la Torre},
  {Franzetti}, {Garilli}, {Halliday}, {Hasinger}, {Iovino}, {Knobel},
  {Koekemoer}, {Kova{\v c}}, {Lamareille}, {Le Borgne}, {Le Brun}, {Maier},
  {Mignoli}, {Pello}, {P{\'e}rez-Montero}, {Ricciardelli}, {Peng}, {Scodeggio},
  {Tanaka}, {Tasca}, {Tresse}, {Vergani}, {Zucca}, {Brusa}, {Cappelluti},
  {Comastri}, {Finoguenov}, {Fu}, {Gilli}, {Hao}, {Ho}, and
  {Salvato}}]{Silverman_et_al_2011}
{Silverman}, J.~D., {Kampczyk}, P., {Jahnke}, K., {Andrae}, R., {Lilly}, S.~J.,
  {Elvis}, M., {Civano}, F., {Mainieri}, V., {Vignali}, C., {Zamorani}, G.,
  {Nair}, P., {Le F{\`e}vre}, O., {de Ravel}, L., {Bardelli}, S., {Bongiorno},
  A., {Bolzonella}, M., {Cappi}, A., {Caputi}, K., {Carollo}, C.~M., {Contini},
  T., {Coppa}, G., {Cucciati}, O., {de la Torre}, S., {Franzetti}, P.,
  {Garilli}, B., {Halliday}, C., {Hasinger}, G., {Iovino}, A., {Knobel}, C.,
  {Koekemoer}, A.~M., {Kova{\v c}}, K., {Lamareille}, F., {Le Borgne}, J.-F.,
  {Le Brun}, V., {Maier}, C., {Mignoli}, M., {Pello}, R., {P{\'e}rez-Montero},
  E., {Ricciardelli}, E., {Peng}, Y., {Scodeggio}, M., {Tanaka}, M., {Tasca},
  L., {Tresse}, L., {Vergani}, D., {Zucca}, E., {Brusa}, M., {Cappelluti}, N.,
  {Comastri}, A., {Finoguenov}, A., {Fu}, H., {Gilli}, R., {Hao}, H., {Ho},
  L.~C., {Salvato}, M., Dec. 2011. {The Impact of Galaxy Interactions on Active
  Galactic Nucleus Activity in zCOSMOS}. \apj 743, 2.

\bibitem[{{Smith} et~al.(2010){Smith}, {Shields}, {Bonning}, {McMullen},
  {Rosario}, and {Salviander}}]{Smith:2010}
{Smith}, K.~L., {Shields}, G.~A., {Bonning}, E.~W., {McMullen}, C.~C.,
  {Rosario}, D.~J., {Salviander}, S., Jun. 2010. {A Search for Binary Active
  Galactic Nuclei: Double-peaked [O III] AGNs in the Sloan Digital Sky Survey}.
  \apj 716, 866--877.

\bibitem[{{Smolcic} et~al.(2015){Smolcic}, {Padovani}, {Delhaize}, {Prandoni},
  {Seymour}, {Jarvis}, {Afonso}, {Magliocchetti}, {Huynh}, {Vaccari}, and
  {Karim}}]{Smolcic2015}
{Smolcic}, V., {Padovani}, P., {Delhaize}, J., {Prandoni}, I., {Seymour}, N.,
  {Jarvis}, M., {Afonso}, J., {Magliocchetti}, M., {Huynh}, M., {Vaccari}, M.,
  {Karim}, A., Apr. 2015. {Exploring AGN Activity over Cosmic Time with the
  SKA}. Advancing Astrophysics with the Square Kilometre Array (AASKA14), 69.

\bibitem[{{Soko{\l}owska} et~al.(2017){Soko{\l}owska}, {Capelo}, {Fall},
  {Mayer}, {Shen}, and {Bonoli}}]{Sokolowska_et_al_2017}
{Soko{\l}owska}, A., {Capelo}, P.~R., {Fall}, S.~M., {Mayer}, L., {Shen}, S.,
  {Bonoli}, S., Feb. 2017. {Galactic Angular Momentum in Cosmological Zoom-in
  Simulations. I. Disk and Bulge Components and the Galaxy-Halo Connection}.
  \apj 835, 289.

\bibitem[{{Solanes} et~al.(2019){Solanes}, {Perea}, {Valent{\'\i}-Rojas}, {del
  Olmo}, {M{\'a}rquez}, {Ramos Almeida}, and {Tous}}]{Solanes_et_al_2019}
{Solanes}, J.~M., {Perea}, J.~D., {Valent{\'\i}-Rojas}, G., {del Olmo}, A.,
  {M{\'a}rquez}, I., {Ramos Almeida}, C., {Tous}, J.~L., Apr 2019. {Intrinsic
  and observed dual AGN fractions from major mergers}. \aap 624, A86.

\bibitem[{{Souza Lima} et~al.(2017){Souza Lima}, {Mayer}, {Capelo}, and
  {Bellovary}}]{SouzaLima_et_al_2017}
{Souza Lima}, R., {Mayer}, L., {Capelo}, P.~R., {Bellovary}, J.~M., Mar. 2017.
  {The Pairing of Accreting Massive Black Holes in Multiphase Circumnuclear
  Disks: the Interplay Between Radiative Cooling, Star Formation, and Feedback
  Processes}. \apj 838, 13.

\bibitem[{{Springel} et~al.(2005){Springel}, {Di Matteo}, and
  {Hernquist}}]{Springel_et_al_2005}
{Springel}, V., {Di Matteo}, T., {Hernquist}, L., Aug. 2005. {Modelling
  feedback from stars and black holes in galaxy mergers}. \mnras 361, 776--794.

\bibitem[{{Springel} and {Hernquist}(2002)}]{Springel_Hernquist_2002}
{Springel}, V., {Hernquist}, L., Jul. 2002. {Cosmological smoothed particle
  hydrodynamics simulations: the entropy equation}. \mnras 333, 649--664.

\bibitem[{{Springel} et~al.(2018){Springel}, {Pakmor}, {Pillepich},
  {Weinberger}, {Nelson}, {Hernquist}, {Vogelsberger}, {Genel}, {Torrey},
  {Marinacci}, and {Naiman}}]{Springel_et_al_2018}
{Springel}, V., {Pakmor}, R., {Pillepich}, A., {Weinberger}, R., {Nelson}, D.,
  {Hernquist}, L., {Vogelsberger}, M., {Genel}, S., {Torrey}, P., {Marinacci},
  F., {Naiman}, J., Mar 2018. {First results from the IllustrisTNG simulations:
  matter and galaxy clustering}. \mnras 475~(1), 676--698.

\bibitem[{{Steidel} et~al.(2014){Steidel}, {Rudie}, {Strom}, {Pettini},
  {Reddy}, {Shapley}, {Trainor}, {Erb}, {Turner}, {Konidaris}, {Kulas}, {Mace},
  {Matthews}, and {McLean}}]{Steidel2014}
{Steidel}, C.~C., {Rudie}, G.~C., {Strom}, A.~L., {Pettini}, M., {Reddy},
  N.~A., {Shapley}, A.~E., {Trainor}, R.~F., {Erb}, D.~K., {Turner}, M.~L.,
  {Konidaris}, N.~P., {Kulas}, K.~R., {Mace}, G., {Matthews}, K., {McLean},
  I.~S., Nov. 2014. {Strong Nebular Line Ratios in the Spectra of z$\sim$2-3
  Star Forming Galaxies: First Results from KBSS-MOSFIRE}. \apj 795, 165.

\bibitem[{{Steinborn} et~al.(2016){Steinborn}, {Dolag}, {Comerford},
  {Hirschmann}, {Remus}, and {Teklu}}]{Steinborn_et_al_2016}
{Steinborn}, L.~K., {Dolag}, K., {Comerford}, J.~M., {Hirschmann}, M., {Remus},
  R.-S., {Teklu}, A.~F., May 2016. {Origin and properties of dual and offset
  active galactic nuclei in a cosmological simulation at z=2}. \mnras 458,
  1013--1028.

\bibitem[{{Steinborn} et~al.(2015){Steinborn}, {Dolag}, {Hirschmann}, {Prieto},
  and {Remus}}]{Steinborn_et_al_2015}
{Steinborn}, L.~K., {Dolag}, K., {Hirschmann}, M., {Prieto}, M.~A., {Remus},
  R.-S., Apr. 2015. {A refined sub-grid model for black hole accretion and AGN
  feedback in large cosmological simulations}. \mnras 448, 1504--1525.

\bibitem[{{Stern} et~al.(2012){Stern}, {Assef}, {Benford}, {Blain}, {Cutri},
  {Dey}, {Eisenhardt}, {Griffith}, {Jarrett}, {Lake}, {Masci}, {Petty},
  {Stanford}, {Tsai}, {Wright}, {Yan}, {Harrison}, and {Madsen}}]{stern2012}
{Stern}, D., {Assef}, R.~J., {Benford}, D.~J., {Blain}, A., {Cutri}, R., {Dey},
  A., {Eisenhardt}, P., {Griffith}, R.~L., {Jarrett}, T.~H., {Lake}, S.,
  {Masci}, F., {Petty}, S., {Stanford}, S.~A., {Tsai}, C.-W., {Wright}, E.~L.,
  {Yan}, L., {Harrison}, F., {Madsen}, K., Jul. 2012. {Mid-infrared Selection
  of Active Galactic Nuclei with the Wide-Field Infrared Survey Explorer. I.
  Characterizing WISE-selected Active Galactic Nuclei in COSMOS}. \apj 753, 30.

\bibitem[{{Stinson} et~al.(2006){Stinson}, {Seth}, {Katz}, {Wadsley},
  {Governato}, and {Quinn}}]{Stinson_et_al_2006}
{Stinson}, G., {Seth}, A., {Katz}, N., {Wadsley}, J., {Governato}, F., {Quinn},
  T., Dec. 2006. {Star formation and feedback in smoothed particle hydrodynamic
  simulations - I. Isolated galaxies}. \mnras 373, 1074--1090.

\bibitem[{{Storchi-Bergmann} et~al.(2017){Storchi-Bergmann}, {Schimoia},
  {Peterson}, {Elvis}, {Denney}, {Eracleous}, and
  {Nemmen}}]{Storchi-Bergmann2017}
{Storchi-Bergmann}, T., {Schimoia}, J.~S., {Peterson}, B.~M., {Elvis}, M.,
  {Denney}, K.~D., {Eracleous}, M., {Nemmen}, R.~S., Feb. 2017. {Double-Peaked
  Profiles: Ubiquitous Signatures of Disks in the Broad Emission Lines of
  Active Galactic Nuclei}. \apj 835, 236.

\bibitem[{{Sun} et~al.(2016){Sun}, {Zhou}, {Hao}, {Jiang}, {Ge}, {Ji}, {Ma},
  {Zhang}, and {Shu}}]{2016ApJ...818...64S}
{Sun}, L., {Zhou}, H., {Hao}, L., {Jiang}, P., {Ge}, J., {Ji}, T., {Ma}, J.,
  {Zhang}, S., {Shu}, X., Feb. 2016. {Keck/ESI Long-slit Spectroscopy of SBS
  1421+511: A Recoiling Quasar Nucleus in an Active Galaxy Pair?} \apj 818, 64.

\bibitem[{{Syer} and {Clarke}(1995)}]{SyerClarke1995}
{Syer}, D., {Clarke}, C.~J., Dec. 1995. {Satellites in discs: regulating the
  accretion luminosity}. \mnras 277, 758--766.

\bibitem[{{Tacconi} et~al.(2013){Tacconi}, {Neri}, {Genzel}, {Combes},
  {Bolatto}, {Cooper}, {Wuyts}, {Bournaud}, {Burkert}, {Comerford}, {Cox},
  {Davis}, {F{\"o}rster Schreiber}, {Garc{\'{\i}}a-Burillo}, {Gracia-Carpio},
  {Lutz}, {Naab}, {Newman}, {Omont}, {Saintonge}, {Shapiro Griffin}, {Shapley},
  {Sternberg}, and {Weiner}}]{Tacconi_et_al_2013}
{Tacconi}, L.~J., {Neri}, R., {Genzel}, R., {Combes}, F., {Bolatto}, A.,
  {Cooper}, M.~C., {Wuyts}, S., {Bournaud}, F., {Burkert}, A., {Comerford}, J.,
  {Cox}, P., {Davis}, M., {F{\"o}rster Schreiber}, N.~M.,
  {Garc{\'{\i}}a-Burillo}, S., {Gracia-Carpio}, J., {Lutz}, D., {Naab}, T.,
  {Newman}, S., {Omont}, A., {Saintonge}, A., {Shapiro Griffin}, K., {Shapley},
  A., {Sternberg}, A., {Weiner}, B., May 2013. {Phibss: Molecular Gas Content
  and Scaling Relations in z \~{} 1-3 Massive, Main-sequence Star-forming
  Galaxies}. \apj 768, 74.

\bibitem[{{Tamanini} et~al.(2016){Tamanini}, {Caprini}, {Barausse}, {Sesana},
  {Klein}, and {Petiteau}}]{2016JCAP...04..002T}
{Tamanini}, N., {Caprini}, C., {Barausse}, E., {Sesana}, A., {Klein}, A.,
  {Petiteau}, A., Apr. 2016. {Science with the space-based interferometer
  eLISA. III: probing the expansion of the universe using gravitational wave
  standard sirens}. \jcap 4, 002.

\bibitem[{{Tamburello} et~al.(2017){Tamburello}, {Capelo}, {Mayer},
  {Bellovary}, and {Wadsley}}]{Tamburello_et_al_2017a}
{Tamburello}, V., {Capelo}, P.~R., {Mayer}, L., {Bellovary}, J.~M., {Wadsley},
  J.~W., Jan. 2017. {Supermassive black hole pairs in clumpy galaxies at high
  redshift: delayed binary formation and concurrent mass growth}. \mnras 464,
  2952--2962.

\bibitem[{{Tamfal} et~al.(2018){Tamfal}, {Capelo}, {Kazantzidis}, {Mayer},
  {Potter}, {Stadel}, and {Widrow}}]{Tamfal_et_al_2018}
{Tamfal}, T., {Capelo}, P.~R., {Kazantzidis}, S., {Mayer}, L., {Potter}, D.,
  {Stadel}, J., {Widrow}, L.~M., Sep. 2018. {Formation of LISA Black Hole
  Binaries in Merging Dwarf Galaxies: The Imprint of Dark Matter}. \apjl 864,
  L19.

\bibitem[{{Tang} and {Grindlay}(2009)}]{Tang2009}
{Tang}, S., {Grindlay}, J., Oct 2009. {The Quasar SDSS J153636.22+044127.0: A
  Double-Peaked Emitter in a Candidate Binary Black Hole System}. \apj 704~(2),
  1189--1194.

\bibitem[{{Tang} et~al.(2018){Tang}, {Haiman}, and {MacFadyen}}]{Tang+2018}
{Tang}, Y., {Haiman}, Z., {MacFadyen}, A., May 2018. {The late inspiral of
  supermassive black hole binaries with circumbinary gas discs in the LISA
  band}. \mnras 476, 2249--2257.

\bibitem[{{Tang} et~al.(2017){Tang}, {MacFadyen}, and {Haiman}}]{Tang+2017}
{Tang}, Y., {MacFadyen}, A., {Haiman}, Z., Aug. 2017. {On the orbital evolution
  of supermassive black hole binaries with circumbinary accretion discs}.
  \mnras 469, 4258--4267.

\bibitem[{{Tavani} et~al.(2018){Tavani}, {Cavaliere}, {Munar-Adrover}, and
  {Argan}}]{tavanietal18}
{Tavani}, M., {Cavaliere}, A., {Munar-Adrover}, P., {Argan}, A., Feb. 2018.
  {The Blazar PG 1553+113 as a Binary System of Supermassive Black Holes}. \apj
  854, 11.

\bibitem[{{Terashima} and {Wilson}(2003)}]{Terashima_Wilson2003}
{Terashima}, Y., {Wilson}, A.~S., Jan. 2003. {Chandra Snapshot Observations of
  Low-Luminosity Active Galactic Nuclei with a Compact Radio Source}. \apj 583,
  145--158.

\bibitem[{{Terquem} and {Papaloizou}(2017)}]{TP2017}
{Terquem}, C., {Papaloizou}, J.~C.~B., Jan. 2017. {On the energy dissipation
  rate at the inner edge of circumbinary discs}. \mnras 464, 2429--2440.

\bibitem[{{Thorne}(1987)}]{1987thyg.book..330T}
{Thorne}, K.~S., 1987. {Gravitational radiation.} pp. 330--458.

\bibitem[{{Tollet} et~al.(2016){Tollet}, {Macci{\`o}}, {Dutton}, {Stinson},
  {Wang}, {Penzo}, {Gutcke}, {Buck}, {Kang}, {Brook}, {Di Cintio}, {Keller},
  and {Wadsley}}]{Tollet_et_al_2016}
{Tollet}, E., {Macci{\`o}}, A.~V., {Dutton}, A.~A., {Stinson}, G.~S., {Wang},
  L., {Penzo}, C., {Gutcke}, T.~A., {Buck}, T., {Kang}, X., {Brook}, C., {Di
  Cintio}, A., {Keller}, B.~W., {Wadsley}, J., Mar. 2016. {NIHAO - IV: core
  creation and destruction in dark matter density profiles across cosmic time}.
  \mnras 456, 3542--3552.

\bibitem[{{Toomre} and {Toomre}(1972)}]{Toomre_Toomre_1972}
{Toomre}, A., {Toomre}, J., Dec. 1972. {Galactic Bridges and Tails}. \apj 178,
  623--666.

\bibitem[{{Tornatore} et~al.(2007){Tornatore}, {Borgani}, {Dolag}, and
  {Matteucci}}]{Tornatore_et_al_2007}
{Tornatore}, L., {Borgani}, S., {Dolag}, K., {Matteucci}, F., Dec. 2007.
  {Chemical enrichment of galaxy clusters from hydrodynamical simulations}.
  \mnras 382, 1050--1072.

\bibitem[{{Tornatore} et~al.(2003){Tornatore}, {Borgani}, {Springel},
  {Matteucci}, {Menci}, and {Murante}}]{Tornatore_et_al_2003}
{Tornatore}, L., {Borgani}, S., {Springel}, V., {Matteucci}, F., {Menci}, N.,
  {Murante}, G., Jul. 2003. {Cooling and heating the intracluster medium in
  hydrodynamical simulations}. \mnras 342, 1025--1040.

\bibitem[{{Torres-Alb{\`a}} et~al.(2018){Torres-Alb{\`a}}, {Iwasawa},
  {D{\'{\i}}az-Santos}, {Charmandaris}, {Ricci}, {Chu}, {Sanders}, {Armus},
  {Barcos-Mu{\~n}oz}, {Evans}, {Howell}, {Inami}, {Linden}, {Medling},
  {Privon}, {U}, and {Yoon}}]{torres-alba_etal2018}
{Torres-Alb{\`a}}, N., {Iwasawa}, K., {D{\'{\i}}az-Santos}, T., {Charmandaris},
  V., {Ricci}, C., {Chu}, J.~K., {Sanders}, D.~B., {Armus}, L.,
  {Barcos-Mu{\~n}oz}, L., {Evans}, A.~S., {Howell}, J.~H., {Inami}, H.,
  {Linden}, S.~T., {Medling}, A.~M., {Privon}, G.~C., {U}, V., {Yoon}, I., Dec.
  2018. {C-GOALS. II. Chandra observations of the lower luminosity sample of
  nearby luminous infrared galaxies in GOALS}. \aap 620, A140.

\bibitem[{{Treister} et~al.(2020){Treister}, {Messias}, {Privon}, {Nagar},
  {Medling}, {U.}, {Bauer}, {Cicone}, {Barcos Munoz}, {Evans},
  {Muller-Sanchez}, {Comerford}, {Armus}, {Chang}, {Koss}, {Venturi},
  {Schawinski}, {Casey}, {Urry}, {Sanders}, {Scoville}, and
  {Sheth}}]{Treister2020}
{Treister}, E., {Messias}, H., {Privon}, G.~C., {Nagar}, N., {Medling}, A.~M.,
  {U.}, V., {Bauer}, F.~E., {Cicone}, C., {Barcos Munoz}, L., {Evans}, A.~S.,
  {Muller-Sanchez}, F., {Comerford}, J.~M., {Armus}, L., {Chang}, C., {Koss},
  M., {Venturi}, G., {Schawinski}, K., {Casey}, C., {Urry}, C.~M., {Sanders},
  D.~B., {Scoville}, N., {Sheth}, K., Jan 2020. {The Molecular Gas in the NGC
  6240 Merging Galaxy System at the Highest Spatial Resolution}. arXiv
  e-prints, arXiv:2001.00601.

\bibitem[{{Treister} et~al.(2012){Treister}, {Schawinski}, {Urry}, and
  {Simmons}}]{treisteretal12}
{Treister}, E., {Schawinski}, K., {Urry}, C.~M., {Simmons}, B.~D., Oct. 2012.
  {Major Galaxy Mergers Only Trigger the Most Luminous Active Galactic Nuclei}.
  \apjl 758, L39.

\bibitem[{{Tremmel} et~al.(2018){Tremmel}, {Governato}, {Volonteri}, {Pontzen},
  and {Quinn}}]{Tremmel_et_al_2018}
{Tremmel}, M., {Governato}, F., {Volonteri}, M., {Pontzen}, A., {Quinn}, T.~R.,
  Apr 2018. {Wandering Supermassive Black Holes in Milky-Way-mass Halos}. \apjl
  857~(2), L22.

\bibitem[{{Tremmel} et~al.(2015){Tremmel}, {Governato}, {Volonteri}, and
  {Quinn}}]{Tremmel_et_al_2015}
{Tremmel}, M., {Governato}, F., {Volonteri}, M., {Quinn}, T.~R., Aug. 2015.
  {Off the beaten path: a new approach to realistically model the orbital decay
  of supermassive black holes in galaxy formation simulations}. \mnras 451,
  1868--1874.

\bibitem[{{Tremmel} et~al.(2017){Tremmel}, {Karcher}, {Governato}, {Volonteri},
  {Quinn}, {Pontzen}, {Anderson}, and {Bellovary}}]{Tremmel_et_al_2017}
{Tremmel}, M., {Karcher}, M., {Governato}, F., {Volonteri}, M., {Quinn}, T.~R.,
  {Pontzen}, A., {Anderson}, L., {Bellovary}, J., Sep. 2017. {The Romulus
  cosmological simulations: a physical approach to the formation, dynamics and
  accretion models of SMBHs}. \mnras 470, 1121--1139.

\bibitem[{{Trump} et~al.(2013){Trump}, {Konidaris}, {Barro}, {Koo}, {Kocevski},
  {Juneau}, {Weiner}, {Faber}, {McLean}, {Yan}, {P{\'e}rez-Gonz{\'a}lez}, and
  {Villar}}]{Trump2013}
{Trump}, J.~R., {Konidaris}, N.~P., {Barro}, G., {Koo}, D.~C., {Kocevski},
  D.~D., {Juneau}, S., {Weiner}, B.~J., {Faber}, S.~M., {McLean}, I.~S., {Yan},
  R., {P{\'e}rez-Gonz{\'a}lez}, P.~G., {Villar}, V., Jan. 2013. {Testing
  Diagnostics of Nuclear Activity and Star Formation in Galaxies at z$>$1}.
  \apjl 763, L6.

\bibitem[{{Trump} et~al.(2015){Trump}, {Sun}, {Zeimann}, {Luck}, {Bridge},
  {Grier}, {Hagen}, {Juneau}, {Montero-Dorta}, {Rosario}, {Brandt},
  {Ciardullo}, and {Schneider}}]{Trump2015}
{Trump}, J.~R., {Sun}, M., {Zeimann}, G.~R., {Luck}, C., {Bridge}, J.~S.,
  {Grier}, C.~J., {Hagen}, A., {Juneau}, S., {Montero-Dorta}, A., {Rosario},
  D.~J., {Brandt}, W.~N., {Ciardullo}, R., {Schneider}, D.~P., Sep. 2015. {The
  Biases of Optical Line-Ratio Selection for Active Galactic Nuclei and the
  Intrinsic Relationship between Black Hole Accretion and Galaxy Star
  Formation}. \apj 811, 26.

\bibitem[{{Tsalmantza} et~al.(2011){Tsalmantza}, {Decarli}, {Dotti}, and
  {Hogg}}]{tsalmantza11}
{Tsalmantza}, P., {Decarli}, R., {Dotti}, M., {Hogg}, D.~W., Sep. 2011. {A
  Systematic Search for Massive Black Hole Binaries in the Sloan Digital Sky
  Survey Spectroscopic Sample}. \apj 738, 20.

\bibitem[{{Urrutia} et~al.(2008){Urrutia}, {Lacy}, and {Becker}}]{Urrutia2008}
{Urrutia}, T., {Lacy}, M., {Becker}, R.~H., Feb 2008. {Evidence for Quasar
  Activity Triggered by Galaxy Mergers in HST Observations of Dust-reddened
  Quasars}. \apj 674~(1), 80--96.

\bibitem[{{Valtonen} and {Pihajoki}(2013)}]{valtonen2013}
{Valtonen}, M., {Pihajoki}, P., Sep. 2013. {A helical jet model for OJ287}.
  \aap 557, A28.

\bibitem[{{Valtonen}(2007)}]{valtonen2007}
{Valtonen}, M.~J., Apr. 2007. {New Orbit Solutions for the Precessing Binary
  Black Hole Model of OJ 287}. \apj 659, 1074--1081.

\bibitem[{{Valtonen} et~al.(2012){Valtonen}, {Ciprini}, and
  {Lehto}}]{valtonen2012}
{Valtonen}, M.~J., {Ciprini}, S., {Lehto}, H.~J., Nov 2012. {On the masses of
  OJ287 black holes}. \mnras 427~(1), 77--83.

\bibitem[{{Valtonen} et~al.(2008){Valtonen}, {Lehto}, {Nilsson}, {Heidt},
  {Takalo}, {Sillanp{\"a}{\"a}}, {Villforth}, {Kidger}, {Poyner}, {Pursimo},
  {Zola}, {Wu}, {Zhou}, {Sadakane}, {Drozdz}, {Koziel}, {Marchev}, {Ogloza},
  {Porowski}, {Siwak}, {Stachowski}, {Winiarski}, {Hentunen}, {Nissinen},
  {Liakos}, and {Dogru}}]{valtonen2008}
{Valtonen}, M.~J., {Lehto}, H.~J., {Nilsson}, K., {Heidt}, J., {Takalo}, L.~O.,
  {Sillanp{\"a}{\"a}}, A., {Villforth}, C., {Kidger}, M., {Poyner}, G.,
  {Pursimo}, T., {Zola}, S., {Wu}, J.-H., {Zhou}, X., {Sadakane}, K., {Drozdz},
  M., {Koziel}, D., {Marchev}, D., {Ogloza}, W., {Porowski}, C., {Siwak}, M.,
  {Stachowski}, G., {Winiarski}, M., {Hentunen}, V.-P., {Nissinen}, M.,
  {Liakos}, A., {Dogru}, S., Apr. 2008. {A massive binary black-hole system in
  OJ287 and a test of general relativity}. \nat 452, 851--853.

\bibitem[{{Van Wassenhove} et~al.(2014){Van Wassenhove}, {Capelo}, {Volonteri},
  {Dotti}, {Bellovary}, {Mayer}, and {Governato}}]{VanWassenhove_et_al_2014}
{Van Wassenhove}, S., {Capelo}, P.~R., {Volonteri}, M., {Dotti}, M.,
  {Bellovary}, J.~M., {Mayer}, L., {Governato}, F., Mar. 2014. {Nuclear coups:
  dynamics of black holes in galaxy mergers}. \mnras 439, 474--487.

\bibitem[{{Van Wassenhove} et~al.(2012){Van Wassenhove}, {Volonteri}, {Mayer},
  {Dotti}, {Bellovary}, and {Callegari}}]{VanWassenhove_et_al_2012}
{Van Wassenhove}, S., {Volonteri}, M., {Mayer}, L., {Dotti}, M., {Bellovary},
  J., {Callegari}, S., Mar. 2012. {Observability of Dual Active Galactic Nuclei
  in Merging Galaxies}. \apjl 748, L7.

\bibitem[{{VanderPlas} and {Ivezi{\'c}}(2015)}]{VanderPlas2015}
{VanderPlas}, J.~T., {Ivezi{\'c}}, {\v Z}., Oct. 2015. {Periodograms for
  Multiband Astronomical Time Series}. \apj 812, 18.

\bibitem[{{Vasiliev} et~al.(2014){Vasiliev}, {Antonini}, and
  {Merritt}}]{vasiliev14}
{Vasiliev}, E., {Antonini}, F., {Merritt}, D., Apr. 2014. {The Final-parsec
  Problem in Nonspherical Galaxies Revisited}. \apj 785, 163.

\bibitem[{{Vasiliev} et~al.(2015){Vasiliev}, {Antonini}, and
  {Merritt}}]{Vasiliev2015}
{Vasiliev}, E., {Antonini}, F., {Merritt}, D., Sep 2015. {The Final-parsec
  Problem in the Collisionless Limit}. \apj 810~(1), 49.

\bibitem[{{Vaughan} et~al.(2016){Vaughan}, {Uttley}, {Markowitz},
  {Huppenkothen}, {Middleton}, {Alston}, {Scargle}, and {Farr}}]{Vaughan2016}
{Vaughan}, S., {Uttley}, P., {Markowitz}, A.~G., {Huppenkothen}, D.,
  {Middleton}, M.~J., {Alston}, W.~N., {Scargle}, J.~D., {Farr}, W.~M., Sep.
  2016. {False periodicities in quasar time-domain surveys}. \mnras 461,
  3145--3152.

\bibitem[{{Veilleux} and {Osterbrock}(1987)}]{Veilleux1987}
{Veilleux}, S., {Osterbrock}, D.~E., Feb. 1987. {Spectral classification of
  emission-line galaxies}. \apjs 63, 295--310.

\bibitem[{{Veilleux} et~al.(2009){Veilleux}, {Rupke}, {Kim}, {Genzel}, {Sturm},
  {Lutz}, {Contursi}, {Schweitzer}, {Tacconi}, {Netzer}, {Sternberg}, {Mihos},
  {Baker}, {Mazzarella}, {Lord}, {Sanders}, {Stockton}, {Joseph}, and
  {Barnes}}]{Veilleux2009}
{Veilleux}, S., {Rupke}, D.~S.~N., {Kim}, D.~C., {Genzel}, R., {Sturm}, E.,
  {Lutz}, D., {Contursi}, A., {Schweitzer}, M., {Tacconi}, L.~J., {Netzer}, H.,
  {Sternberg}, A., {Mihos}, J.~C., {Baker}, A.~J., {Mazzarella}, J.~M., {Lord},
  S., {Sanders}, D.~B., {Stockton}, A., {Joseph}, R.~D., {Barnes}, J.~E., Jun
  2009. {Spitzer Quasar and Ulirg Evolution Study (QUEST). IV. Comparison of 1
  Jy Ultraluminous Infrared Galaxies with Palomar-Green Quasars}. \apjs
  182~(2), 628--666.

\bibitem[{{Verbiest} et~al.(2016){Verbiest}, {Lentati}, {Hobbs}, {van
  Haasteren}, {Demorest}, {Janssen}, {Wang}, {Desvignes}, {Caballero}, {Keith},
  {Champion}, {Arzoumanian}, {Babak}, {Bassa}, {Bhat}, {Brazier}, {Brem},
  {Burgay}, {Burke-Spolaor}, {Chamberlin}, {Chatterjee}, {Christy}, {Cognard},
  {Cordes}, {Dai}, {Dolch}, {Ellis}, {Ferdman}, {Fonseca}, {Gair},
  {Garver-Daniels}, {Gentile}, {Gonzalez}, {Graikou}, {Guillemot}, {Hessels},
  {Jones}, {Karuppusamy}, {Kerr}, {Kramer}, {Lam}, {Lasky}, {Lassus},
  {Lazarus}, {Lazio}, {Lee}, {Levin}, {Liu}, {Lynch}, {Lyne}, {Mckee},
  {McLaughlin}, {McWilliams}, {Madison}, {Manchester}, {Mingarelli}, {Nice},
  {Os{\l}owski}, {Palliyaguru}, {Pennucci}, {Perera}, {Perrodin}, {Possenti},
  {Petiteau}, {Ransom}, {Reardon}, {Rosado}, {Sanidas}, {Sesana}, {Shaifullah},
  {Shannon}, {Siemens}, {Simon}, {Smits}, {Spiewak}, {Stairs}, {Stappers},
  {Stinebring}, {Stovall}, {Swiggum}, {Taylor}, {Theureau}, {Tiburzi},
  {Toomey}, {Vallisneri}, {van Straten}, {Vecchio}, {Wang}, {Wen}, {You},
  {Zhu}, and {Zhu}}]{2016MNRAS.458.1267V}
{Verbiest}, J.~P.~W., {Lentati}, L., {Hobbs}, G., {van Haasteren}, R.,
  {Demorest}, P.~B., {Janssen}, G.~H., {Wang}, J.-B., {Desvignes}, G.,
  {Caballero}, R.~N., {Keith}, M.~J., {Champion}, D.~J., {Arzoumanian}, Z.,
  {Babak}, S., {Bassa}, C.~G., {Bhat}, N.~D.~R., {Brazier}, A., {Brem}, P.,
  {Burgay}, M., {Burke-Spolaor}, S., {Chamberlin}, S.~J., {Chatterjee}, S.,
  {Christy}, B., {Cognard}, I., {Cordes}, J.~M., {Dai}, S., {Dolch}, T.,
  {Ellis}, J.~A., {Ferdman}, R.~D., {Fonseca}, E., {Gair}, J.~R.,
  {Garver-Daniels}, N.~E., {Gentile}, P., {Gonzalez}, M.~E., {Graikou}, E.,
  {Guillemot}, L., {Hessels}, J.~W.~T., {Jones}, G., {Karuppusamy}, R., {Kerr},
  M., {Kramer}, M., {Lam}, M.~T., {Lasky}, P.~D., {Lassus}, A., {Lazarus}, P.,
  {Lazio}, T.~J.~W., {Lee}, K.~J., {Levin}, L., {Liu}, K., {Lynch}, R.~S.,
  {Lyne}, A.~G., {Mckee}, J., {McLaughlin}, M.~A., {McWilliams}, S.~T.,
  {Madison}, D.~R., {Manchester}, R.~N., {Mingarelli}, C.~M.~F., {Nice}, D.~J.,
  {Os{\l}owski}, S., {Palliyaguru}, N.~T., {Pennucci}, T.~T., {Perera},
  B.~B.~P., {Perrodin}, D., {Possenti}, A., {Petiteau}, A., {Ransom}, S.~M.,
  {Reardon}, D., {Rosado}, P.~A., {Sanidas}, S.~A., {Sesana}, A., {Shaifullah},
  G., {Shannon}, R.~M., {Siemens}, X., {Simon}, J., {Smits}, R., {Spiewak}, R.,
  {Stairs}, I.~H., {Stappers}, B.~W., {Stinebring}, D.~R., {Stovall}, K.,
  {Swiggum}, J.~K., {Taylor}, S.~R., {Theureau}, G., {Tiburzi}, C., {Toomey},
  L., {Vallisneri}, M., {van Straten}, W., {Vecchio}, A., {Wang}, Y., {Wen},
  L., {You}, X.~P., {Zhu}, W.~W., {Zhu}, X.-J., May 2016. {The International
  Pulsar Timing Array: First data release}. \mnras 458, 1267--1288.

\bibitem[{{Vignali} et~al.(2018){Vignali}, {Piconcelli}, {Perna}, {Hennawi},
  {Gilli}, {Comastri}, {Zamorani}, {Dotti}, and {Mathur}}]{Vignali2018}
{Vignali}, C., {Piconcelli}, E., {Perna}, M., {Hennawi}, J., {Gilli}, R.,
  {Comastri}, A., {Zamorani}, G., {Dotti}, M., {Mathur}, S., Jun. 2018.
  {Probing black hole accretion in quasar pairs at high redshift}. \mnras 477,
  780--790.

\bibitem[{{Villforth} and {Hamann}(2015)}]{2015AJ....149...92V}
{Villforth}, C., {Hamann}, F., Mar. 2015. {The Host Galaxies and Narrow-Line
  Regions of Four Double-Peaked [OIII] AGNs}. \aj 149, 92.

\bibitem[{{Vogelsberger} et~al.(2013){Vogelsberger}, {Genel}, {Sijacki},
  {Torrey}, {Springel}, and {Hernquist}}]{Vogelsberger_et_al_2013}
{Vogelsberger}, M., {Genel}, S., {Sijacki}, D., {Torrey}, P., {Springel}, V.,
  {Hernquist}, L., Dec. 2013. {A model for cosmological simulations of galaxy
  formation physics}. \mnras 436, 3031--3067.

\bibitem[{{Vogelsberger} et~al.(2014){Vogelsberger}, {Genel}, {Springel},
  {Torrey}, {Sijacki}, {Xu}, {Snyder}, {Nelson}, and
  {Hernquist}}]{Vogelsberger_et_al_2014}
{Vogelsberger}, M., {Genel}, S., {Springel}, V., {Torrey}, P., {Sijacki}, D.,
  {Xu}, D., {Snyder}, G., {Nelson}, D., {Hernquist}, L., Oct. 2014.
  {Introducing the Illustris Project: simulating the coevolution of dark and
  visible matter in the Universe}. \mnras 444, 1518--1547.

\bibitem[{{Voges} et~al.(1999){Voges}, {Aschenbach}, {Boller},
  {Br{\"a}uninger}, {Briel}, {Burkert}, {Dennerl}, {Englhauser}, {Gruber},
  {Haberl}, {Hartner}, {Hasinger}, {K{\"u}rster}, {Pfeffermann}, {Pietsch},
  {Predehl}, {Rosso}, {Schmitt}, {Tr{\"u}mper}, and {Zimmermann}}]{Voges1999}
{Voges}, W., {Aschenbach}, B., {Boller}, T., {Br{\"a}uninger}, H., {Briel}, U.,
  {Burkert}, W., {Dennerl}, K., {Englhauser}, J., {Gruber}, R., {Haberl}, F.,
  {Hartner}, G., {Hasinger}, G., {K{\"u}rster}, M., {Pfeffermann}, E.,
  {Pietsch}, W., {Predehl}, P., {Rosso}, C., {Schmitt}, J.~H.~M.~M.,
  {Tr{\"u}mper}, J., {Zimmermann}, H.~U., Sep. 1999. {The ROSAT all-sky survey
  bright source catalogue}. \aap 349, 389--405.

\bibitem[{{Volonteri} et~al.(2015{\natexlab{a}}){Volonteri}, {Capelo},
  {Netzer}, {Bellovary}, {Dotti}, and {Governato}}]{Volonteri_et_al_2015a}
{Volonteri}, M., {Capelo}, P.~R., {Netzer}, H., {Bellovary}, J., {Dotti}, M.,
  {Governato}, F., Sep. 2015{\natexlab{a}}. {Black hole accretion versus star
  formation rate: theory confronts observations}. \mnras 452, L6--L10.

\bibitem[{{Volonteri} et~al.(2015{\natexlab{b}}){Volonteri}, {Capelo},
  {Netzer}, {Bellovary}, {Dotti}, and {Governato}}]{Volonteri_et_al_2015b}
{Volonteri}, M., {Capelo}, P.~R., {Netzer}, H., {Bellovary}, J., {Dotti}, M.,
  {Governato}, F., May 2015{\natexlab{b}}. {Growing black holes and galaxies:
  black hole accretion versus star formation rate}. \mnras 449, 1470--1485.

\bibitem[{{Volonteri} et~al.(2016){Volonteri}, {Dubois}, {Pichon}, and
  {Devriendt}}]{Volonteri_et_al_2016}
{Volonteri}, M., {Dubois}, Y., {Pichon}, C., {Devriendt}, J., Aug. 2016. {The
  cosmic evolution of massive black holes in the Horizon-AGN simulation}.
  \mnras 460, 2979--2996.

\bibitem[{{Volonteri} et~al.(2003){Volonteri}, {Haardt}, and
  {Madau}}]{2003ApJ...582..559V}
{Volonteri}, M., {Haardt}, F., {Madau}, P., Jan 2003. {The Assembly and Merging
  History of Supermassive Black Holes in Hierarchical Models of Galaxy
  Formation}. \apj 582~(2), 559--573.

\bibitem[{{Volonteri} et~al.(2009){Volonteri}, {Miller}, and
  {Dotti}}]{Volonteri09}
{Volonteri}, M., {Miller}, J.~M., {Dotti}, M., Sep. 2009. {Sub-Parsec
  Supermassive Binary Quasars: Expectations at z $<$ 1}. \apjl 703, L86--L89.

\bibitem[{{Volonteri} and {Perna}(2005)}]{2005MNRAS.358..913V}
{Volonteri}, M., {Perna}, R., Apr 2005. {Dynamical evolution of
  intermediate-mass black holes and their observable signatures in the nearby
  Universe}. \mnras 358~(3), 913--922.

\bibitem[{{Wang} et~al.(2009){Wang}, {Chen}, {Hu}, {Mao}, {Zhang}, and
  {Bian}}]{Wang:2009}
{Wang}, J.-M., {Chen}, Y.-M., {Hu}, C., {Mao}, W.-M., {Zhang}, S., {Bian},
  W.-H., Nov. 2009. {Active Galactic Nuclei with Double-Peaked Narrow Lines:
  Are they Dual Active Galactic Nuclei?} \apjl 705, L76--L80.

\bibitem[{{Wang} et~al.(2017){Wang}, {Greene}, {Ju}, {Rafikov}, {Ruan}, and
  {Schneider}}]{wang17}
{Wang}, L., {Greene}, J.~E., {Ju}, W., {Rafikov}, R.~R., {Ruan}, J.~J.,
  {Schneider}, D.~P., Jan. 2017. {Searching for Binary Supermassive Black Holes
  via Variable Broad Emission Line Shifts: Low Binary Fraction}. \apj 834, 129.

\bibitem[{{Ward}(1997)}]{Ward1997}
{Ward}, W.~R., Apr. 1997. {Protoplanet Migration by Nebula Tides}. \icarus 126,
  261--281.

\bibitem[{{Weinberger} et~al.(2017){Weinberger}, {Springel}, {Hernquist},
  {Pillepich}, {Marinacci}, {Pakmor}, {Nelson}, {Genel}, {Vogelsberger},
  {Naiman}, and {Torrey}}]{Weinberger_et_al_2017}
{Weinberger}, R., {Springel}, V., {Hernquist}, L., {Pillepich}, A.,
  {Marinacci}, F., {Pakmor}, R., {Nelson}, D., {Genel}, S., {Vogelsberger}, M.,
  {Naiman}, J., {Torrey}, P., Mar. 2017. {Simulating galaxy formation with
  black hole driven thermal and kinetic feedback}. \mnras 465, 3291--3308.

\bibitem[{{Weinberger} et~al.(2018){Weinberger}, {Springel}, {Pakmor},
  {Nelson}, {Genel}, {Pillepich}, {Vogelsberger}, {Marinacci}, {Naiman},
  {Torrey}, and {Hernquist}}]{Weinberger_et_al_2018}
{Weinberger}, R., {Springel}, V., {Pakmor}, R., {Nelson}, D., {Genel}, S.,
  {Pillepich}, A., {Vogelsberger}, M., {Marinacci}, F., {Naiman}, J., {Torrey},
  P., {Hernquist}, L., Sep 2018. {Supermassive black holes and their feedback
  effects in the IllustrisTNG simulation}. \mnras 479~(3), 4056--4072.

\bibitem[{{Weston} et~al.(2017){Weston}, {McIntosh}, {Brodwin}, {Mann},
  {Cooper}, {McConnell}, and {Nielsen}}]{Weston2017}
{Weston}, M.~E., {McIntosh}, D.~H., {Brodwin}, M., {Mann}, J., {Cooper}, A.,
  {McConnell}, A., {Nielsen}, J.~L., Feb. 2017. {Incidence of WISE -selected
  obscured AGNs in major mergers and interactions from the SDSS}. \mnras 464,
  3882--3906.

\bibitem[{{Wills} and {Browne}(1986)}]{Wills1986}
{Wills}, B.~J., {Browne}, I.~W.~A., Mar. 1986. {Relativistic beaming and quasar
  emission lines}. \apj 302, 56--63.

\bibitem[{{Wisnioski} et~al.(2015){Wisnioski}, {F{\"o}rster Schreiber},
  {Wuyts}, {Wuyts}, {Bandara}, {Wilman}, {Genzel}, {Bender}, {Davies},
  {Fossati}, {Lang}, {Mendel}, {Beifiori}, {Brammer}, {Chan}, {Fabricius},
  {Fudamoto}, {Kulkarni}, {Kurk}, {Lutz}, {Nelson}, {Momcheva}, {Rosario},
  {Saglia}, {Seitz}, {Tacconi}, and {van Dokkum}}]{Wisnioski_et_al_2015}
{Wisnioski}, E., {F{\"o}rster Schreiber}, N.~M., {Wuyts}, S., {Wuyts}, E.,
  {Bandara}, K., {Wilman}, D., {Genzel}, R., {Bender}, R., {Davies}, R.,
  {Fossati}, M., {Lang}, P., {Mendel}, J.~T., {Beifiori}, A., {Brammer}, G.,
  {Chan}, J., {Fabricius}, M., {Fudamoto}, Y., {Kulkarni}, S., {Kurk}, J.,
  {Lutz}, D., {Nelson}, E.~J., {Momcheva}, I., {Rosario}, D., {Saglia}, R.,
  {Seitz}, S., {Tacconi}, L.~J., {van Dokkum}, P.~G., Feb. 2015. {The
  KMOS$^{3D}$ Survey: Design, First Results, and the Evolution of Galaxy
  Kinematics from 0.7 $\leq$ z $\leq$ 2.7}. \apj 799, 209.

\bibitem[{{Wizinowich} et~al.(2006){Wizinowich}, {Chin}, {Johansson},
  {Kellner}, {Lafon}, {Le Mignant}, {Neyman}, {Stomski}, {Summers}, {Sumner},
  and {van Dam}}]{WizinowichCJ06}
{Wizinowich}, P.~L., {Chin}, J., {Johansson}, E., {Kellner}, S., {Lafon}, R.,
  {Le Mignant}, D., {Neyman}, C., {Stomski}, P., {Summers}, D., {Sumner}, R.,
  {van Dam}, M., Jun. 2006. {Adaptive optics developments at Keck Observatory}.
  In: Society of Photo-Optical Instrumentation Engineers (SPIE) Conference
  Series. Vol. 6272 of \procspie. p. 627209.

\bibitem[{{Woo} et~al.(2014){Woo}, {Cho}, {Husemann}, {Komossa}, {Park}, and
  {Bennert}}]{Woo:2014}
{Woo}, J.-H., {Cho}, H., {Husemann}, B., {Komossa}, S., {Park}, D., {Bennert},
  V.~N., Jan. 2014. {A sub-kpc-scale binary active galactic nucleus with double
  narrow-line regions}. \mnras 437, 32--37.

\bibitem[{{Wright}(2006)}]{Wright06}
{Wright}, E.~L., Dec. 2006. {A Cosmology Calculator for the World Wide Web}.
  \pasp 118, 1711--1715.

\bibitem[{{Wright} et~al.(2010){Wright}, {Eisenhardt}, {Mainzer}, {Ressler},
  {Cutri}, {Jarrett}, {Kirkpatrick}, {Padgett}, {McMillan}, {Skrutskie},
  {Stanford}, {Cohen}, {Walker}, {Mather}, {Leisawitz}, {Gautier}, {McLean},
  {Benford}, {Lonsdale}, {Blain}, {Mendez}, {Irace}, {Duval}, {Liu}, {Royer},
  {Heinrichsen}, {Howard}, {Shannon}, {Kendall}, {Walsh}, {Larsen}, {Cardon},
  {Schick}, {Schwalm}, {Abid}, {Fabinsky}, {Naes}, and {Tsai}}]{Wright2010}
{Wright}, E.~L., {Eisenhardt}, P.~R.~M., {Mainzer}, A.~K., {Ressler}, M.~E.,
  {Cutri}, R.~M., {Jarrett}, T., {Kirkpatrick}, J.~D., {Padgett}, D.,
  {McMillan}, R.~S., {Skrutskie}, M., {Stanford}, S.~A., {Cohen}, M., {Walker},
  R.~G., {Mather}, J.~C., {Leisawitz}, D., {Gautier}, III, T.~N., {McLean}, I.,
  {Benford}, D., {Lonsdale}, C.~J., {Blain}, A., {Mendez}, B., {Irace}, W.~R.,
  {Duval}, V., {Liu}, F., {Royer}, D., {Heinrichsen}, I., {Howard}, J.,
  {Shannon}, M., {Kendall}, M., {Walsh}, A.~L., {Larsen}, M., {Cardon}, J.~G.,
  {Schick}, S., {Schwalm}, M., {Abid}, M., {Fabinsky}, B., {Naes}, L., {Tsai},
  C.-W., Dec. 2010. {The Wide-field Infrared Survey Explorer (WISE): Mission
  Description and Initial On-orbit Performance}. \aj 140, 1868--1881.

\bibitem[{{Wright} et~al.(1998){Wright}, {McHardy}, and {Abraham}}]{wright1998}
{Wright}, S.~C., {McHardy}, I.~M., {Abraham}, R.~G., Apr 1998. {Host galaxies
  of the optically violently variable quasars PKS 0736+017, OJ 287 and LB
  2136}. \mnras 295~(4), 799--812.

\bibitem[{{Wrobel} and {Laor}(2009)}]{wrobel2009}
{Wrobel}, J.~M., {Laor}, A., Jul. 2009. {Discovery of Radio Emission from the
  Quasar SDSS J1536+0441, a Candidate Binary Black Hole System}. Astrophysical
  Journal Letters 699, L22--L25.

\bibitem[{{Wrobel} et~al.(2014){Wrobel}, {Walker}, and {Fu}}]{wrobel2014}
{Wrobel}, J.~M., {Walker}, R.~C., {Fu}, H., Sep. 2014. {Evidence from the Very
  Long Baseline Array that J1502SE/SW are Double Hotspots, not a Supermassive
  Binary Black Hole}. Astrophysical Journal Letters 792, L8.

\bibitem[{{Wylezalek} et~al.(2018){Wylezalek}, {Zakamska}, {Greene}, {Riffel},
  {Drory}, {Andrews}, {Merloni}, and {Thomas}}]{Wylezalek2018}
{Wylezalek}, D., {Zakamska}, N.~L., {Greene}, J.~E., {Riffel}, R.~A., {Drory},
  N., {Andrews}, B.~H., {Merloni}, A., {Thomas}, D., Feb. 2018. {SDSS-IV MaNGA:
  identification of active galactic nuclei in optical integral field unit
  surveys}. \mnras 474, 1499--1514.

\bibitem[{{Xu} and {Komossa}(2009)}]{Xue:2009}
{Xu}, D., {Komossa}, S., Nov. 2009. {Narrow Double-Peaked Emission Lines of
  SDSS J131642.90+175332.5: Signature of a Single or a Binary AGN in a Merger,
  Jet-Cloud Interaction, or Unusual Narrow-Line Region Geometry}. \apjl 705,
  L20--L24.

\bibitem[{{Yamada} et~al.(2018){Yamada}, {Ueda}, {Oda}, {Tanimoto}, {Imanishi},
  {Terashima}, and {Ricci}}]{yamada_etal2018}
{Yamada}, S., {Ueda}, Y., {Oda}, S., {Tanimoto}, A., {Imanishi}, M.,
  {Terashima}, Y., {Ricci}, C., May 2018. {Broadband X-Ray Spectral Analysis of
  the Double-nucleus Luminous Infrared Galaxy Mrk 463}. \apj 858, 106.

\bibitem[{{Yang} et~al.(2019){Yang}, {Ge}, and {Lu}}]{Yang_et_al_2019}
{Yang}, C., {Ge}, J., {Lu}, Y., Dec 2019. {Investigating the co-evolution of
  massive black holes in dual active galactic nuclei and their host galaxies
  via galaxy merger simulations}. Science China Physics, Mechanics, and
  Astronomy 62~(12), 129511.

\bibitem[{{York} et~al.(2000){York}, {Adelman}, {Anderson}, {Anderson},
  {Annis}, {Bahcall}, {Bakken}, {Barkhouser}, {Bastian}, {Berman}, {Boroski},
  {Bracker}, {Briegel}, {Briggs}, {Brinkmann}, {Brunner}, {Burles}, {Carey},
  {Carr}, {Castander}, {Chen}, {Colestock}, {Connolly}, {Crocker}, {Csabai},
  {Czarapata}, {Davis}, {Doi}, {Dombeck}, {Eisenstein}, {Ellman}, {Elms},
  {Evans}, {Fan}, {Federwitz}, {Fiscelli}, {Friedman}, {Frieman}, {Fukugita},
  {Gillespie}, {Gunn}, {Gurbani}, {de Haas}, {Haldeman}, {Harris}, {Hayes},
  {Heckman}, {Hennessy}, {Hindsley}, {Holm}, {Holmgren}, {Huang}, {Hull},
  {Husby}, {Ichikawa}, {Ichikawa}, {Ivezi{\'c}}, {Kent}, {Kim}, {Kinney},
  {Klaene}, {Kleinman}, {Kleinman}, {Knapp}, {Korienek}, {Kron}, {Kunszt},
  {Lamb}, {Lee}, {Leger}, {Limmongkol}, {Lindenmeyer}, {Long}, {Loomis},
  {Loveday}, {Lucinio}, {Lupton}, {MacKinnon}, {Mannery}, {Mantsch}, {Margon},
  {McGehee}, {McKay}, {Meiksin}, {Merelli}, {Monet}, {Munn}, {Narayanan},
  {Nash}, {Neilsen}, {Neswold}, {Newberg}, {Nichol}, {Nicinski}, {Nonino},
  {Okada}, {Okamura}, {Ostriker}, {Owen}, {Pauls}, {Peoples}, {Peterson},
  {Petravick}, {Pier}, {Pope}, {Pordes}, {Prosapio}, {Rechenmacher}, {Quinn},
  {Richards}, {Richmond}, {Rivetta}, {Rockosi}, {Ruthmansdorfer}, {Sandford},
  {Schlegel}, {Schneider}, {Sekiguchi}, {Sergey}, {Shimasaku}, {Siegmund},
  {Smee}, {Smith}, {Snedden}, {Stone}, {Stoughton}, {Strauss}, {Stubbs},
  {SubbaRao}, {Szalay}, {Szapudi}, {Szokoly}, {Thakar}, {Tremonti}, {Tucker},
  {Uomoto}, {Vanden Berk}, {Vogeley}, {Waddell}, {Wang}, {Watanabe},
  {Weinberg}, {Yanny}, {Yasuda}, and {SDSS Collaboration}}]{YorkAA00}
{York}, D.~G., {Adelman}, J., {Anderson}, Jr., J.~E., {Anderson}, S.~F.,
  {Annis}, J., {Bahcall}, N.~A., {Bakken}, J.~A., {Barkhouser}, R., {Bastian},
  S., {Berman}, E., {Boroski}, W.~N., {Bracker}, S., {Briegel}, C., {Briggs},
  J.~W., {Brinkmann}, J., {Brunner}, R., {Burles}, S., {Carey}, L., {Carr},
  M.~A., {Castander}, F.~J., {Chen}, B., {Colestock}, P.~L., {Connolly}, A.~J.,
  {Crocker}, J.~H., {Csabai}, I., {Czarapata}, P.~C., {Davis}, J.~E., {Doi},
  M., {Dombeck}, T., {Eisenstein}, D., {Ellman}, N., {Elms}, B.~R., {Evans},
  M.~L., {Fan}, X., {Federwitz}, G.~R., {Fiscelli}, L., {Friedman}, S.,
  {Frieman}, J.~A., {Fukugita}, M., {Gillespie}, B., {Gunn}, J.~E., {Gurbani},
  V.~K., {de Haas}, E., {Haldeman}, M., {Harris}, F.~H., {Hayes}, J.,
  {Heckman}, T.~M., {Hennessy}, G.~S., {Hindsley}, R.~B., {Holm}, S.,
  {Holmgren}, D.~J., {Huang}, C.-h., {Hull}, C., {Husby}, D., {Ichikawa},
  S.-I., {Ichikawa}, T., {Ivezi{\'c}}, {\v Z}., {Kent}, S., {Kim}, R.~S.~J.,
  {Kinney}, E., {Klaene}, M., {Kleinman}, A.~N., {Kleinman}, S., {Knapp},
  G.~R., {Korienek}, J., {Kron}, R.~G., {Kunszt}, P.~Z., {Lamb}, D.~Q., {Lee},
  B., {Leger}, R.~F., {Limmongkol}, S., {Lindenmeyer}, C., {Long}, D.~C.,
  {Loomis}, C., {Loveday}, J., {Lucinio}, R., {Lupton}, R.~H., {MacKinnon}, B.,
  {Mannery}, E.~J., {Mantsch}, P.~M., {Margon}, B., {McGehee}, P., {McKay},
  T.~A., {Meiksin}, A., {Merelli}, A., {Monet}, D.~G., {Munn}, J.~A.,
  {Narayanan}, V.~K., {Nash}, T., {Neilsen}, E., {Neswold}, R., {Newberg},
  H.~J., {Nichol}, R.~C., {Nicinski}, T., {Nonino}, M., {Okada}, N., {Okamura},
  S., {Ostriker}, J.~P., {Owen}, R., {Pauls}, A.~G., {Peoples}, J., {Peterson},
  R.~L., {Petravick}, D., {Pier}, J.~R., {Pope}, A., {Pordes}, R., {Prosapio},
  A., {Rechenmacher}, R., {Quinn}, T.~R., {Richards}, G.~T., {Richmond}, M.~W.,
  {Rivetta}, C.~H., {Rockosi}, C.~M., {Ruthmansdorfer}, K., {Sandford}, D.,
  {Schlegel}, D.~J., {Schneider}, D.~P., {Sekiguchi}, M., {Sergey}, G.,
  {Shimasaku}, K., {Siegmund}, W.~A., {Smee}, S., {Smith}, J.~A., {Snedden},
  S., {Stone}, R., {Stoughton}, C., {Strauss}, M.~A., {Stubbs}, C., {SubbaRao},
  M., {Szalay}, A.~S., {Szapudi}, I., {Szokoly}, G.~P., {Thakar}, A.~R.,
  {Tremonti}, C., {Tucker}, D.~L., {Uomoto}, A., {Vanden Berk}, D., {Vogeley},
  M.~S., {Waddell}, P., {Wang}, S.-i., {Watanabe}, M., {Weinberg}, D.~H.,
  {Yanny}, B., {Yasuda}, N., {SDSS Collaboration}, Sep. 2000. {The Sloan
  Digital Sky Survey: Technical Summary}. \aj 120, 1579--1587.

\bibitem[{{Younger} et~al.(2008){Younger}, {Hopkins}, {Cox}, and
  {Hernquist}}]{Younger_et_al_2008}
{Younger}, J.~D., {Hopkins}, P.~F., {Cox}, T.~J., {Hernquist}, L., Oct. 2008.
  {The Self-Regulated Growth of Supermassive Black Holes}. \apj 686, 815--828.

\bibitem[{{Yu}(2002)}]{yu02}
{Yu}, Q., Apr. 2002. {Evolution of massive binary black holes}. \mnras 331,
  935--958.

\bibitem[{{Yu} et~al.(2005){Yu}, {Lu}, and {Kauffmann}}]{yu05}
{Yu}, Q., {Lu}, Y., {Kauffmann}, G., Dec. 2005. {Evolution of Accretion Disks
  around Massive Black Holes: Constraints from the Demography of Active
  Galactic Nuclei}. \apj 634, 901--909.

\bibitem[{{Yuan} et~al.(2016){Yuan}, {Komossa}, {Zhang}, {Feng}, {Ling},
  {Zhao}, {Zhang}, {Osborne}, {O'Brien}, {Willingale}, and
  {Lapington}}]{yuan2016}
{Yuan}, W., {Komossa}, S., {Zhang}, C., {Feng}, H., {Ling}, Z.~X., {Zhao},
  D.~H., {Zhang}, S.~N., {Osborne}, J.~P., {O'Brien}, P., {Willingale}, R.,
  {Lapington}, J., Feb 2016. {Detecting tidal disruption events of massive
  black holes in normal galaxies with the Einstein Probe}. In: {Meiron}, Y.,
  {Li}, S., {Liu}, F.~K., {Spurzem}, R. (Eds.), Star Clusters and Black Holes
  in Galaxies across Cosmic Time. Vol. 312 of IAU Symposium. pp. 68--70.

\bibitem[{{Zabludoff} et~al.(1996){Zabludoff}, {Zaritsky}, {Lin}, {Tucker},
  {Hashimoto}, {Shectman}, {Oemler}, and {Kirshner}}]{Zabludoff_et_al_1996}
{Zabludoff}, A.~I., {Zaritsky}, D., {Lin}, H., {Tucker}, D., {Hashimoto}, Y.,
  {Shectman}, S.~A., {Oemler}, A., {Kirshner}, R.~P., Jul 1996. {The
  Environment of ``E+A'' Galaxies}. \apj 466, 104.

\bibitem[{{Zheng} et~al.(2016){Zheng}, {Butler}, {Shen}, {Jiang}, {Wang},
  {Chen}, and {Cuadra}}]{Zheng2016}
{Zheng}, Z.-Y., {Butler}, N.~R., {Shen}, Y., {Jiang}, L., {Wang}, J.-X.,
  {Chen}, X., {Cuadra}, J., Aug 2016. {SDSS J0159+0105: A Radio-Quiet Quasar
  with a Centi-Parsec Supermassive Black Hole Binary Candidate}. \apj 827~(1),
  56.

\end{thebibliography}

\end{document}